\documentclass[twocolumn]{aastex63}

\usepackage{threeparttable}
\usepackage{multirow}
\usepackage{interval}
\usepackage{scalerel}

\usepackage{hyperref}
\usepackage{nameref}

\usepackage{natbib}

\usepackage{savesym}
\savesymbol{tablenum}
\usepackage{siunitx,booktabs}
\restoresymbol{SIX}{tablenum}
\newcommand{\Msun}{\mbox{$\mathrm{M}_{\odot}$}}
\newcommand{\Lsun}{\mbox{$\mathrm{L}_{\odot}$}}
\newcommand{\Mi}{\mbox{$M_{\rm i}$}}
\newcommand{\MTO}{\mbox{$M_{\rm TO}$}}
\newcommand{\Mup}{\mbox{$M_{\rm up}$}}
\newcommand{\Mf}{\mbox{$M_{\rm f}$}}
\newcommand{\Mc}{\mbox{$M_{\rm c}$}}

\newcommand{\jks}{\mbox{$J\!-\!K_{\rm s}$}}

\newcommand{\J}{\mbox{$J$}}
\newcommand{\h}{\mbox{$H$}}
\newcommand{\ks}{\mbox{$K_{\rm s}$}}
\newcommand{\Mks}{\mbox{$M_{K_{\rm s}}$}}
\newcommand{\WRP}{\mbox{$W_{\rm BR}$}}
\newcommand{\WJK}{\mbox{$W_{\rm JK_s}$}}
\newcommand{\gaia}{\textit{Gaia}}

\newcommand{\cprime}{C$^{\prime}$}
\newcommand{\gbp}{G_{\scaleto{\rm BP}{4.5pt}}}
\newcommand{\grp}{G_{\scaleto{\rm RP}{4.5pt}}}

\newcommand{\beq}[1]{\vspace{-0.5mm} \begin{equation} \label{#1}}
\newcommand{\beqa}{\begin{eqnarray}}
\newcommand{\eeq}{\end{equation}}\vspace{-0.2mm}
\newcommand{\eeqa}{\end{eqnarray}}
\newcounter{subeqn}
\newcommand{\bsubeqa}[2]{\beqa\label{#1}\nonumber\eeqa\vspace*{-12.5mm}\renewcommand{\theequation}{\arabic{equation}\alph{subeqn}}\setcounter{subeqn}{1}\beqa\label{#2}}
\newcommand{\nsubeqn}[1]{\addtocounter{subeqn}{1}\addtocounter{equation}{-1}\label{#1}}
\newcommand{\esubeqa}{\eeqa\renewcommand{\theequation}{\arabic{equation}}}

\defcitealias{Mowlavi_etal_18}{GDR2-LPV}
\defcitealias{Groenewegen21}{G21}
\defcitealias{Lindegren_etal_21}{L21}

\received{September, 2021}
\revised{xxx, 2021}
\accepted{\today}
\submitjournal{ApJ}
\shorttitle{AGB stars in open clusters}
\shortauthors{Marigo et al.}
\begin{document}
\title{A Fresh Look at AGB stars in Galactic Open Clusters with \gaia: \\Impact on Stellar Models and the Initial-Final Mass Relation}

\correspondingauthor{Paola Marigo}
\email{paola.marigo@unipd.it}

\author[0000-0002-9137-0773]{Paola Marigo}
\affiliation{Department of Physics and Astronomy G. Galilei,
University of Padova, Vicolo dell'Osservatorio 3, I-35122, Padova, Italy}
\author[0000-0002-9480-8400]{Diego Bossini}
\affiliation{Instituto de Astrofísica e Ciências do Espaço, Universidade do Porto CAUP, Rua das Estrelas, PT4150-762 Porto, Portugal} 
\author[0000-0002-1429-2388]{Michele Trabucchi}
\affiliation{Astronomy Department, Geneva University 
Ch. Pegasi 51, CH-1290 Versoix, Switzerland}
\author[0000-0002-3867-9966]{Francesco Addari}
\affiliation{SISSA, via Bonomea 265, I-34136 Trieste, Italy} 
\author[0000-0002-6301-3269]{L\'eo Girardi}
\affiliation{INAF-Osservatorio Astronomico di Padova  Vicolo dell’Osservatorio 5, I-35122 Padova, Italy}
\author[0000-0001-7453-9947]{Jeffrey Cummings}
\affiliation{Department of Astronomy, Indiana University, 727 E 3rd Street, Bloomington, IN 47405, USA}
\author[0000-0002-9300-7409]{Giada Pastorelli}
\affiliation{STScI, 3700 San Martin Drive, Baltimore, MD 21218, USA}
\author[0000-0002-0834-5092]{Piero Dal tio}
\affiliation{Department of Physics and Astronomy G. Galilei,
University of Padova, Vicolo dell'Osservatorio 3, I-35122, Padova, Italy}
\author[0000-0002-6213-6988]{Guglielmo Costa}
\affiliation{Department of Physics and Astronomy G. Galilei,
University of Padova, Vicolo dell'Osservatorio 3, I-35122, Padova, Italy}
\author[0000-0002-5434-1973]{Alessandro Bressan}
\affiliation{SISSA, via Bonomea 265, I-34136 Trieste, Italy}



\begin{abstract}
Benefiting from the \gaia\ second and early third  releases of photometric and astrometric data we examine the population of asymptotic giant branch (AGB) stars that appear in the fields of intermediate-age and young open star clusters.
We identify 49 AGB star candidates, brighter than the tip of the red giant branch, with a good-to-high cluster membership probability. Among them we find 19 TP-AGB stars with known  spectral type: 4 M stars, 3 MS/S stars and 12 C stars.
By combining observations, stellar models, and radiative transfer calculations that include the effect of circumstellar dust, we characterize each star in terms of initial mass, luminosity, mass-loss rate, core mass, period and mode of pulsation.
The information collected helps us  shed light on the TP-AGB evolution at solar-like  metallicity, placing constraints on the third dredge-up process, the initial masses of carbon stars, stellar winds, and the initial-final mass relation (IFMR).
In particular, we find that two bright carbon stars,  MSB 75 and BM IV 90, members of the clusters NGC 7789 and NGC 2660 (with similar ages of $\simeq 1.2-1.6$ Gyr and initial masses $ 2.1 \ga \Mi/\Msun \ga 1.9$), have  unusually high core  masses, $\Mc \approx 0.67-0.7\,\Msun$. These results support the findings of a recent work \citep{Marigo_etal_20} that identified a kink in the IFMR,  which  interrupts its monotonic trend just at the same initial masses.  
Finally, we investigate 
two competing scenarios to explain the \Mc\ data: the role of stellar winds in single-star evolution, and binary interactions through the blue-straggler channel.
\end{abstract}



\section{Introduction} 
The final stages of the evolution of low and intermediate mass stars, the so-called thermally-pulsing asymptotic giant branch (TP-AGB)  phase, are important drivers of the evolution of galaxies, contributing to their integrated light \citep[e.g.,][]{Bruzual_07, Maraston_etal_06} and chemical enrichment in the form of gas and dust expelled by stellar winds \citep[e.g.,][]{Ventura_etal_18, Slemer_etal_17, Nanni_etal_14, Cristallo_etal_11, FerrarottiGail_06, KarakasLattanzio_02, Marigo_01}.
The complexity of the physical processes that characterize the TP-AGB phase makes stellar modeling particularly difficult and the predictions are sometimes very uncertain. For example, the predicted efficiency of the third dredge-up (3DU) or of the hot-bottom burning (HBB) process still does not find an agreement among the different evolution codes, due to the rough treatment of convection and mixing, and the sensitivity to numerical details
\citep[e.g.,][]{Wagstaff_etal_20, Marigo_etal_13, VenturaDantona_05, FrostLattanzio_96}

In this context, observations play a key role, as they can place constraints on processes whose physics are still poorly defined \citep[][]{Marigo_15}. Moreover, through observations of AGB stars in different environments it is possible to study their dependence on metallicity \citep[][]{Marigo_etal_07}.

The Magellanic Clouds are fundamental calibrators of the intermediate-metallicity regime ($-0.2 \la {\rm [Fe/H]} \la -0.4$) thanks to their large population of AGB stars both in the field
\citep{Boyer_etal_11, Meixner_etal_06, Blum_etal_06}, and in massive globular clusters \citep{FrogelMouldBlanco_90, Maraston05, Pessev_etal_08, Noel_etal_13}.
Various works have focused on Magellanic Clouds in the attempt to calibrate mass loss, third dredge-up and hot-bottom burning either through the population synthesis technique \citep{Pastorelli_etal_20, Pastorelli_etal_19, GirardiMarigo_07, Marigo_etal_99}, or
using single AGB tracks \citep{Wagstaff_etal_20,Ventura_etal_15, DellAgli_etal_15, Kamath_etal_12,Kamath_etal_10,Lebzelter_etal_08, Ventura_etal_00}, or with radiative transfer calculations across  dusty circumstellar envelopes \citep{Nanni_etal_19,Nanni_etal_18,GroenewegenSloan_18,GullieuszikGroenewegen_12,Sargent_etal_11, Srinivasan_etal_11,  Groenewegen_etal_09}.
The Magellanic Clouds have also been serving as ideal laboratories for studying the variability of AGB stars \citep[e.g.][]{Feast_etal_1989,Wood_etal_1999,Wood_2000,Cioni_etal_2003,Whitelock_etal_2008,Soszynski_etal_2009_LMC,Soszynski_etal_2011_SMC,Soszynski_etal_2013,Riebel_etal_2015,Ita_etal_2018,Goldman_etal_2019}, and in particular for modeling their pulsation \citep[e.g.][]{Wood_Sebo_1996,Wood_etal_1999,Soszynski_etal_2007,Wood_15,Trabucchi_etal_17,Trabucchi_etal_2019,Trabucchi_etal_2021}.
At low metallicities $({\rm [Fe/H]} \la -0.9)$, the calibration has been extended to
AGB star samples in dwarf galaxies  \citep{Rosenfield_etal_14, Girardi_etal_10} from the ANGST survey \citep{Dalcanton_etal_09}.

In the solar-like metallicity regime, a relevant source of information comes from the Andromeda galaxy (M31) thanks to the extraordinary  photometric data provided by the PHAT survey \citep{Dalcanton_etal_12, Boyer_etal_13}, which also made it possible to identify 2753 star clusters with $\simeq 294$ candidate AGB stars
\citep{Girardi_etal_20}. Another valuable tool is the semi-empirical IFMR  which links the initial mass of low- and intermediate-mass stars with the final mass of white dwarfs left behind on their death \citep[e.g.,][]{Barnett_etal_21,Cummings_etal_18, El-Badry_etal_18, Salaris_etal_09}. 
Recent studies have shown the great potential of the IFMR in placing constraints on mass loss, third dredge-up, and hot-bottom burning of the progenitor stars \citep[e.g.,][]{Althaus_etal_21,Marigo_etal_20, Kalirai_etal_14}. In the IFMR context, it is worth mentioning the work by \citet{Fragkou_etal_19} who measured the central star mass of a planetary nebula, member of the open cluster Andrews–Lindsay 1.

The solar-metallicity calibration based on Milky Way AGB stars has so far been hampered by the severe uncertainties on their distances, and  by the fact that the TP-AGB population in star clusters is very small, except for some notable exceptions, such as the old globular clusters 47 Tuc \citep{LebzelterWood_05, McDonald_etal_11a, Momany_etal_12, Lebzelter_etal_14} and $\omega$ Cen \citep{Boyer_etal_08,McDonald_etal_09,McDonald_etal_11b}.

The situation is now remarkably improving thanks to the advent of the \gaia\ satellite and its data releases, in particular \citet[][\gaia\ DR2]{Gaia_DR2_18} and \citet[][\gaia\ EDR3]{Gaia_EDR3_21}. Despite some intrinsic  issues related to the variability and surface dynamics of AGB stars \citep{Chiavassa_etal_18}, we can rely on much better parallaxes than ever, which paves the way for detailed studies also in the Milky Way. When combined with near-infrared (NIR) data, the full-sky, multi-epoch optical photometry provided by \gaia\ is especially promising for characterizing the chemistry and evolutionary stage of evolved red giants, in particular AGB stars \citep{Lebzelter_etal_18,Lebzelter_etal_19}. Recently, \citet{Abia_etal_20} derived the luminosity function and the kinematic properties of a sample of 210 field carbon stars in the solar neighborhood, having measured parallaxes with relative errors of less than 20\%.

A limitation in using field stars is that the ages, and therefore the initial masses of the progenitors, are not easily known.
This weakness can potentially be overcome if AGB stars belong to star clusters, which are routinely dated with stellar isochrones.
In this respect, Milky Way open clusters are potentially suitable targets, as they cover a wide age range ($\approx 0.001 - 10$ Gyr) where AGB stars can be found.
It is dutiful to recall the early efforts made in the past to visually identify AGB stars on photographic plates and report them as probable cluster members: some of the carbon stars analyzed in this work were discovered roughly 50 years ago!
The interested reader may refer to the papers of  \citet{GaustadConti_71, HartwickHesser_71, HartwickHesser_73, CatchpoleFeast_73, Kalinowsli_etal_74, JorgensenWesterlund_88, EggenIben_91}.

Since those first studies until today, AGB stars in open clusters have been largely neglected due to two main problems: the small-number statistics (in the best cases no more than one or two TP-AGB stars are expected per cluster, and very often none), and the difficulty to assess reliable cluster membership.

With \gaia\ this last limitation has been  overcome and we now have large catalogs available that provide astrometric data and membership probabilities for the stars of many open clusters \citep{Monteiro_etal_20,Castro-Ginard_etal_20,Ferreira_etal_20,Cantat-Gaudin_Anders_20, LiuPang_19, Castro-Ginard_etal_19,Cantat-Gaudin_etal_18}.
Using the data from \gaia\ DR2, \citet{Pal_Worthey_21} recently identified 9 carbon stars, likely members of open clusters covering a large age range, $\simeq 0.03-3.20$ Gyr, and studied their frequency normalized to the integrated luminosity of the host cluster, $N_{\rm C}/L_{\rm V}$, as a function of the turn-off mass $\MTO$, or equally of the cluster age.
Their study indicates that the normalized frequency of carbon stars peaks at $\MTO \simeq 1.7\,\Msun$, and drops to zero for $\MTO < 1.24\,\Msun$. This piece of information is useful to constrain the AGB contribution to the integrated light of galaxies.

Spurred by the new perspective offered by \gaia\ (DR2 and  EDR3), in this study we aim to carry out a systematic analysis of AGB stars in open clusters and to characterize them as much as possible by exploiting the observational data (parallax and proper motion, spectral type, spectral energy distribution, light curve and pulsation period) and coupling them with stellar models and radiative transfer calculations that include the effect of circumstellar dust.

The paper is structured as follows. 
In Sect.~\ref{sect_brightsample} we select a sample of bright near-infrared evolved giants and supergiants, likely members of open clusters, using membership data and cluster ages from \gaia\ DR2-based catalogs \citep{Cantat-Gaudin_Anders_20, Cantat-Gaudin_etal_20}, and the  diagnostic \gaia-2MASS diagram, originally designed by \citet{Lebzelter_etal_18}. 
Starting from this sample, in Sect.~\ref{sect_AGBcandidate} we extract the candidate AGB stars by means of suitable photometric and age criteria, and discuss their main characteristics as a function of spectral type (M, S, C) and initial stellar mass. Sect.~\ref{ssec_mscstars} reviews the main information available for the 19 identified TP-AGB stars with assigned spectral type. We add, whenever possible, variability information and new period estimates from the analysis of available light curves.
In Sect.~\ref{ssec_plx} we introduce and discuss the AGB star parallaxes from \gaia\ EDR3, and apply two different prescriptions for the zero-point correction, namely \citet{Lindegren_etal_21}, and \citealt{Groenewegen21}. We use the new \gaia\ EDR3 astrometric and kinematic data (including the zero-point corrections) to re-evaluate the cluster membership of the TP-AGB stars with known spectroscopic type. The procedure is described in Sect.~\ref{ssec_membedr3}.
By using the individual distances, obtained from the parallax inversion, and the extinction $A_V$ from the catalogs of cluster ages, we fit the spectral energy distribution (SED) of each TP-AGB star by means of radiative transfer models that account for the presence of circumstellar dust. The details are given in Sect.~\ref{ssec_sed}. From the SED fitting we derive the bolometric luminosity and the present-day dust-mass loss rate.
From the luminosity  we infer the current mass of the core, \Mc, by means of TP-AGB stellar models available in the literature.
Sect.~\ref{ssec_cmlr} provides a full description of the method.
The entire Sect.~\ref{sect_results} is devoted to present and analyze the results of this study, addressing several aspects of the TP-AGB evolution (e.g, onset of the 3DU and carbon star formation, stellar winds and dust production, long-period variability). In Sect.~\ref{ssec_ifmr} we compare the values of the current core mass inferred from the luminosity with the initial-final mass relation of white dwarfs in the Milky Way. 
In Sect.~\ref{ssec_highmc} we investigate two competitive scenarios for the formation of carbon stars with high \Mc\ in old open clusters: the single stellar evolution mode and the blue straggler channel.
Finally, Sect.~\ref{sec_conclusion} closes the paper.

\section{Identification of near-infrared bright stars}
\label{sect_brightsample}
To investigate the  near-infrared bright stellar populations  in Galactic  open clusters we start with the catalog of \citet{Cantat-Gaudin_Anders_20}, which provides main parameters and lists of star members for 1481 open star clusters based on the second \gaia\ data release \citep[DR2,][]{Gaia_DR2_18}. Each star in the catalog is assigned the \gaia\ DR2  photometry, the distance from parallax inversion, and its membership probability.

Then, we select the brightest near-infrared stars by adopting the following four criteria:
1) stars have 2MASS photometry in the \J, \h, \ks\ pass-bands \citep{Cutri_etal_03}, in addition to the \gaia\ ones;
2) they have an absolute magnitude $\Mks < -5$; 
3) they belong to the associated clusters with a membership probability $p\ge 0.5$;
4) the age of the host cluster is estimated from recent studies,  mostly based on \gaia\ DR2  \citep{Dias_etal_21,Cantat-Gaudin_Anders_20, Bossini_etal_19, Cummings_etal_18}.

To cross-match \gaia\ DR2 and 2MASS catalogs we adopt a search radius of 3$\arcsec$, which avoids spurious duplicates.
The absolute  magnitude of each star $\Mks$ is obtained by subtracting the true distance modulus from the apparent magnitude \ks\ and then applying the extinction correction $A_{k_s}$. In general, the same is done for any relevant pass-band $i$ using the transformation coefficients $A_i/A_V$\footnote{The transformation coefficients for the different filters are derived from the extinction curve of \citet{Cardelli_etal_89} and \citet{ODonnell_James_94} with a total-to-selective extinction ratio of $R_V=3.1$.}, provided that the visual extinction $A_V$ is known.  This latter is taken to be the same $A_V$ of the host cluster, available from the age-estimation studies based on the best-fit isochrone technique \citep{Dias_etal_21,Cantat-Gaudin_Anders_20, Bossini_etal_19, Cummings_etal_18}. We note that some stars may be also affected by circumstellar extinction  due the presence of dust shells; this aspect will be discussed in Sect.~\ref{ssec_sed}.

We consider stars brighter than $\Mks = -5$.
This value is roughly 2 magnitudes fainter than the tip  of the red giant branch (RGB) at solar-like metallicities, $\Mks^{\rm RGBtip} \simeq -7$ \citep{Freedman_etal_20}.
We end up with a sample of 543 stars that satisfy the four constraints listed above.
To obtain a more informative characterization of the selected stars we search for those of known spectroscopic type with the aid of the \texttt{SIMBAD} astronomical database \citep{simbad}.
In total the spectroscopic sample contains 191 objects.
Among them we find 14 O stars, 26 B stars, 11 A stars, 16 F stars, 18 G stars, 54 K stars, 4 Wolf-Rayet stars, 35 M stars,  1 MS star, 2 S stars, 9  C stars, and 1 peculiar star, classified as LBV.
We consider 2 additional carbon stars, not present in the catalog of \citet{Cantat-Gaudin_Anders_20}. More details about these objects are provided in Sect.~\ref{sect_AGBcandidate}.

\subsection{The \gaia-2MASS diagram}
\label{sect_gaia2mass}
\begin{figure*}[ht!]
\begin{center}
\begin{minipage}{0.48\textwidth}
\resizebox{\hsize}{!}{\includegraphics{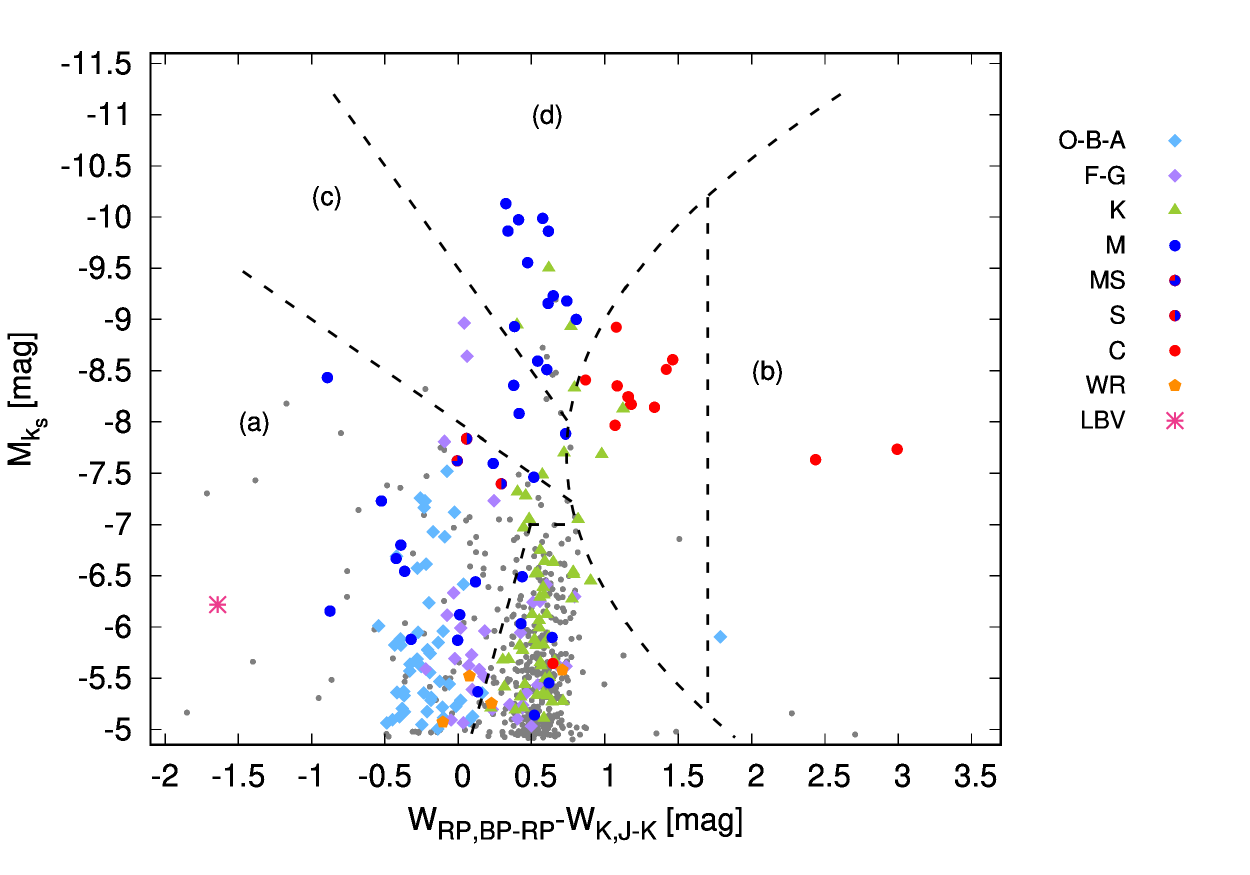}}
\end{minipage}
\begin{minipage}{0.48\textwidth}
\resizebox{\hsize}{!}{\includegraphics{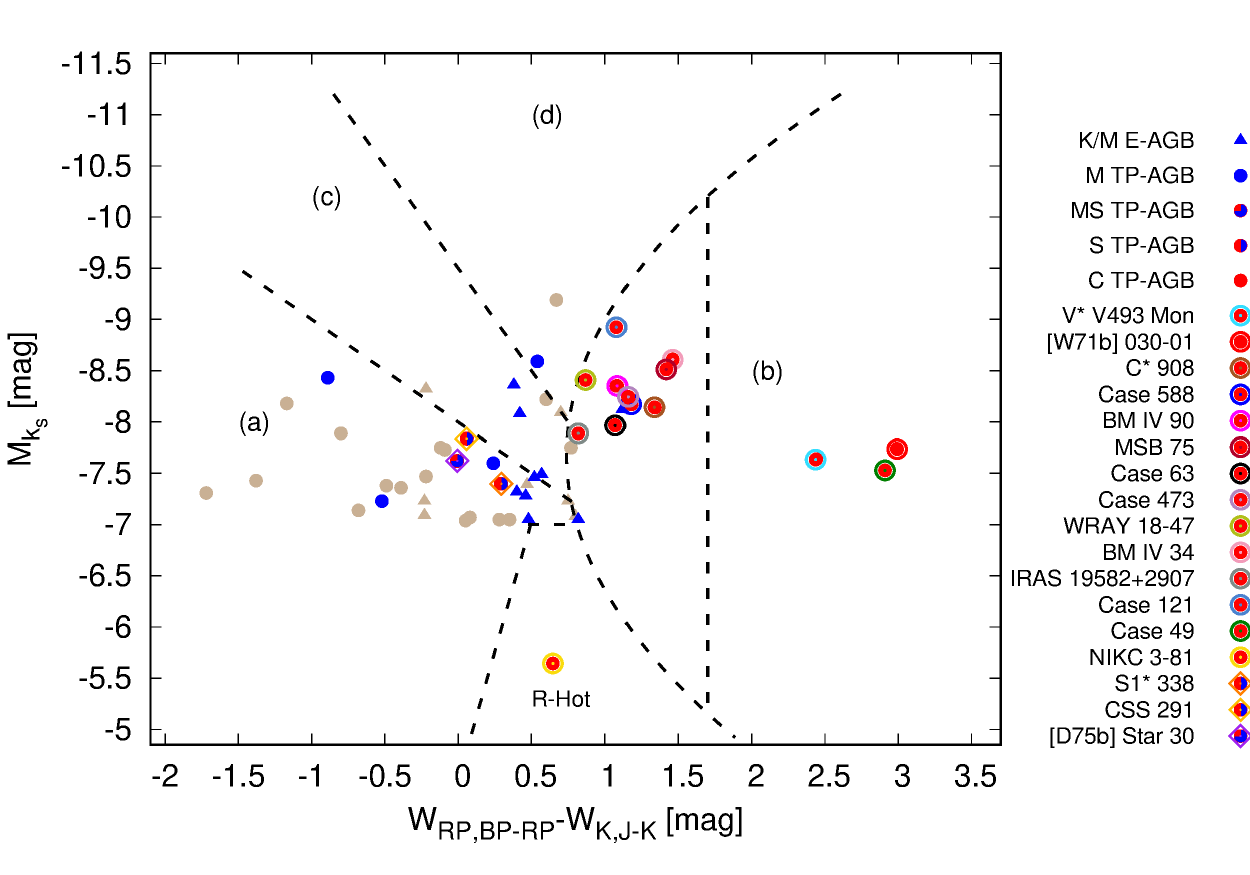}}
\end{minipage}
\begin{minipage}{0.48\textwidth}
\resizebox{\hsize}{!}{\includegraphics{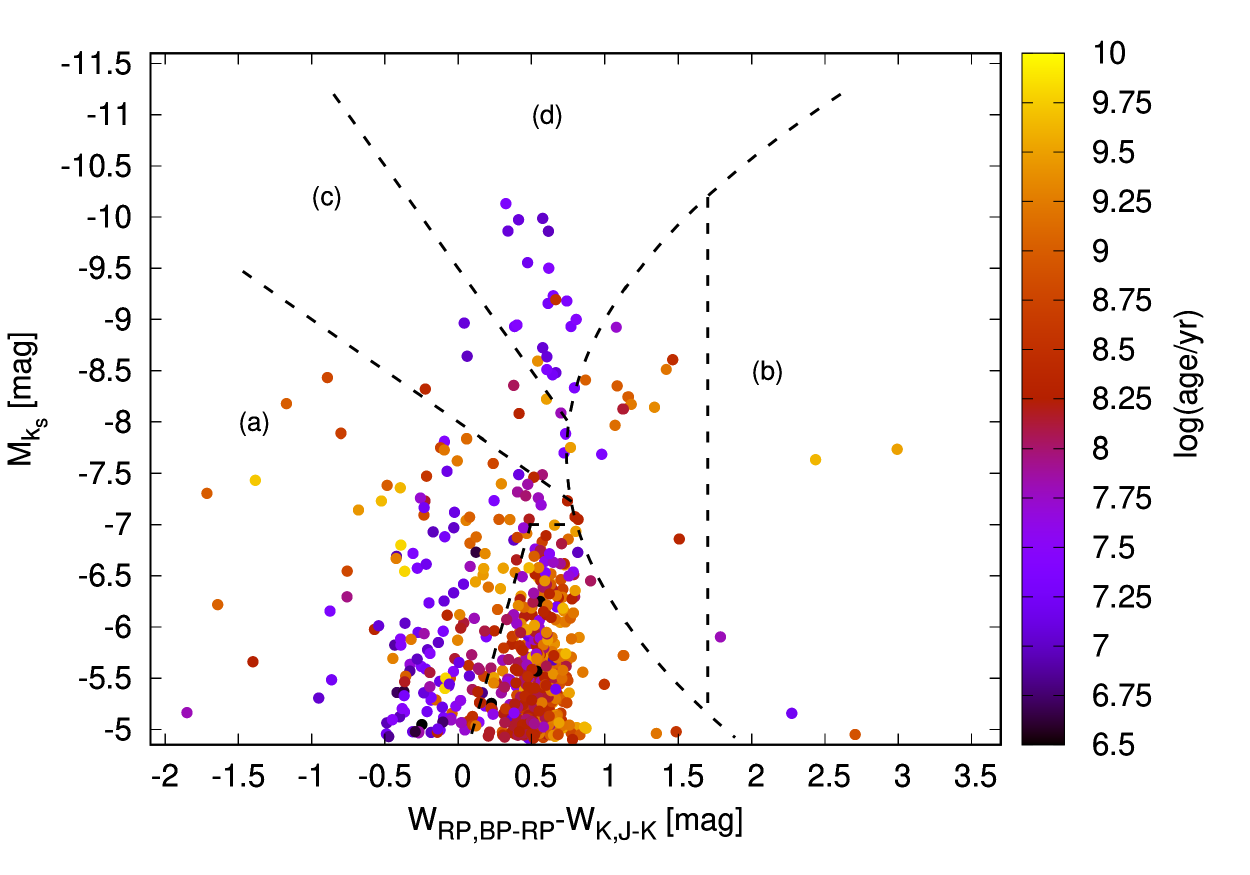}}
\end{minipage}
\begin{minipage}{0.48\textwidth}
\resizebox{\hsize}{!}{\includegraphics{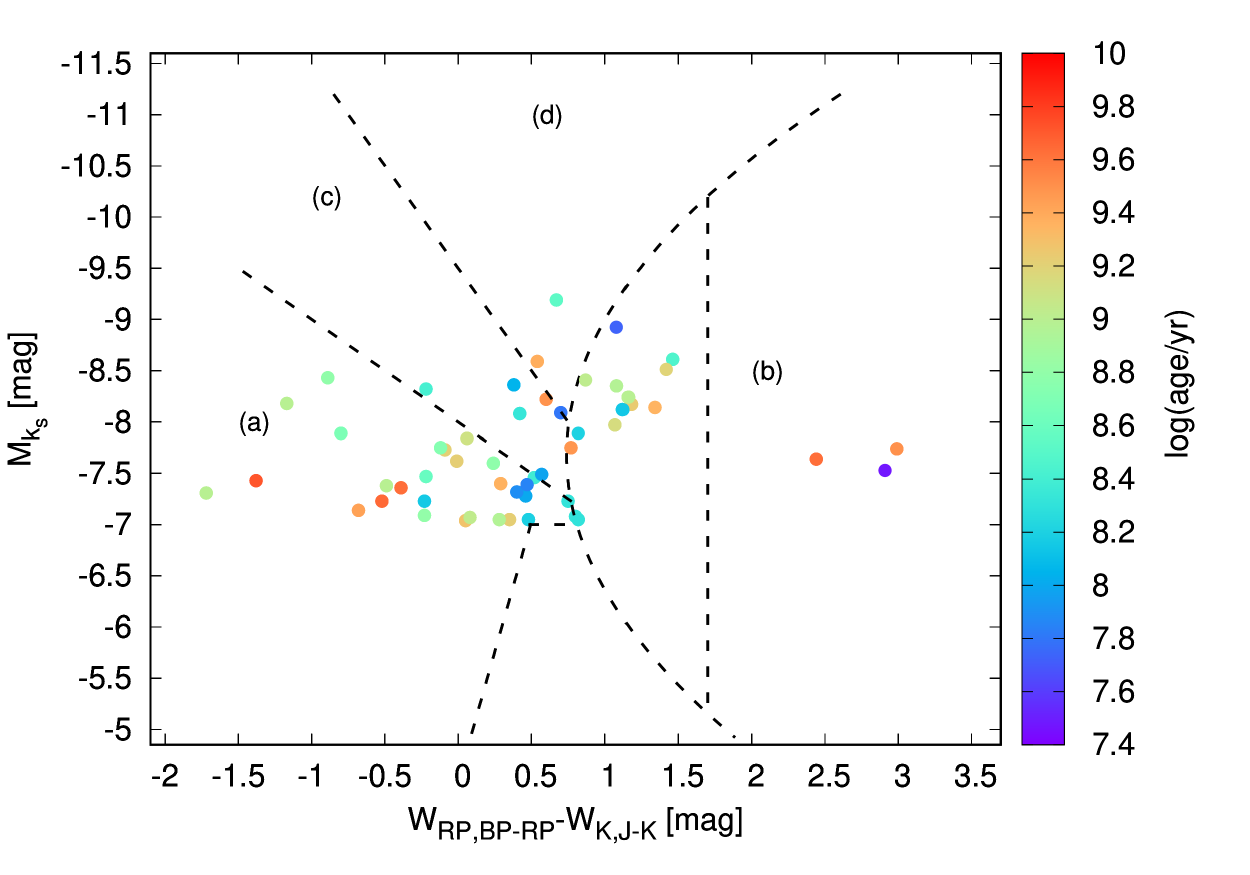}}
\end{minipage}
\begin{minipage}{0.48\textwidth}
\resizebox{\hsize}{!}{\includegraphics{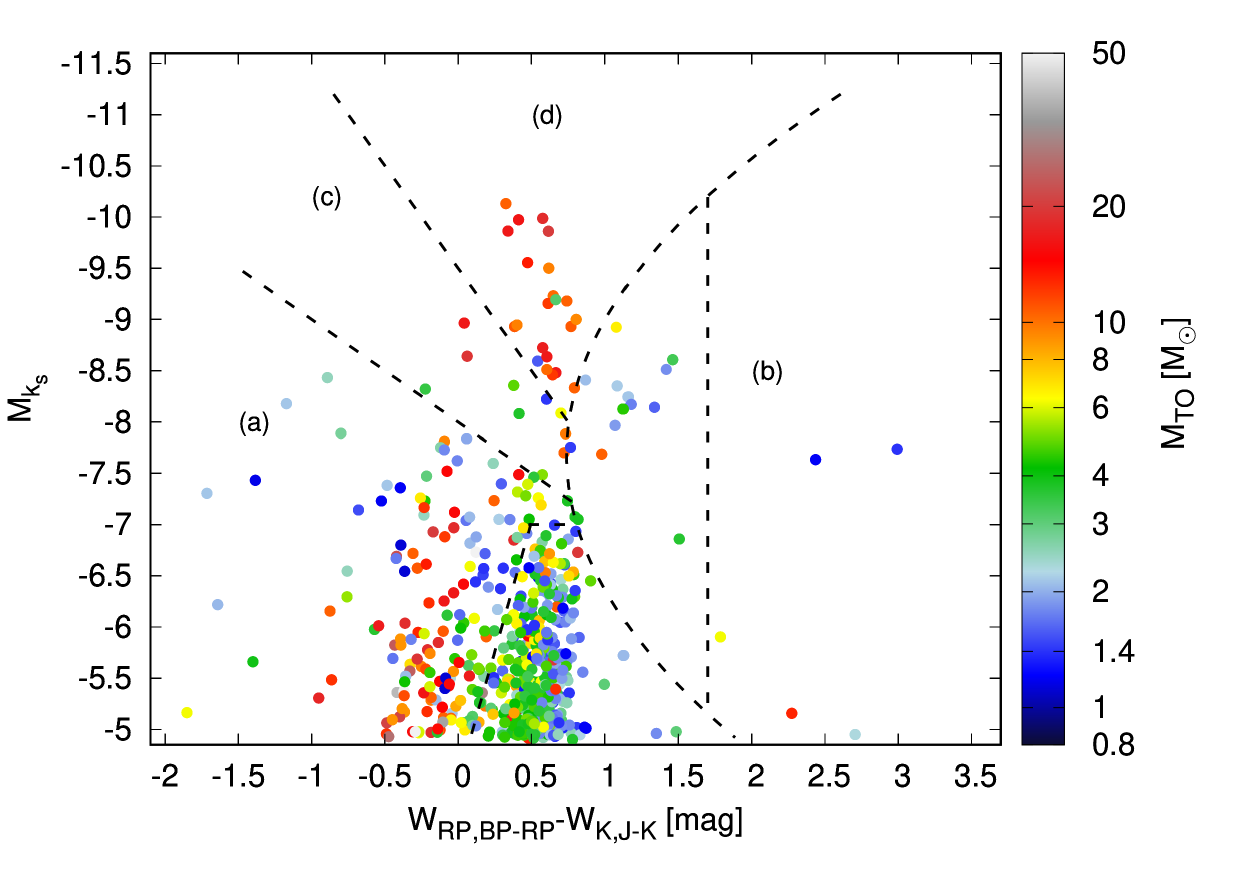}}
\end{minipage}
\begin{minipage}{0.48\textwidth}
\resizebox{\hsize}{!}{\includegraphics{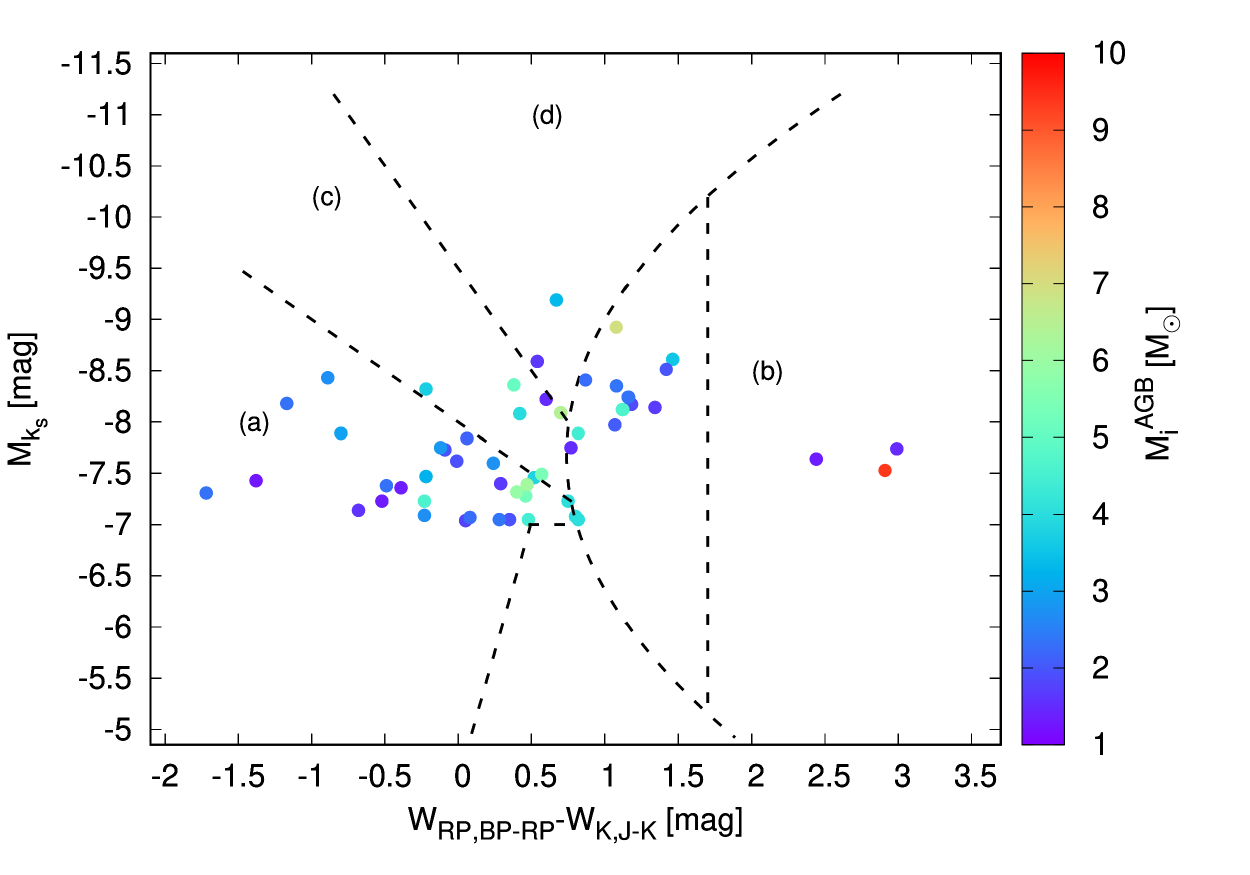}}
\end{minipage}
\caption{Near-infrared bright  stars of open clusters in the \gaia-2MASS diagram. The absolute $M_{K_s}$ magnitude is corrected for interstellar extinction, derived from $A_V$ of each cluster.
Dashed lines show the approximate boundaries of main stellar branches, defined by \citep{Lebzelter_etal_18}.
\emph{Left panels}: all stars from the catalog of \citet{Cantat-Gaudin_Anders_20} with cluster membership probability ($p\ge 0.5$). The stars are marked according to their spectral type if known (otherwise they are indicated by grey dots), age and turn-off mass of the host cluster. 
\emph{Right panels}: AGB star candidates in open clusters, brighter than the RGB tip. From top to bottom the stars are color-coded as a function of the spectral type (when unknown the gray color is used), cluster's age, and initial mass. The top panel distinguishes between E-AGB (triangles) and TP-AGB (circles) stars, based on \texttt{PARSEC} stellar isochrones. It also includes the full list of MS, S, and C stars. See the text for more details.} 
\label{fig_lebz}
\end{center}
\end{figure*}

Once the cross-identification and dereddening operations are completed, we construct the \gaia-2MASS diagram (see Fig.~\ref{fig_lebz}), originally designed by \citet[][hereinafter L18, L19]{Lebzelter_etal_18, Lebzelter_etal_19}, that is especially suitable to highlight the presence of AGB stars. In this diagram the \ks\ magnitude is correlated with a particular combination of \gaia\ and 2MASS pass-bands, through the quantity \WRP-\WJK. Here \WRP\ and \WJK\ are reddening-free Wesenheit functions \citep{Madore_1982,Soszynski_etal_2005}, defined as $\WRP = G_{\mathrm{RP}} -1.3\times(G_{\mathrm {BP}} - G_{\mathrm{RP}})$  and $\WJK = \ks - 0.686\times(J-\ks)$, respectively.
L18 first demonstrated that this diagram is a powerful tool to analyze and identify sub-classes of AGB stars, as a function of chemical type and initial  mass.

Figure~\ref{fig_lebz} shows three versions of the same \gaia-2MASS diagram, in which the stars are flagged by their spectroscopic type, age and turn-off mass of the host cluster.
To guide the analysis we mark with letters the regions of the four main stellar branches as introduced by L18 (see their figures 1-3), namely:
(a) low-mass O-rich AGB stars, RGB and faint AGB stars; (b) carbon stars; (c) intermediate-mass O-rich AGB stars; (d) red super giants (RSG) and massive O-rich AGB stars.

In this respect, we should note that some important differences exist between the stellar populations examined by L18, L19 and our study.
In the original L18, L19 works the data mainly refer to long-period variables (LPVs) in the Magellanic Clouds (LMC), identified with \gaia\ DR2 data, while in this study we consider {\em all} stars of Galactic open clusters satisfying the four criteria given in Sect.~\ref{sect_brightsample}. As a consequence, the \gaia-2MASS diagrams of Fig.~\ref{fig_lebz} are populated also by stars not belonging to the AGB star population.
Let us discuss the main features of each region, as a function of spectral type (top panel), age (middle panel) and turn-off mass (hereafter also \MTO; bottom panel).

\subsubsection{Region (a)}
The main differences between L18 and this study show up in the mixture of stars that populate the region (a).
In our diagram we identify a group of O-B-A stars, mostly of I luminosity class (light blue diamonds in the top-left panel). They draw a sort of vertical sequence  leftward of the main branch at $\WRP-\WJK \approx 0.5$, where most of the K stars (green triangles) and some faint M stars (blue circles) are found. The O-B-A stars are associated to young open clusters, with  $7.0 \la  \log({\rm age/yr}) \la 7.5$, likely corresponding to core H-burning massive stars of initial mass $\Mi > 10\, \Msun$ (bottom panel). Slightly to the right of the O-B-A type group  we find F-G stars (purple diamonds) of older ages, with $\log({\rm age/yr}) > 8$. These should mainly correspond to core He-burning stars. Intermediate-age K stars, with $8 \la \log({\rm age/yr}) \la 9$, appear to populate the main branch at $\WRP-\WJK \approx 0.5$ as E-AGB stars, together with the old RGB K stars, with $\log({\rm age/yr}) > 9 $. Finally, we note that the region (a) hosts also M-type stars distributed over a wide age range, typically with $\log({\rm age/yr}) > 8 $. These stars are mainly O-rich AGB star candidates.
Among them there are  3 stars of spectral type S and MS, which are particularly interesting since they witness the occurrence of the 3DU.
A more in-depth discussion about AGB stars in the region (a) is given in Sect.~\ref{sect_AGBcandidate}.

In addition to the groups of stars just mentioned, we find  a few peculiar objects in region (a).
 We identify a faint carbon star, NIKC 3-81, located at $\WRP-\WJK \simeq 0.65$  and $\Mks \simeq -5.63$ on the main branch of RGB K stars.
Following the recent work by \citet{Abia_etal_20} on field Galactic carbon stars, we suggest that NIKC 3-81 might belong to the class of R-hot carbon stars. These objects are faint carbon stars, without s-process enhancement, whose origin is still a matter of debate. One hypothesis is that the carbon enrichment occurs during the He-flash at the tip of the RGB, but other channels are also invoked \citep[see][and references therein for a thorough discussion]{Abia_etal_20}.
The star NIKC 3-81 has an age of 2.7 Gyr, and $\Mi \simeq 1.53 \, \Msun$ assuming it is in a post-He-flash stage. 

At the bottom of the same sequence 
there are five Wolf-Rayet stars (orange pentagons),  belonging to young open clusters (Hogg 15, NGC 6231, NGC 6871, Negueruela 1, and Teutsch 127), with $6.3 \la  \log({\rm age/yr}) \la 7.2$. In particular, the star HD 311884, member of Hogg 15, is assigned a spectral type WN6+O5V with an age of 2.2 Myr, and hence it is consistent with a very massive progenitor, $\Mi \ga 100\, \Msun$. Following recent stellar evolution calculations, this object may end up its life as a pulsation pair-instability supernova  \citep{Goswami_etal_21, Costa_etal_21, Chen_etal_15}.

 On the left side of the region (a), at $\WRP-\WJK \simeq -1.64$ and $\Mks \simeq -6.22$, there is an intriguing object, MR 35, (asterisk), which is attributed the spectral type of Luminous Blue Variable (LBV), typical of very massive and bright stars near the Humphreys-Davidson limit \citep{HDlimit}.
 This classification appears in conflict with the age of about 1.1 Gyr of the host cluster Teutsch 143a, which has $\MTO \simeq 2\, \Msun$. A closer inspection reveals that this star is included in the atlas of post-AGB stars and planetary nebulae selected from IRAS point source catalogue \citep{Suarez_etal_06}. MR 35 is classified as a peculiar star (neither a genuine planetary nebula, nor a post-AGB star),  with an optical spectrum showing a few emission lines including  H${\alpha}$. It is suggested to be a very young proto-planetary nebula. This configuration better matches with the intermediate age of MR 35.
 
\subsubsection{Region (b)}
It should correspond to the domain of intrinsic carbon stars. In fact, we count 11 spectroscopically identified C stars in this region, which have a good probability of being cluster members.
We devote an extensive analysis of their properties in Sect.~\ref{sect_AGBcandidate}. They span an age range from about 0.2 Gyr to 4.3 Gyr. We note that only 2 carbon stars have $\WRP-\WJK > 1.7$, the approximate boundary beyond which stars are expected to experience significant mass loss driven by carbonaceous dust. The remaining 9 carbon stars are located to the left of this limit where dust should be inefficiently  produced. We will deepen this aspect with the aid of radiative transfer models in Sect.~\ref{ssec_sed}.

\subsubsection{Regions (c) and (d)}
The region (c) should contain intermediate-mass O-rich AGB stars. 
Though it appears barely populated, we find three  stars consistent with this expectation, having  $8.0 \la \log({\rm age/yr}) \la 8.4$, and initial masses $5.0 \ga \Mi/\Msun \ga 3.5$. They are likely E-AGB stars \citep[based on \texttt{PARSEC} stellar isochrones of][triangles]{Pastorelli_etal_20}. Two of them have spectral type K (blue triangles in the top-right panel of Fig.~\ref{fig_lebz}).

M-type red supergiants, evolved from massive stars with $7.0 \la \log({\rm age}) \la 7.8$, are detected close to the bright edge of the region (c), and mostly in region (d) where they draw a nearly vertical branch. This finding is fully consistent with 
the analysis of L18. In the same region we identify a younger bright star (at $\WRP-\WJK \simeq 0.66$ and $\Mks \simeq -9.19$) with an age of 0.34 Gyr. This object, 2MASS J07570972-2553064, is also a candidate TP-AGB star (with $M_{\rm i}^{\rm AGB} \simeq 3.3\, \Msun$), perhaps undergoing a mild HBB.

\section{AGB stars in open  clusters}
\label{sect_AGBcandidate}
From the sample of bright near-infrared stars we aim to extract a sub-sample of AGB star candidates. This is done by combining the \gaia-2MASS diagram with the age estimates of the clusters.
The AGB star candidates are selected according to the following criteria:
1) they must be brighter than the RGB tip, i.e. $\Mks <-7$; 
2) they must be older than about 100 Myr, corresponding to an upper limit of the initial post-main sequence mass of about 6 \Msun; 
3) they are assigned a spectral type among K, M, MS, S, C, when known.

The lower age limit is chosen on the basis of stellar evolution models. At solar metallicity, \texttt{PARSEC} models predict a maximum mass for the development of a degenerate carbon-oxygen core, $\Mup \simeq 6\,\Msun$, corresponding to a main-sequence lifetime of roughly 100 Myr.
Stars with $\Mi < \Mup$ experience the standard double-shell AGB phase, that includes both the Early-AGB (E-AGB) and the TP-AGB phases.
In total, we identified 49 candidate AGB stars: 17 are expected to be on the E-AGB and the remaining 32 are interpreted as evolving on the TP-AGB. This distinction is made using the information stored in the \texttt{PARSEC} isochrones filtered through 2MASS pass-bands.

The main properties of the host clusters and the \gaia\ DR2 and 2MASS photometry of the individual AGB candidates are summarized in  Table~\ref{tab_agbcand}. 
For about half of the AGB sample (24 sources) we also know the spectral type from the \texttt{SIMBAD} database: 5 stars are of K type, 7 stars are of M type, 1 star is of MS type, 2 stars are of S type, and 8 stars are of type C.
The group of spectroscopic C stars is complemented with 2 additional sources, namely: the TP-AGB stars MSB 75 and Wray 18-47, not included in the catalog of \citet{Cantat-Gaudin_Anders_20}. Analyzed with \gaia\ EDR3 data (see Sect.~\ref{ssec_membedr3}), they turn out to be likely members of the open clusters NGC 7789 and NGC 2533, respectively.

The data for the AGB stars with known spectral type are summarized  in Table~\ref{tab_specagb}. For comparison we report the age of their associated cluster from two different sources, namely:
\citet{Cantat-Gaudin_etal_20}, and \citet{Dias_etal_21}. Both catalogs use \gaia\ DR2 photometric data and membership information.

All AGB star candidates are plotted in the \gaia-2MASS diagram (Fig.~\ref{fig_lebz}). 
In Sect.~\ref{sect_gaia2mass} we already analyzed the main features of the different regions of the diagram introduced by L18 and L19. 
Here we focus on some specific aspects  of the AGB population that still deserve to be properly taken into consideration.

\subsection{O-rich AGB stars}
L18 pointed out that in the region (a) the branch extending towards negative $\WRP-\WJK$ primarily contains low-mass O-rich AGB stars (with $M_{\rm i}^{\rm AGB} \ga 2\, \Msun$), while more massive objects are found in regions (c) and (d).
In our sample the region (a) contains a few stars, with ages exceeding 1 Gyr, that are consistent with that indication, though also younger, and more massive ($M_{\rm i}^{\rm AGB} \ga 3\,\Msun $) stars are present. This is a substantial difference with respect to the Magellanic  Clouds populations analyzed by L18 and L19.
 In practice, we do not detect a clear separation between low- and intermediate-mass stars in regions (a) and (c), though our conclusions might be conditioned by the small size of the sample. However, this does not seem to be the case. 
 In fact, the lack of a well-defined gap between branches (a) and (c) stands out quite clearly if we consider the rich LPVs population in the Milky Way observed with \gaia\ DR2, as well as the corresponding synthetic populations that include  AGB stars with solar-like metallicity (Pastorelli et al., in preparation).  As discussed in that study, the disappearance of the gap between regions (a) and (c) is driven by a metallicity effect, mainly reflecting i) the reduction of the initial-mass range for carbon-star formation at increasing metallicity, and  ii) the specific behaviour of spectra for O-rich AGB stars as a function of effective temperature and metallicity (Pastorelli et al., in preparation).

Quite interesting is the case of the three stars of spectral type S and MS detected in region (a). To our knowledge this is the first time that  AGB stars of this spectral type are confirmed to be members of open clusters. An early study \citep{Jorgensen_85} adopted a statistical approach to examine 15 S stars found along the line of sight of open clusters, but concluded that such configuration could be just explained as the result of a random field distribution of stars. Thanks to \gaia\ data, this negative conclusion has now been importantly revised.
We note that the 1 MS and 2 S stars are located in the vicinity of the bright side of the region (c), supporting the predictions of evolutionary models for the TP-AGB phase in which repeated 3DU  events increase the photospheric C/O approaching unity \citep[][more details in Pastorelli et al., in prep.]{Marigo_etal_13}. 

By using the estimated ages of the host clusters we can characterize the stellar progenitors: the two S stars, S1$^\ast$ 338 and CSS 291, belong  to the old open clusters BH 55 and Tombaugh 1, and correspond to initial masses $M_{\rm i}^{\rm AGB} \simeq 1.65\, \Msun$ and $M_{\rm i}^{\rm AGB} \simeq 2.13\, \Msun$, respectively. The only MS star, [D75b] Star 30, member of the cluster  NGC 1798, is associated with $M_{\rm i}^{\rm AGB} \simeq 1.93\, \Msun$. 
These objects may set useful constraints on the onset of the 3DU, as discussed in Sect.~\ref{sect_results},
\nameref{para_sstars}.

Moving to intermediate-mass stars, we  identify in regions (c) and (d) few AGB stars with initial masses $M_{\rm i}^{\rm AGB} > 3 \, \Msun$, which are in principle compatible with the occurrence of HBB. However, we note that they
are not particularly bright ($\ks > -9.5$) and their evolutionary stage is expected to correspond to the E-AGB.  Therefore these objects still have to enter the TP-AGB phase. 

\subsection{Carbon stars}
We find 10 spectroscopically confirmed carbon stars, all falling inside region (b), as expected from the studies by L18, L19 and \citet{Abia_etal_20}. Among them, 8 carbon stars derive from the \gaia\ DR2 catalog of \citet{Cantat-Gaudin_Anders_20}, having a membership probability $p\ge 0.5$.  We add 2 stars, MSB 75 and Wray 18-47, likely members of the open clusters NGC 7789 and NGC 2533, according to a new membership analysis based on \gaia\ EDR3 (Sect.~\ref{ssec_membedr3}).

\citet{Pal_Worthey_21} recently analyzed the frequency of carbon stars in open clusters using the data from \gaia\ DR2. They obtained a list of 9 objects possibly associated to star clusters (their table 2), of which 6 are in common with our 10-star sample, namely: V493 Mon, C* 908, V* BI Per\footnote{In this work V* BI Per is indicated also with the name Case 63.}, BM IV 90, Case 473, and BM IV 34. The remaining 3 carbon stars considered by \citet[][namely IRAS 19582+2907, Case 121, Case 49]{Pal_Worthey_21}  are not included in our initial selection as their cluster membership probability is rather low, $p=0.4$, according to \citet{Cantat-Gaudin_etal_20}.
However, for the purpose of comparison we add these 3 doubtful cases  at the bottom of Table~\ref{tab_specagb} and in Fig.~\ref{fig_lebz}.
We just note that these carbon stars are related to relatively young star clusters, with $29 \la \mathrm{(age/Myr)} \la 158$ and turn-off masses in the range $8.97 \ga (\MTO/\Msun) \ga 4.20$.
We analyze the cluster membership of these sources in  Sect.~\ref{ssec_membedr3}, and the evolutionary status in Sect.~\ref{sect_results}.

\begin{table*}
\centering
\begin{threeparttable}
\tiny
\caption{Open cluster parameters and properties of AGB star candidate members, based on \gaia\ DR2.}
\label{tab_agbcand}
\begin{tabular*}{\textwidth}{@{\extracolsep{\fill}}ccccccccccccc@{}}
\hline\hline
\multicolumn{1}{c}{2MASS ID} &
\multicolumn{1}{c}{Cluster} &
\multicolumn{1}{c}{$\mathrm{\log(age)}$} & 
\multicolumn{1}{c}{$M_{\rm TO}$} & 
\multicolumn{1}{c}{$D$} &
\multicolumn{1}{c}{$A_V$} &
\multicolumn{1}{c}{$p$} &
\multicolumn{1}{c}{$(\jks)_0$} &
\multicolumn{1}{c}{$\WRP-\WJK$} &
\multicolumn{1}{c}{$\Mks$} &
\multicolumn{1}{c}{$\Mi^{\rm AGB}$}&
\multicolumn{1}{c}{stage}\\
\hline
\noalign{\smallskip}
\multicolumn{1}{c}{} &
\multicolumn{1}{c}{} &
\multicolumn{1}{c}{dex} & 
\multicolumn{1}{c}{[$\Msun$]} & 
\multicolumn{1}{c}{[pc]} &
\multicolumn{1}{c}{[mag]} &
\multicolumn{1}{c}{} &
\multicolumn{1}{c}{[mag]} &
\multicolumn{1}{c}{[mag]} &
\multicolumn{1}{c}{[mag]} &
\multicolumn{1}{c}{$[\Msun]$}&
\multicolumn{1}{c}{}\\
\hline
\noalign{\smallskip}
  $09551489-5633003$ &            FSR 1521 &  9.72 &  1.16 & 4933 &  2.72 & 0.7 &   1.40 &  $-1.38$ &  $-7.43$ &   1.25 & TP-AGB\\ 
  $23475728+6835426$ &             King 11 &  9.65 &  1.18 & 3097 &  2.67 & 0.8 &   1.17 &  $-0.52$ &  $-7.23$ &   1.31 & TP-AGB\\ 
  $07263860-4739437$ &          Melotte 66 &  9.63 &  1.19 & 4830 &  0.25 & 0.9 &   1.28 &  $-0.39$ &  $-7.36$ &   1.33 & TP-AGB\\ 
  $06363268+0925393$ &          Trumpler 5 &  9.63 &  1.19 & 3047 &  1.20 & 0.7 &   2.94 &  $+2.44$ &  $-7.64$ &   1.33 & TP-AGB\\ 
  $08310560-3838107$ &            Pismis 3 &  9.50 &  1.40 & 2349 &  2.35 & 0.9 &   3.17 &  $+2.99$ &  $-7.74$ &   1.47 & TP-AGB\\ 
  $06525082+1655289$ &         Berkeley 29 &  9.49 &  1.42 & 12604 &  0.24 & 0.5 &   1.06 &  $+0.60$ &  $-8.22$ &   1.48 & TP-AGB\\ 
  $10593792-5936032$ &         Teutsch 106 &  9.48 &  1.44 & 7184 &  1.74 & 0.6 &   1.17 &  $+0.77$ &  $-7.75$ &   1.50 & TP-AGB\\ 
  $21030493+4021370$ &         Berkeley 54 &  9.43 &  1.51 & 6680 &  2.24 & 1.0 &   1.27 &  $-0.68$ &  $-7.14$ &   1.56 & TP-AGB\\ 
  $07002883-0014204$ &         Berkeley 34 &  9.38 &  1.56 & 7609 &  1.64 & 1.0 &   0.77 &  $+0.54$ &  $-8.59$ &   1.62 & TP-AGB\\ 
  $08561346-3930429$ &               BH 55 &  9.36 &  1.59 & 4783 &  1.22 & 0.8 &   1.26 &  $+0.29$ &  $-7.40$ &   1.65 & TP-AGB\\ 
  $07494578-1715141$ &         Ruprecht 37 &  9.35 &  1.60 & 4526 &  0.30 & 0.5 &   1.80 &  $+1.34$ &  $-8.14$ &   1.66 & TP-AGB\\ 
  $05482764+0715292$ &        Collinder 74 &  9.28 &  1.69 & 2498 &  0.85 & 0.7 &   1.30 &  $+0.05$ &  $-7.04$ &   1.76 & TP-AGB\\ 
  $06090764+0436414$ &              Dias 2 &  9.24 &  1.74 & 4009 &  0.47 & 0.9 &   2.05 &  $+1.18$ &  $-8.17$ &   1.84 & TP-AGB\\ 
  $05113360+4740468$ &            NGC 1798 &  9.22 &  1.77 & 5124 &  1.12 & 0.6 &   1.24 &  $-0.01$ &  $-7.62$ &   1.93 & TP-AGB\\ 
  $08003039-1047532$ &            NGC 2506 &  9.22 &  1.77 & 3191 &  0.09 & 1.0 &   1.18 &  $+0.35$ &  $-7.05$ &   1.93 & TP-AGB\\ 
  $07030677-2050128$ &          Tombaugh 2 &  9.21 &  1.79 & 9316 &  0.83 & 0.7 &   1.33 &  $-0.09$ &  $-7.73$ &   1.94 & TP-AGB\\ 
  $03325580+5244137$ &          Berkeley 9 &  9.14 &  1.89 & 1814 &  2.69 & 1.0 &   1.49 &  $+1.07$ &  $-7.97$ &   2.06 & TP-AGB\\ 
  $07001366-2033294$ &          Tombaugh 1 &  9.10 &  1.95 & 2554 &  0.66 & 0.8 &   1.19 &  $+0.06$ &  $-7.84$ &   2.13 & TP-AGB\\ 
  $10092236-5718402$ &            FSR 1530 &  9.03 &  2.06 & 5421 &  3.90 & 1.0 &   2.08 &  $+0.08$ &  $-7.07$ &   2.26 & TP-AGB \\ 
  $20541769+5039128$ &         Berkeley 53 &  8.99 &  2.13 & 3415 &  4.42 & 0.7 &   1.28 &  $-1.72$ &  $-7.31$ &   2.33 & TP-AGB\\ 
  $20553632+5100498$ &         Berkeley 53 &  8.99 &  2.13 & 3415 &  4.42 & 0.9 &   1.44 &  $-1.17$ &  $-8.18$ &   2.33 & TP-AGB\\ 
  $20560804+5104316$ &         Berkeley 53 &  8.99 &  2.13 & 3415 &  4.42 & 0.6 &   1.56 &  $+1.16$ &  $-8.24$ &   2.33 & TP-AGB\\ 
  $20560894+5059071$ &         Berkeley 53 &  8.99 &  2.13 & 3415 &  4.42 & 1.0 &   1.17 &  $-0.49$ &  $-7.38$ &   2.33 & TP-AGB\\ 
  $08423384-4712252$ &            NGC 2660 &  8.97 &  2.17 & 2788 &  1.19 & 1.0 &   1.70 &  $+1.08$ &  $-8.35$ &   2.36 & TP-AGB\\ 
  $20104694+4110170$ &             IC 1311 &  8.96 &  2.18 & 6167 &  1.34 & 0.7 &   1.13 &  $+0.28$ &  $-7.05$ &   2.38 & TP-AGB\\ 
  $21104859+4832062$ &         Berkeley 91 &  8.80 &  2.49 & 6057 &  4.04 & 0.7 &   1.09 &  $-0.23$ &  $-7.09$ &   2.70 & E-AGB \\ 
  $09264885-5114107$ &               BH 67 &  8.79 &  2.51 & 8167 &  3.00 & 0.8 &   1.35 &  $-0.89$ &  $-8.43$ &   2.72 & TP-AGB \\ 
  $00161695+5958115$ &   Juchert Saloran 1 &  8.78 &  2.53 & 5383 &  2.82 & 0.9 &   1.20 &  $+0.24$ &  $-7.60$ &   2.74 & TP-AGB\\ 
  $14573523-6235349$ &        Ruprecht 112 &  8.73 &  2.64 & 2733 &  2.07 & 0.7 &   1.12 &  $-0.12$ &  $-7.75$ &   2.85 & TP-AGB\\ 
  $14543493-6233265$ &        Ruprecht 112 &  8.73 &  2.64 & 2733 &  2.07 & 0.8 &   1.22 &  $+7.68$ &  $-7.33$ &   2.85 & E-AGB\\ 
  $19474773+2328064$ &            FSR 0154 &  8.68 &  2.76 & 3594 &  3.52 & 0.6 &   1.24 &  $-0.80$ &  $-7.89$ &   2.97 & TP-AGB\\ 
  $17532024-2721185$ &          Czernik 37 &  8.58 &  3.00 & 2314 &  3.43 & 0.9 &   1.14 &  $-0.22$ &  $-7.47$ &   3.21 & TP-AGB\\ 
  $07570972-2553064$ &         Ruprecht 42 &  8.53 &  3.13 & 6560 &  0.96 & 0.5 &   0.98 &  $+0.67$ &  $-9.19$ &   3.34 & TP-AGB\\ 
  $07444463-2824077$ &          Haffner 14 &  8.46 &  3.33 & 4108 &  1.52 & 0.6 &   1.44 &  $+1.46$ &  $-8.61$ &   3.54 & TP-AGB \\ 
  $01292302+6316584$ &             NGC 559 &  8.41 &  3.48 & 2884 &  2.19 & 0.9 &   1.17 &  $-0.22$ &  $-8.32$ &   3.69 & E-AGB\\ 
  $07593754-6035134$ &            NGC 2516 &  8.38 &  3.57 &  423 &  0.11 & 0.6 &   1.09 &  $+0.52$ &  $-7.46$ &   3.79 & E-AGB\\ 
  $10190419-5625014$ &               BH 92 &  8.33 &  3.74 & 2384 &  1.33 & 0.9 &   1.13 &  $+0.42$ &  $-8.08$ &   3.95 & E-AGB\\ 
  $07082701-1312329$ &            NGC 2345 &  8.32 &  3.77 & 2663 &  1.04 & 0.9 &   1.08 &  $+0.80$ &  $-7.08$ &   3.98 & E-AGB\\ 
  $13193042-6454484$ &        Ruprecht 107 &  8.31 &  3.80 & 3464 &  1.01 & 0.9 &   1.14 &  $+0.75$ &  $-7.23$ &   4.02 & E-AGB\\ 
  $14350882-5633453$ &            NGC 5662 &  8.30 &  3.84 &  761 &  0.60 & 0.8 &   1.19 &  $+0.82$ &  $-7.05$ &   4.05 & E-AGB\\ 
  $23244485+6120384$ &            NGC 7654 &  8.19 &  4.24 & 1653 &  1.85 & 1.0 &   0.79 &  $+0.48$ &  $-7.05$ &   4.46 & E-AGB\\ 
  $01331401+6041111$ &         Gulliver 16 &  8.15 &  4.40 & 4886 &  1.39 & 0.9 &  -0.13 &  $-0.23$ &  $-7.23$ &   4.63 & E-AGB\\ 
  $10473842-5728027$ &         Ruprecht 91 &  8.14 &  4.44 & 1070 &  0.19 & 0.6 &   1.28 &  $+1.12$ &  $-8.12$ &   4.67 & E-AGB\\ 
  $09490713-5434130$ &         Ruprecht 83 &  8.05 &  4.84 & 3706 &  2.38 & 0.5 &   1.10 &  $+0.38$ &  $-8.36$ &   5.07 & E-AGB\\ 
  $01574017+6013079$ &             NGC 743 &  8.00 &  5.07 & 1169 &  1.40 & 0.7 &   1.00 &  $+0.46$ &  $-7.28$ &   5.32 & E-AGB\\ 
  $10384498-5910584$ &               BH 99 &  7.98 &  5.17 &  467 &  0.34 & 1.0 &   0.92 &  $+0.57$ &  $-7.49$ &   5.42 & E-AGB\\ 
  $16553781-3930177$ &            NGC 6242 &  7.89 &  5.65 & 1246 &  1.03 & 1.0 &   0.76 &  $+0.40$ &  $-7.32$ &   5.92 & E-AGB\\ 
  $18332658-1025317$ &            NGC 6649 &  7.85 &  5.89 & 2124 &  3.90 & 0.8 &   0.94 &  $+0.47$ &  $-7.39$ &   6.15 & E-AGB\\ 
  $05241683+4218238$ &              SAI 47 &  7.80 &  6.20 & 4683 &  1.40 & 0.7 &   1.01 &  $+0.70$ &  $-8.09$ &   6.45 & E-AGB\\      
\hline

\noalign{\smallskip}
\end{tabular*}
\footnotesize{{\bf Notes:} The table is sorted by cluster's age decreasing downwards. From left to right the columns show: the star's 2MASS identifier (1), the cluster's name (2), age (3), turn-off mass \MTO\ (4), distance $D$ (5), visual extinction $A_V$ (6), the star's membership probability $p$ (7), intrinsic $(\jks)_0$ color (8), the difference of \gaia-2MASS Wasenheit functions (9),  the absolute deredenned \Mks\ magnitude (10), the initial mass $M_{\rm i}^{\rm AGB}$ (11) and the evolutionary stage (12).
Cluster's parameters (age, $A_V$ and distance $D$) are taken from \citet{Cantat-Gaudin_etal_20}, while the turn-off mass \MTO\ is extracted from the library of \texttt{PARSEC} stellar isochrones, assuming solar metallicity.
For each star, the membership probability $p$ is from \citet{Cantat-Gaudin_Anders_20}, intrinsic $(J-\ks)_0$ color, absolute magnitude $\Mks$, initial mass $M_{\rm i}^{\rm AGB}$, and evolutionary stage (E-AGB or TP-AGB) are derived from the library of \texttt{PARSEC} stellar isochrones.}
\end{threeparttable}
\end{table*}

\begin{table*}
\centering
\begin{threeparttable}
\setlength{\tabcolsep}{2pt}
\tiny
\caption{Main properties of spectroscopically identified AGB stars, based on \gaia\ DR2.}
\label{tab_specagb}
\begin{tabular*}{\textwidth}{@{\extracolsep{\fill}}ccccccccccccc@{}}
\hline\hline\multicolumn{4}{c}{} &
\multicolumn{2}{c}{Cantat-G. 20} &
\multicolumn{2}{c}{Dias 21}\\
\cmidrule(lr){5-6}\cmidrule(lr){7-8}
\noalign{\smallskip}
\multicolumn{1}{c}{2MASS ID} &
\multicolumn{1}{c}{SIMBAD} &
\multicolumn{1}{c}{Cluster} &
\multicolumn{1}{c}{$p$} &
\multicolumn{1}{c}{$\mathrm{\log(age)\tnote{\bf a}}$} & 
\multicolumn{1}{c}{$M_{\rm i}^{\rm AGB}$} & 
\multicolumn{1}{c}{$\mathrm{\log(age)\tnote{\bf b}}$} & 
\multicolumn{1}{c}{$M_{\rm i}^{\rm AGB}$} & 
\multicolumn{1}{c}{$(\jks)_0$} &
\multicolumn{1}{c}{$\WRP-\WJK$} &
\multicolumn{1}{c}{$\Mks$} &
\multicolumn{1}{c}{type}&
\multicolumn{1}{c}{stage}\\
\hline
\noalign{\smallskip}
\multicolumn{1}{c}{} &
\multicolumn{1}{c}{} &
\multicolumn{1}{c}{} &
\multicolumn{1}{c}{} &
\multicolumn{1}{c}{dex} &
\multicolumn{1}{c}{[$\Msun$]} & 
\multicolumn{1}{c}{dex} &
\multicolumn{1}{c}{[$\Msun$]} & 
\multicolumn{1}{c}{[mag]} &
\multicolumn{1}{c}{[mag]} &
\multicolumn{1}{c}{[mag]} &
\multicolumn{1}{c}{}&
\multicolumn{1}{c}{}\\
\hline
\noalign{\smallskip}
$23475728+6835426$ & IRAS 23455+6819 & King 11 & 0.8 & 9.65 & 1.31 & 9.55 & 1.41 & 1.17 & $-0.52$ & $-7.23$ & M & TP-AGB \\
$06363268+0925393$ & V* V493 Mon & Trumpler 5 & 0.7 & 9.63 & 1.33 & 9.54 & 1.42 & 2.94 & $+2.44$ & $-7.64$ & C  & TP-AGB \\
$08310560-3838107$ & [W71b] 030-01 & Pismis 3 & 0.9 & 9.50 & 1.47 & 9.00 & 2.32 & 3.17 & $+2.99$ & $-7.74$ & C  & TP-AGB \\
$07002883-0014204$ & HD 292921 & Berkeley 34 & 1.0 & 9.38 & 1.62 & - & - & 0.77 & $+0.54$ & $-8.59$ & M  & TP-AGB \\
$08561346-3930429$ & S1* 338 & BH 55 & 0.8 & 9.36 & 1.65 & 9.40 & 1.60 & 1.26 & $+0.29$ & $-7.40$ & S  & TP-AGB \\
$07494578-1715141$ & C* 908 & Ruprecht 37 & 0.5 & 9.35 & 1.66 & 9.49 & 1.49 & 1.80 & $+1.34$ & $-8.14$ & C  & TP-AGB \\
$06090764+0436414$ & Case 588 & Dias 2 & 0.9 & 9.24 & 1.84 & 9.11 & 2.12 & 2.05 & $+1.18$ & $-8.17$ & C  & TP-AGB \\
$05113360+4740468$ & [D75b] Star 30 & NGC 1798 & 0.6 & 9.22 & 1.93 & 9.14 & 2.07 & 1.24 & $-0.01$ & $-7.62$ & MS  & TP-AGB \\
$08423384-4712252$ & BM IV 90 & NGC 2660\tnote{\bf c} & 1.0 & 9.21 & 1.94 & 9.08 & 2.17 & 1.70 & $+1.08$ & $-8.35$ & C  & TP-AGB \\
$05113360+4740468$ & MSB 75 & NGC 7789 & \multicolumn{1}{r}{--\tnote{\bf d}} & 9.19 & 1.97 & 9.21 & 1.94 & 1.90 & $+1.42$ & $-8.51$ & C  & TP-AGB \\
$03325580+5244137$ & V* Case 63 & Berkeley 9 & 1.0 & 9.14 & 2.06 & 9.18 & 1.99 & 1.49 & $+1.07$ & $-7.97$ & C  & TP-AGB \\
$07001366-2033294$ & CSS 291 & Tombaugh 1 & 0.8 & 9.10 & 2.13 & 9.11 & 2.12 & 1.19 & $+0.06$ & $-7.84$ & S  & TP-AGB \\
$20560804+5104316$ & Case 473 & Berkeley 53 & 0.6 & 8.99 & 2.33 & - & - & 1.56 & $+1.16$ & $-8.24$ & C  & TP-AGB \\
$08070513-2947435$ & Wray 18-47 & NGC 2533 & \multicolumn{1}{r}{--\tnote{\bf d}} & 8.85 & 2.59 & 8.87 & 2.56 & 1.62 & $+0.87$ & $-8.41$ & C  & TP-AGB \\
$09264885-5114107$ & IRAS 09251-5101 & BH 67 & 0.8 & 8.79 & 2.72 & - & - & 1.35 & $-0.89$ & $-8.43$ & M  & TP-AGB \\
$00161695+5958115$ & - & Juchert Saloran 1 & 0.9 & 8.78 & 2.74 & 8.81 & 2.69 & 1.20 & $+0.24$ & $-7.60$ & M  & TP-AGB \\
$07444463-2824077$ & BM IV 34 & Haffner 14 & 0.6 & 8.46 & 3.54 & 8.56 & 3.25 & 1.44 & $+1.46$ & $-8.61$ & C  & TP-AGB \\
$07593754-6035134$ & V* V460 Car & NGC 2516 & 0.6 & 8.38 & 3.79 & 8.44 & 3.59 & 1.09 & $+0.52$ & $-7.46$ & M  & E-AGB \\
$10190419-5625014$ & HD 300666 & BH 92 & 0.9 & 8.33 & 3.95 & 7.46 & 9.02 & 1.13 & $+0.42$ & $-8.08$ & M  & E-AGB \\
$14350882-5633453$ & HD 127753 & NGC 5662 & 0.8 & 8.30 & 4.05 & 8.02 & 5.20 & 1.19 & $+0.82$ & $-7.05$ & K  & E-AGB \\
$23244485+6120384$ & BD+60 2534 & NGC 7654 & 1.0 & 8.19 & 4.46 & 7.72 & 6.74 & 0.79 & $+0.48$ & $-7.05$ & K  & E-AGB \\
$10473842-5728027$ & HD 93662 & Ruprecht 91 & 0.6 & 8.14 & 4.67 & 8.05 & 5.07 & 1.28 & $+1.12$ & $-8.12$ & K  & E-AGB \\
$09490713-5434130$ & IRAS 09473-5420 & Ruprecht 83 & 0.5 & 8.05 & 5.07 & 8.19 & 4.48 & 1.10 & $+0.38$ & $-8.36$ & M  & E-AGB \\
$01574017+6013079$ & HD 11800 & NGC 743 & 0.7 & 8.00 & 5.32 & 8.04 & 5.11 & 1.00 & $+0.46$ & $-7.28$ & K  & E-AGB \\
$10384498-5910584$ & * t02 Car & BH 99 & 1.0 & 7.98 & 5.42 & 7.79 & 6.23 & 0.92 & $+0.57$ & $-7.49$ & K  & E-AGB \\
$16553781-3930177$ & HD 152524 & NGC 6242 & 1.0 & 7.89 & 5.92 & 7.77 & 6.37 & 0.76 & $+0.4$ & $-7.32$ & K  & E-AGB \\
\hline
\noalign{\smallskip}
\multicolumn{11}{c}{Doubtful cases}\\
\noalign{\smallskip}
\hline
$20001359+2915413$ & IRAS 19582+2907 & FSR 0172 & 0.4 & 8.20 & 4.42 & - & - & 1.84 & $+0.82$ & $-7.89$ & C  & TP-AGB \\
$05500040+2216113$ & Case 121 & Berkeley 72 & 0.4 & 7.73 & 6.96 & 8.73 & 2.86 & 1.56 & $+1.27$ & $-8.92$ & C  & S-/TP-AGB\tnote{(e)} \\
$01443756+6049533$ & Case 49 & NGC 663 & 0.4 & 7.47 & 9.37 & 7.40 & 9.74 & 2.85 & $+2.91$ & $-7.53$ & C  & S-AGB\tnote{(e)}  \\
\noalign{\smallskip}
\hline
\end{tabular*}
\footnotesize{{\bf Notes:} Similar to Table~\ref{tab_agbcand}, but restricted to AGB star candidates (\texttt{SIMBAD} designation in column 2) with known spectroscopic type (column 12). Note that compared to Table~\ref{tab_agbcand}, here we include 5 additional C stars: MSB 75, Wray 18-47, IRAS 19582+2907, Case 121, and Case 49.  The first two stars are not present in the original catalog of \citet{Cantat-Gaudin_Anders_20}.
The latter three objects are doubtful cases as their \gaia\ DR2-based cluster membership probability is rather low, $p=0.4$.  See the text for more explanation.
\begin{tablenotes}[flushleft]
\item[(a)] Cluster ages from \citet{Cantat-Gaudin_etal_20} with $M_{\rm i}^{\rm AGB}$ derived accordingly (next column).
\item[(b)] Cluster ages from \citet{Dias_etal_21}  with $M_{\rm i}^{\rm AGB}$ derived accordingly (next column). A dash (-) means that the cluster is not present in the catalog.
\item[(c)] Age and $M_{\rm i}^{\rm AGB}$ for NGC 2660 are reported according to the Bayesian analysis of \citet{Jeffery_etal_16} and the recent work of \citet{Rain_etal_21}, in place of the \citet{Cantat-Gaudin_etal_20} estimate. The reason of this choice is related to the dual morphology of the red clump of core He-burning stars, as explained in Sect.~\ref{ssec_mscstars}. 
\item[(d)] Membership probability absent from  the \gaia\ DR2 catalog of \citet{Cantat-Gaudin_Anders_20}. The inclusion in the list of C-star cluster members follows from our new analysis based on \gaia\ EDR3 (Sect.~\ref{ssec_membedr3}).
\item[(e)] S-AGB stands for Super-AGB.
\end{tablenotes}
}
\end{threeparttable}
\end{table*}

\section{Basic information on individual M, S and C stars}
\label{ssec_mscstars}
Table~\ref{tab_specagb} summarizes a few relevant properties of the AGB stars with known spectroscopic type. Clusters ages are extracted from two recent works, both based on \gaia\ DR2, namely: \citet{Cantat-Gaudin_etal_20} and \citet{Dias_etal_21}.
The analysis that follows is limited to the stars that are expected to be on the TP-AGB phase, according to the \texttt{PARSEC} evolutionary tracks.
Some of the evolved giants in our sample, the carbon stars in particular, were already discovered and associated to open clusters by early studies in the past century. 
Here  we review the relevant literature about these historical TP-AGB  stars and include basic data for other, less studied, M, S and C stars. The information about long-period variability, including our new analysis, is provided in Sect.~\ref{ssec_var} and in Tables~\ref{tab:obsvar} and ~\ref{tab:obsvar2}.
\paragraph{M stars: IRAS 23455+6819, HD 292921, IRAS 09251-5101, 2MASS J00161695+5958115}
We identify 4 M-type stars in the sample that are expected to be on the TP-AGB phase. Two of them are low-mass stars (IRAS 23455+6819 and HD 292921) with $M_{\rm i}^{\rm AGB}\simeq 1.3\,\Msun \,{\rm and}\, 1.6\,\Msun$,  and the other two (IRAS 09251-5101 and 2MASS J00161695 + 5958115) are intermediate-mass stars with similar initial mass, $M_{\rm i}^{\rm AGB}\simeq 2.7\,\Msun$. To our knowledge there is no specific literature on these stars, except for some information on their variability, which we find in the \gaia\ DR2 catalog dedicated to LPVs \citep{Mowlavi_etal_18}. More details can be found in Sect.~\ref{ssec_var}.

\paragraph{S stars: [D75b] Star 30, S1$^\ast$ 338, CSS 291 }
The MS star [D75b] Star 30 is a  member of the cluster NGC 1798, with an estimated age of $\simeq 1.66$ Gyr \citep{Cantat-Gaudin_etal_20}, and a slightly sub-solar metallicity [Fe/H]$=-0.18$ \citep{Carrera_etal_19}. The progenitor initial mass is $M_{\rm i}^{\rm AGB} \simeq 1.79\, \Msun$. This star has experienced a few dredge-up episodes and should become a carbon star according to current TP-AGB stellar models \citep{Cristallo_etal_11, Marigo_etal_13}.
The S star S1$^\ast$ 338  belongs to the old cluster BH 55, which is assigned an age of $\simeq 2$ Gyr \citep{Cantat-Gaudin_etal_20}. A spectroscopic measure of [Fe/H] is missing for this cluster, so we assume it to be solar. Under this assumption, its S-star witnesses the occurrence of the 3DU at $M_{\rm i}^{\rm AGB} \simeq 1.60-1.65\,\Msun$.
A younger S star, CSS 291,  is associated to the star cluster Tombaugh 1 with an estimated age of $\simeq 1.26$ Gyr. Also for this cluster no metallicity estimate is available. The initial mass of the progenitor is $M_{\rm i}^{\rm AGB}  \simeq 2.1 \, \Msun$ at solar metallicity. The spectral type is S4/2, which would correspond to C/O$\simeq 0.95$ according to the classification of \citet{Keenan_80}.

\paragraph{V$^\ast$ V493 Mon}
This carbon star belongs to the old open cluster Trumpler 5, with an estimated age of $\simeq 4.3$ Gyr \citep{Cantat-Gaudin_etal_20},or $\simeq $ 3.5 Gyr \citep{Dias_etal_21}. The initial mass of the progenitor is $M_{\rm i}^{\rm AGB}  \simeq 1.3-1.4\, \Msun$.
We note that this cluster has a relatively low metallicity, [Fe/H] $\simeq -0.4$ according to the APOGEE and GALAH spectroscopic surveys \citep{Carrera_etal_19}. Therefore, V$^\ast$ V493 Mon cannot be used to set constraints of carbon star formation at solar metallicity, rather it is useful to characterize an intermediate metallicity regime, similar to that of the Small Magellanic Cloud.
In this respect, we note that $M_{\rm i}^{\rm AGB}  \simeq 1.3-1.4\, \Msun$ is broadly consistent with the calibration of the 3DU, in particular in terms of the minimum mass for carbon star formation, carried out by \citet{Pastorelli_etal_19} from a detailed analysis of the AGB star populations in the Small Magellanic Cloud. 

\paragraph{[W71b] 030-01} 
It is a carbon star belonging to the intermediate-age open cluster Pismis 3, with an estimated age of $\simeq 3.2$ Gyr \citep{Cantat-Gaudin_etal_20}, or $\simeq 1$ Gyr \citep{Dias_etal_21}.
Following the former age indication the initial mass of the progenitor is $M_{\rm i}^{\rm AGB} \simeq 1.5\,\Msun$, which lies towards the low-mass end of the interval for C star formation. Instead, adopting the latter age, the progenitor is more massive with $M_{\rm i}^{\rm AGB} \simeq 2.3\,\Msun$.
The metal content of the host cluster is not available from spectroscopic measurements. Indirect estimates of the metallicity  were derived by  reproducing the morphology of color-magnitude diagrams with traditional best-isochrone fitting methods. The results suggested metallicities ranging from subsolar \citep{CarraroOrtolani_94} to solar \citep{Salaris_etal_04}, but it should be noted that these works predate the significant revision of the Sun metallicity \citep{AGSS09,Caffau_etal_11}.

\paragraph{C$^\ast$ 908} This star has no genuine spectral classification, but it is considered a possible carbon star and is listed in the General Catalog of Galactic Carbon  Stars \citep{Alksnis_etal_01}, as already noted by \citet{Pal_Worthey_21}. Its photometric properties match those of carbon stars very well: it has a red near-IR color ($\jks_0 \simeq 1.8$) and in the \gaia-2MASS diagram it is placed inside the region b) where all TP-AGB stars of spectral type C are found.  C$^\ast$ 908 is a likely member of the cluster Ruprecht 37, which is assigned an age of $\simeq 2.2$ Gyr \citep{Cantat-Gaudin_etal_20}, or $\simeq 3.1$ Gyr \citep{Dias_etal_21}. Taking into account both age estimates the initial mass of the progenitor should lie in the range $1.5 \la M_{\rm i}^{\rm AGB} /\Msun \la 1.7\Msun$. The spectroscopic metallicity of the cluster is unknown. By using the \gaia\ DR2 astrometric data for individual stars  and matching the observed CMD with synthetic simple stellar populations, \citet[][]{Piatti_etal_19} derived for Ruprecth 37 an age of $\simeq 4.5 \pm 0.5$ Gyr and a metallicity [Fe/H]$=-0.32\pm 0.24$.

\paragraph{MSB~75, BM IV 90, Case 588}
 The carbon star MSB~75 was first pointed out  by \citet{GaustadConti_71} as a likely member of the intermediate-age open cluster NGC~7789, on the basis of the good consistency between its radial velocity and the  measurements for a few K giants.  NGC 7789 has an estimated age of $\simeq 1.6$ Gyr \citep{Cantat-Gaudin_etal_20, Cummings_etal_18}. The metallicity is almost solar, [Fe/H]$\simeq +0.06$ \citep{Carrera_etal_19},
 or [Fe/H]$\simeq -0.06$ \citep{Zhong_etal_20}.
 
 The carbon star BM IV 90 was identified within 1$\arcmin$ of the center of the cluster NGC 2660 by \citet{HartwickHesser_73, HartwickHesser_71}, while its evolutionary status was discussed by \citet{EggenIben_91}. 
 NGC 2660 is almost coeval of NGC 7789, with an age of $\simeq 1.6$ Gyr \citep{Rain_etal_21, Jeffery_etal_16}. We note that \citet{Cantat-Gaudin_etal_20} derived a younger age of $\approx 0.93$ Gyr, using a neural network approach, but we tend to disfavor the \citet{Cantat-Gaudin_etal_20} solution for the reasons provided below.
 
 The two clusters NGC 7789 and NGC 2660 exhibit a similar peculiar morphology of the red clump, that is the locus of core-He burning stars in color-magnitude diagrams. As extensively discussed by \citet[][see also \citealt{Girardi_etal_10}]{Girardi_etal_09}, the so-called dual clump, that is  the coexistence of a faint extension, slightly to the blue of the main concentration of clump stars, is a likely clue that both clusters host, at the same time,  stars of low-mass which undergo the helium flash at the RGB tip, and those just massive enough to avoid electron degeneracy in the helium core.
 An age of $\simeq 1.6$ Gyr derived by \citet{Rain_etal_21, Jeffery_etal_16} is nicely consistent with this interpretation, while a younger age \citep{Cantat-Gaudin_etal_20} hardly accounts for the observed dual clump morphology.

The carbon star Case 588 belongs to the cluster Dias 2, whose parameters (age, distance and extinction) were first derived by \citet{Tadross_09} on the base of 2MASS photometric data \citep{Cutri_etal_03}. 
Following the systematic revision based on \gaia\ DR2 of \citet{Cantat-Gaudin_etal_20}, Dias 2 is assigned an age of $\simeq 1.74$ Gyr.  We note that the isochrone fitting of the  \gaia\ color-magnitude diagram relies on a small number of stars (44 stars have a membership probability $p > 0.7$), and therefore significant uncertainties may affect this age derivation. Anyhow, it is reassuring that this value is in broad agreement with the age ($\simeq 1.86$ Gyr) independently derived by \citet[][MWSC \# 0756]{Kharchenko_etal_13} from UBV and 2MASS data, and the recent estimate of $\simeq 1.3$ Gyr by \citet{Dias_etal_21}.
The initial mass of the progenitor should be in the range $1.8 \la M_{\rm i}^{\rm AGB}/\Msun \la 2.0$.

It follows that the carbon stars MSB 75, BM VI 90, and Case 588 sample the region close to the transition mass $M_{\rm HeF}$, between the classes of low- and intermediate-mass stars. 
In addition, these carbon stars are  extremely interesting objects, as their progenitors fall
in the initial mass range where a kink in the initial-final mass relation has been recently detected \citep{Marigo_etal_20}. 
This point will be further addressed in  Sect.~\ref{ssec_ifmr}.

\paragraph{Case 63}
This carbon star is a likely member of the cluster
Berkeley 9, with an estimated age of $\simeq 1.4-1.5$ Gyr. The progenitor initial mass is $M_{\rm i}^{\rm AGB} \simeq 2.0-2.1\, \Msun$ that is close or just above the transition mass $M_{\rm HeF}$.  The metallicity of the cluster is slightly subsolar, [Fe/H]$\simeq -0.17$ \citep{Carrera_etal_19}.

\paragraph{Case 473}
It is a carbon star associated to the cluster Berkeley 53, with an age of $\simeq 0.98$ Gyr. The initial mass of the progenitor
is $M_{\rm i}^{\rm AGB}  \simeq 2.3 \, \Msun$.
The metallicity of the cluster is close to solar, with [Fe/H]$\simeq -0.02$ according to the spectroscopic estimate from  APOGEE and GALAH surveys \citep{Carrera_etal_19}. 

\paragraph{Wray 18-47} This carbon star was identified in the field of the open cluster NGC 2533 by \citet{JorgensenWesterlund_88}. Not present in \gaia\ DR2 catalog of \citet{Cantat-Gaudin_Anders_20}, its cluster membership probability turns out relatively good ($p \simeq 0.6$) using the \gaia\ EDR3. 
The age of NGC 2533 is  $\simeq 0.71$ Gyr following \citet{Cantat-Gaudin_etal_20}, fully consistent with the study of \citet{Dias_etal_21}, as well as of \citet{Siegel_etal_19} based on  the Swift UVOT Stars Survey. The initial mass of the progenitor is $M_{\rm i}^{\rm AGB}  \simeq 2.6\, \Msun$.

\paragraph{BM IV 34}
It is a carbon star associated to the relatively young star cluster Haffner 14 \citep{JorgensenWesterlund_88,Groenewegen_etal_95}, with an estimated age of $\simeq 288$ Myr \citep{Cantat-Gaudin_etal_20}, or  $\simeq 195$ Myr \citep{Bossini_etal_19}, $\simeq 362$ Myr \citep{Dias_etal_21}. The relevance of this star lies in the fact that the initial mass of the progenitor is relatively high, $\Mi^{\rm AGB} \simeq 3.3-4.0\, \Msun$ and therefore places constraints on the maximum mass for the formation of  carbon stars at solar metallicity during the TP-AGB phase.

\paragraph{IRAS 19582+2907, Case 121, Case 49}
These 3 carbon stars have a relatively low membership probability, $p=0.4$, according to \citet{Cantat-Gaudin_Anders_20}.
We will further investigate this aspect with  \gaia\ EDR3  in Sect.~\ref{ssec_membedr3}.
Looking at the ages of the parent clusters \citep{Cantat-Gaudin_etal_20}, it turns out that these carbon stars belong to  young clusters and their initial masses suggest them as plausible candidates for the occurrence of HBB, either in the standard TP-AGB phase (IRAS 19582+2907: $M_{\rm i} \simeq 4.4\,\Msun$) or during the Super-AGB evolution 
(Case 121: $M_{\rm i} \simeq 7 \,\Msun$, Case 69: $M_{\rm i} \simeq 9.7\,\Msun$). We note that \citet{Dias_etal_21} derive an older age for Berkeley 72, so that the progenitor mass of Case 121 becomes $\simeq 2.9\,\Msun$, which brings it back into a normal TP-AGB evolution. A more detailed discussion of these stars will be carried out in Sect.~\ref{sect_results}.

\subsection{Variability}
\label{ssec_var}

Variability is a common feature of AGB stars that can be used to better constrain their physical properties. For all TP-AGB sources listed in Table~\ref{tab_specagb} (with the exception of Case 49, see Sect.~\ref{sect_evprop}) we searched for variability information in the General Catalog of Variable Stars \citep[GCVS,][]{Samus_etal_17}. We found the variability type of five stars, but only for one star (MSB 75) did we find an estimate of the period.

Therefore we extended the search to three ongoing surveys dedicated to time-domain astronomy, namely \gaia\ \citep[see][for information about the variability processing of \gaia\ DR2 data]{Holl_etal_2018}, the All-Sky Automated Survey for SuperNovae \citep[ASAS-SN,][]{Shappee_etal_2014}, and the Zwicky Transient Facility \citep[ZTF,][]{Bellm_etal_2019}. Each one of these surveys provide access to photometric time series for a large number of variable stars, as well as catalogs of variability properties from the processing of the light curves.

As a first step, we crossmatched our list of 19 sources with the variability catalog provided in the framework of each one of the three surveys. The \gaia\ DR2 catalog of LPV candidates \citep{Mowlavi_etal_18} provides information for more than 150\,000 sources, among which we found seven of the stars in our sample (six with a period). ASAS-SN is an automated, ground-based visual survey aimed at the detection of transient phenomena that also collected a large amount of observations of variable stars. The ASAS-SN catalog of variable stars \citep[][and references therein]{Jayasinghe_etal_2019} includes more than 660\,000 objects automatically processed and classified. In this catalog we found 16 of the stars in our sample, and found a period estimate for ten of them\footnote{
    We integrated data for a few stars with more recent information from the corresponding dedicated pages at the ASAS-SN Sky Patrol website (\url{https://asas-sn.osu.edu/}).
}. ZTF is a ground-based wide-field survey for transients observing at visual and IR wavelengths. \citet{Chen_etal_2020_ZTF} describe the ZTF catalog of periodic variable stars and its content of more than 780\,000 sources, in which we found a match for four stars, all of them with period information. We note that \citet{Chen_etal_2020_ZTF} provide an additional catalog of more than 1\,380\,000 `suspected' variables, with which we matched two further sources from our sample. However we disregarded these matches as the reported periods are shorter than one day.

\begin{table*}
\centering
\begin{threeparttable}
\scriptsize
\caption{Period and variability type for the sources in our sample according to the stellar variability catalogs from \gaia\ DR2, ASAS-SN, ZTF, as well as to the GCVS. The final column lists the period adopted in this work.}
\label{tab:obsvar}
\begin{tabular*}{\textwidth}{@{\extracolsep{\fill}}ccccccccccl@{}}
\hline
\hline
 & & \gaia\ DR2 & \multicolumn{2}{c}{ASAS-SN} & \multicolumn{2}{c}{ZTF} & \multicolumn{2}{c}{GCVS} & \multicolumn{2}{c}{adopted} \\
\cmidrule(lr){3-3}\cmidrule(lr){4-5}\cmidrule(lr){6-7}\cmidrule(lr){8-9}\cmidrule(lr){10-11}
Star & GCVS Name & $P$ & $P^{(a)}$ & Type & $P$ & Type & $P^{(a)}$ & Type & $P$ & Ref.$^{(b)}$ \\
\hline
 & & [days] & [days] & & [days] & & [days] & & [days] & \\
\hline 
V$^{\ast}$ V493 Mon     & V$^{\ast}$ V493 Mon & 317.7 &     0 &   L & 288.3 & Mira &     0 & SRB: & 432.4 & Z$^{(c)}$ \\
$[$W71b$]$ 030-01       &                   - & 393.3 & 460.0 &  SR &     - &    - &     - &    - & 460.0 & A         \\
C$^{\ast}$ 908          &                   - &     - & 517.7 &  SR &     - &    - &     - &    - & 235.9 & A$^{(c)}$ \\
Case 588                &                   - &     0 & 539.0 &  SR &     - &    - &     - &    - & 539.0 & A         \\
BM IV 90                & V$^{\ast}$   GV Vel &     - &     0 &   L &     - &    - &     0 &   LB & 438.2 & A$^{(c)}$ \\
MSB 75                  & V$^{\ast}$ V532 Cas & 432.5 & 449.2 &  SR &     - &    - & 450.0 &  SRA & 432.5 & G         \\
Case 63                 & V$^{\ast}$   BI Per &     - &     0 &   L &     - &    - &     0 &   LB & 163.4 & A$^{(c)}$ \\
Case 473                &                   - &     - &     0 & YSO &     - &    - &     - &    - & 358.4 & Z$^{(c)}$ \\
WRAY 18-47              & V$^{\ast}$ V521 Pup & 368.9 & 277.9 &  SR &     - &    - &     0 &  SR: & 368.9 & G         \\
BM IV 34                &                   - &     - & 136.7 &   L &     - &    - &     - &    - & 136.7 & A         \\
IRAS 19582+2907         &                   - &     - &     - &   - & 201.1 &   SR &     - &    - & 364.6 & Z$^{(c)}$ \\
Case 121                &                   - &     - & 228.2 &  SR &     - &    - &     - &    - & 228.2 & A         \\
S1$^{\ast}$ 338         &                   - &     - & 43.28 &  SR &     - &    - &     - &    - & 43.28 & A         \\
$[$D75b$]$ Star 30      &                   - &     - &     - &   - &     - &    - &     - &    - & 76.36 & A$^{(c)}$ \\
CSS 291                 &                   - &     - &     0 &   L &     - &    - &     - &    - & 95.52 & A$^{(c)}$ \\
IRAS 23455+6819         &                   - & 604.1 &     0 &   L & 248.1 &   SR &     - &    - & 85.02 & Z$^{(c)}$ \\
HD 292921               &                   - &     - &     - &   - &     - &    - &     - &    - &     - & -         \\
IRAS 09251-5101         &                   - & 148.9 & 144.4 &  SR &     - &    - &     - &    - & 144.4 & A         \\
2MASS J00161695+5958115 &                   - &     - & 801.7 &  SR & 118.1 &   SR &     - &    - & 95.01 & A$^{(c)}$ \\
\hline
\noalign{\smallskip}
\end{tabular*}
\footnotesize{{\bf{Notes:}}
\begin{tablenotes}[flushleft]
\item{(a)} Zero-day periods under the ASAS-SN or GCVS column indicate that the source is included in the catalog but its period is not reported. 
\item{(b)} indicates whether the adopted period was taken from \gaia\ DR2 (G), ASAS-SN (A) or ZTF (Z).
\item{(c)} means that the adopted period was recomputed from a time series produced by the survey rather than taken from the corresponding variability catalog.
\end{tablenotes}
}
\end{threeparttable}
\end{table*}

The periods and variability type we retrieved are summarized in Table~\ref{tab:obsvar}. We found at least one match for 17 sources, but four of them have no period estimate. There are two sources, HD 292921 and $[$W71b$]$ 030-01, the we could not find in any of the examined catalogs. For seven stars we found information in more than one of the catalogs, that in most cases are not entirely in agreement. For instance, the two periods reported for 2MASS J00161695+5958115 differ by more than a factor seven, while three different variability types are reported for V$^{\ast}$ V493 Mon. Such discrepancies can be easily explained in terms of the nature of the observed stars and of the differences in cadence and coverage of the three surveys. Indeed, LPVs can be multi-periodic, in which case different observational strategies can lead to the detection of distinct periods. Even for mono-periodic variables, cycle-to-cycle variations can cause discrepancies associated with the epochs of observation. Moreover, all three catalogs adopted here have been produced by means of automated pipelines, which inevitably leads to spurious results when the number of processed objects is large.

This prompted us to re-examine the light curves of each individual star. Not only this allowed us to select the most reasonable value of the period, but we could also find data for sources not present in the catalogs, or for which a period is not given.

\begin{figure*}[ht!]
\centering
\resizebox{\hsize}{!}{\includegraphics{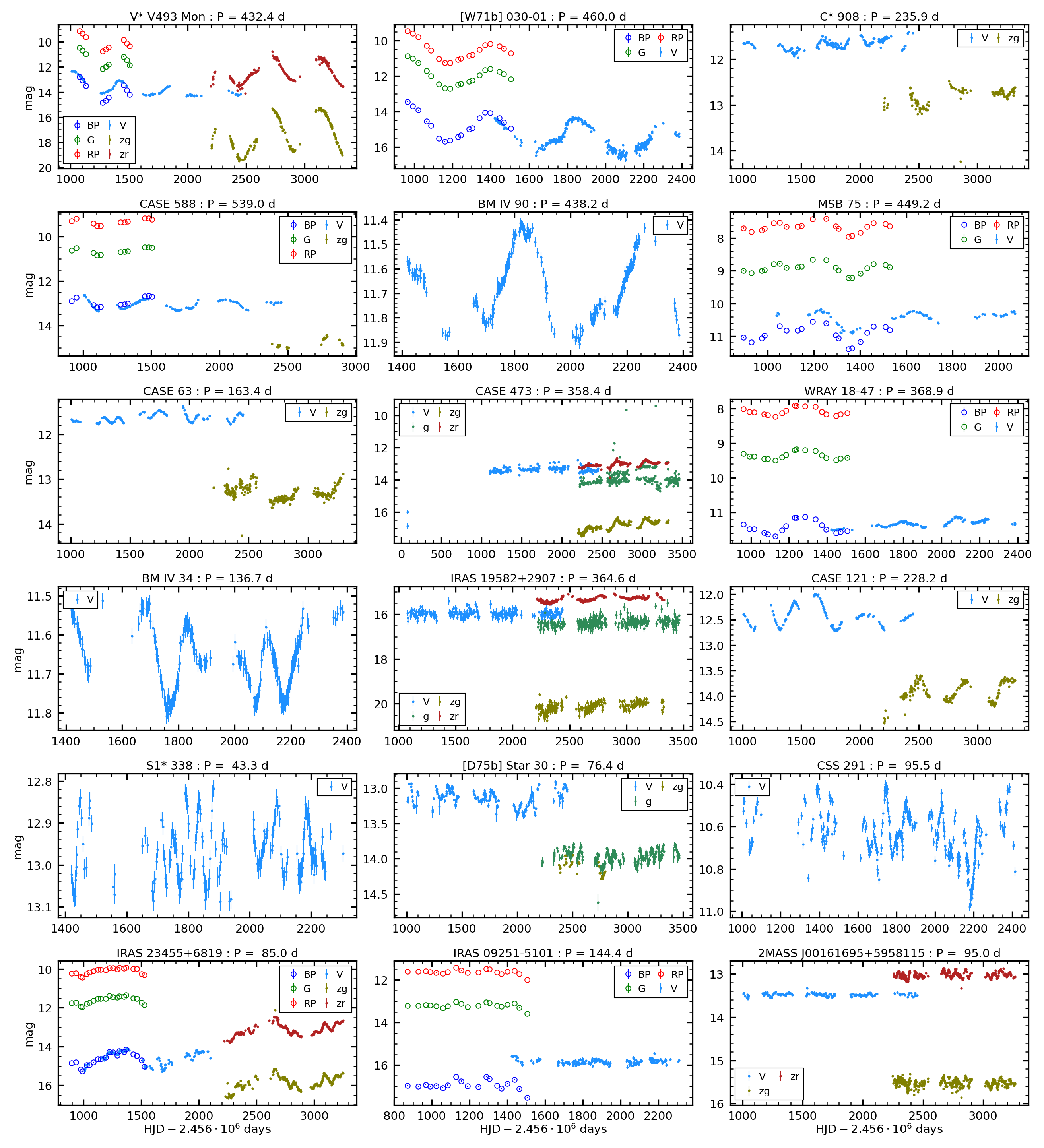}}
\caption{Light curves of the 18 sources for which we determined a period in filters of \gaia\ ($G$, $\gbp$, $\grp$), ASAS-SN ($V$, $g$), and ZTF ($g$, $r$).}
\label{fig_lcs}
\end{figure*}

Using the ASAS-SN Sky Patrol service\footnote{\url{https://asas-sn.osu.edu/}}, that gives almost real-time access to ASAS-SN observations, we were able to retrieve $V$-band light curves for all sources in our sample (including the three stars not reported in the ASAS-SN variable stars catalog). For a few sources we also retrieved time series in the $g$ band. Similarly, by accessing the ZTF data products \citep{Masci_etal_2019_ZTFdata} through the NASA/IPAC Infrared Science Archive (IRSA\footnote{\url{https://irsa.ipac.caltech.edu/frontpage/}}), we retrieved $g$-band light curves for the four stars included in the ZTF variable stars catalog, as well as for six additional sources. In some cases we also retrieved time series in the $r$ band. The retrieved light curves are displayed in Fig.~\ref{fig_lcs}.

For each time series we computed the Lomb-Scargle periodogram \citep[e.g.][and references therein]{vanderplas_2018} to determine the variability period, then subtracted the best-fit sine model from the data and repeated the process in order to check for the presence of possible secondary periods. A higher-order Fourier series model was adopted instead in some cases, selected by visual inspection, in which the light curve displays a clearly asymmetric shape (a common feature of Miras and related LPVs).

Many sources present long-term trends, not necessarily periodic, that are nonetheless fitted by the periodic model. Whether or not these are long secondary periods is hard to say given the relatively short observational baselines of the data. Regardless of the nature of these features, their subtraction from the time series during the first step of our processing effectively acts as a detrending, allowing for a better characterization of the true period of each star. We disregard all periods associated with such trends and consider only the periods over shorter timescales. We were not able to identify secondary periods that can be realistically attributed to pulsation, so we consider a single period for each star.

For each source we selected the most realistic period by considering the quality of the light curve in each band of each survey (photometric errors, time sampling and coverage), and by visually inspecting the best-fitting models and the Fourier transform of the window function of the time series. Then we compared our periods with those reported in the literature catalogs. If the literature periods are reasonably similar to the values we obtained, we adopted the former. Otherwise, we employed the periods we derived. We were not able to determine a reliable period for the source HD 292921, for which we also did not find a period in literature, so we exclude it from further analysis. We caution the reader that this does not necessarily mean that HD 292921 is not variable.

For 12 stars (most of the sample) we ended up relying on ASAS-SN data for the period, that was recomputed from the light curve in half of the cases. We adopted ZTF data for four sources, in all cases using the value of period computed by us. Indeed, we found significant differences between the four periods published in the ZTF catalog and the values we derived for the same sources using ZTF time series. Such discrepancies were expected as the catalog of \citet{Chen_etal_2020_ZTF} is based on the ZTF data release 2, while we employed significantly longer time series from the more recent data release 7\footnote{\url{https://www.ztf.caltech.edu/page/dr7}}. Only for two stars we adopted the period published in \gaia\ DR2, whose time series in most cases have insufficient coverage or sampling. We note, however, that the \gaia\ DR2 $\gbp$-band light curves match remarkably well the ones in the ASAS-SN $V$-band over the overlapping time interval, an agreement that we often took as confirmation of the results we derived from ASAS-SN data.

Besides the impact of differences in coverage and cadence associated with the three surveys, we did not identify any evident systematic trend associated with the differences between the periods we derived and the ones we found in literature. Two relatively bright sources (MSB 75 and CSS 291) are close to the saturation limit of ASAS-SN (as discussed below), but no other source is at risk of saturation for the surveys we considered.

Below we provide an overview of the results and variability properties of each star.

\paragraph{V$^{\ast}$ V493 Mon} The shape and regularity of the ZTF light curves, from which we derived the adopted period of 432 days, suggest this object is a Mira, despite it being reported as a SRB star in the GCVS and classified as irregular by ASAS-SN. Overall, the derived light curves suggest this star may be subject to long-term variations.

\paragraph{$[$W71b$]$ 030-01} This star shows an asymmetric light curve, typical of Miras, that can be appreciated in the ASAS-SN time series. The latter covers two full cycles with good sampling, and the period we computed from it is in good agreement with the automatically-derived value from the ASAS-SN catalog of 460 days, that we adopted.

\paragraph{C$^{\ast}$ 908} Most likely a semi-regular variable. The processing of both the ASAS-SN and the $g$-band ZTF light curves highlighted a long-term trend and a periodicity about half that reported in the ASAS-SN catalog. We adopted the 236 day period we extracted from the ASAS-SN time series.

\paragraph{Case 588} Our analysis is in agreement with the results from the ASAS-SN catalog, from which we take the adopted period of 539 days. Despite its low amplitude, that would be consistent with it being classifies as SR, this star displays a relatively long period and regular light curve, being thus similar to Miras.

\paragraph{BM IV 90} Despite being reported in the ASAS-SN catalog as an irregular variable (with no period), this star shows a clear periodicity in its $V$-band light curve, from which we derive a period of 438 days. It has a rather small amplitude ($\Delta V\lesssim0.5$ mag), but a rather regular light curve with no clear sign of multi-periodicity.

\paragraph{MSB 75} The \gaia\ DR2 light curves show an asymmetric shape with a secondary maximum that is not evident in the ASAS-SN time series due to its cadence. We note that the latter light curve is flagged with a saturation warning, yet its periodicity and compatibility with \gaia\ data are evident. When modelling the data of both surveys using a two-component Fourier series we obtain a period consistent with the value of 433 days published in \gaia\ DR2, that we adopt. The amplitude and regularity of the light curve are consistent with the GCVS classification as a SRA variable.

\paragraph{Case 63} While not highly regular, the $V$-band light curve of this source clearly shows a periodicity that we constrain to 163 days. This period is not evident in the $g$-band time series from ZTF, that is rather suggestive of a longer-timescale variability. The latter could be a secondary pulsation period, which would be consistent with identifying this star as a semiregular variable.

\paragraph{Case 473} This source is classified as a young stellar object in the ASAS-SN catalog, with no reported period. This classification is unlikely given the data presented in the previous section and that this is a spectroscopically confirmed C-rich star. From ZTF photometry we identified a period of 358 days, that we adopted, corresponding to asymmetric variability. While close to the one-year alias, visual inspection of the light curve suggest this period to be realistic. 

\paragraph{WRAY 18-47} Our analysis is consistent with the \gaia\ DR2 period of 369 days, that we adopt. Since the period is close to 1 yr, the coverage of the ASAS-SN time series is insufficient do confidently derive a period estimate.

\paragraph{BM IV 34} This star shows a clear semiregular behavior with an evident periodicity. We adopt the ASAS-SN estimate of 137 days, consistent with our analysis.

\paragraph{IRAS 19582+2907} ASAS-SN time series for this source do not show any sign of clear periodicity. The ZTF $r$-band light curve, having smaller errors, allows for the derivation of a period of 365 days. The evident $\sim0.5$ mag fluctuations are evidence that this period is unlikely to be attributed to an alias.

\paragraph{Case 121} The periodicity of this star is evident from both the ASAS-SN anf ZTF light curves. Our analysis of the former is consistent with the published period of 228 days, that we adopt. We note, however, that the periodogram for the ZTF time series shows a strong peak at 395 days that could be a secondary pulsation period. While this would be consistent with this star being a bi-periodic semiregular variable, this latter period is rather uncertain.

\paragraph{S1$^{\ast}$ 338} This is the shortest-period star in our sample (43 days adopted from ASAS-SN), and has a $V$-band amplitude of only $\sim0.2$ mag. These features are consistent with this star being a semiregular variable pulsating in an overtone mode. Other pulsation modes are likely active in this star, but the coverage and sampling of the light curve does not allow them to be reliably constrained.

\paragraph{$[$D75b$]$ Star 30} We found no published period for this source, that indeed has quite an irregular light curve. Upon subtracting long-term trends, analysis of the three available time series (ASAS-SN $V$ and $g$ band, ZTF $g$ band) lead to consistent results, and we adopt a 76 days period derived from the ASAS-SN $g$-band light curve. This star is probably a semiregular variable.

\paragraph{CSS 291} Similar to S1$^{\ast}$ 338, this appears to be a semiregular variable pulsating in one or more overtone modes. In absence of published periods, we derive a value of 96 days from the ASAS-SN $V$-band time series. This source is flagged with a saturation warning, and indeed is close to the ASAS-SN saturation limit of $\sim$10-11 mag. However, the ASAS-SN aperture photometry pipeline is capable of recovering saturated sources up to $\sim7-8$ mag, suggesting that this light curve is likely reliable.

\paragraph{IRAS 23455+6819} The combination of \gaia\ DR2, ASAS-SN and ZTF light curves clearly highlights the presence of two periods with remarkably different timescales in this source. The longer period, easily detected as the primary variability in any automated pipeline, is likely a long secondary period. From the $r$-band ZTF light curve we derive a value of 631 days for it, in fairly good agreement with the value published in \gaia\ DR2. For the shorter period, more likely associated with pulsation, we derive a value of 85 days from the ZTF data, in good agreement with the result from the ASAS-SN light curve. The sampling of the \gaia\ DR2 time series is insufficient to constrain this period.

\paragraph{HD 292921} For this source we only found a time series in the ASAS-SN $V$-band, characterized by relatively large photometric errors. The period derived from the periodogram is uncertain, so we do not adopt it here.

\paragraph{IRAS 09251-5101} The good agreement between the periods published in \gaia\ DR2 and ASAS-SN for this source suggests they are correct, although the poor sampling in the time series of the former and relatively large uncertainties in the light curves from of the latter make it difficult to confirm their validity. We adopt the 144 day period from ASAS-SN.

\paragraph{2MASS J00161695+5958115} Visual inspection of the light curves available for this source suggests that its period can be better constrained from ZTF data, from which we derived our adopted value of 95 days. This star is probably a semiregular variable.

\subsection{\gaia\ EDR3 parallaxes and zero-point correction}
\label{ssec_plx}
The \gaia\ space mission marks an unprecedented progress in the precision of astrometric parameters of stars.
Compared to other types of stars, the parallaxes, $\pi$, of AGB stars may be affected by greater uncertainties for two main reasons. First, the errors can be amplified by the photocentric variability caused by complex surface convection structures \citep{Chiavassa_etal_18}.
Second, assuming a constant mean color to compute the parallaxes \citep{Lindegren_etal_18, Lindegren_etal_21a} may introduce an unknown color bias which, in principle, could be substantial for pulsating AGB stars with large color variations over a cycle \citep{Platais_etal_03}. 
It is therefore possible that these inconsistencies may affect not only the precision, but also the accuracy of the astrometric solution for AGB star parallaxes.

With these caveats in mind, we adopt the \gaia\ EDR3 parallaxes to derive the distances of the 29 AGB star candidates in our spectroscopic sample (see Table~\ref{tab_specagb}).
The majority of them -- 69\%, 86\%, 93 \% -- have \gaia\ EDR3 parallax uncertainty, $\epsilon(\pi)/\pi < 10\%\,, 15\%\,, 20\%$, where $\epsilon(\pi)$ is the tabulated error.
The goodness of the astrometric solution can be checked with the aid of  the renormalised unit weight error (RUWE), which should be $\simeq 1$ for a well-behaved \gaia\ EDR3 source \citep{Lindegren_etal_21a}. Most of the stars in our spectroscopic sample (86\%) have $\mathrm{RUWE}<1.1$ and the maximum value, $\mathrm{RUWE}=1.24$, is associated to the variable M star V$^\ast$ V460 Car. These indicators, therefore, support the good quality of the astrometric data, bearing in mind the issues described above. 

To complete the discussion on AGB star parallaxes, another aspect is noteworthy.
Recent studies have shown that the \gaia\ EDR3 parallaxes are affected by a systematic offset, the magnitude of which primarily depends on the position in the sky, magnitude, and
colour \citep{Groenewegen21,Lindegren_etal_21,Zinn21,Bhardwaj_etal_21,Stassun_etal_21,Huang_etal21}.
Following a standard formalism \citep[see, e.g., ][]{Groenewegen21}, the parallax zero-point offset, ZP, is defined through
\begin{equation}
\pi_{\rm t} = \pi_0 - \mathrm{ZP}\,,     
\end{equation}
where $\pi_0$ is the observed parallax listed in the published catalog, and $\pi_{\rm t}$ denotes the true parallax corrected for the ZP offset.
The ZP offset  is typically negative; quasars have a median \gaia\ EDR3 parallax of $\approx -17\, \mu$as. 
\citet{Lindegren_etal_21} provided the community with a useful tool to compute the parallax ZP as a function of object parameters, which has been used and tested extensively in various works. A few studies have 
pointed out that the model of \citet{Lindegren_etal_21} could produce an over-correction, in the sense that the ZP are too much negative and therefore the corrected parallaxes are too large \citep{Riess_etal21, Zinn21, Huang_etal21,Groenewegen21}.
This occurs for classical Cepheids \citep{Riess_etal21}, RGB stars \citep{Zinn21}, and RR-Lyrae variables \citep{Bhardwaj_etal_21}.
In light of this, a few different formulations have been proposed that either correct Lindegren's prescription \citep[e.g.,][]{Zinn21}, or offer alternative approaches \citep[e.g.,][]{Groenewegen21}. In any case it should be noted that the source samples used for the ZP calibration generally do not adequately cover the characteristic intervals of  AGB stars for some parameters (for example, the effective wavenumber $\nu_{\rm eff}$, and the color $BP-RP$).
In general, AGB stars have smaller $\nu_{\rm eff}$ and redder $BP-RP$ colors  compared with those of the calibrating objects. It follows that the ZP corrections have not yet been properly validated for AGB stars and must be taken with some caution.

\begin{figure}[ht!]
\centering
\resizebox{0.7\hsize}{!}{\includegraphics{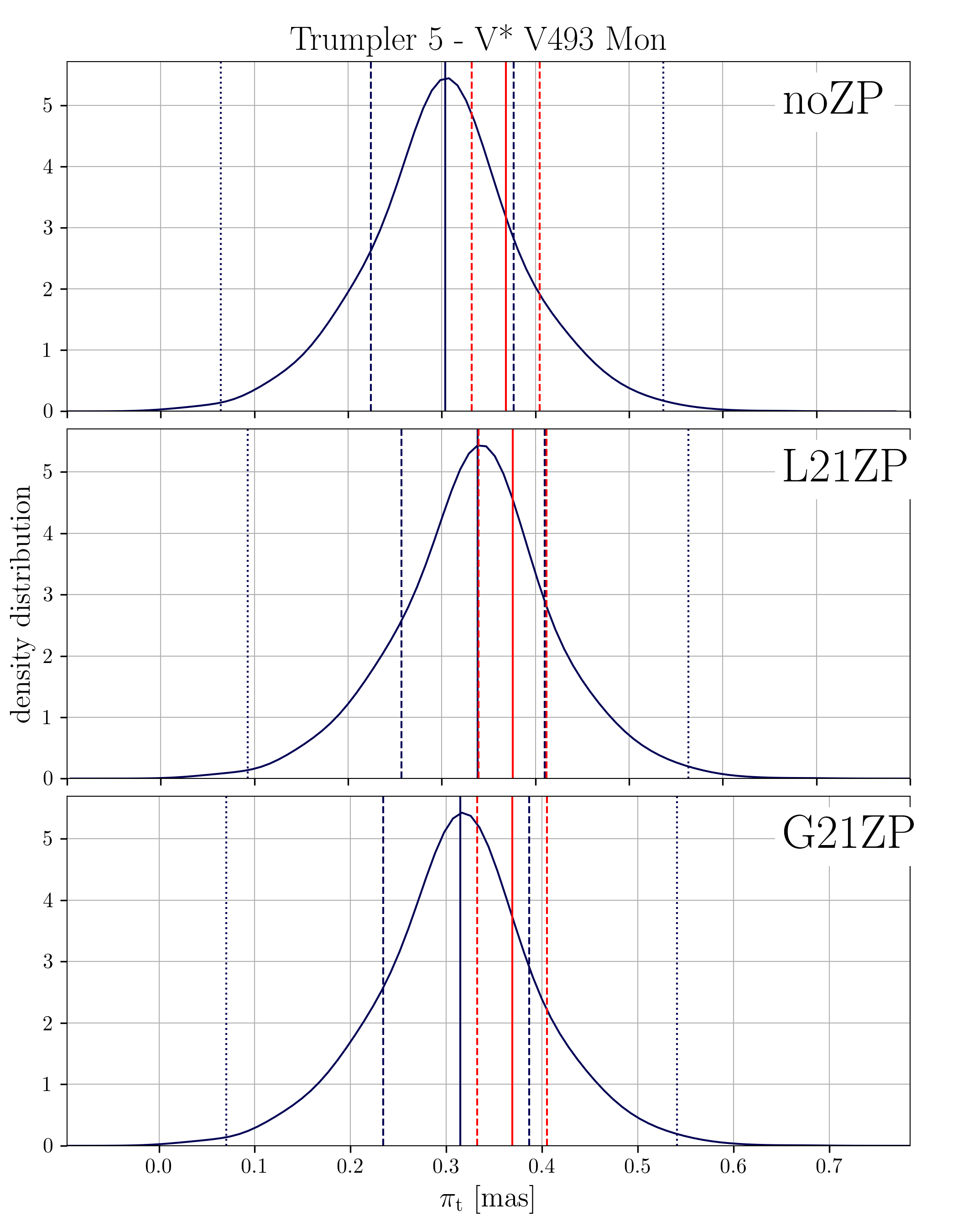}}
\caption{Effect of the \gaia\ EDR3 parallax zero-point corrections considered in this work (no ZP correction, \citealt{Lindegren_etal_21}, and \citealt{Groenewegen21}) applied to the carbon star star V$^\ast$ V493 Mon and its candidate hosting cluster Trumpler 5. 
The blue curves describe the parallax distributions of the cluster members in form of the density distribution function (calculated using the Kernel Density Estimation method - KDE), normalized by the total area.
The solid, dashed, and dotted blue lines mark, respectively,  the median, the 68$^{\rm th}$ and the 95$^{\rm th}$ confidence intervals of the distributions. 
The parallax central value and the $\pm 1-\sigma$ errors  for  V$^\ast$ V493 Mon are also shown (red solid and dashed lines).
}
\label{fig_ex_ZP_comp}
\end{figure}

In consideration of the above, in this work we have decided to check three options, namely: no ZP correction (noZP case), and the ZP corrections according to \citet[][L21ZP case]{Lindegren_etal_21} and \citet[][G21ZP case]{Groenewegen21}. 
As to the L21ZP case, we used the python script provided by the authors, while to implement the G21ZP correction we  followed the instructions given by the author at the end of the paper. In this latter case, we used his models 30-34 to extract the spatial correction at $G=20$ mag for the 5  HEALPix levels (from 0 to 4) and chose the highest level for which the number of calibrating sources, $n_{\rm obj}$, is not less than 100. As discussed by \citet{Groenewegen21},  forty objects or more are required to maintain a good signal-to-noise and get robust results. With the condition $n_{\rm obj} > 100$, most stars are attributed HEALPix levels 2 or 3 and not higher. To estimate the magnitude correction to the parallax we adopted his equation 6. Finally, the total ZP is given by the sum of the spatial and magnitude corrections. Figure \ref{fig_ex_ZP_comp} shows an example of the effect produced by applying the ZP parallax corrections to the carbon star V$^\ast$ V493 Mon and the stars of its candidate hosting cluster Trumpler 5.

\begin{figure*}[ht!]
\centering
\begin{minipage}{0.80\textwidth}
\resizebox{\hsize}{!}{\includegraphics{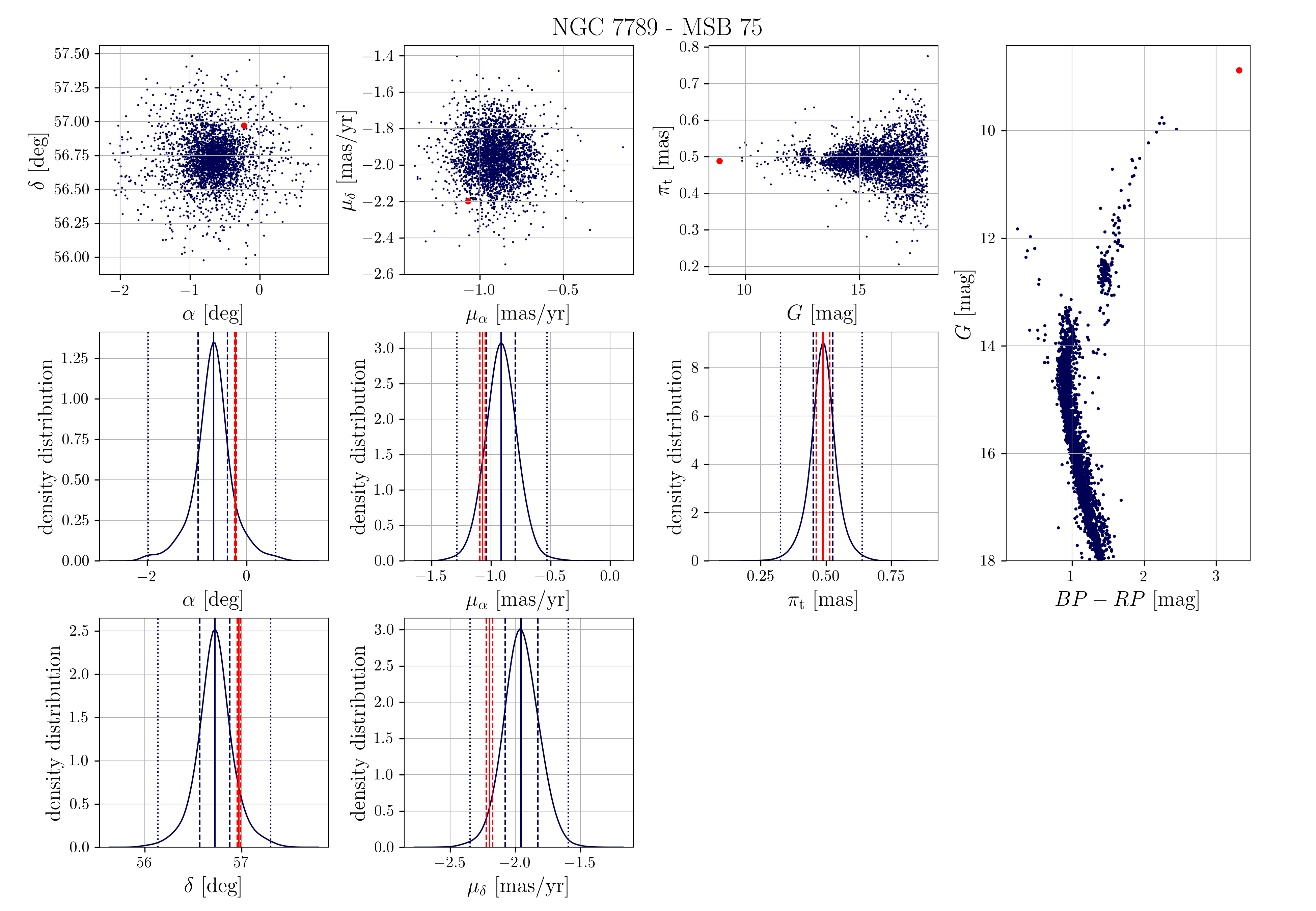}}
\end{minipage}
\caption{Visual example of membership assessment for the target star MSB 75 with respect to the cluster NGC 7789. 
The blue dots in the upper panels refer to the cluster members (with $p > 0.7$) according to \citet{Cantat-Gaudin_etal_20}, while the red dot marks the target star. 
The blue curves in the center and lower panels are the density distribution functions of each astrometric parameter (calculated using KDE method, similar to the one adopted in Figure \ref{fig_ex_ZP_comp}), with the solid, dashed, and dotted blue lines indicating  the median, the 68$^{\rm th}$, and the 95$^{\rm th}$ percentiles, respectively. 
The red lines denote the central values (solid lines) and the 1$-\sigma$ errors (dash lines) of MSB 75 parameters. Our analysis indicates that MSB 75 is a member of NGC 7789.
The bottom-right panel shows the CMD using the three optical \gaia\ pass-bands.      
The parameters of each star are updated to \gaia\ EDR3 and the parallaxes corrected for the zero-point offset (in this example using  \citet{Groenewegen21}).}
\label{fig_exm_memb}
\end{figure*}

\subsection{Cluster membership revisited with \gaia\ EDR3}
\label{ssec_membedr3}
To assess the cluster membership of the TP-AGB stars, we exploit the astrometric data from both \gaia\ DR2 and EDR3.
Our analysis is applied only to the sources with known spectral type and classified as TP-AGB or Super-AGB stars (Table~\ref{tab_specagb}), as we are primarily interested in the final stages of evolution when surface carbon enrichment due to the 3DU can cause the spectral type change along the sequence M $\rightarrow$ S $\rightarrow$ C.
We end up with a sample of 18 TP-AGB stars and 2 potential Super-AGB stars.
We add also the C star NIKC 3-81 which is not a genuine TP-AGB star, but possibily a R-hot star. In total we examine 21 stars.

For each cluster we first consider the most representative stars, which are those with membership probability $p > 0.7$ according to \citet[][to which we refer for all details]{Cantat-Gaudin_etal_20}, on the base of a statistical analysis of parallax $\pi$ and  proper motions ($\mu_{\alpha}$ and $\mu_{\delta}$) performed with the UPMASK code \citep{Krone_etal_14, Cantat-Gaudin_etal_18}.
Using the  same sample of stars we then re-determine the global properties of the clusters considering the updated astrometry from \gaia\ EDR3. 
We then derive the median $\alpha$, $\delta$, $\mu_{\alpha}$, $\mu_{\delta}$, and $\omega$ of each cluster and assess their 68\% and 99\% confidence levels (C.L.), accordingly with the $68^{\rm th}$ and $95^{\rm th}$ percentiles of each parameter distribution of the members. 
Figure~\ref{fig_exm_memb} provides an example of the method applied to the C star V$^\ast$ V493 Mon.

The evaluation of the target star membership is obtained by comparing parameter-by-parameter each individual star with the clusters C.L. (also accounting for the errors on the target star parameters).
We consider members only the stars for which \emph{all} the parameters are within the 99\% C.L. of the cluster distributions.   
The results are reported in Table~\ref{tab_stars} in Appendix~\ref{sec_app}. We list the target stars together with the C.L. they belong to with respect to the parent cluster. 
Specifically, we indicate if a star lies  within the 68\% C.L. (i.e. when all the parameters are within this range), or it is found within the 99\% C.L. (i.e. when at least one parameter is outside the 68\% C.L. but still inside the 99\% C.L.), or it is rejected (when at least one parameter is outside the 99\% C.L.).

For all the stars and parent clusters for which radial velocity data are available, we also confirm that the two measurements are compatible. However, given the low number of cluster members with measured radial velocities, we do not consider this test stringent for the final membership evaluation. 

The same analysis for each star is repeated three times (Sect.~\ref{ssec_plx}), adopting  the \gaia\ EDR3 parallaxes without ZP correction (noZP case), and with the ZP corrections of \citet[][L21ZP case]{Lindegren_etal_21} and \citet[][G21ZP case]{Groenewegen21}. Although the parallax distributions of the host clusters and the parallaxes of the target stars change somewhat, the results of the membership assessment always remain in agreement.
Our analysis indicates that out of 21 stars examined, 20 are members of the associated cluster, with the exception of Case 49 whose membership to the  cluster NGC 663 is rejected.

\subsection{SED fitting: bolometric luminosity and other parameters}
\label{ssec_sed}
The bolometric luminosity is estimated by fitting the observed spectral energy distribution (SED), which typically includes photometry data from \gaia\ EDR3 \citep{Gaia_EDR3_21}, 2MASS \citep{Cutri_etal_03}, WISE \citep{wise14}, MSX \citep{Egan_etal_03}, AKARI \citep{Tshihara_etal_10}, and IRAS \citep{IRAS_88}.

A potential issue with the SED fitting of TP-AGB stars comes from their periodic large-amplitude variability, especially at visual wavelengths. However, the \gaia\ EDR3 photometry we adopt is expected to be well representative of the mean stellar luminosity as it consists of mean magnitudes computed over time series longer than (or at least comparable with) the variability timescales of TP-AGB stars \citep[$\sim$1000 days,][]{Gaia_EDR3_21}. The variability amplitude decreases rapidly in the IR \citep{Ita_etal_2021,Iwanek_etal_2021}, so that even single-epoch survey data can be considered reasonably safe to use.

The SED fitting is performed with the Virtual Observatory SED Analyzer \citep[VOSA; ][]{vosa08}.
The results are reported in Table~\ref{tab_sedfit}.
In VOSA the synthetic photometry is derived from the GRAMS grid of  theoretical spectra for M  and C stars \citep{Sargent_etal_11,Srinivasan_etal_11}. These models account for the radiative transfer of the photospheric emission across the circumstellar dust shell produced by a mass-losing AGB star. The starting hydrostatic  atmospheres are taken from the PHOENIX models for M stars \citep{Kucinskas_etal_05}, and from the COMA models for C stars \citep{Aringer_etal_09}.

Each theoretical spectrum in the grid  is characterized by a combination of input parameters, namely: stellar mass $M$, surface gravity $g=GM/R^2$ ($G$ is the gravitation constant and $R$ is the photospheric radius),  effective temperature $T_{\rm eff}$, and photospheric C/O ratio (only for carbon stars). 

During the SED fitting procedure these parameters are allowed to vary within selected ranges. In our analysis we used the entire grid of GRAMS models for $g$, $M$, $T_{\rm eff}$, and C/O.  To match the observed spectrum we need also to specify the distance $D$ of the source and the visual extinction $A_V$. 
Distance and associated uncertainty are derived by inverting the \gaia\ EDR3 parallax, possibly  corrected for the zero-point \citep{Lindegren_etal_21, Groenewegen21}.
The visual extinction $A_V$ is obtained from the same catalogs we use to date the host clusters. In the fitting $A_V$  is let vary within a range of $\pm 0.2$ mag around the central value.

To single out the synthetic photometry that best reproduces the observed data  we choose the VOSA option of a reduced $\chi^2$ statistical test.  In this way the observational errors associated to the different pass-bands are used to weigh the importance of each photometric point when calculating the $\chi^2$ final value for each model.
In addition to $A_V$, $g$, $M$, $T_{\rm eff}$, and C/O, the best-fit model returns other important quantities that contribute to characterize the AGB star, namely: the dust mass-loss rate, $\dot M_{\rm d}$, the inner radius of the dust shell $r_{\rm in}$, and the optical depth at 11.3 microns, $\tau_{11.3}$, for C-rich stars, or the optical depth at 10 microns, $\tau_{10}$, for O-rich stars.
We note that there are no appropriately calculated spectra for S stars; we treat these cases in the same way as for O-rich stars. For the 3 stars of type MS and S, the quality of the fit is still good. 

We set the VOSA option that provides the fit parameter uncertainties using a statistical approach. The best fit is obtained for 100 virtual SEDs by applying a gaussian random noise (proportional to the observational error) to each photometric point.
Then, for each parameter, the reported uncertainty is given by the standard deviation of the derived  distribution, if its value is larger that half the grid step for the parameter. Otherwise, the reported uncertainty is just half the grid step. 

The  best  fitting  model  is  then used  to  infer  the  total  observed flux $F_{\rm obs}$ of the star. Finally, the bolometric luminosity is obtained through the relation $L=4 \pi D^2 F_{\rm obs}$. It is important to note that the uncertainty in luminosity, $\sigma_L$, is obtained through standard error propagation, that is
\begin{equation}
   \frac{\sigma_L}{L}  = 2\frac{\sigma_D}{D} + \frac{\sigma _{F_{\rm obs}}}{F_{\rm obs}}
    \label{eq_deltal}
\end{equation}
Taking the uncorrected \gaia\ EDR3 parallaxes (noZP case) the typical flux uncertainties are $\sigma_{F_{\rm obs}}/{F_{\rm obs}}\simeq 1-7 \%$.
As to the distance error, we find that the large majority (85\%) of the 20 analyzed stars ${\sigma_D}/{D} < 15 \%$, and the largest uncertainty is ${\sigma_D}/{D} \simeq 32 \%$.
The majority (65\%) of the 20 stars analyzed have  $\sigma_L/{L} < 25 \%$, and only 2 stars have $\sigma_{\sigma_L}/{L} \ga 50 \%$.  Similar results apply for the L21ZP and G21ZP cases.
It follows that in all cases under consideration the distance uncertainty, $\sigma_D$, dominates the error budget for $L$.

\begin{figure*}[h!]
\begin{center}
\begin{minipage}{0.23\textwidth}
\resizebox{\hsize}{!}{\includegraphics{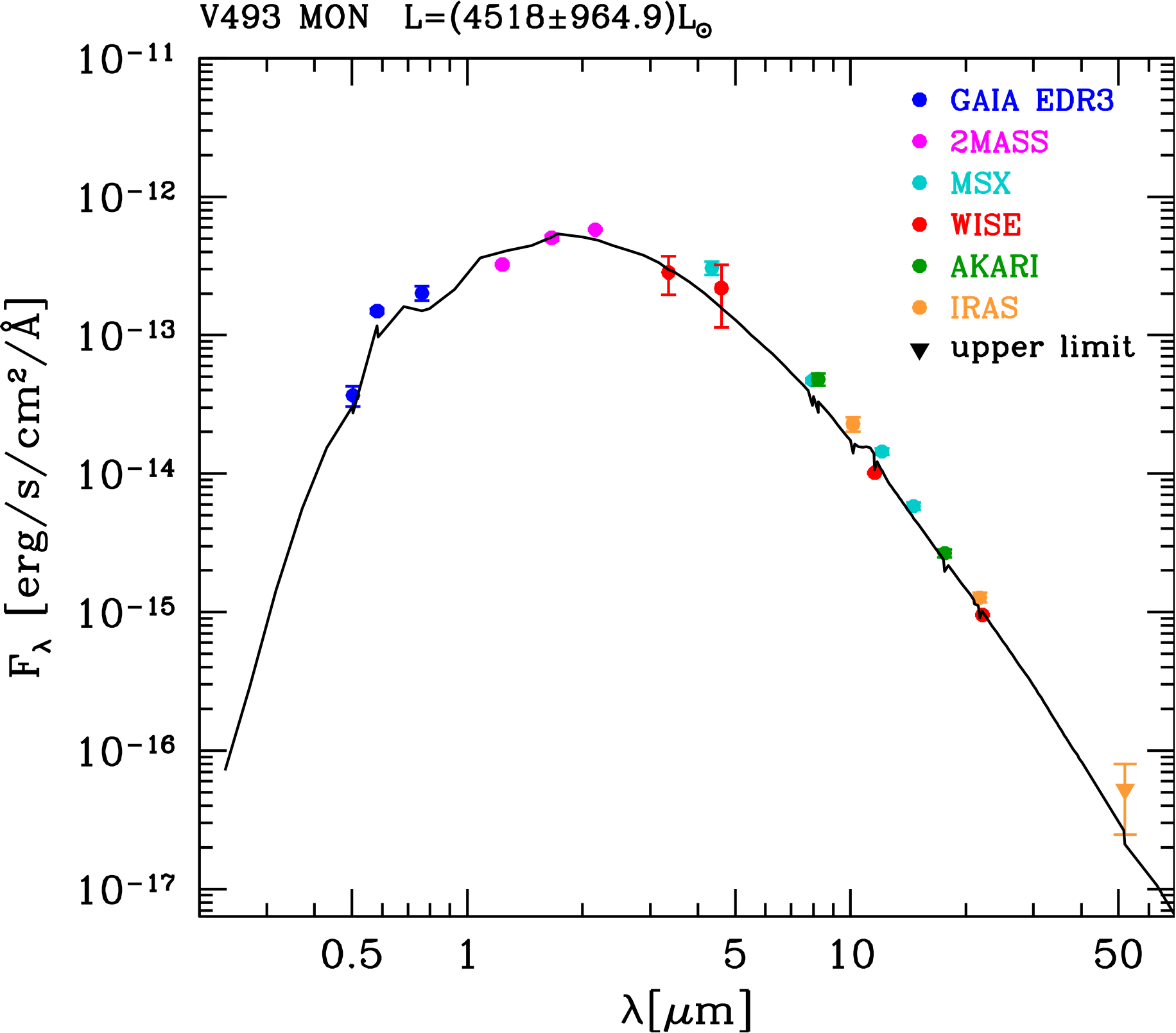}}
\end{minipage}
\begin{minipage}{0.23\textwidth}
\resizebox{\hsize}{!}{\includegraphics{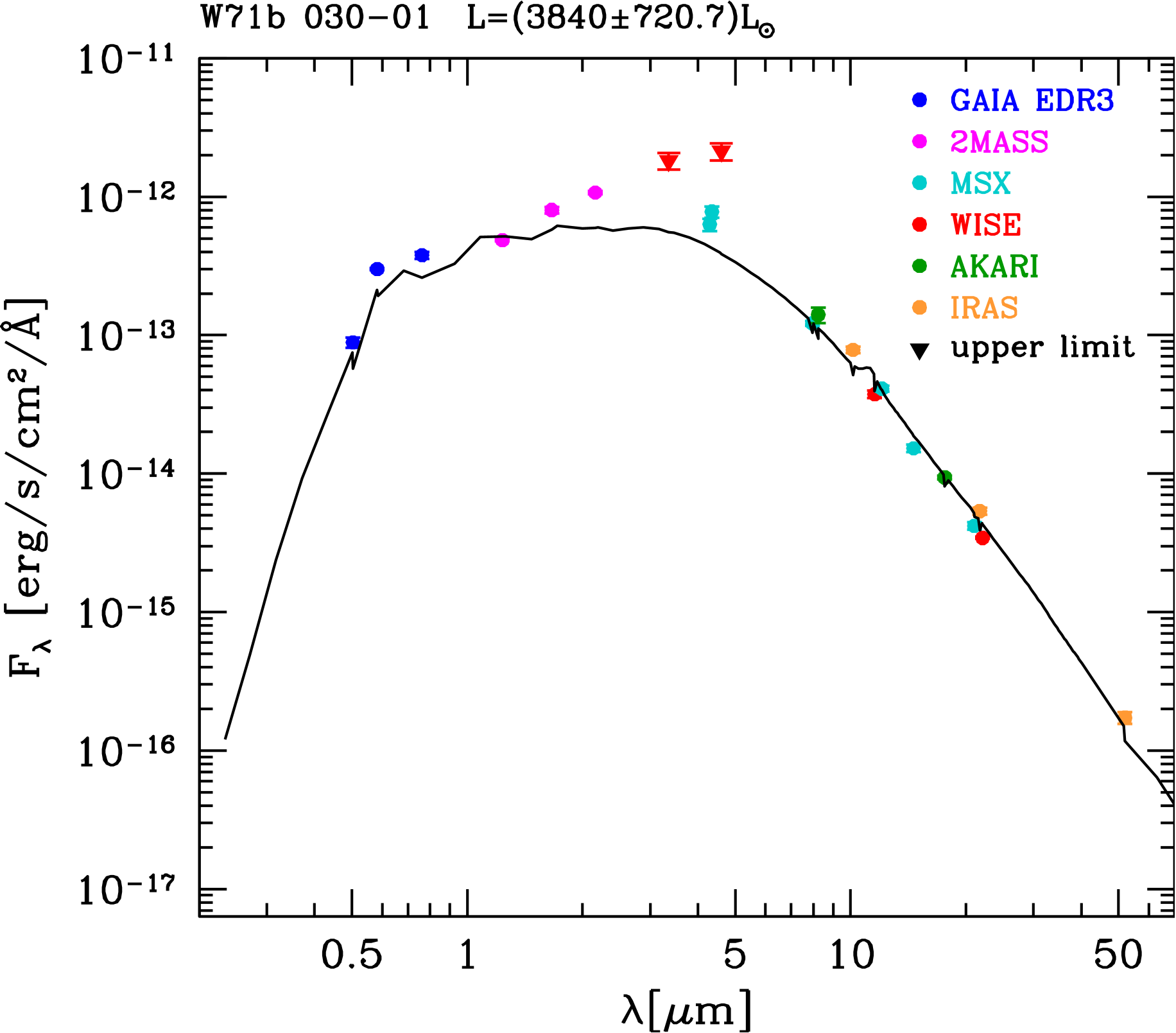}}
\end{minipage}
\begin{minipage}{0.23\textwidth}
\resizebox{\hsize}{!}{\includegraphics{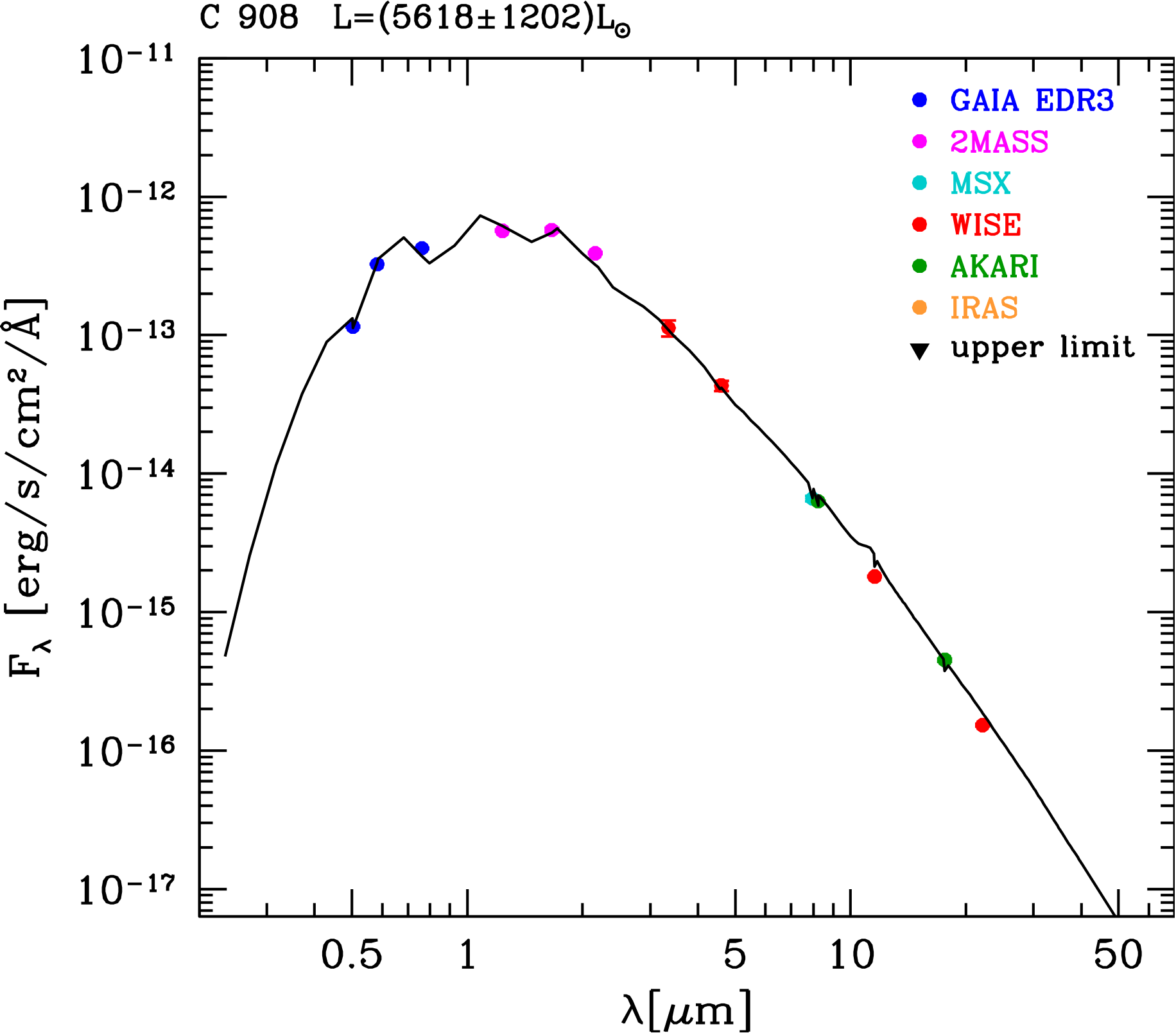}}
\end{minipage}
\begin{minipage}{0.23\textwidth}
\resizebox{\hsize}{!}{\includegraphics{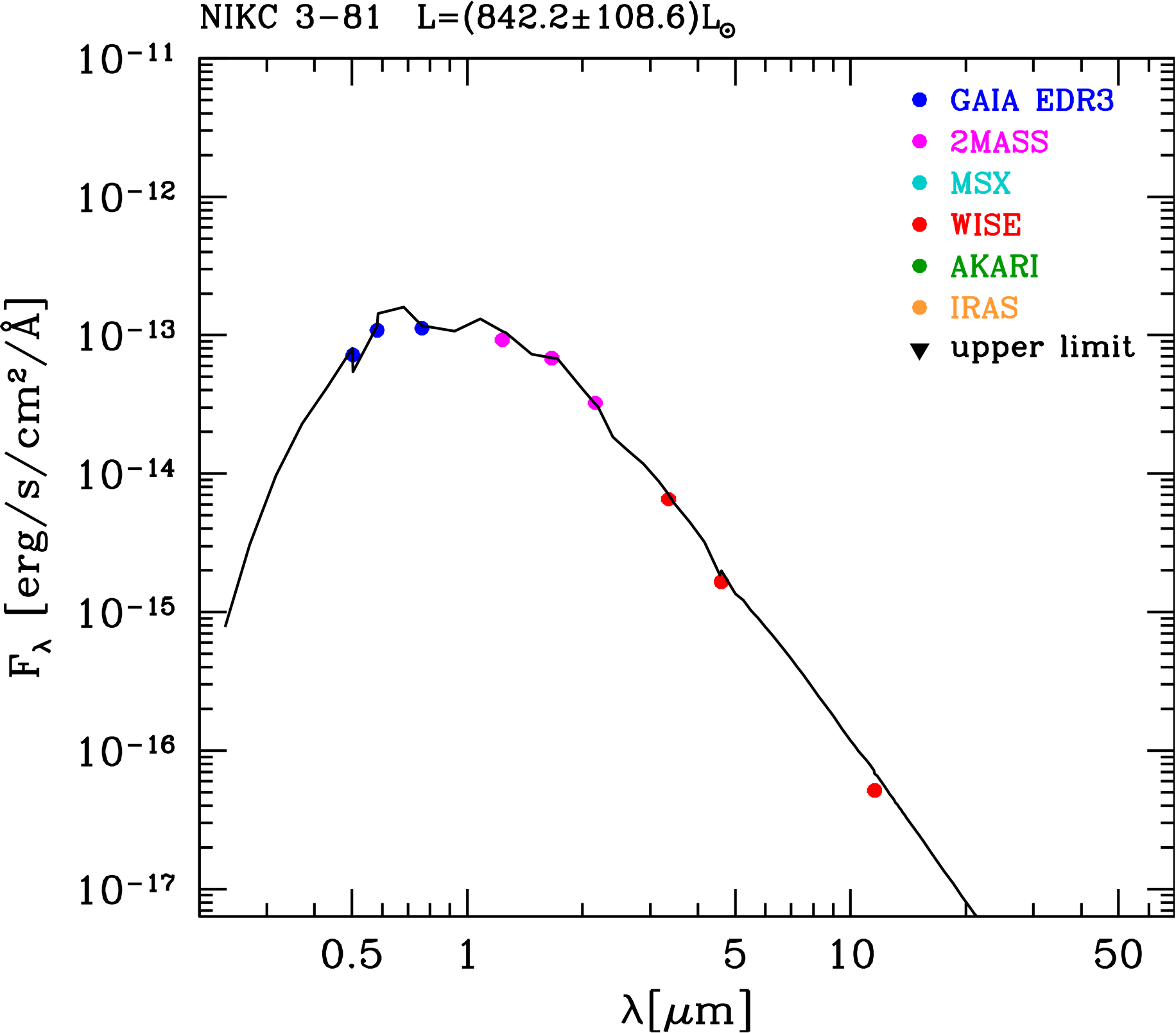}}
\end{minipage}
\begin{minipage}{0.23\textwidth}
\resizebox{\hsize}{!}{\includegraphics{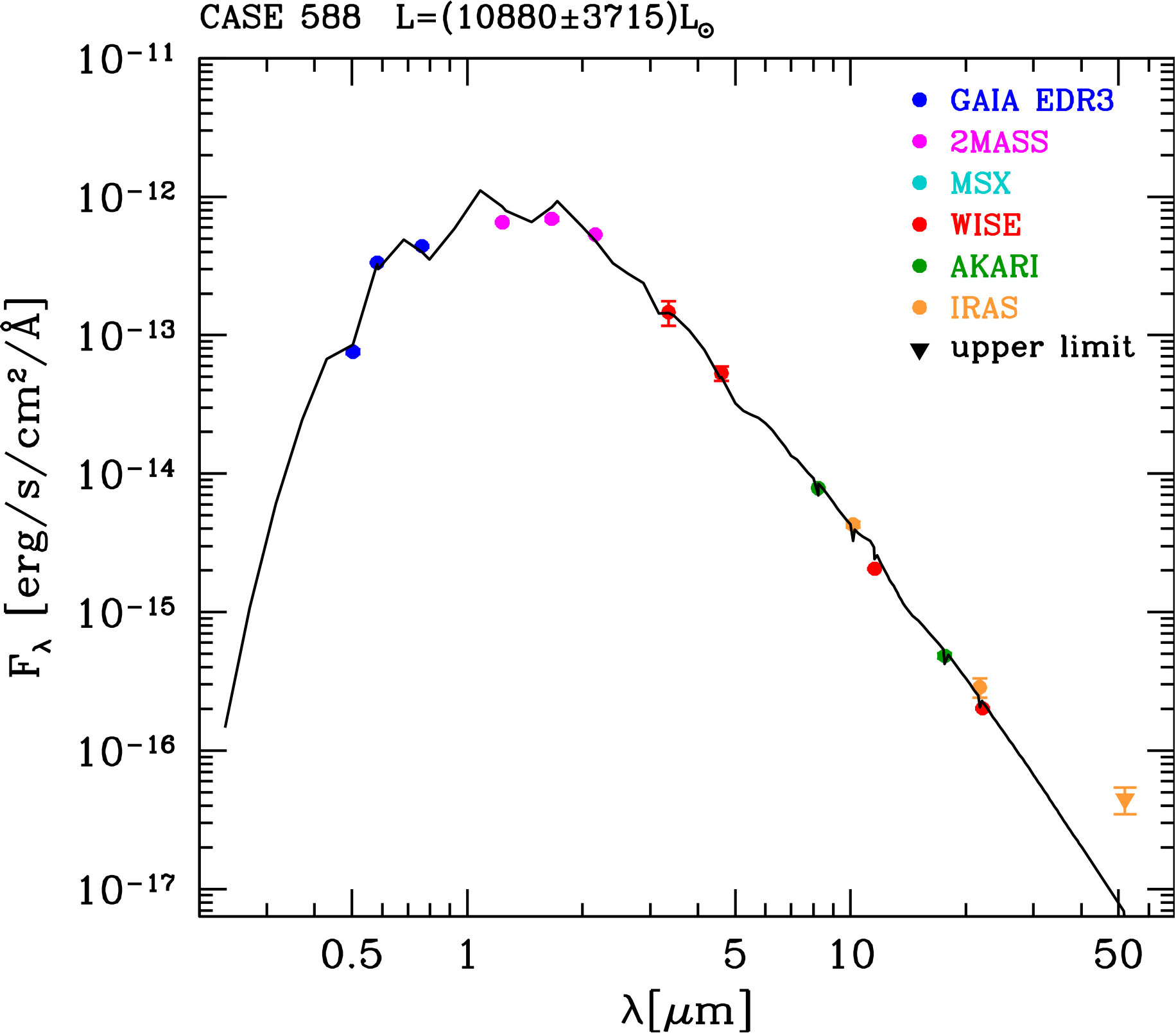}}
\end{minipage}
\begin{minipage}{0.23\textwidth}
\resizebox{\hsize}{!}{\includegraphics{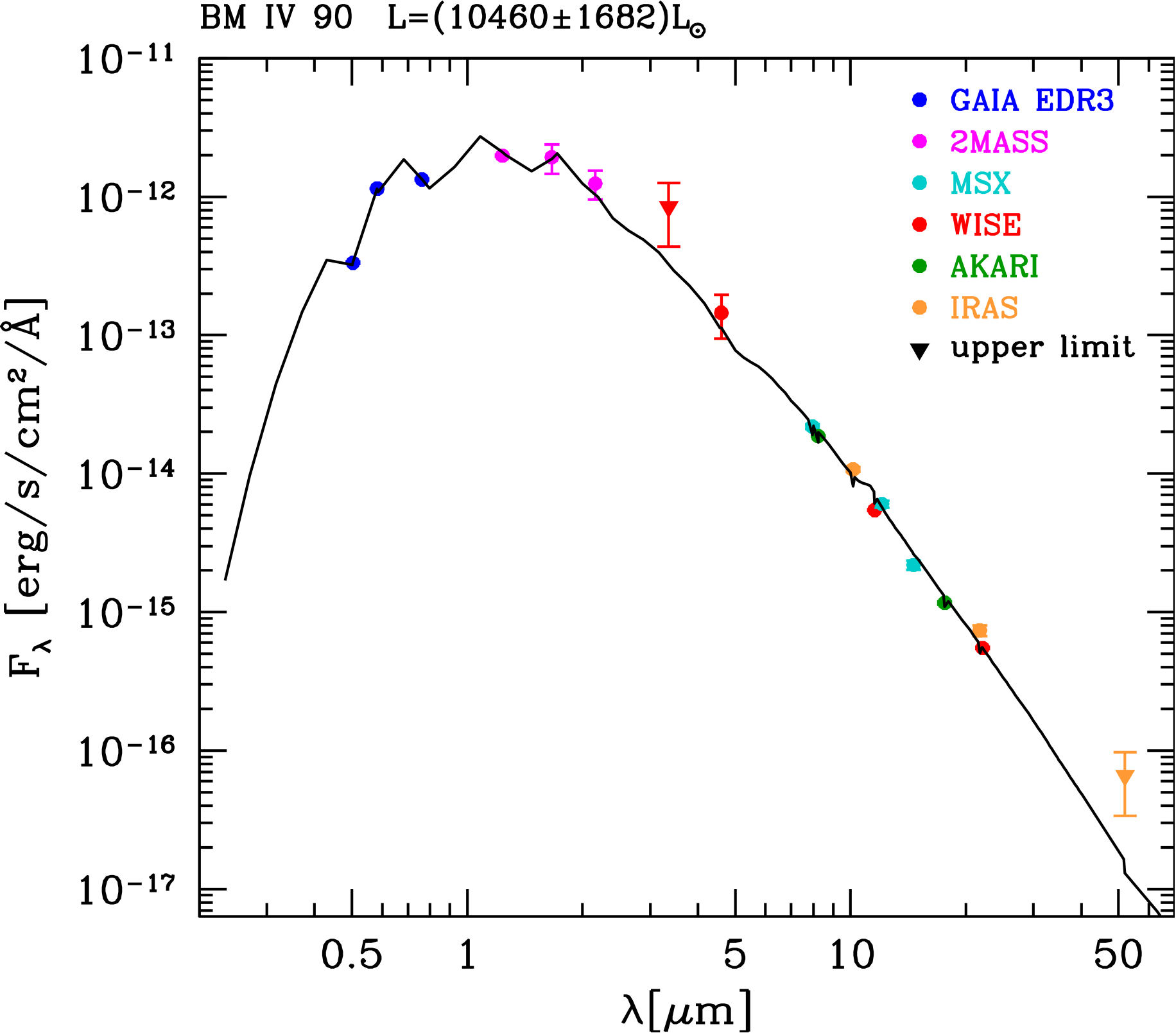}}
\end{minipage}
\begin{minipage}{0.23\textwidth}
\resizebox{\hsize}{!}{\includegraphics{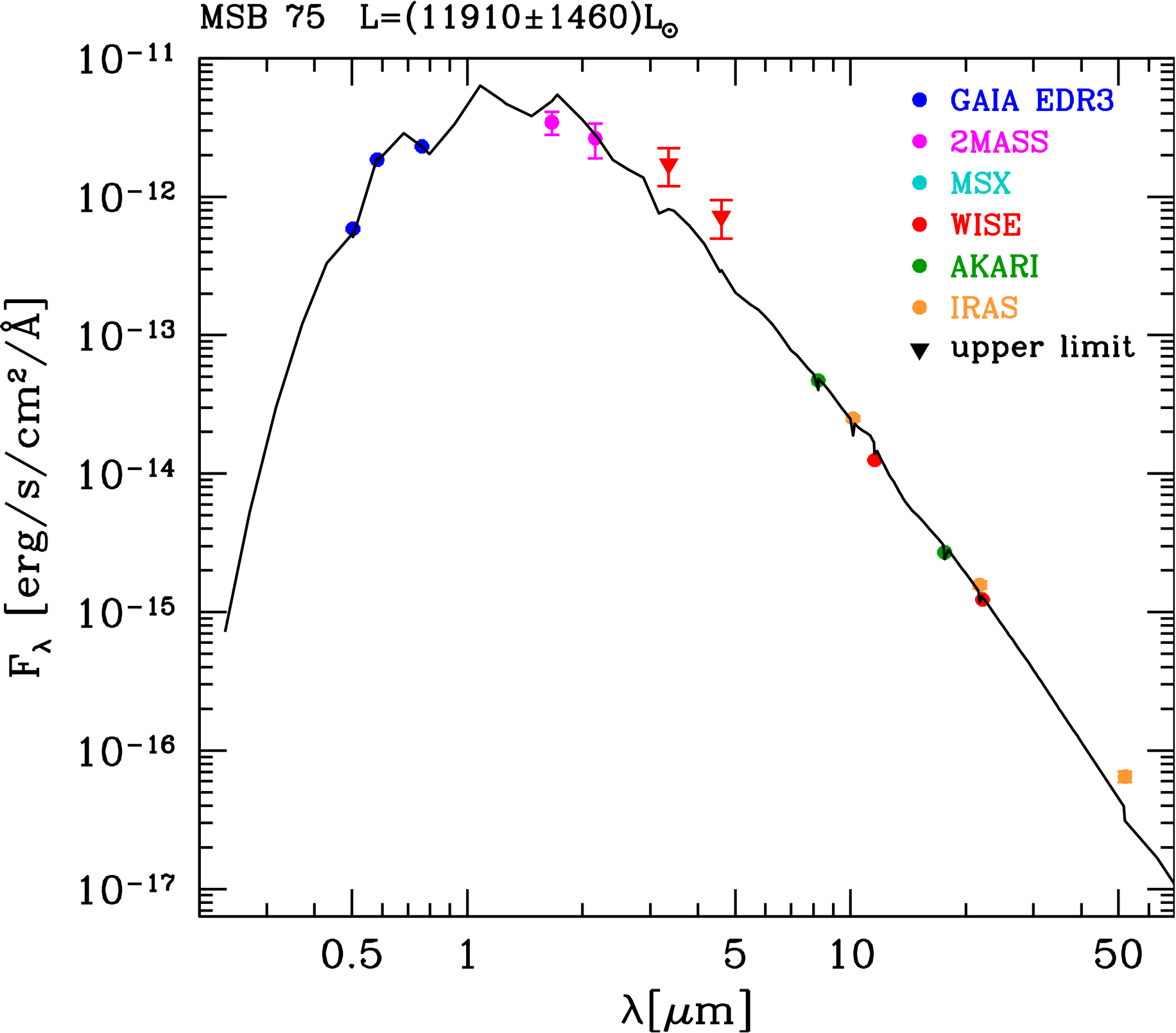}} 
\end{minipage}
\begin{minipage}{0.23\textwidth}
\resizebox{\hsize}{!}{\includegraphics{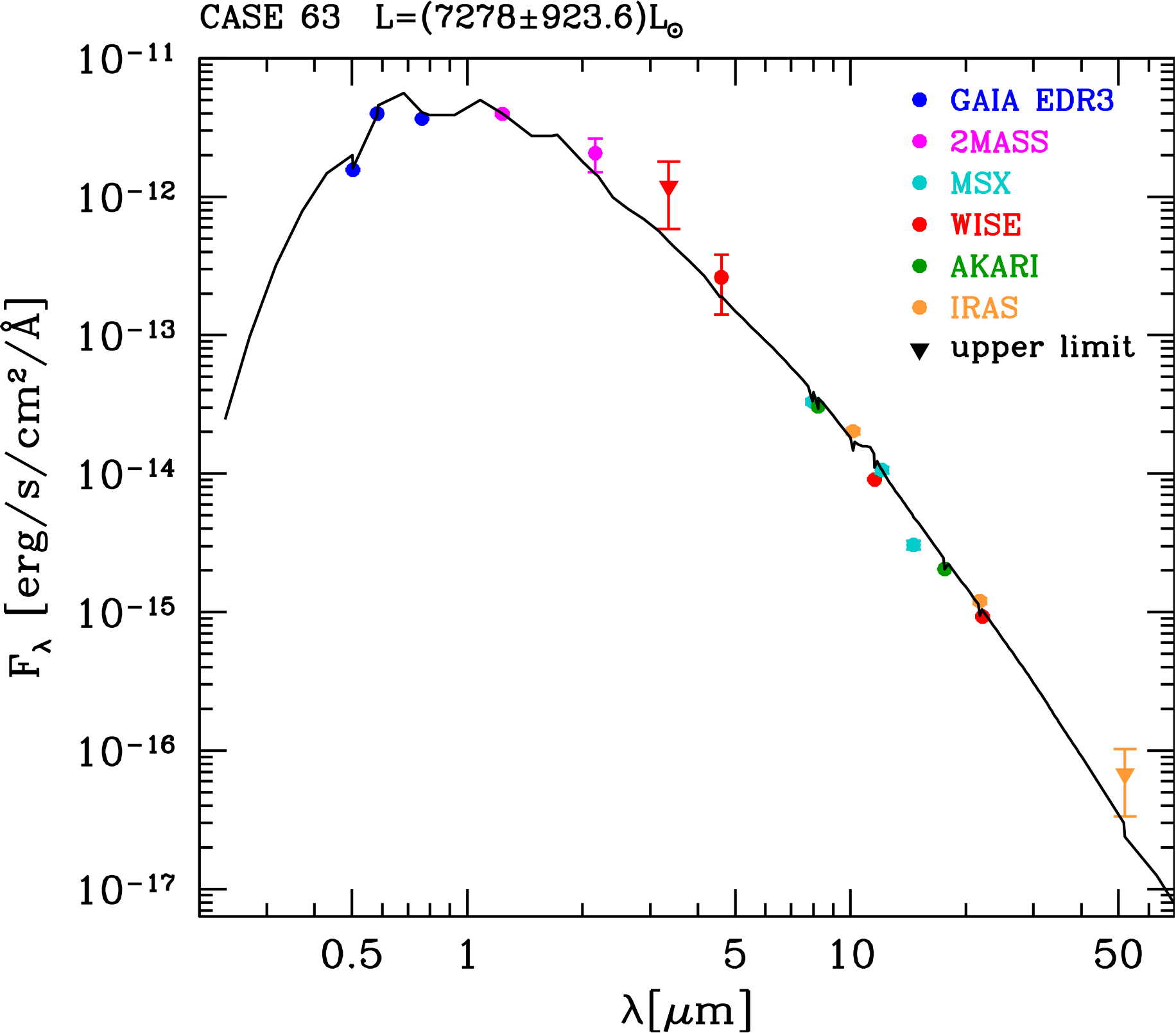}}
\end{minipage}
\begin{minipage}{0.23\textwidth}
\resizebox{\hsize}{!}{\includegraphics{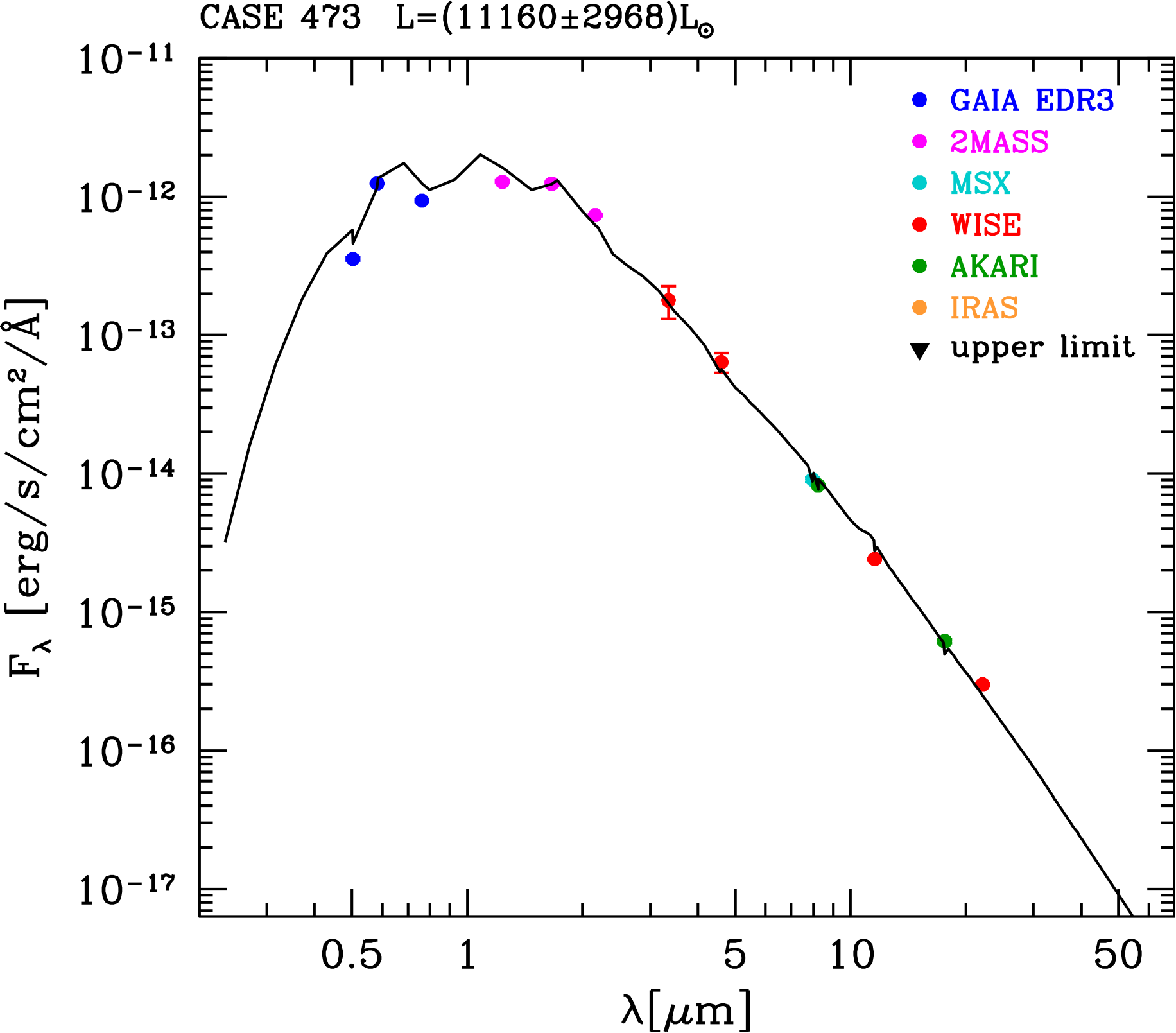}}
\end{minipage}
\begin{minipage}{0.23\textwidth}
\resizebox{\hsize}{!}{\includegraphics{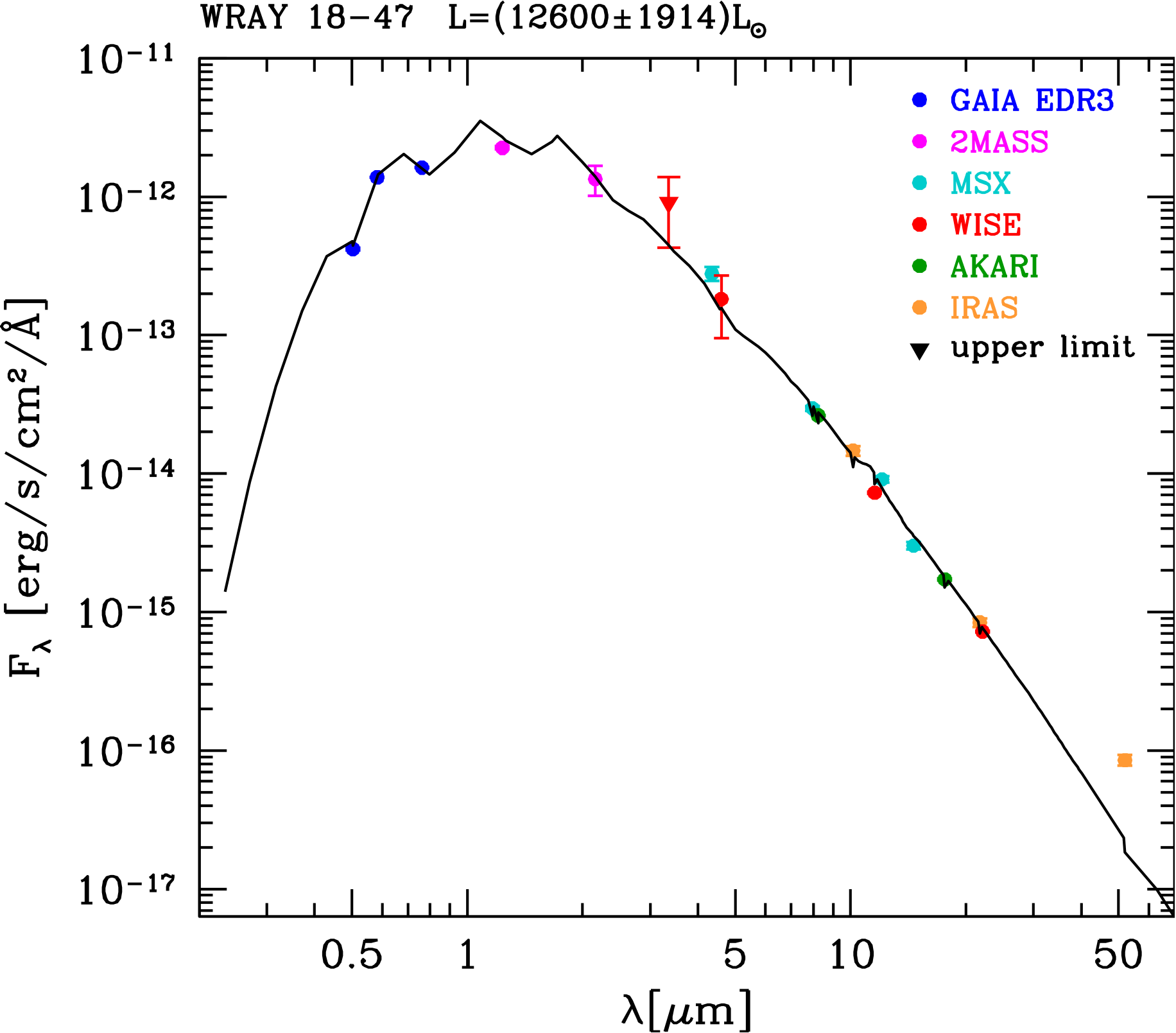}}
\end{minipage}
\begin{minipage}{0.23\textwidth}
\resizebox{\hsize}{!}{\includegraphics{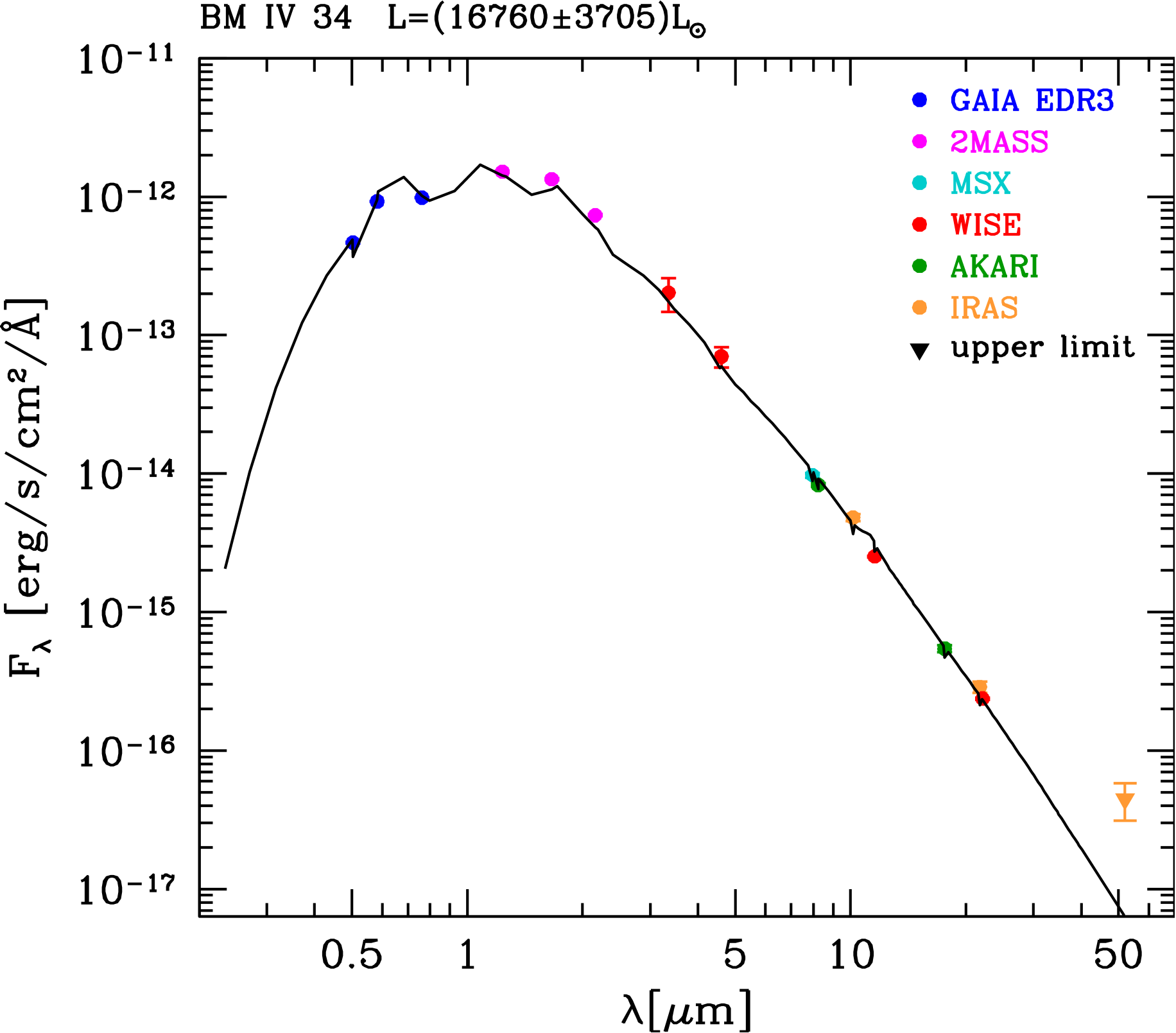}}
\end{minipage}
\begin{minipage}{0.23\textwidth}
\resizebox{\hsize}{!}{\includegraphics{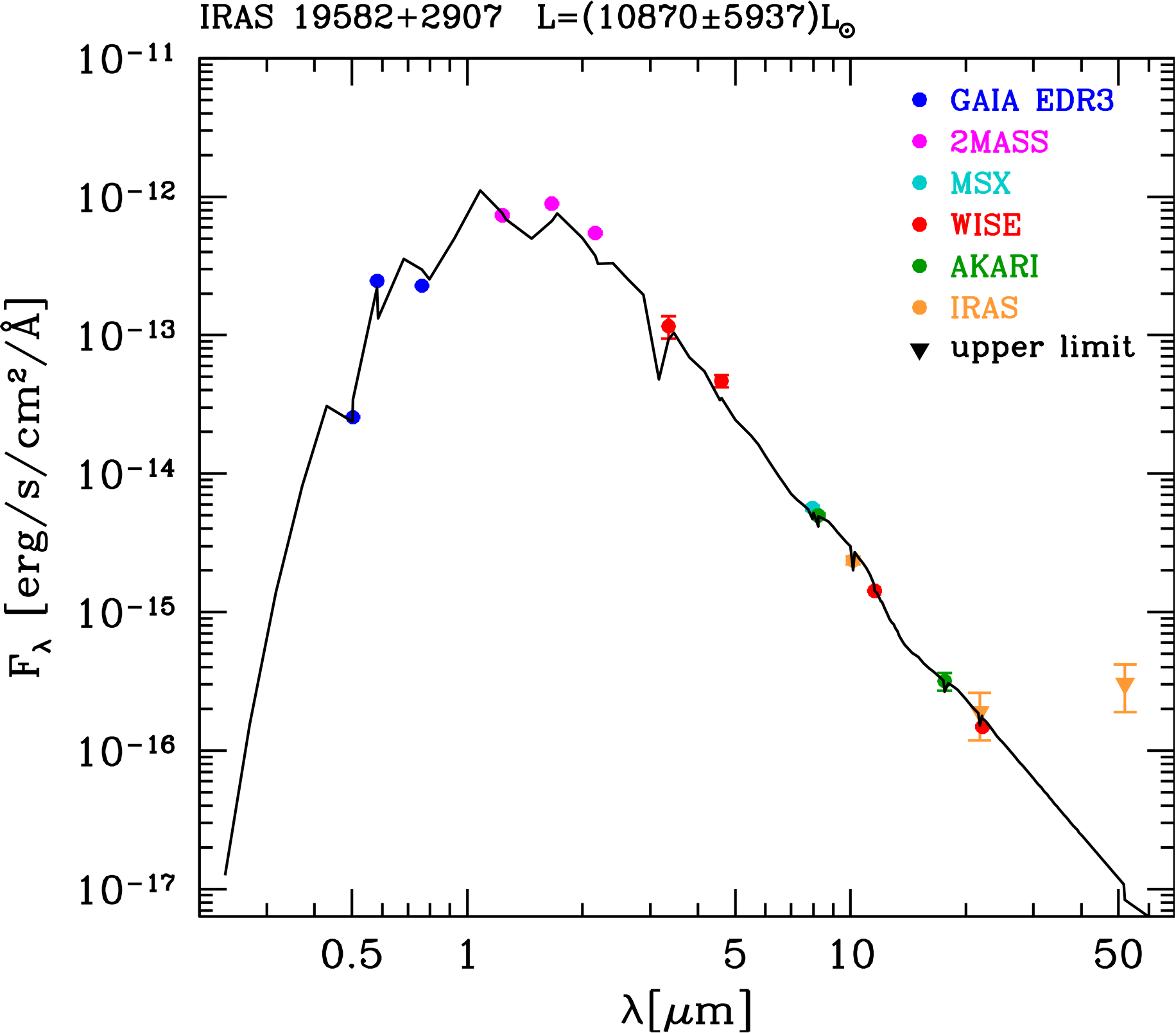}}
\end{minipage}
\begin{minipage}{0.23\textwidth}
\resizebox{\hsize}{!}{\includegraphics{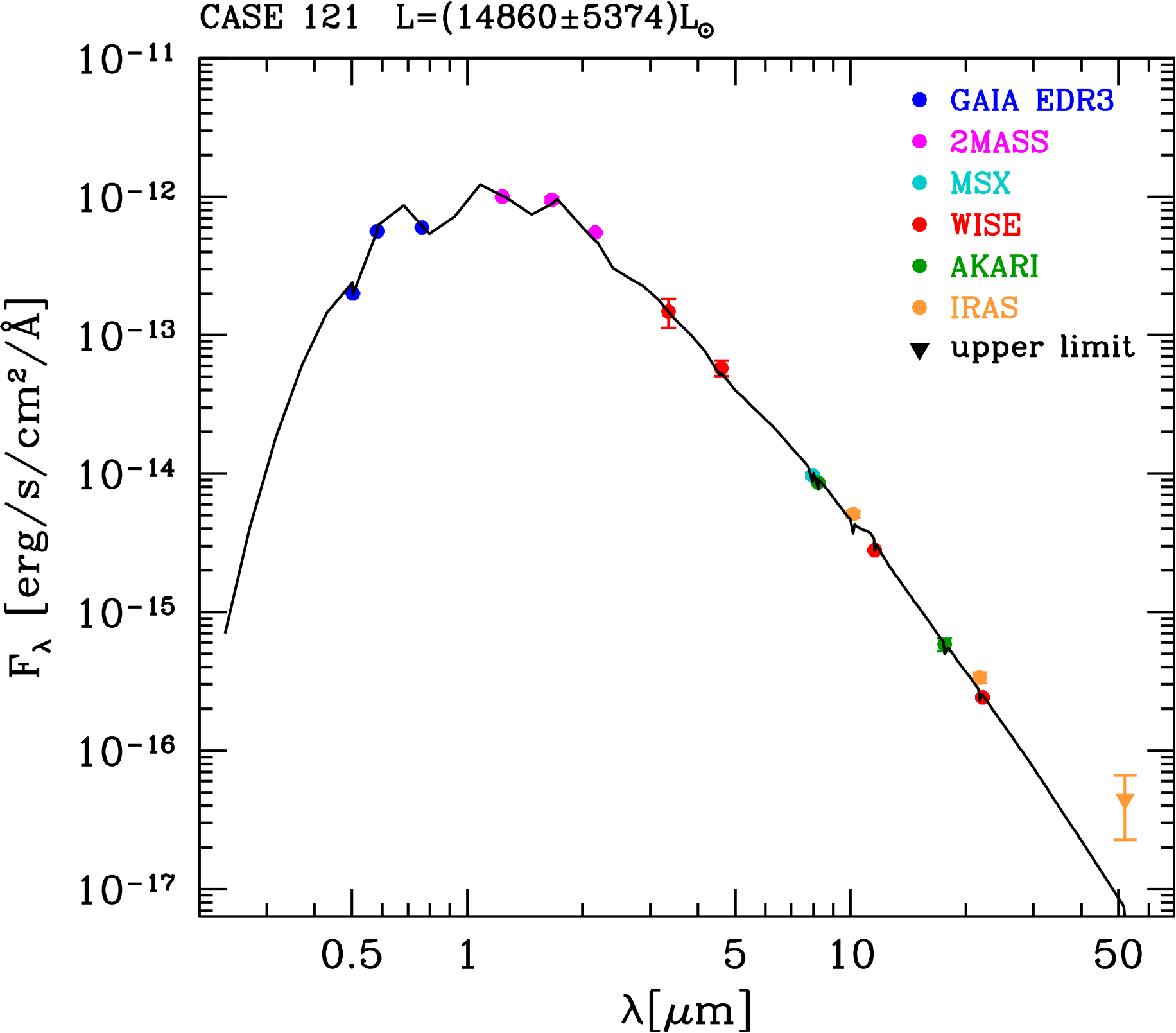}}
\end{minipage}
\begin{minipage}{0.23\textwidth}
\resizebox{\hsize}{!}{\includegraphics{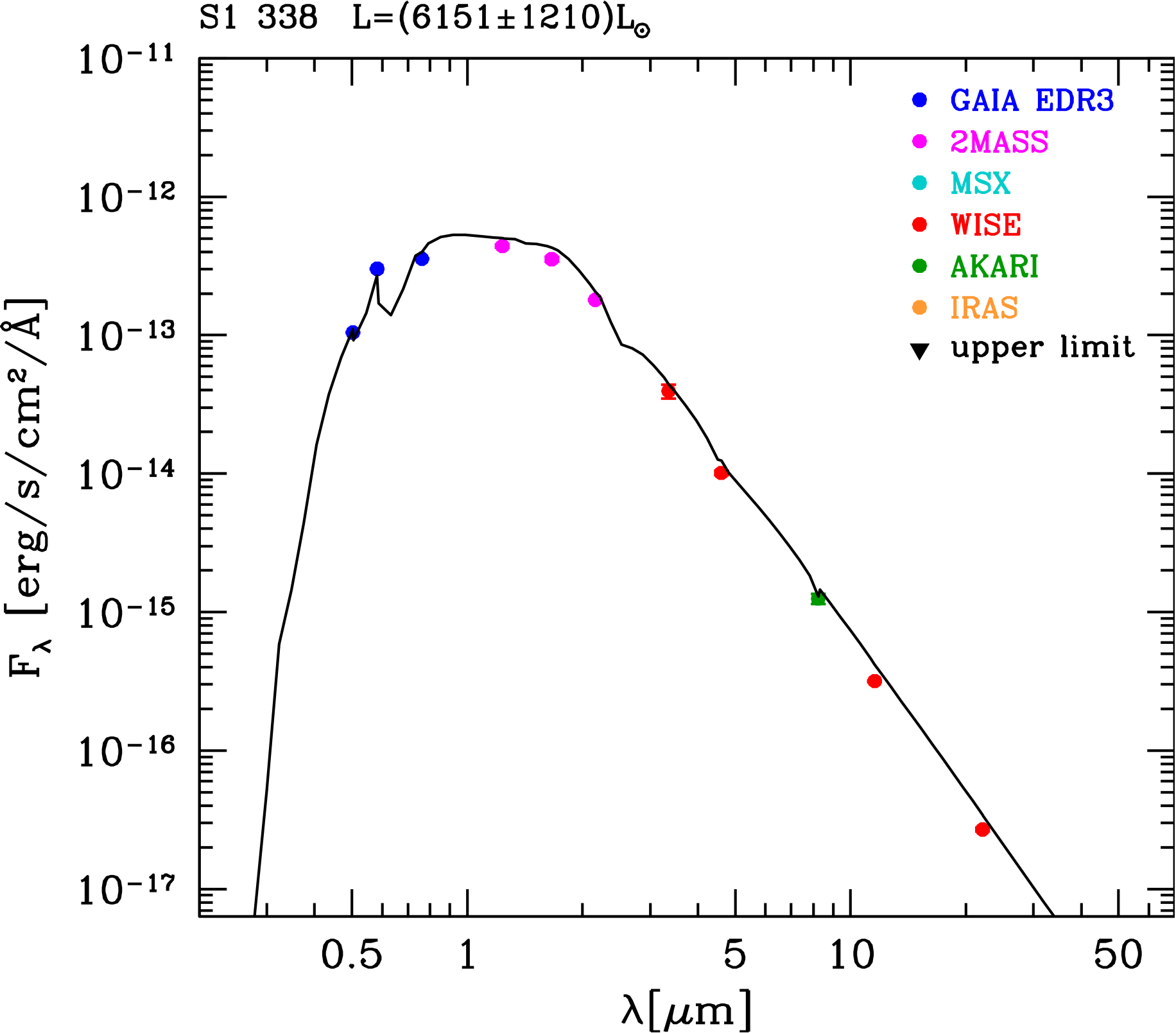}}
\end{minipage}
\begin{minipage}{0.23\textwidth}
\resizebox{\hsize}{!}{\includegraphics{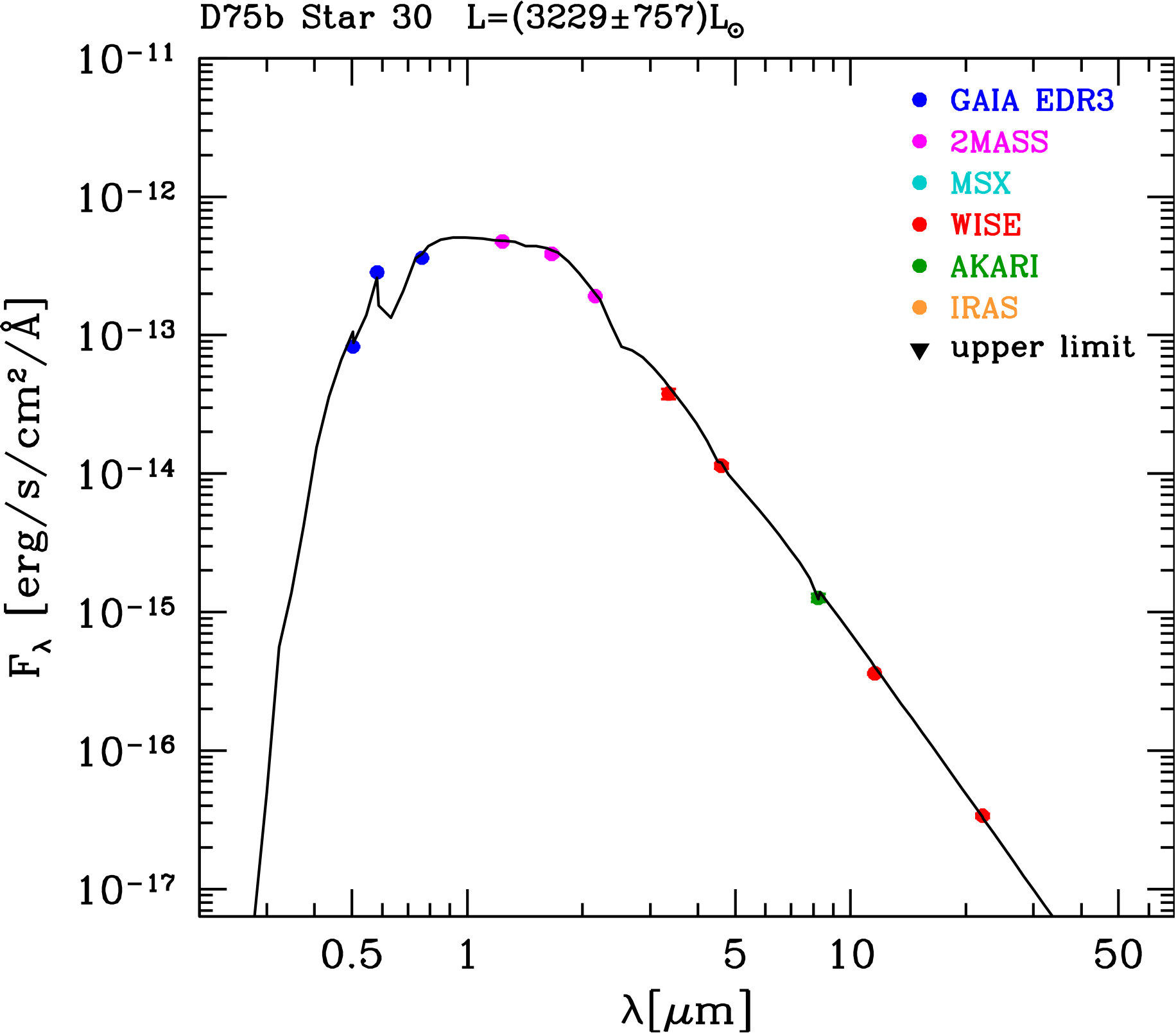}}
\end{minipage}
\begin{minipage}{0.23\textwidth}
\resizebox{\hsize}{!}{\includegraphics{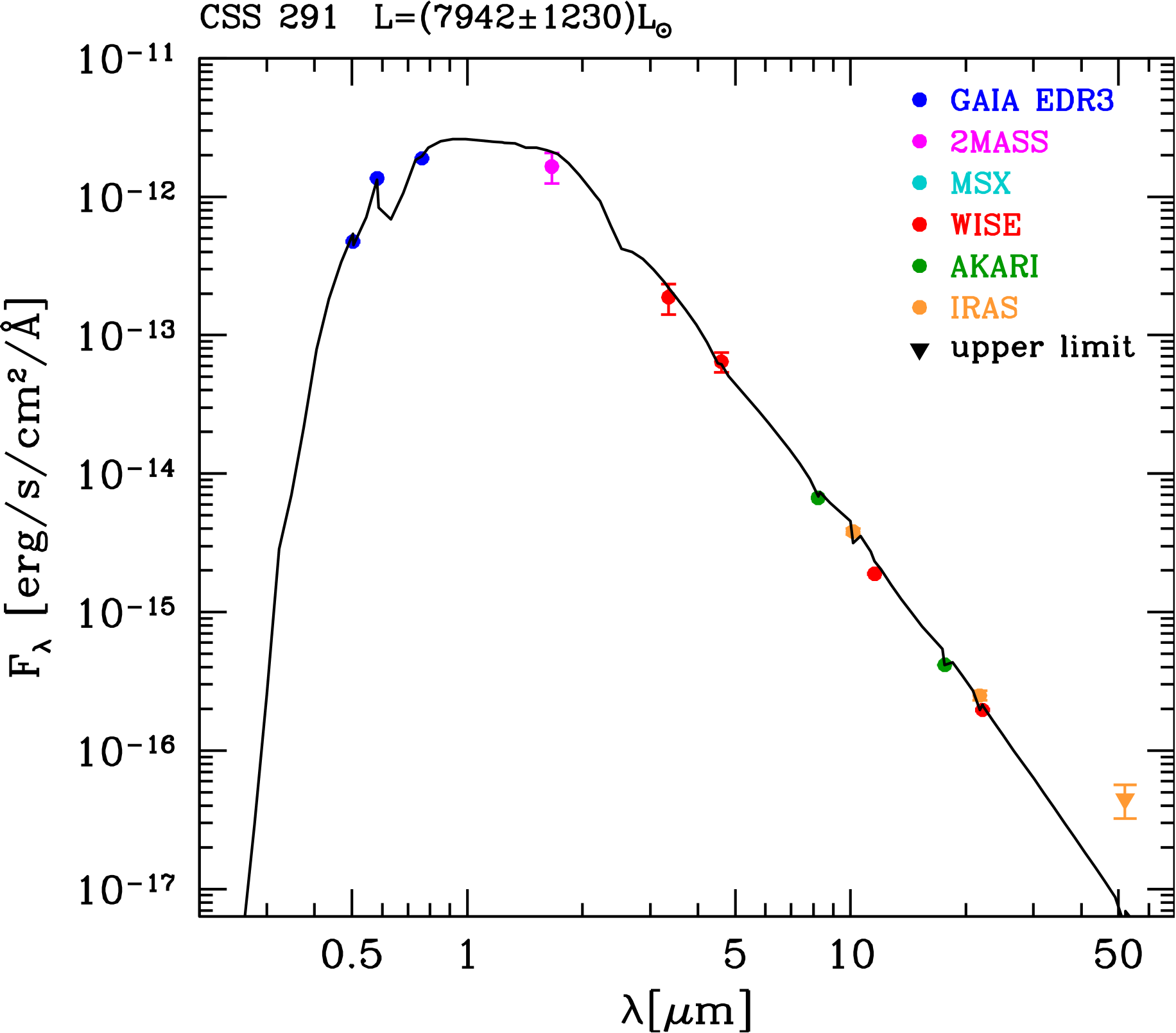}}
\end{minipage}
\begin{minipage}{0.23\textwidth}
\resizebox{\hsize}{!}{\includegraphics{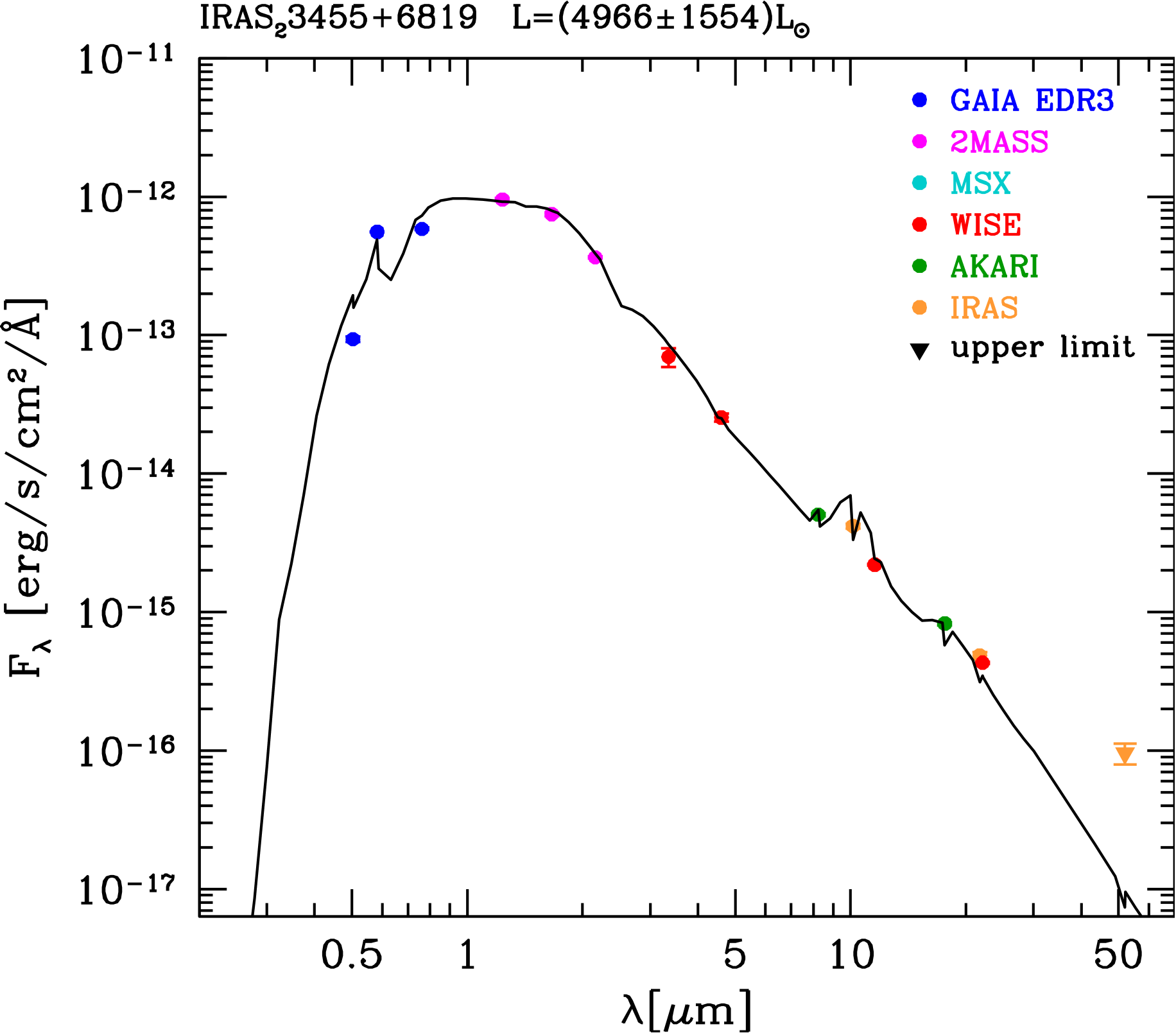}}
\end{minipage}
\begin{minipage}{0.23\textwidth}
\resizebox{\hsize}{!}{\includegraphics{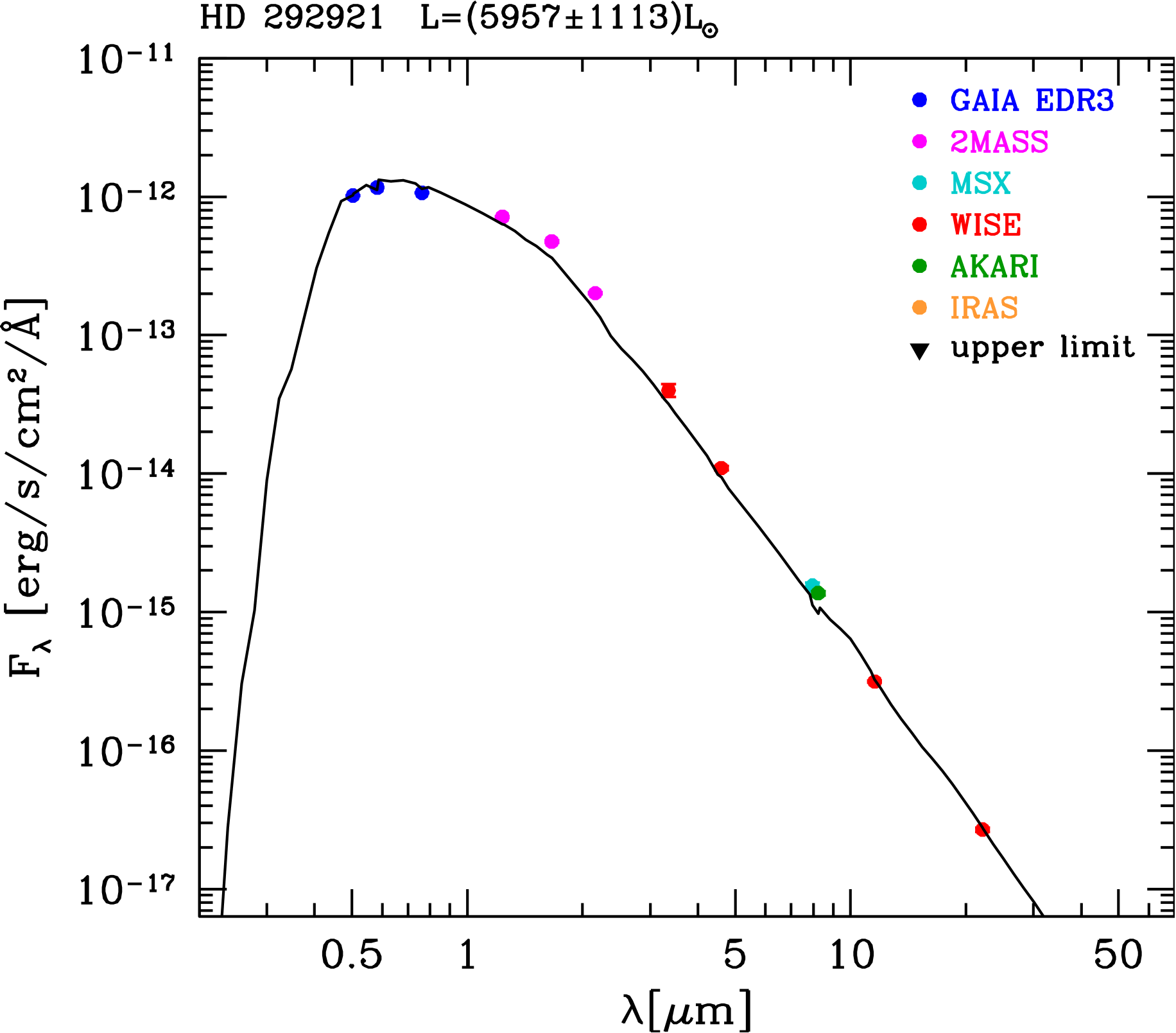}}
\end{minipage}
\begin{minipage}{0.23\textwidth}
\resizebox{\hsize}{!}{\includegraphics{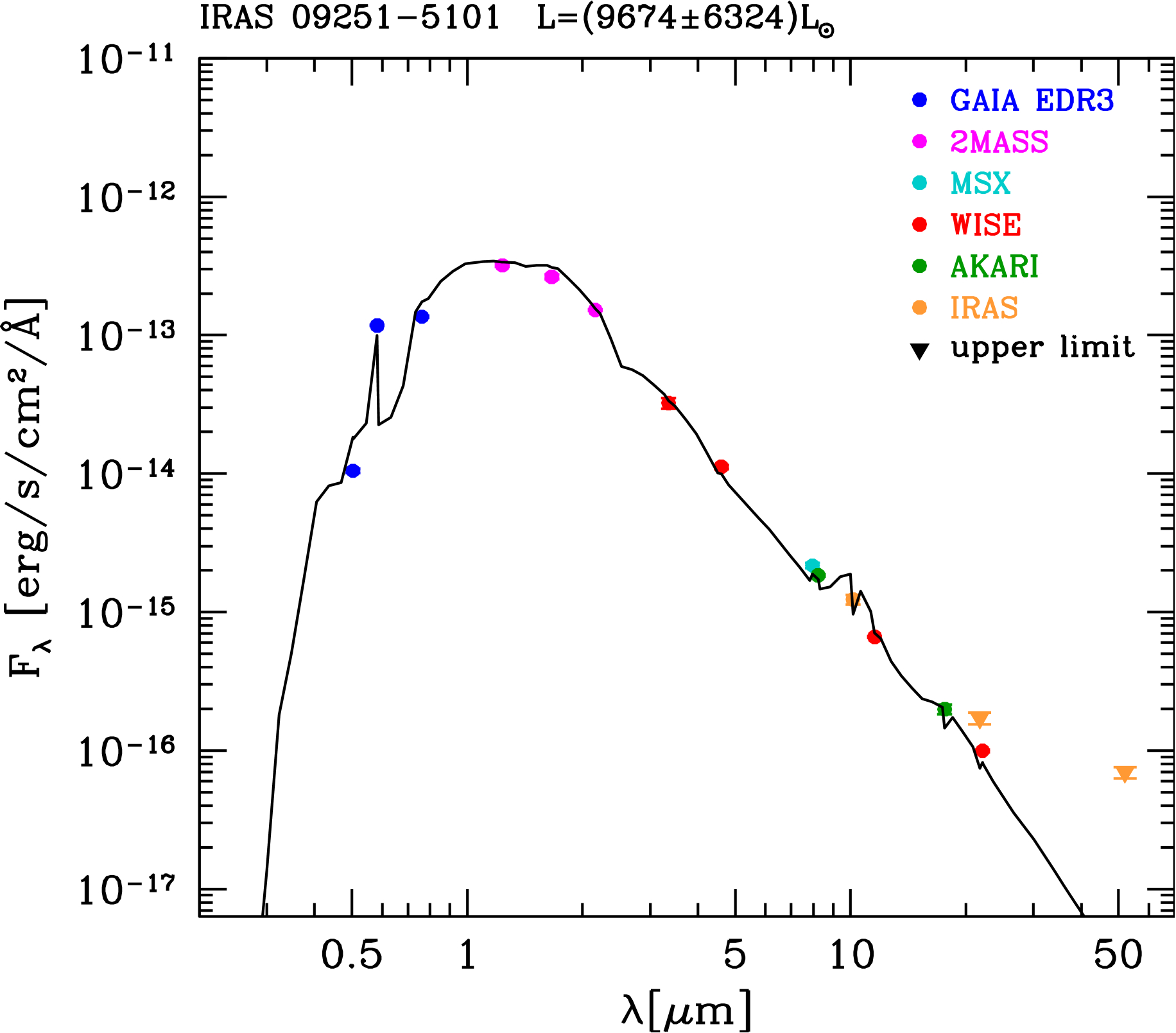}}
\end{minipage}
\begin{minipage}{0.23\textwidth}
\resizebox{\hsize}{!}{\includegraphics{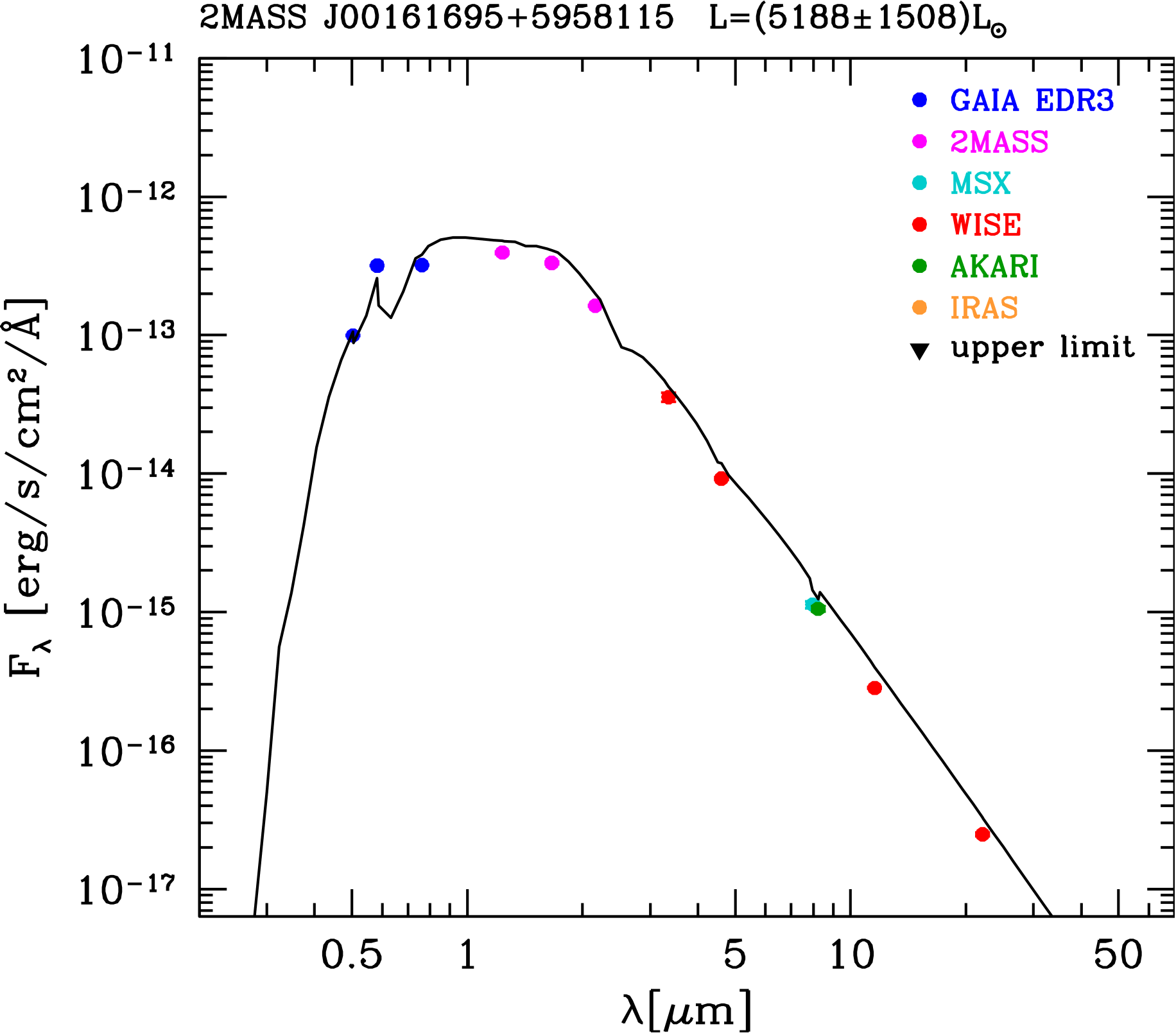}}
\end{minipage}\\
\caption{SED fitting of TP-AGB stars of type M, S and C 
in Milky Way open clusters. Interstellar extinction $A_V$ is taken from the catalog of \citet{Dias_etal_21}. 
Each star is placed at the distance given by its own \gaia\ EDR3 parallax (no ZP correction). Photometric data (with error bars) from various sources are displayed with coloured symbols. Upper limits (denoted with filled triangles) are not considered in the fits. Note that all sources are consistent with being TP-AGB stars, except for NIKC 3-81 which is probably a R-hot star, placed below the RGB tip.}
\label{fig_sedstars}
\end{center}
\end{figure*}

\subsection{Evolutionary models: current core mass}
\label{ssec_cmlr}
The bolometric luminosity derived from the SED fitting is used to estimate the current mass of the core, $\Mc$, for each TP-AGB star of spectroscopic type M, S, and C in our sample. 
This is done by using  the predictions of TP-AGB stellar models. We adopt the core mass-luminosity (CMLR) relation, that describes a direct proportionality between the mass of the core \Mc\ and  the quiescent luminosity $L_{\rm CMLR}$ sustained by the H-burning shell (when the He-shell is almost extinguished).
The existence of a positive correlation between the luminosity of a shell-burning
star and the mass of its degenerate core -- almost independent of the stellar mass, as is the case for RGB and TP-AGB stars --, was explained and extensively investigated in several works carried out in the past \citep{Eggleton_67,Paczynski_70, Tuchman_etal_83, BoothroydSackmann_88}.

\begin{figure}[h!]
\centering
\resizebox{\hsize}{!}{\includegraphics{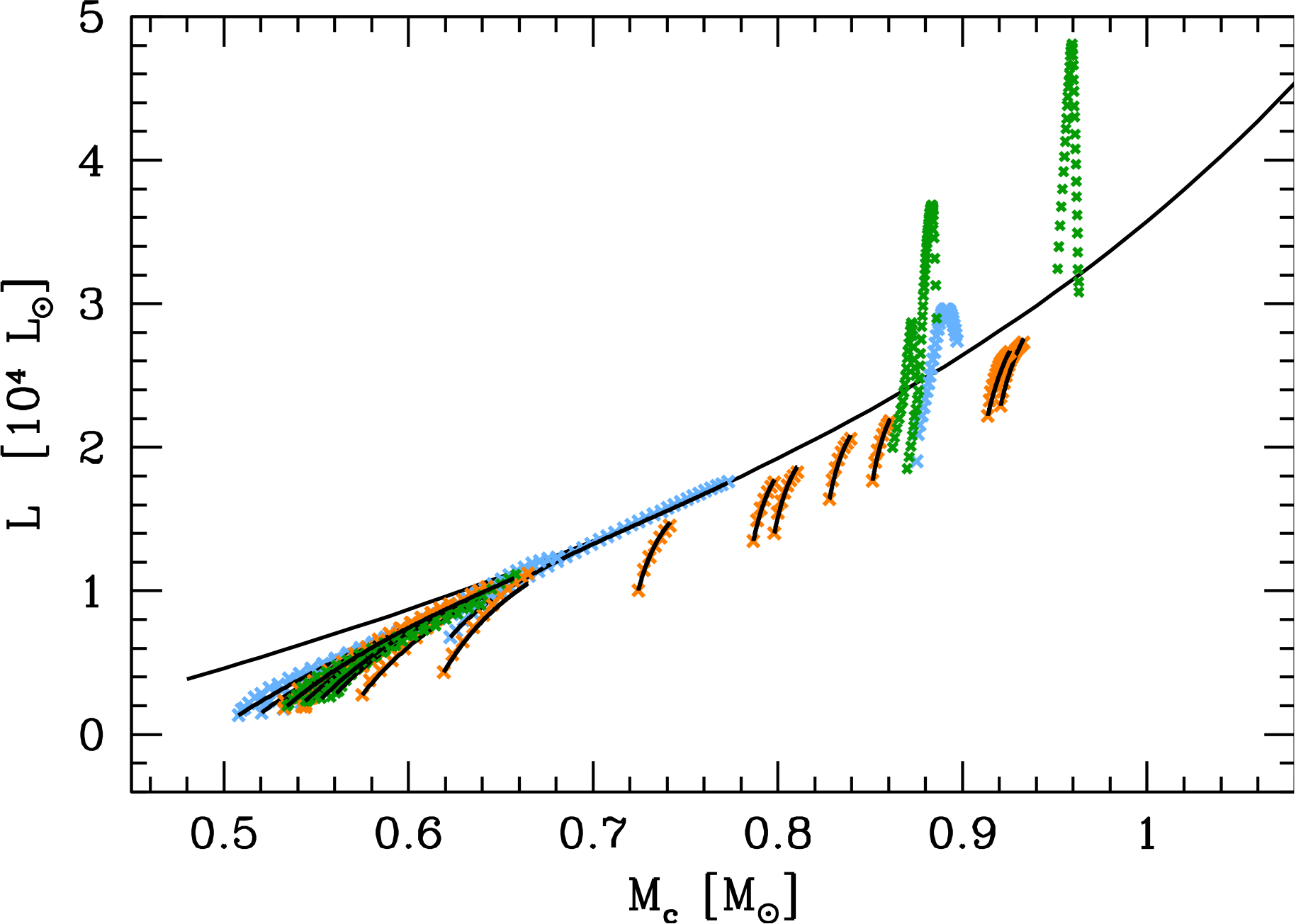}}
\caption{Core mass - luminosity relation of TP-AGB stars. TP-AGB model sequences (crosses) show the evolution of the core mass and luminosity during the quiescent stages of the TP cycles, for a few values of the initial stellar mass in the interval $1.0 \la M_{\rm i}^{\rm AGB}/\Msun \la 6.5$, and metallicities close to the solar value ($0.01 \le Z \le 0.02$). Each model (cross) is taken at the pre-flash luminosity maximum, just before the occurrence of a He-shell thermal instability. Various sources are included, namely:  \citet[][green]{KarakasLattanzio_02}; \citet[][orange - \texttt{FRUITY} database]{Cristallo_etal_11};   \citet[][blue - models computed with the \texttt{MESA} code]{Addari_2020}.
The solid black line running over the entire \Mc\ range is an analytic fit to the asymptotic core mass - luminosity relation for $Z=0.014$, given by Eq.~(\ref{eq-Lqa}+\ref{eq-Lqb1}).  Overplotted to the TP-AGB models we show the predictions (black lines) obtained with the synthetic CMLR (Eq.~\ref{eq-Lqa}+\ref{eq-Lqb1}+\ref{eq-Lqb2}), including the pre-pulses and assuming the same $M_{\rm c, 1TP}$, $L_{\rm 1TP}$, $Z$, and  \Mc, as in the stellar models.
}
\label{fig_cmlr}
\end{figure}

As we see in Fig.~\ref{fig_cmlr}, if we exclude the first few TPs of each sequence (the so-called subluminous pre-pulses), the CMLR relation is well defined by TP-AGB stellar models over the range $0.5 \la \Mc/\Msun \la 0.8$, and a general good agreement is found among stellar models computed with different codes.
Conversely, at larger core masses, $\Mc > 0.8 \,\Msun$, the CMLR breaks down due to the possible occurrence of HBB, which makes the models overluminous \citep{BoothroydSackmann_91, Blocker_91}. Moreover, the extreme dependence  of HBB on the adopted treatment of convection in the models causes a significant dispersion of the  predictions for the luminosity at the same \Mc.
Note, for example, the large differences between the \texttt{FRUITY} models (orange) and those computed by \citet[][green]{KarakasLattanzio_02} at $\Mc \ga 0.85\, \Msun$.
We also recall that these high-core mass stars are expected to recover the CMLR  towards the end of the TP-AGB evolution, when the drastic reduction of the envelope mass by stellar winds makes HBB extinguish \citep{VassiliadisWood_93, Marigo_etal_13}. This effect produces the bell-shape luminosity curves exhibited by some model sequences with $\Mc > 0.8 \,\Msun$ of Fig.~\ref{fig_cmlr}. Another possible violation of the CMLR towards higher luminosity may be caused by very deep 3DU episodes, when the efficiency is high \citep{Herwig_etal_98}.

We have paid particular attention to considering all these aspects. First, we do not need to worry about the HBB effect as the  luminosities estimated for our stars with $M_{\rm i}^{\rm AGB}\ga  3-4\, \Msun$ lie in the range $8\,000 \la L/\Lsun \la 15\,000$, and therefore they are well below the typical values expected when the process is operating ($L > 20\,000\, \Lsun$; see Fig.~\ref{fig_cmlr}).
Second, to homogeneously estimate \Mc\ for all our stars  we derive a synthetic fit formula for the CMLR, an approach that seems appropriate considering the well-defined and regular behavior of the models, as shown in Fig.~\ref{fig_cmlr} for $\Mc < 0.8\,\Msun$. 
The quiescent luminosity is expressed as a function of the current core  mass, \Mc,  the core mass at the first thermal pulse, $M_{\rm c,1TP}$, and the metallicity $Z$:
\bsubeqa{eq-Lq}{eq-Lqa}
L_{\rm CMLR} \!=\! & & (25\,667 + 481\, z)(\Mc - 0.446)    \\
\nsubeqn{eq-Lqb1}
& + & 10^{2.684 \,+\, 1.643M_{\rm c} } \\
\nsubeqn{eq-Lqb2}
& - & 10^{\,\xi \,-\, (M_{\rm c,1TP} - 0.446) (M_{\rm c} - M_{\rm c,1TP})/ 0.012 }\, ,
\label{eq_cmlr}
\esubeqa
where $z=Z/Z_{\odot}$ and $Z_{\odot}=0.014$. The relation is valid for  $0.50 \la \Mc/\Msun \la 0.95$  and $0.01 \la Z \la 0.02$, in absence of HBB. The metallicity range ($\mathrm { -0.15 \la [Fe/H] \la +0.15 }$) is suitable for the open clusters analyzed in this work, except for Trumpler 5   ( $\mathrm{Fe/H}\simeq -0.37$) and its carbon stars V$^\ast$ V493 Mon that we treat separately.

The functional form for $L_{\rm CMLR}$ is essentially the same as the one proposed by \citet[][their equation 5]{WagenhuberGroenewegen_98}, but with some important changes.
First, we derive new numerical coefficients by fitting the pre-flash luminosity for a high number (291) of thermal pulses taken from the large \texttt{FRUITY}\footnote{Full-network Repository of Updated Isotopic Tables and Yields: http://fruity.oa-teramo.inaf.it/} database.
We select all TP-AGB models  with $1.5\le M_{\rm i}^{\rm AGB}/\Msun \le 6.0$ and  metallicity $0.01\le Z \le 0.02$. We also add some additional TP-AGB tracks (132 TPs) computed with the \texttt{MESA} code \citep{Paxton_etal_11} for $M_{\rm i}^{\rm AGB}=1.7,\, 1.9, \,2.0, \,2.5,\, 3.0\, \Msun$ and  $Z=0.014$  \citep{Addari_2020}, and the TP-AGB tracks (152 TPs) with $1.0 \le M_{\rm i}^{\rm AGB}/\Msun \le 2.5\, \Msun$  and $Z=0.012,\,0.02$ published by \citet{KarakasLattanzio_02}.
In total our reference data-set includes 575 models with known \Mc, $M_{\rm c,1TP}$, $Z$, and $L$.
A standard chi-square minimization technique is adopted to obtain the coefficients.
We note that at these solar-like metallicities the most massive \texttt{FRUITY} models do not show evident overluminosity effects due to the occurrence of HBB. This circumstance turns out to be useful as it allows us to map the CMLR at high \Mc\ (where other models predict a strong HBB), which is  necessary if we deal with massive TP-AGB stars at the end of their evolution when HBB is off.

The terms (\ref{eq-Lqa}) and (\ref{eq-Lqb1}) describe the asymptotic behavior when thermal pulses have reached the full-amplitude regime, while the negative term (\ref{eq-Lqb2}) provides a correction to account for the first sub-luminous pulses.  This phase is particularly important for low-mass stars with $\Mc \la 0.65\, \Msun$, as their brightening rate is 
rather slow and they reach the asymptotic regime after a relatively high number of thermal pulses. Neglecting the sub-luminous pulses would therefore lead to underestimating \Mc\ for these stars.
We emphasize again that the analytic relation of Eq.(3) is meant to describe the behavior of the quiescent luminosity of a TP-AGB star in absence of HBB and very efficient 3DU.

The CMLR is characterized by a mild dependence on metallicity: at increasing $Z$ the luminosity is somewhat higher at given \Mc.
In the term~(\ref{eq-Lqa}) we include linear dependence with the metallicity, expressed through the scaling factor $z=Z/Z_{\odot}$, where the reference solar metallicity is $Z_{\odot}=0.014$ (in place of $Z_{\odot}=0.02$ as in \citet{WagenhuberGroenewegen_98}).

The parameter $\xi$ in Eq.~(\ref{eq-Lqb2}) deserves some comments. It has a precise physical meaning as it determines the luminosity, $L_{\rm 1TP}$, at the first thermal pulse.
In fact, if we set $\Mc=M_{\rm c,1TP}$ in Eq.~(3) it is clear that $10^\xi$ measures the luminosity difference between the CMLR in the asymptotic regime (\ref{eq-Lqa} + \ref{eq-Lqb1}) and the first thermal pulse. The best fitting returns a global mean value $\xi=3.596$.
However, if we know the values $M_{\rm c, 1TP}$ and $L_{\rm 1TP}$  of a specific TP-AGB model, it is straightforward to derive the corresponding parameter $\xi$. 
In this case $\xi$ is not a constant, but it varies as a function of the TP-AGB model under consideration.

We checked the accuracy of our fit relation for $L_{\rm CMLR}$ against the reference data-set of TP-AGB models at solar-like metallicities, as illustrated in Fig.~\ref{fig_cmlr}. 
The performance is quite good: not only the relation reproduces satisfactorily the asymptotic behavior, but also recovers well the initial turn-on phase of the pre-pulses.
This can be appreciated by comparing  the various TP-AGB model sequences (crosses) with the synthetic CMLR (black lines), assuming the same $M_{\rm c,1TP}$, $L_{\rm 1TP}$  and metallicity $Z$ of the stellar models.
As we can see, the formula reproduces the behavior of the different tracks very well, including the initial turn-on part.
Considering the whole set of $575$ luminosity points, the median relative error of the fit formula is $\simeq 4\%$, while  for 70\%, 80\%, 90\% of the cases the synthetic predictions for $L$ deviate from the models less than 7\%, 9\%, 14\%.

In this work, however, we employ the fit formula in the opposite way, that is to infer \Mc\ starting from a known luminosity value.
In brief, to estimate \Mc\ from a given observed luminosity, $L_{\rm obs}$, we proceed as follows. 
For each TP-AGB star member of an open cluster with known age and metallicity, its initial mass $M_{\rm i}^{\rm AGB}$ is obtained from the corresponding \texttt{PARSEC} isochrone. Then, we extract $M_{\rm c,1TP}$ and $L_{\rm 1TP}$ from the \texttt{PARSEC} stellar evolutionary track of initial mass $M_{\rm i}^{\rm AGB}$ and metallicity $Z$. Finally, with the aid of a standard root-finding technique we solve the equation $L_{\rm obs}-L_{\rm CMLR}(\Mc, M_{\rm c,1TP},L_{\rm 1TP,},Z)=0$ and get \Mc.  We can also compute the error on \Mc\ due to the uncertainty on $L_{\rm obs}$, obtained with Eq.~(\ref{eq_deltal}), which depends on both flux and distance errors. Adopting Eq.~(3) for $L_{\rm CMLR}$, it is straightforward to convert $\sigma_L$ into the statistical uncertainty of the core mass, $\sigma_{M_{\rm c, CMLR}}$, by using the standard error propagation law.
We find that for the noZP case $\sigma_{M_{\rm c, CMLR}}$ varies in the range $\simeq 0.006-0.125\, \Msun$, with a median value of $\simeq 0.023\, \Msun$. Among the 19 TP-AGB stars examined, 13 (16) have $\sigma_{M_{\rm c, CMLR}} \le 0.03, (0.05)\, \Msun$. The largest uncertainties with $\sigma_{M_{\rm c, CMLR}} \ga 0.1\, \Msun$ apply to the C star IRAS 19582+2907 and the M star IRAS 09251-5101. The other two cases, L21ZP and G21ZP, show somewhat different results for individual objects, but the order of magnitude and the typical range for $\sigma_{M_{\rm c, CMLR}}$ remain the same.

At this point it is worth analyzing the accuracy of the method to infer \Mc.
Clearly, intrinsic differences in luminosity and core mass exist among different sets of TP-AGB models, but in the mass range not affected by the occurrence of HBB they appear to be small.
\begin{table}
\begin{center}
\begin{threeparttable}
\scriptsize
\caption{Comparison of \Mc\ predicted by various sets of full TP-AGB models and with the synthetic CMLR.}
\label{tab_cmlr}
\noindent 
\begin{tabular}{@{}cccc}
\hline\hline
\multicolumn{1}{c}{} & FRUITY & K02 & MESA \\
\hline
\multicolumn{1}{r}{$L_{\rm 1TP}\,[\Lsun]$} & 
2346 & 1982 & 2208 \\
\multicolumn{1}{r}{$M_{\rm c, 1TP}\,[\Msun]$} & 
0.533 &   0.534 &   0.542 \\
\multicolumn{1}{c}{$Z$} & 
0.014 &   0.02 &   0.014 \\
\hline
\multicolumn{1}{c}{$L\,[\Lsun]$} & 
\multicolumn{3}{c}{$M_{\rm c}\,[\Msun]$} \\
\hline
4000 &   0.552 &    0.550 &   0.550 \\ 
6000 &   0.576 &    0.580 &   0.577 \\ 
8000 &   0.604 &    0.615 &   0.610 \\ 
\hline
\multicolumn{4}{c}{Results with the fit relation of Eq.~(3)} \\
\hline
4000 &   0.552 &    0.554 &   0.558 \\ 
6000 &   0.578 &    0.579 &   0.581 \\ 
8000 &   0.609 &    0.609 &   0.610 \\
\hline
\end{tabular}
\noindent 
\footnotesize{{\bf Notes:} Full TP-AGB models are from \citet[][\texttt{FRUITY}]{Cristallo_etal_11}, \citet[][K02]{KarakasLattanzio_02}, and \citet[][\texttt{MESA}]{Addari_2020}.
}
\end{threeparttable}
\end{center}
\end{table}
For example, let us consider a TP-AGB model with $M_{\rm i}^{\rm AGB}=2.5\,\Msun$ and compare the results from \citet[][K02]{KarakasLattanzio_02} for $Z=0.02$,  \texttt{FRUITY} and \texttt{MESA} for $Z=0.014$.
Taking a small grid of luminositiy, $L= 4\,000,6\,000,8\,000 \,\Lsun$, the different  models\footnote{In all cases a linear interpolation in $L$ is adopted to get \Mc.} predict the values of \Mc\ reported in Table~\ref{tab_cmlr}.
Overall, the differences in \Mc\ are modest, ranging within $ 0.002-0.01\, \Msun$. 
Using our fit relation with $M_{\rm c, 1TP}$ and $L_{\rm 1TP }$ extracted from the three TP-AGB models, we also obtain very similar \Mc, the deviations being of the order of  few $10^{-3}\, \Msun$ in all cases.
From these tests we are therefore confident that the formula (Eq.~3) is a reliable tool to derive \Mc\ of the TP-AGB stars in our sample.

Comparing the dispersion in \Mc\ produced by different TP-AGB models at fixed luminosity with $\sigma_{M_{\rm c, CMLR}}$ obtained for the observed stars (Eq.~\ref{eq_deltal}), it turns out that in many cases the latter uncertainty  -- due to distance and flux errors -- is larger (see Table~\ref{tab_sedfit}). In other words, differences in \Mc\ among TP-AGB models at solar-like metallicities are mostly washed away by the observational errors.

\begin{figure*}[ht!]
\begin{center}
\resizebox{0.9\textwidth}{!}{\includegraphics{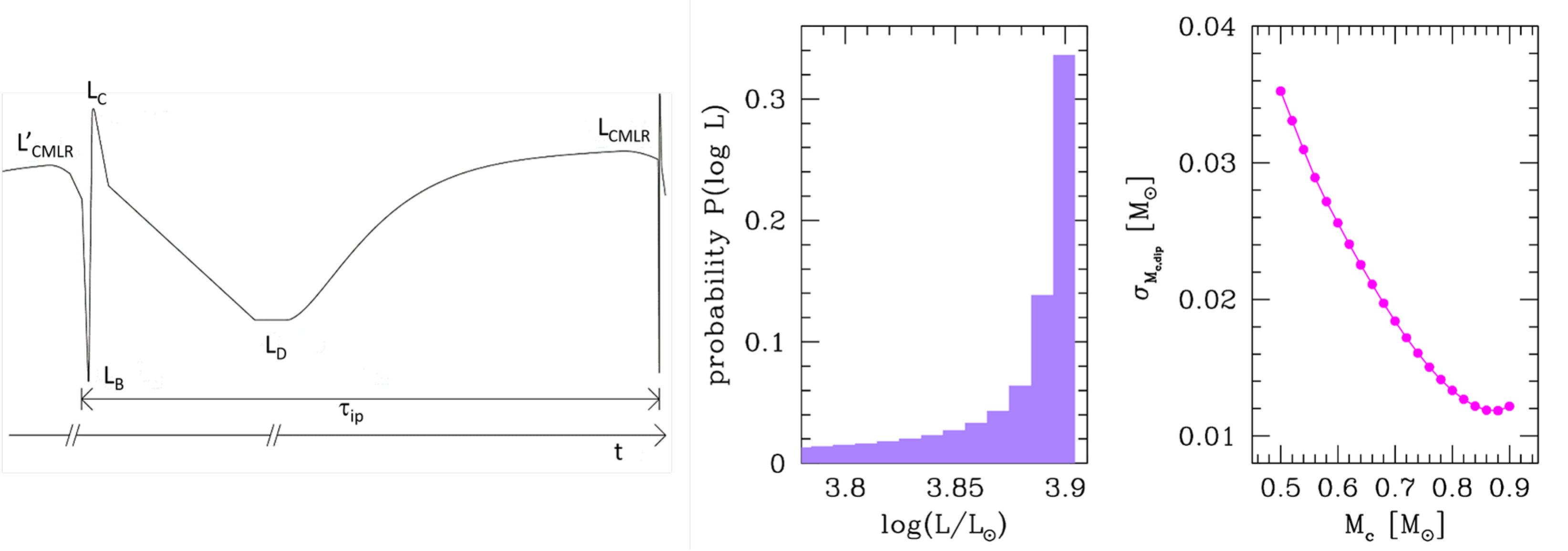}}
\caption{Effects of the slow luminosity dip driven by thermal pulses.
\emph{Left panel}: sketch of the luminosity evolution as a function of time between two consecutive thermal pulses, with interpulse period $\tau_{\rm ip}$.
The labels indicate a few relevant stages:
the quiescent luminosity $L_{\rm CMLR}$, 
the rapid dip $L_{\rm B}$ and peak $L_{\rm C}$, and the slow dip $L_{\rm D}$. The time axis is stretched around the the first TP, from  $L'_{\rm CMLR}$ to $L_{\rm D}$. These short-lived stages  typically covers $\la 0.1$ of the interpulse period.
\emph{Middle panel}: probability distribution (purple histogram) of the luminosity along the slow dip
from $L_{\rm D}$ to $L_{\rm CMLR}$. The width of the luminosity bins is set to 0.01 dex.
We consider a TP-AGB star with solar metallicity, $\Mc=0.6\, \Msun$,  and $L_{\rm CMLR} \simeq 8000 \, \Lsun$, and $L_{\rm D}\simeq 3400\,\Lsun$.
\emph{Right panel}: statistical (positive only) uncertainty  of the core mass derived from the CMLR, due to the probability that a star intercepts the observed luminosity while evolving on a slow dip of a TP at higher \Mc. The uncertainty is shown as a function of the core mass in the range $0.5 \le \Mc/\Msun \le 0.9$. The corresponding  quiescent luminosity in the asymptotic regime  spans the interval 
$4\,600\la L_{\rm CMLR}/\Lsun \la 26\,400$. 
See the text for more details.
}
\label{fig_ldip}
\end{center}
\end{figure*}

A further effect is related to the changes in luminosity caused by the occurrence of thermal pulses (see left panel of Fig.~\ref{fig_ldip}).
The stars in our sample, in principle, may not lie on the CMLR for this reason \citep{Marigo_etal_99,BoothroydSackmann_88}.
The probability and extent of these deviations from the CMLR were thoroughly quantified \citep{WoodZarro_81,BoothroydSackmann_88, WagenhuberGroenewegen_98}.  Finding a star above the CMLR is always quite unlikely (with a probability $\simeq 1-2$\%), as the luminosity rise (up to $L_{\rm C}$ in Fig.~\ref{fig_ldip})  when a TP occurs is extremely short-lived. Instead, the probability increases appreciably  to observe TP-AGB stars below the CMLR as they are evolving on the slow luminosity dip following a thermal pulse (from $L_{\rm D}$ to $L_{\rm CMLR}$), when the nuclear activity of the H-burning shell starts recovering during the power-down phase of a He-shell flash. At the luminosity minimum, $L_{\rm D}$, models predict that the nuclear luminosities produced by the H- and He-burning shells are comparable, $L_{\rm H} \simeq L_{\rm He} \simeq 0.5\, L_{\rm D}$.
The maximum extent of $L_{\rm D}$ below the CMLR can be as much as $\approx 1$ mag, for the models with the lowest core mass ($\Mc \approx 0.5-0.6\, \Msun$). But what really matters is the time spent on these fainter stages with respect to the total duration of the pulse cycle, since this controls the probability of observing a star below the CMLR. An example  of the probability distribution of the luminosity on the slow dip is given in the top-right panel of Fig.~\ref{fig_ldip}.

Using the predictions of TP-AGB models \citep[e.g.,][]{BoothroydSackmann_88, WagenhuberGroenewegen_98} it turns out that
the fraction of the pulse cycle that occurs on the slow luminosity dip  becomes smaller at increasing core mass. For example, at solar metallicity,  the probability of finding a star with $\Mc \simeq 0.55, 0.65, 0.75 \,\Msun$ and a luminosity between $L_{\rm CMLR}$ down to $\Delta\log(L)=0.1$ dex below (i.e. within 0.25 mag fainter), is equal to 57\%, 67\%, 77\% respectively. It means that  stars with higher \Mc\ stay statistically closer to the CMLR (in absence of HBB).

Given these arguments, we can now ask what is the order of magnitude of the error we commit in using the CMLR to derive \Mc, that is under the assumption  that the observed luminosity, $L_{\rm obs}$, lies exactly on the CMLR. The answer is that the CMLR provides a lower limit to the true \Mc, as there is always a finite probability that $L_{\rm obs}$ is intercepted while the star is evolving in the slow luminosity dip of a pulse cycle at higher \Mc. 
 
To quantify the associated statistical error we proceed as follows. 
Let us denote with $M_{\rm c, CMLR}$ the core mass predicted by the CMLR for $L_{\rm CMLR}=L_{\rm obs}$. First, we note that the maximum range of \Mc, compatible with the observed $L_{\rm obs}$, goes from $M_{\rm c, CMLR}$ to the maximum core mass, $M_{\rm c, D}$, at which the minimum luminosity, reached during the flash-driven variations, equals the observed luminosity, that is  $L_{\rm D}=L_{\rm obs}$. Second, we compute the probability density function that  a star with  $M_{\rm c, CMLR} \le \Mc \le M_{\rm c, D}$ attains $L_{\rm obs}$ during a luminosity dip. Then,  we estimate the standard deviation of this distribution, $\sigma_{M_{\rm c}, {\rm dip}}$, by extracting the value of the core mass, $\tilde M_{\rm c}$, such that the probability to intercept $L_{\rm obs}$ from all luminosity dips with $M_{\rm c, CMLR} \le \Mc \le \tilde M_{\rm c}$ is equal to 68 \%. Finally we define $\sigma_{M_{\rm c}, {\rm dip}} =\tilde M_{\rm c}-M_{\rm c, CMLR}$ as the statistical uncertainty on the estimation of \Mc\ due to the flash-driven luminosity variations, for a given $L_{\rm obs}$.

We repeated the procedure just described over a range of quiescent luminosities corresponding to $0.5 \le \Mc/\Msun \le 0.9$. The results are shown in the right panel of Fig.~\ref{fig_ldip}. We find that $\sigma_{M_{\rm c}, {\rm dip}}$ decreases at increasing \Mc\ (or equivalently $L$), and  typically ranges from $\approx 0.035 \, \Msun$ to $\approx 0.01\, \Msun$. The inverse correlation of  $\sigma_{M_{\rm c}, {\rm dip}}$ with \Mc\ can be  explained as the result of the time spent in the slow luminosity dip, which is shorter at higher \Mc. 

The final estimate of the core mass is computed as follows
\begin{equation}
\Mc = M_{\rm c, CMLR} - \sigma_{M_{\rm c, CMLR}} + \sqrt{\sigma_{M_{\rm c, CMLR}}^2 + \sigma_{M_{\rm c}, {\rm dip}}^2}\, .
\end{equation}
Note that the uncertainty due to the slow luminosity dip, $\sigma_{M_{\rm c}, {\rm dip}}$, is only positive and it is combined with  $\sigma_{M_{\rm c, CMLR}}$ through the standard summation in quadrature. The results for the core mass of the TP-AGB stars of the spectroscopic sample are reported in Tables~\ref{tab_sedfit} and \ref{tab_sedfit:D21}.
As expected, we see that the asymmetry of the  error-bars increases for the stars with $\Mc \la 0.6\, \Msun$ (such as V$^\ast$ V493 Mon, [W71b] 030-31, C$^\ast$ 908,  S1$^\ast$ 338, [D75b] Star 30,  IRAS 23455-6819, HD 292921), as these objects have a higher probability to be observed while they are evolving on the slow luminosity dip, which corresponds to a larger $\sigma_{M_{\rm c}, {\rm dip}}$. 

\begin{table*}
\centering
\begin{threeparttable}
\tiny
\caption{Results from SED fitting and TP-AGB models. Cluster ages and $A_V$ from \citet{Cantat-Gaudin_etal_20}.}
\label{tab_sedfit}
\begin{tabular*}{\textwidth}{@{\extracolsep{\fill}}cccccccccccc@{}}
\noalign{\smallskip}
\hline\hline
\noalign{\smallskip}
\multicolumn{1}{c}{star} & 
\multicolumn{1}{c}{$\chi^2$} & 
\multicolumn{1}{c}{$A_V$\tnote{(b)}} &
\multicolumn{1}{c}{$\dot M_{\rm dust}$} & 
\multicolumn{1}{c}{$\tau_{11.3} /\tau_{10} $} & 
\multicolumn{1}{c}{$M_{\rm i}^{\rm AGB}$\tnote{(c)}} &
\multicolumn{1}{c}{$D$} &
\multicolumn{1}{c}{$L$} &
\multicolumn{1}{c}{$\Mc$} \\
 & & \multicolumn{1}{c}{[mag]} & 
\multicolumn{1}{c}{[${M_{\odot}\, {\rm yr}^{-1}}$]}  & & \multicolumn{1}{c}{[${M_{\odot}}$]} &
\multicolumn{1}{c}{[pc]} &
\multicolumn{1}{c}{[${L_{\odot}}$]} &
\multicolumn{1}{c}{[${M_{\odot}}$]}\\
\noalign{\smallskip}
\hline
\noalign{\smallskip}
              V* V493 Mon &  4.431E+01 & $ 1.38\pm 0.10 $     & 1.19E-09 & 2.00E-01 & $1.328\pm 0.107$     & $ 2714\pm 268  $     & $ 4091\pm 867  $     & $0.553^{+0.034}_{-0.011}$      \\
\multicolumn{6}{c}{}                               & $ 2661\pm 258  $     & $ 3933\pm 819  $     & $0.550^{+0.034}_{-0.010}$      \\
\multicolumn{6}{c}{}                               & $ 2711\pm 268  $     & $ 4081\pm 864  $     & $0.553^{+0.034}_{-0.011}$      \\
$\mathrm{[W71b]\;030-31}$ &  5.822E+01 & $ 2.53\pm 0.10 $     & 3.37E-09 & 2.00E-01 & $1.473\pm 0.120$     & $ 1790\pm 107  $     & $ 3520\pm 500  $     & $0.545^{+0.032}_{-0.006}$      \\
\multicolumn{6}{c}{}                               & $ 1788\pm 106  $     & $ 3512\pm 498  $     & $0.545^{+0.032}_{-0.006}$      \\
\multicolumn{6}{c}{}                               & $ 1698\pm 96   $     & $ 3167\pm 430  $     & $0.538^{+0.035}_{-0.005}$      \\
                   C* 908 &  1.549E+02 & $ 0.36\pm 0.09 $     & 1.69E-10 & 4.00E-02 & $1.660\pm 0.155$     & $ 3831\pm 361  $     & $ 5355\pm 1133 $     & $0.578^{+0.034}_{-0.015}$      \\
\multicolumn{6}{c}{}                               & $ 3293\pm 267  $     & $ 3956\pm 732  $     & $0.553^{+0.034}_{-0.009}$      \\
\multicolumn{6}{c}{}                               & $ 3735\pm 343  $     & $ 5088\pm 1052 $     & $0.573^{+0.033}_{-0.014}$      \\
                 Case 588 &  1.715E+02 & $ 0.65\pm 0.10 $     & 2.83E-10 & 2.48E-02 & $1.834\pm 0.229$     & $ 4669\pm 748  $     & $ 9293\pm 3316 $     & $0.652^{+0.055}_{-0.050}$      \\
\multicolumn{6}{c}{}                               & $ 3706\pm 471  $     & $ 5854\pm 1702 $     & $0.591^{+0.037}_{-0.024}$      \\
\multicolumn{6}{c}{}                               & $ 4999\pm 857  $     & $10657\pm 4043 $     & $0.675^{+0.065}_{-0.061}$      \\
                 BM IV 90 &  3.418E+01 & $ 1.35\pm 0.12 $     & 1.59E-10 & 1.96E-02 & $1.941\pm 0.224$     & $ 2884\pm 211  $     & $10421\pm 1874 $     & $0.681^{+0.034}_{-0.028}$      \\
\multicolumn{6}{c}{}                               & $ 2640\pm 177  $     & $ 8730\pm 1463 $     & $0.652^{+0.032}_{-0.022}$      \\
\multicolumn{6}{c}{}                               & $ 2853\pm 207  $     & $10196\pm 1819 $     & $0.677^{+0.035}_{-0.027}$      \\
                   MBS 75 &  1.315E+02 & $ 1.01\pm 0.10 $     & 1.89E-10 & 2.38E-02 & $1.977\pm 0.229$     & $ 2050\pm 107  $     & $11473\pm 2028 $     & $0.696^{+0.036}_{-0.030}$      \\
\multicolumn{6}{c}{}                               & $ 1871\pm 89   $     & $ 9551\pm 1602 $     & $0.663^{+0.032}_{-0.024}$      \\
\multicolumn{6}{c}{}                               & $ 2047\pm 107  $     & $11432\pm 2021 $     & $0.695^{+0.036}_{-0.030}$      \\
                  Case 63 &  6.298E+01 & $ 2.63\pm 0.07 $     & 1.97E-10 & 1.84E-02 & $2.063\pm 0.229$     & $ 1745\pm 95   $     & $ 6887\pm 907  $     & $0.611^{+0.029}_{-0.013}$      \\
\multicolumn{6}{c}{}                               & $ 1594\pm 80   $     & $ 5750\pm 705  $     & $0.590^{+0.030}_{-0.010}$      \\
\multicolumn{6}{c}{}                               & $ 1719\pm 92   $     & $ 6686\pm 872  $     & $0.607^{+0.029}_{-0.013}$      \\
                 Case 473 &  2.577E+02 & $ 4.26\pm 0.04 $     & 7.87E-11 & 8.35E-03 & $2.326\pm 0.189$     & $ 3648\pm 443  $     & $11162\pm 2908 $     & $0.673^{+0.050}_{-0.045}$      \\
\multicolumn{6}{c}{}                               & $ 3414\pm 388  $     & $ 9773\pm 2394 $     & $0.650^{+0.043}_{-0.036}$      \\
\multicolumn{6}{c}{}                               & $ 3367\pm 378  $     & $ 9509\pm 2300 $     & $0.646^{+0.041}_{-0.034}$      \\
               WRAY 18-47 &  5.362E+01 & $ 0.98\pm 0.09 $     & 2.12E-10 & 3.41E-02 & $2.596\pm 0.212$     & $ 2782\pm 173  $     & $10958\pm 1661 $     & $0.665^{+0.033}_{-0.025}$      \\
\multicolumn{6}{c}{}                               & $ 2521\pm 142  $     & $ 8999\pm 1257 $     & $0.634^{+0.030}_{-0.017}$      \\
\multicolumn{6}{c}{}                               & $ 2452\pm 134  $     & $ 8518\pm 1163 $     & $0.627^{+0.029}_{-0.015}$      \\
                 BM IV 34 &  9.637E+01 & $ 1.60\pm 0.11 $     & 8.44E-11 & 1.56E-02 & $3.540\pm 0.309$     & $ 4531\pm 433  $     & $15757\pm 3551 $     & $0.770^{+0.028}_{-0.023}$      \\
\multicolumn{6}{c}{}                               & $ 3848\pm 312  $     & $11366\pm 2235 $\tnote{(a)}     & $0.659^{+0.052}_{-0.046}$      \\
\multicolumn{6}{c}{}                               & $ 3733\pm 294  $     & $10695\pm 2051 $\tnote{(a)}     & $0.645^{+0.049}_{-0.044}$      \\
          IRAS 19582+2907 &  6.266E+02 & $ 3.94\pm 0.05 $     & 4.21E-11 & 9.94E-03 & $4.420\pm 0.419$     & $ 4938\pm 1288 $     & $10866\pm 5905 $\tnote{(a)}     & $0.648^{+0.127}_{-0.125}$      \\
\multicolumn{6}{c}{}                               & $ 3743\pm 740  $     & $ 6243\pm 2604 $\tnote{(a)}     & $0.542^{+0.072}_{-0.065}$      \\
\multicolumn{6}{c}{}                               & $ 4139\pm 905  $     & $ 7635\pm 3504 $\tnote{(a)}     & $0.576^{+0.089}_{-0.083}$      \\
                 Case 121 &  1.917E+01 & $ 1.63\pm 0.04 $     & 1.27E-10 & 2.08E-02 & $6.960\pm 0.800$     & $ 5028\pm 870  $     & $14148\pm 5075 $\tnote{(a)}     & $0.714^{+0.097}_{-0.095}$      \\
\multicolumn{6}{c}{}                               & $ 3856\pm 512  $     & $ 8323\pm 2314 $\tnote{(a)}     & $0.592^{+0.061}_{-0.054}$      \\
\multicolumn{6}{c}{}                               & $ 4454\pm 682  $     & $11103\pm 3543 $\tnote{(a)}     & $0.653^{+0.078}_{-0.074}$      \\
                  S1* 338 &  2.009E+02 & $ 1.22\pm 0.05 $     & 1.67E-12 & 2.31E-04 & $1.646\pm 0.150$     & $ 4708\pm 454  $     & $ 5012\pm 1107 $     & $0.572^{+0.033}_{-0.014}$      \\
\multicolumn{6}{c}{}                               & $ 4314\pm 382  $     & $ 4209\pm 862  $     & $0.558^{+0.034}_{-0.011}$      \\
\multicolumn{6}{c}{}                               & $ 4265\pm 373  $     & $ 4114\pm 835  $     & $0.556^{+0.034}_{-0.010}$      \\
$\mathrm{[D75b]\;Star\,30}$ &  4.446E+02 & $ 1.30\pm 0.02 $     & 6.70E-12 & 3.45E-04 & $1.927\pm 0.224$     & $ 3427\pm 375  $     & $ 2988\pm 689  $     & $0.539^{+0.036}_{-0.010}$      \\
\multicolumn{6}{c}{}                               & $ 3044\pm 296  $     & $ 2358\pm 486  $     & $0.523^{+0.034}_{-0.007}$      \\
\multicolumn{6}{c}{}                               & $ 3215\pm 330  $     & $ 2630\pm 571  $     & $0.530^{+0.035}_{-0.008}$      \\
                  CSS 291 &  8.390E+01 & $ 0.84\pm 0.09 $     & 5.74E-11 & 2.39E-03 & $2.130\pm 0.178$     & $ 2392\pm 141  $     & $ 7954\pm 1371 $     & $0.626^{+0.032}_{-0.020}$      \\
\multicolumn{6}{c}{}                               & $ 2190\pm 118  $     & $ 6668\pm 1082 $     & $0.604^{+0.030}_{-0.015}$      \\
\multicolumn{6}{c}{}                               & $ 2214\pm 121  $     & $ 6814\pm 1113 $     & $0.606^{+0.031}_{-0.016}$      \\
          IRAS 23455+6819 &  1.530E+02 & $ 2.85\pm 0.00 $     & 9.19E-10 & 2.56E-02 & $1.310\pm 0.100$     & $ 3044\pm 445  $     & $ 4872\pm 1480 $     & $0.567^{+0.035}_{-0.019}$      \\
\multicolumn{6}{c}{}                               & $ 2816\pm 381  $     & $ 4167\pm 1174 $     & $0.554^{+0.035}_{-0.014}$      \\
\multicolumn{6}{c}{}                               & $ 2679\pm 345  $     & $ 3773\pm 1014 $     & $0.547^{+0.034}_{-0.012}$      \\
                HD 292921 &  1.299E+02 & $ 1.48\pm 0.00 $     & 2.11E-11 & 5.59E-04 & $1.620\pm 0.144$     & $ 4054\pm 340  $     & $ 5981\pm 1134 $     & $0.588^{+0.032}_{-0.015}$      \\
\multicolumn{6}{c}{}                               & $ 3549\pm 261  $     & $ 4584\pm 774  $     & $0.563^{+0.031}_{-0.010}$      \\
\multicolumn{6}{c}{}                               & $ 3925\pm 319  $     & $ 5608\pm 1034 $     & $0.581^{+0.030}_{-0.014}$      \\
          IRAS 09251-5101 &  1.127E+02 & $ 2.86\pm 0.00 $     & 4.29E-10 & 2.55E-02 & $2.721\pm 0.221$     & $ 6978\pm 2211 $     & $ 9650\pm 6325 $     & $0.647^{+0.081}_{-0.078}$      \\
\multicolumn{6}{c}{}                               & $ 4928\pm 1103 $     & $ 4813\pm 2258 $\tnote{(a)}     & $0.505^{+0.069}_{-0.059}$      \\
\multicolumn{6}{c}{}                               & $ 6213\pm 1753 $     & $ 7651\pm 4483 $     & $0.620^{+0.050}_{-0.044}$      \\
  2MASS J00161695+5958115 &  9.889E+02 & $ 2.62\pm 0.01 $     & 1.67E-12 & 1.05E-04 & $2.743\pm 0.222$     & $ 4439\pm 609  $     & $ 4924\pm 1381 $    & $0.508^{+0.051}_{-0.036}$      \\
\multicolumn{6}{c}{}                               & $ 4190\pm 542  $     & $ 4387\pm 1163 $\tnote{(a)}     & $0.494^{+0.046}_{-0.031}$      \\
\multicolumn{6}{c}{}                               & $ 3952\pm 483  $     & $ 3903\pm 977  $\tnote{(a)}     & $0.481^{+0.042}_{-0.026}$      \\

\\\noalign{\smallskip}
\hline
\noalign{\smallskip}
\end{tabular*}
\tiny{{\bf Notes:} For each star the distance, luminosity and core mass are given for 3 choices of \gaia\ EDR3 parallax ZP correction (from top to bottom): noZP, L21ZP, and G21ZP. See the text for more details.
\\
\begin{tablenotes}[flushleft]
\item[(a)] Luminosity below the predicted onset of the TP-AGB phase. The core mass is estimated from the asymptotic CMLR (Eq. (\ref{eq-Lqa} + \ref{eq-Lqb1})).
\item[(b)] In the SED fitting the visual extinction is adjusted within the range $(A_V\pm 0.2)$ mag.
\item[(c)] The initial mass is obtained from the  \texttt{PARSEC} stellar isochrone of the cluster's age; the error bar corresponds to an uncertainty of $\pm 0.1$ dex 
in $\mathrm{\log(age)}$.
\end{tablenotes}
}
\end{threeparttable}
\end{table*}

\begin{table*}
\centering
\begin{threeparttable}
\tiny
\caption{Results from SED fitting and TP-AGB models. Cluster ages and $A_V$ from \citet{Dias_etal_21}.}
\label{tab_sedfit:D21}
\begin{tabular*}{\textwidth}{@{\extracolsep{\fill}}cccccccccccc@{}}
\noalign{\smallskip}
\hline\hline
\noalign{\smallskip}
\multicolumn{1}{c}{star} & 
\multicolumn{1}{c}{$\chi^2$} & 
\multicolumn{1}{c}{$A_V$\tnote{(b)}} &
\multicolumn{1}{c}{$\dot M_{\rm dust}$} & 
\multicolumn{1}{c}{$\tau_{11.3} /\tau_{10} $} & 
\multicolumn{1}{c}{$M_{\rm i}^{\rm AGB}$\tnote{(c)}} &
\multicolumn{1}{c}{$D$} &
\multicolumn{1}{c}{$L$} &
\multicolumn{1}{c}{$\Mc$} \\
 & & \multicolumn{1}{c}{[mag]} & 
\multicolumn{1}{c}{[${M_{\odot}\, {\rm yr}^{-1}}$]}  & & \multicolumn{1}{c}{[${M_{\odot}}$]} &
\multicolumn{1}{c}{[pc]} &
\multicolumn{1}{c}{[${L_{\odot}}$]} &
\multicolumn{1}{c}{[${M_{\odot}}$]}\\
\noalign{\smallskip}
\hline
\noalign{\smallskip}
              V* V493 Mon &  4.086E+01 & $ 2.00\pm 0.12 $     & 1.27E-09 & 2.00E-01 & $1.423\pm 0.123$     & $ 2714\pm 268  $     & $ 4518\pm 965  $     & $0.563^{+0.031}_{-0.012}$      \\
\multicolumn{6}{c}{}                               & $ 2661\pm 258  $     & $ 4343\pm 911  $     & $0.560^{+0.035}_{-0.011}$      \\
\multicolumn{6}{c}{}                               & $ 2711\pm 268  $     & $ 4506\pm 962  $     & $0.562^{+0.031}_{-0.012}$      \\
$\mathrm{[W71b]\;030-31}$ &  1.024E+02 & $ 3.11\pm 0.07 $     & 1.97E-09 & 2.00E-01 & $2.315\pm 0.188$     & $ 1790\pm 107  $     & $ 3840\pm 721  $     & $0.550^{+0.033}_{-0.009}$      \\
\multicolumn{6}{c}{}                               & $ 1788\pm 106  $     & $ 3831\pm 717  $     & $0.550^{+0.033}_{-0.009}$      \\
\multicolumn{6}{c}{}                               & $ 1698\pm 96   $     & $ 3455\pm 626  $     & $0.543^{+0.032}_{-0.007}$      \\
                   C* 908 &  1.474E+02 & $ 0.57\pm 0.12 $     & 1.69E-10 & 4.00E-02 & $1.491\pm 0.122$     & $ 3831\pm 361  $     & $ 5618\pm 1202 $     & $0.581^{+0.031}_{-0.015}$      \\
\multicolumn{6}{c}{}                               & $ 3293\pm 267  $     & $ 4151\pm 778  $     & $0.556^{+0.034}_{-0.009}$      \\
\multicolumn{6}{c}{}                               & $ 3735\pm 343  $     & $ 5338\pm 1117 $     & $0.577^{+0.034}_{-0.014}$      \\
                 Case 588 &  1.414E+02 & $ 1.09\pm 0.08 $     & 1.50E-10 & 2.00E-02 & $2.119\pm 0.177$     & $ 4669\pm 748  $     & $10880\pm 3715 $     & $0.677^{+0.061}_{-0.056}$      \\
\multicolumn{6}{c}{}                               & $ 3706\pm 471  $     & $ 6855\pm 1887 $     & $0.608^{+0.038}_{-0.027}$      \\
\multicolumn{6}{c}{}                               & $ 4999\pm 857  $     & $12477\pm 4542 $     & $0.704^{+0.072}_{-0.069}$      \\
                 BM IV 90 &  3.394E+01 & $ 1.40\pm 0.05 $     & 1.59E-10 & 1.94E-02 & $2.169\pm 0.181$     & $ 2884\pm 211  $     & $10464\pm 1682 $     & $0.667^{+0.034}_{-0.026}$      \\
\multicolumn{6}{c}{}                               & $ 2640\pm 177  $     & $ 8766\pm 1302 $     & $0.638^{+0.032}_{-0.019}$      \\
\multicolumn{6}{c}{}                               & $ 2853\pm 207  $     & $10238\pm 1632 $     & $0.663^{+0.033}_{-0.025}$      \\
                   MBS 75 &  1.204E+02 & $ 1.20\pm 0.08 $     & 1.50E-10 & 2.02E-02 & $1.943\pm 0.224$     & $ 2050\pm 107  $     & $11909\pm 1460 $     & $0.706^{+0.028}_{-0.021}$      \\
\multicolumn{6}{c}{}                               & $ 1871\pm 89   $     & $ 9913\pm 1127 $     & $0.672^{+0.027}_{-0.017}$      \\
\multicolumn{6}{c}{}                               & $ 2047\pm 107  $     & $11866\pm 1455 $     & $0.705^{+0.028}_{-0.021}$      \\
                  Case 63 &  7.246E+01 & $ 2.81\pm 0.03 $     & 1.97E-10 & 1.82E-02 & $1.992\pm 0.232$     & $ 1745\pm 95   $     & $ 7278\pm 924  $     & $0.621^{+0.028}_{-0.013}$      \\
\multicolumn{6}{c}{}                               & $ 1594\pm 80   $     & $ 6077\pm 716  $     & $0.599^{+0.030}_{-0.010}$      \\
\multicolumn{6}{c}{}                               & $ 1719\pm 92   $     & $ 7066\pm 888  $     & $0.617^{+0.030}_{-0.013}$      \\
                 Case 473 &  2.577E+02 & $ 4.26\pm 0.07 $     & 7.87E-11 & 8.35E-03 & $2.326\pm 0.189$     & $ 3648\pm 443  $     & $11162\pm 2968 $     & $0.673^{+0.051}_{-0.046}$      \\
\multicolumn{6}{c}{}                               & $ 3414\pm 388  $     & $ 9773\pm 2447 $     & $0.650^{+0.044}_{-0.037}$      \\
\multicolumn{6}{c}{}                               & $ 3367\pm 378  $     & $ 9509\pm 2351 $     & $0.646^{+0.042}_{-0.035}$      \\
               WRAY 18-47 &  5.634E+01 & $ 1.35\pm 0.12 $     & 1.59E-10 & 2.14E-02 & $2.553\pm 0.209$     & $ 2782\pm 173  $     & $12599\pm 1914 $     & $0.693^{+0.037}_{-0.031}$      \\
\multicolumn{6}{c}{}                               & $ 2521\pm 142  $     & $10347\pm 1448 $     & $0.655^{+0.032}_{-0.021}$      \\
\multicolumn{6}{c}{}                               & $ 2452\pm 134  $     & $ 9794\pm 1341 $     & $0.647^{+0.030}_{-0.019}$      \\
                 BM IV 34 &  9.945E+01 & $ 1.82\pm 0.07 $     & 8.44E-11 & 1.62E-02 & $3.252\pm 0.279$     & $ 4531\pm 433  $     & $16764\pm 3705 $     & $0.766^{+0.052}_{-0.050}$      \\
\multicolumn{6}{c}{}                               & $ 3848\pm 312  $     & $12093\pm 2325 $     & $0.716^{+0.024}_{-0.014}$      \\
\multicolumn{6}{c}{}                               & $ 3733\pm 294  $     & $11379\pm 2132 $     & $0.710^{+0.022}_{-0.012}$      \\
          IRAS 19582+2907 &  6.266E+02 & $ 3.94\pm 0.06 $     & 4.21E-11 & 9.94E-03 & $2.089\pm 0.175$     & $ 4938\pm 1288 $     & $10866\pm 5937 $     & $0.678^{+0.092}_{-0.090}$      \\
\multicolumn{6}{c}{}                               & $ 3743\pm 740  $     & $ 6243\pm 2622 $     & $0.598^{+0.047}_{-0.037}$      \\
\multicolumn{6}{c}{}                               & $ 4139\pm 905  $     & $ 7635\pm 3526 $     & $0.622^{+0.057}_{-0.051}$      \\
                 Case 121 &  2.643E+01 & $ 1.85\pm 0.03 $     & 1.27E-10 & 2.55E-02 & $2.858\pm 0.232$     & $ 5028\pm 870  $     & $14862\pm 5374 $     & $0.731^{+0.088}_{-0.087}$      \\
\multicolumn{6}{c}{}                               & $ 3856\pm 512  $     & $ 8743\pm 2456 $     & $0.643^{+0.032}_{-0.022}$      \\
\multicolumn{6}{c}{}                               & $ 4454\pm 682  $     & $11663\pm 3755 $     & $0.681^{+0.053}_{-0.049}$      \\
                  S1* 338 &  5.090E+02 & $ 1.59\pm 0.01 $     & 1.67E-12 & 1.04E-04 & $1.600\pm 0.140$     & $ 4708\pm 454  $     & $ 6151\pm 1210 $     & $0.590^{+0.032}_{-0.016}$      \\
\multicolumn{6}{c}{}                               & $ 4314\pm 382  $     & $ 5165\pm 933  $     & $0.573^{+0.032}_{-0.012}$      \\
\multicolumn{6}{c}{}                               & $ 4265\pm 373  $     & $ 5048\pm 902  $     & $0.571^{+0.032}_{-0.012}$      \\
$\mathrm{[D75b]\;Star30}$ &  4.141E+02 & $ 1.44\pm 0.03 $     & 1.67E-12 & 1.56E-04 & $2.068\pm 0.211$     & $ 3427\pm 375  $     & $ 3229\pm 757  $     & $0.540^{+0.033}_{-0.010}$      \\
\multicolumn{6}{c}{}                               & $ 3044\pm 296  $     & $ 2548\pm 535  $     & $0.525^{+0.034}_{-0.007}$      \\
\multicolumn{6}{c}{}                               & $ 3215\pm 330  $     & $ 2843\pm 627  $     & $0.532^{+0.035}_{-0.009}$      \\
                  CSS 291 &  8.374E+01 & $ 0.83\pm 0.00 $     & 5.74E-11 & 1.97E-03 & $2.119\pm 0.177$     & $ 2392\pm 141  $     & $ 7942\pm 1230 $     & $0.626^{+0.030}_{-0.018}$      \\
\multicolumn{6}{c}{}                               & $ 2190\pm 118  $     & $ 6658\pm 964  $     & $0.604^{+0.029}_{-0.014}$      \\
\multicolumn{6}{c}{}                               & $ 2214\pm 121  $     & $ 6803\pm 993  $     & $0.607^{+0.030}_{-0.014}$      \\
          IRAS 23455+6819 &  1.510E+02 & $ 2.90\pm 0.00 $     & 9.19E-10 & 2.56E-02 & $1.410\pm 0.122$     & $ 3044\pm 445  $     & $ 4966\pm 1554 $     & $0.571^{+0.036}_{-0.019}$      \\
\multicolumn{6}{c}{}                               & $ 2816\pm 381  $     & $ 4249\pm 1235 $     & $0.558^{+0.036}_{-0.015}$      \\
\multicolumn{6}{c}{}                               & $ 2679\pm 345  $     & $ 3847\pm 1068 $     & $0.551^{+0.035}_{-0.013}$      \\
                HD 292921 &  1.460E+02 & $ 1.48\pm 0.00 $     & 2.11E-11 & 5.90E-04 & $1.620\pm 0.144$     & $ 4054\pm 340  $     & $ 5957\pm 1113 $     & $0.587^{+0.031}_{-0.015}$      \\
\multicolumn{6}{c}{}                               & $ 3549\pm 261  $     & $ 4566\pm 758  $     & $0.563^{+0.031}_{-0.010}$      \\
\multicolumn{6}{c}{}                               & $ 3925\pm 319  $     & $ 5586\pm 1014 $     & $0.581^{+0.030}_{-0.013}$      \\
          IRAS 09251-5101 &  1.371E+02 & $ 2.86\pm 0.00 $     & 4.29E-10 & 2.56E-02 & $2.721\pm 0.221$     & $ 6978\pm 2211 $     & $ 9674\pm 6324 $     & $0.647^{+0.082}_{-0.078}$      \\
\multicolumn{6}{c}{}                               & $ 4928\pm 1103 $     & $ 4825\pm 2256 $     & $0.506^{+0.069}_{-0.059}$      \\
\multicolumn{6}{c}{}                               & $ 6213\pm 1753 $     & $ 7669\pm 4480 $     & $0.620^{+0.051}_{-0.044}$      \\
  2MASS J00161695+5958115 &  1.166E+03 & $ 2.72\pm 0.09 $     & 1.67E-12 & 1.02E-04 & $2.689\pm 0.218$     & $ 4439\pm 609  $     & $ 5188\pm 1508 $     & $0.515^{+0.054}_{-0.039}$      \\
\multicolumn{6}{c}{}                               & $ 4190\pm 542  $     & $ 4622\pm 1272 $\tnote{(a)}     & $0.500^{+0.049}_{-0.033}$      \\
\multicolumn{6}{c}{}                               & $ 3952\pm 483  $     & $ 4112\pm 1072 $ \tnote{(a)}    & $0.487^{+0.044}_{-0.029}$      \\

\\\noalign{\smallskip}
\hline
\noalign{\smallskip}
\end{tabular*}
\tiny{{\bf Notes:} For each star the distance, luminosity and core mass are given for 3 choices of \gaia\ EDR3 parallax ZP correction (from top to bottom): noZP, L21ZP, and G21ZP. See the text for more details.
\\
\begin{tablenotes}[flushleft]
\item[(a)] Luminosity below the predicted onset of the TP-AGB phase. 
The core mass is estimated from the asymptotic CMLR (Eq. (\ref{eq-Lqa} + \ref{eq-Lqb1})).
\item[(b)] In the SED fitting the visual extinction is adjusted within the range $(A_V\pm 0.2)$ mag.
\item[(c)] The initial mass is obtained from the  \texttt{PARSEC} stellar isochrone of the cluster's age; the error bar corresponds to an uncertainty of $\pm 0.1$ dex 
in $\mathrm{\log(age)}$.
\end{tablenotes}
}
\end{threeparttable}
\end{table*}

\section{Analysis of the results} 
\label{sect_results}
Our preliminary study provides us with a lot of information for each star which, combined together, can help us try to sketch a general view of the TP-AGB phase at solar-like metallicity. There is an important aspect that also needs to be highlighted: the homogeneity of the stellar models. The same PARSEC stellar tracks and isochrones \citep{Bressan_etal_12} were used to derive the ages of the clusters \citep{Dias_etal_21,Cantat-Gaudin_etal_20} and hence $M_{\rm i}^{\rm AGB}$, to extract the conditions at the first thermal pulse $M_{\rm c, 1TP}$ and $L_{\rm 1TP}$,  and to estimate the initial masses of the white dwarf progenitors that define the initial-final mass relation \citep{Cummings_etal_18, Marigo_etal_20}. Clearly, this does not safeguard the work from possible systematic errors, but guarantees the internal self-consistency of the results.

\begin{figure*}[ht!]
\centering
\begin{minipage}{0.48\textwidth}
\resizebox{\hsize}{!}{\includegraphics{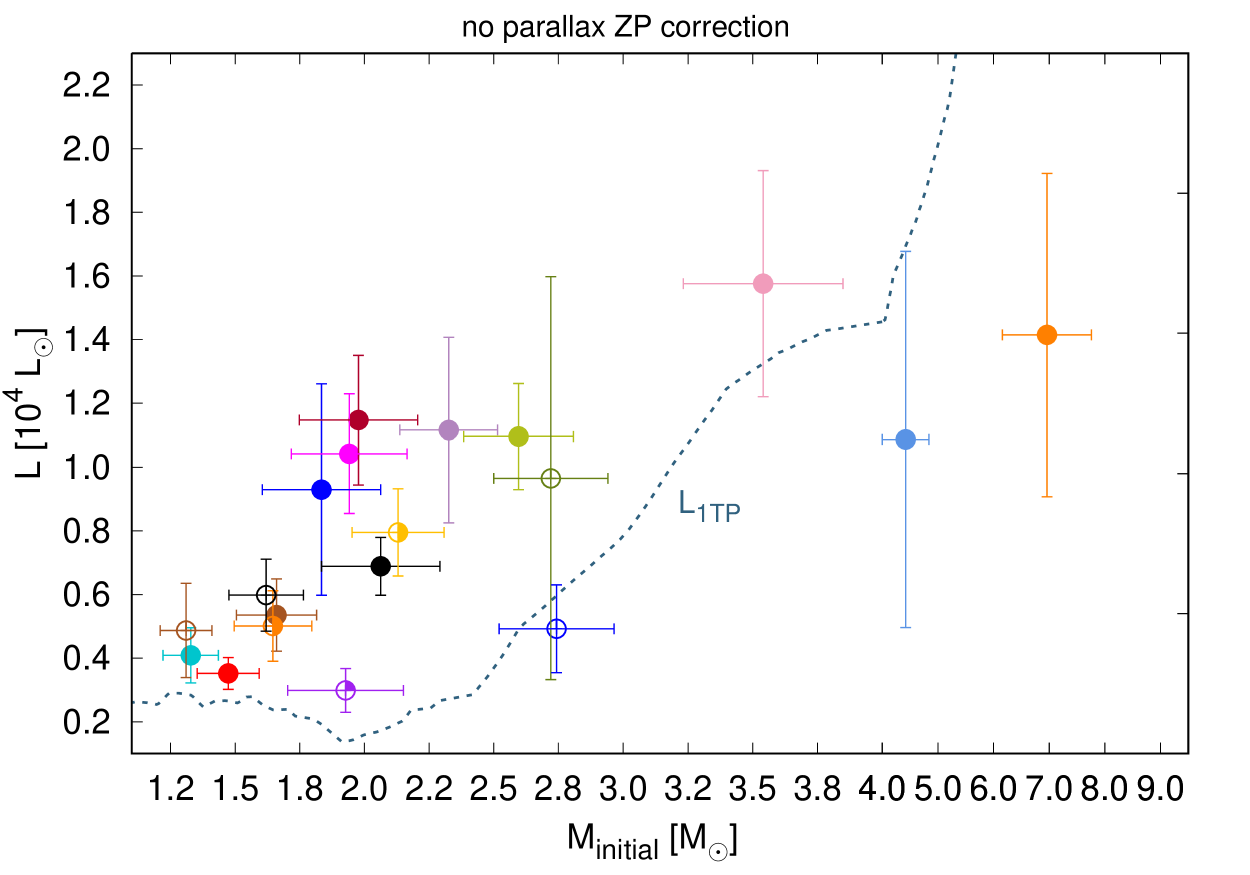}}
\end{minipage}
\begin{minipage}{0.48\textwidth}
\resizebox{\hsize}{!}{\includegraphics{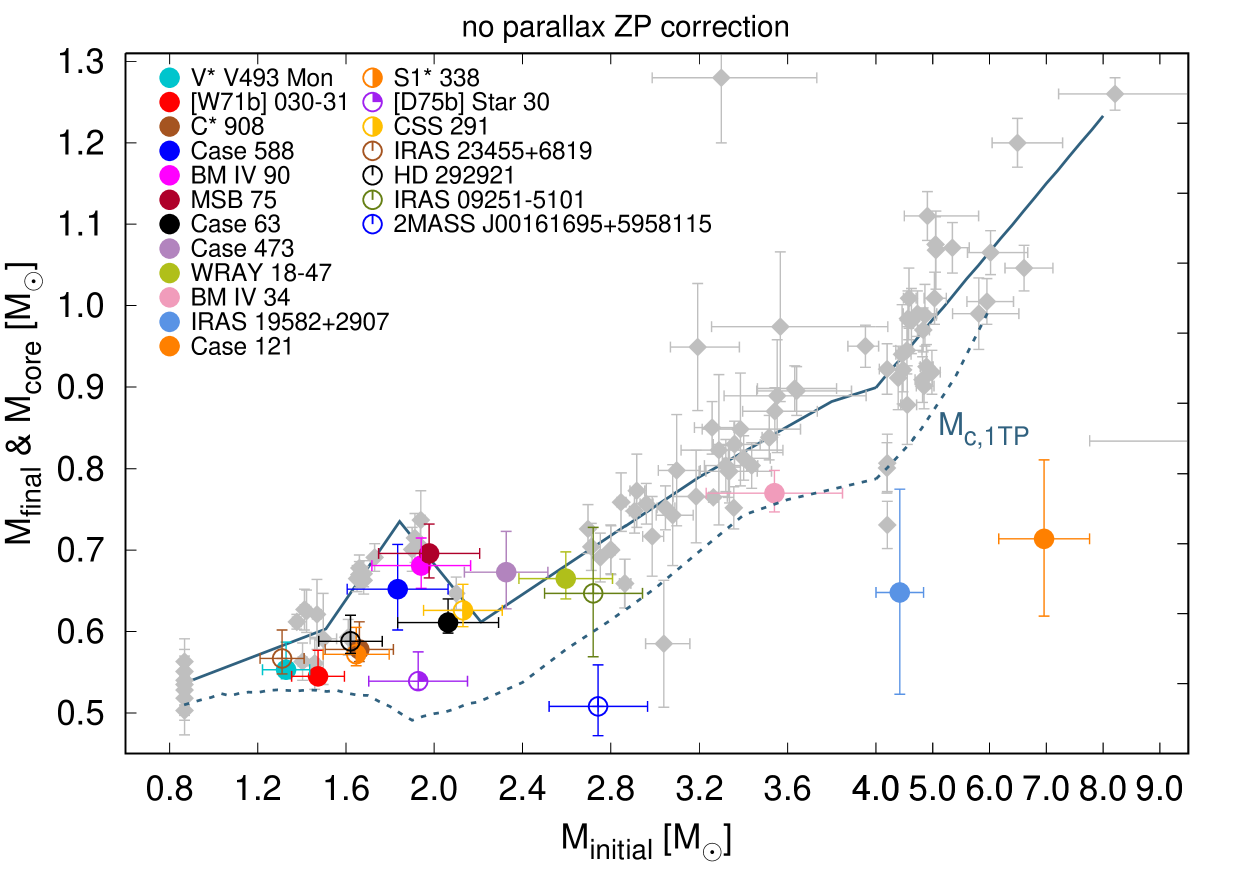}}
\end{minipage}
\begin{minipage}{0.48\textwidth}
\resizebox{\hsize}{!}{\includegraphics{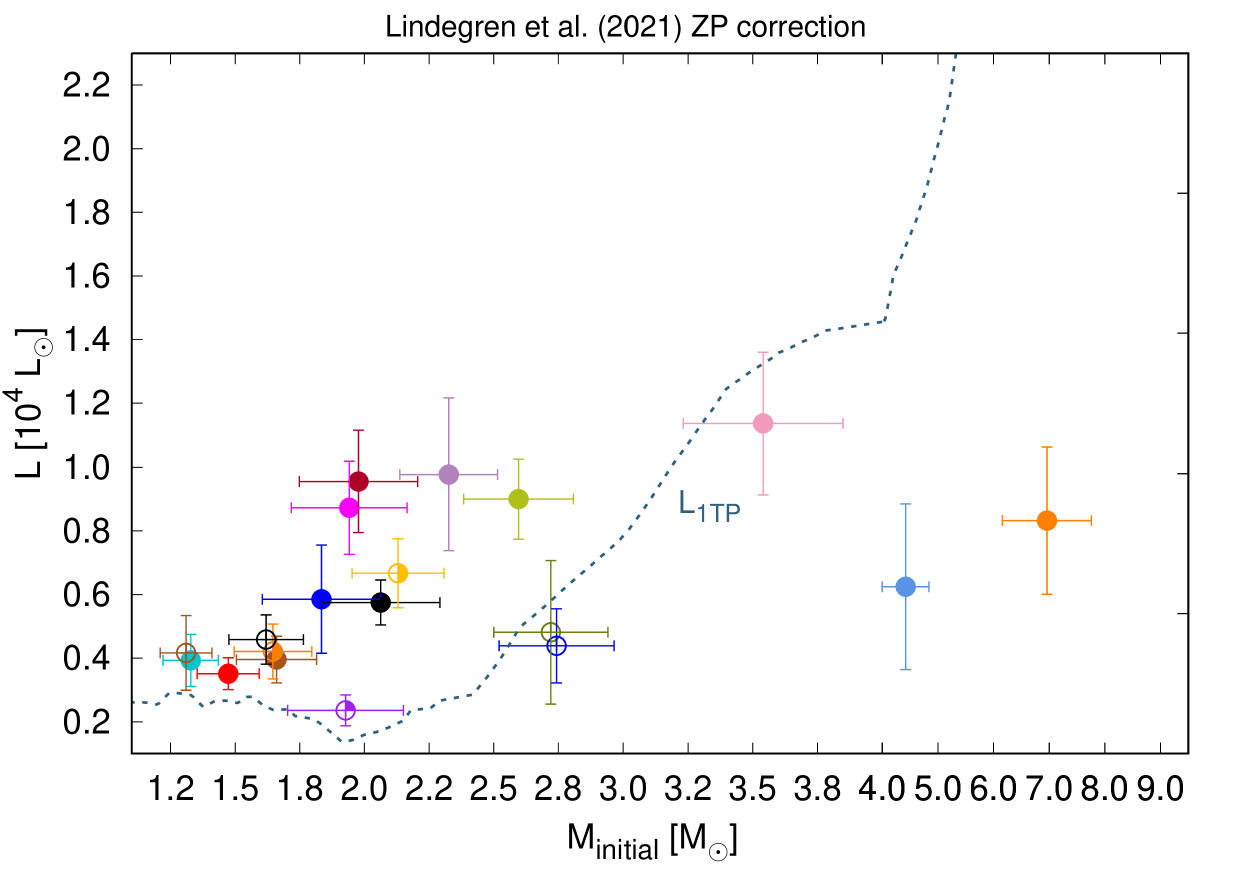}}
\end{minipage}
\begin{minipage}{0.48\textwidth}
\resizebox{\hsize}{!}{\includegraphics{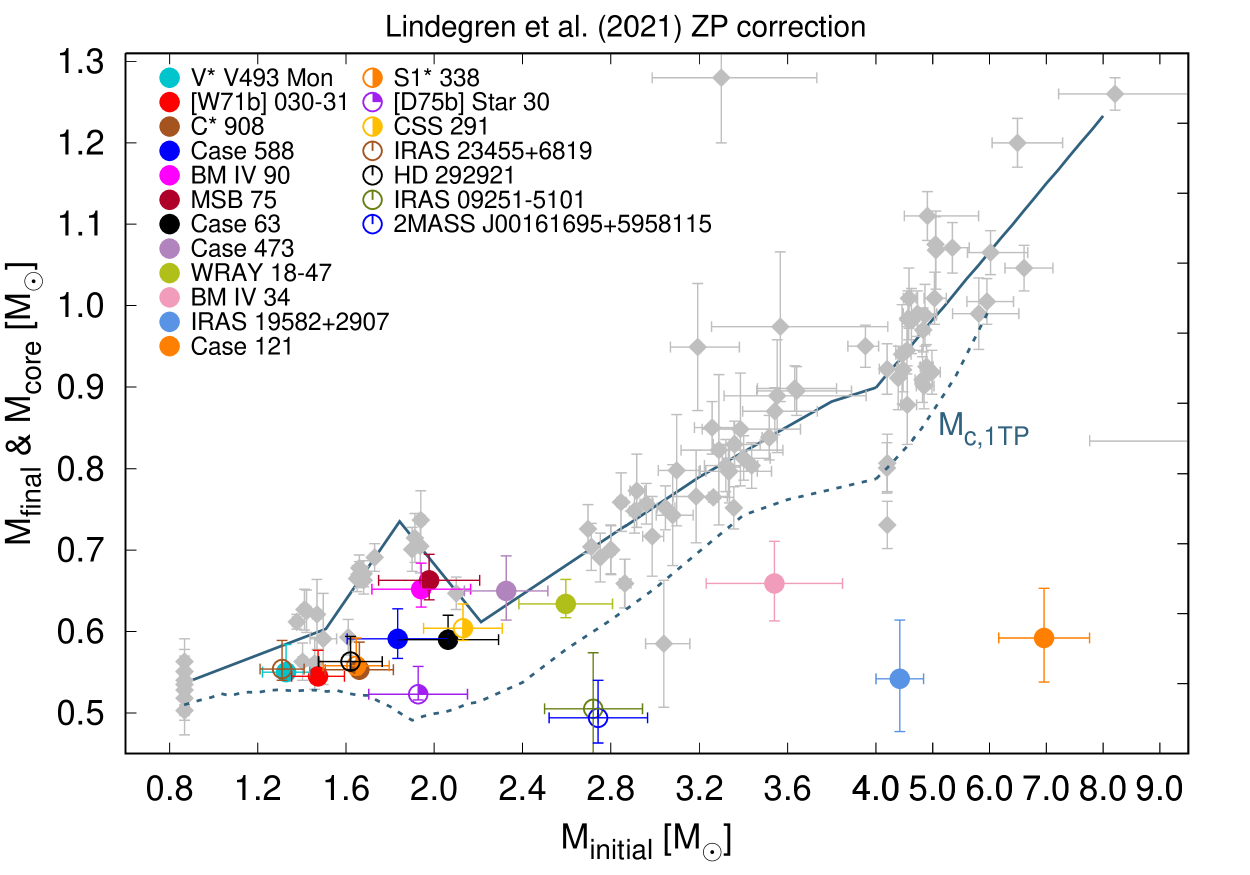}}
\end{minipage}
\begin{minipage}{0.48\textwidth}
\resizebox{\hsize}{!}{\includegraphics{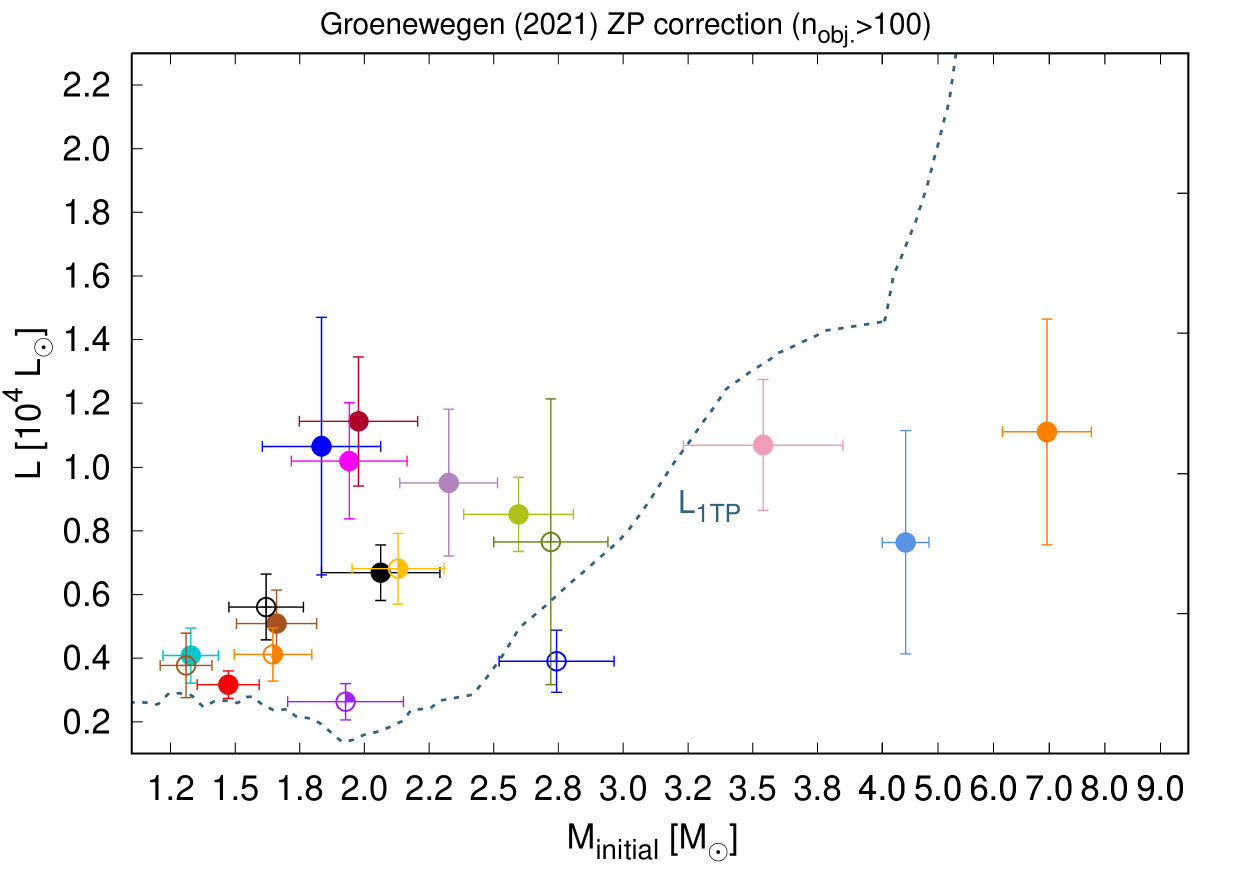}}
\end{minipage}
\begin{minipage}{0.48\textwidth}
\resizebox{\hsize}{!}{\includegraphics{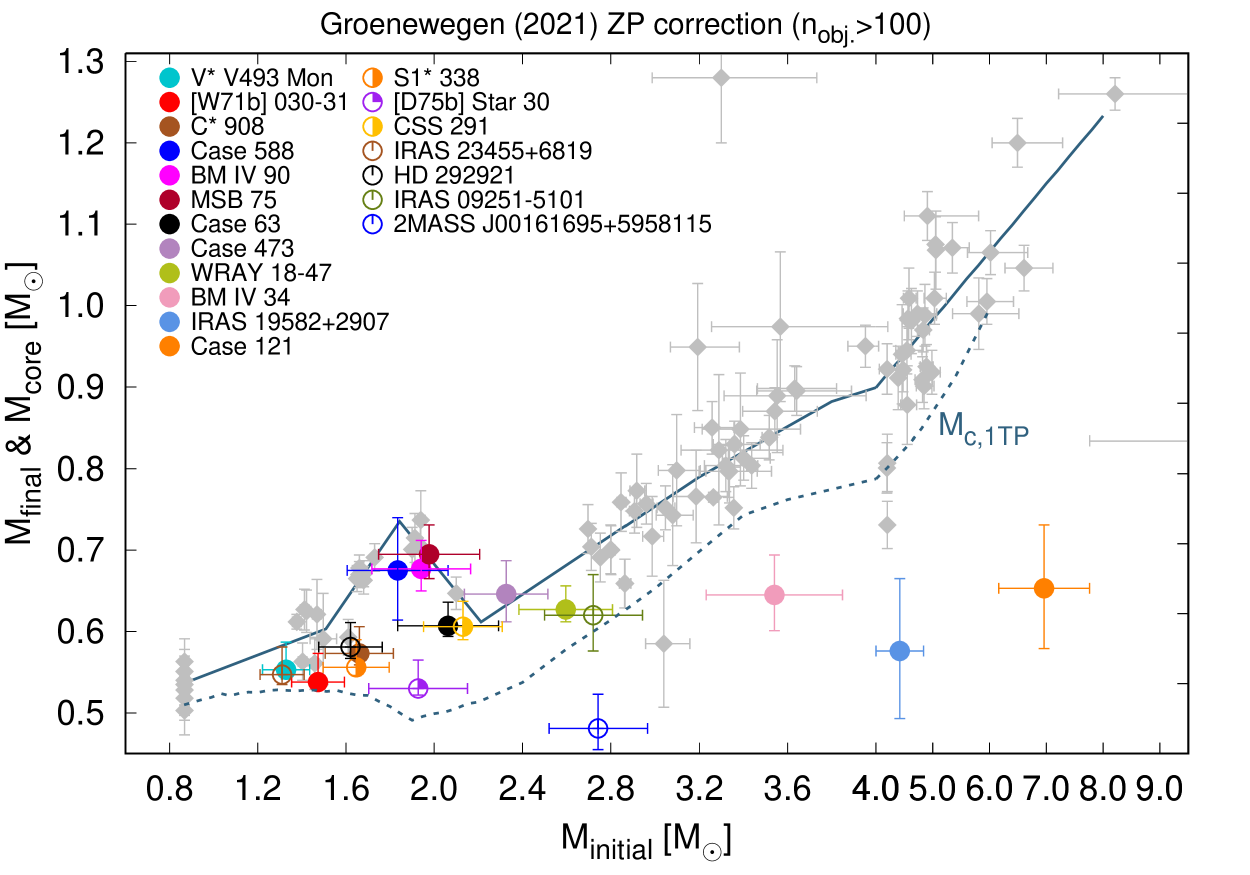}} 
\end{minipage}
\caption{Luminosities and core masses of TP-AGB stars in open clusters
 as a function of the initial stellar mass. The M, S, and C stars (reported in the legenda) are marked with colored symbols and error bars.
 For comparison we over-plot the luminosity  and core mass at the first thermal pulse, $L_{\rm 1TP}$ and $M_{\rm c,1TP}$,  predicted by the \texttt{PARSEC} stellar models at solar metallicity (dashed lines). 
 The cluster ages and visual extinctions are taken from the work of \citet{Cantat-Gaudin_etal_20}. Moving downward the panels show the results for three prescriptions of the \gaia\ EDR3 parallax zero-point (ZP), namely: no ZP correction (top),  \citet[][middle]{Lindegren_etal_21}, and \citet[][bottom]{Groenewegen21}.
Note that the $X$-axis is stretched over the range $0.8\le M_{\rm i}^{\rm AGB}/\Msun \le 4.0$.
\emph{Left panels}: Bolometric luminosties derived from the fitting of the spectral energy distributions. 
\emph{Right panels}: The initial-final mass relation of white dwarfs in the Milky Way is compared to the current core masses of TP-AGB stars in open clusters.  The semi-empirical IFMR (gray diamonds with error bars) is taken from \citet{Marigo_etal_20} and \citet{Cummings_etal_18}. The solid line is a fit to the IFMR data. }
\label{fig_lum_ifmr}
\end{figure*}

\begin{figure*}[ht!]
\centering
\begin{minipage}{0.48\textwidth}
\resizebox{\hsize}{!}{\includegraphics{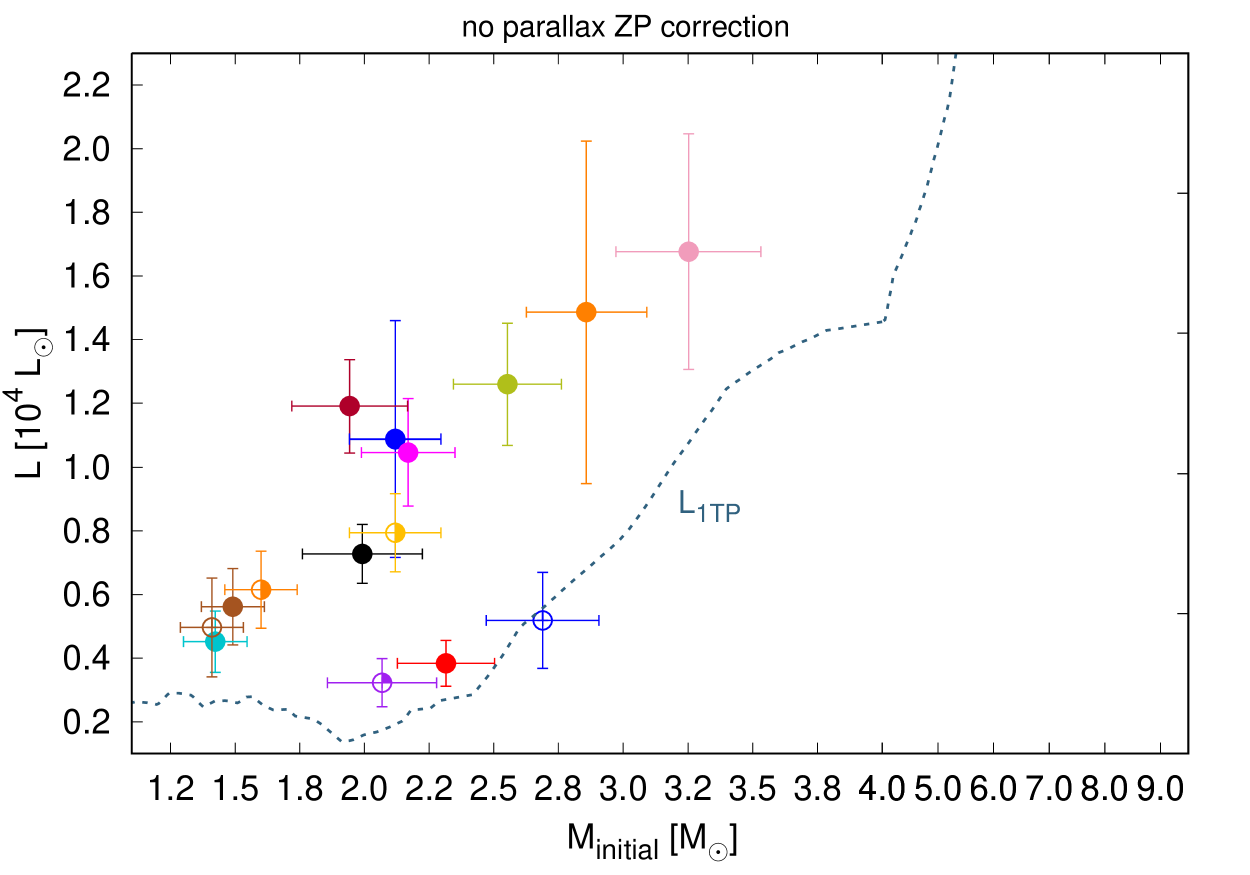}}
\end{minipage}
\begin{minipage}{0.48\textwidth}
\resizebox{\hsize}{!}{\includegraphics{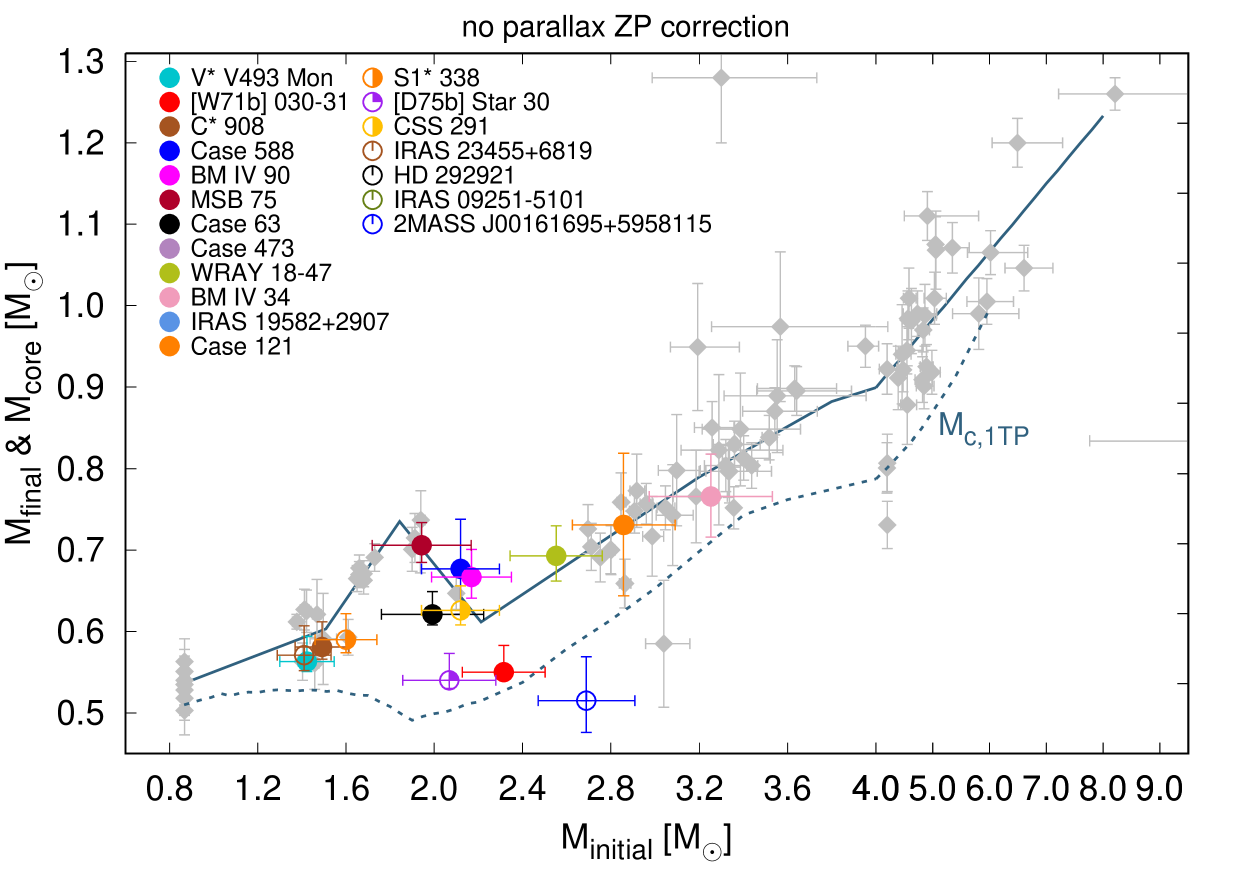}}
\end{minipage}
\begin{minipage}{0.48\textwidth}
\resizebox{\hsize}{!}{\includegraphics{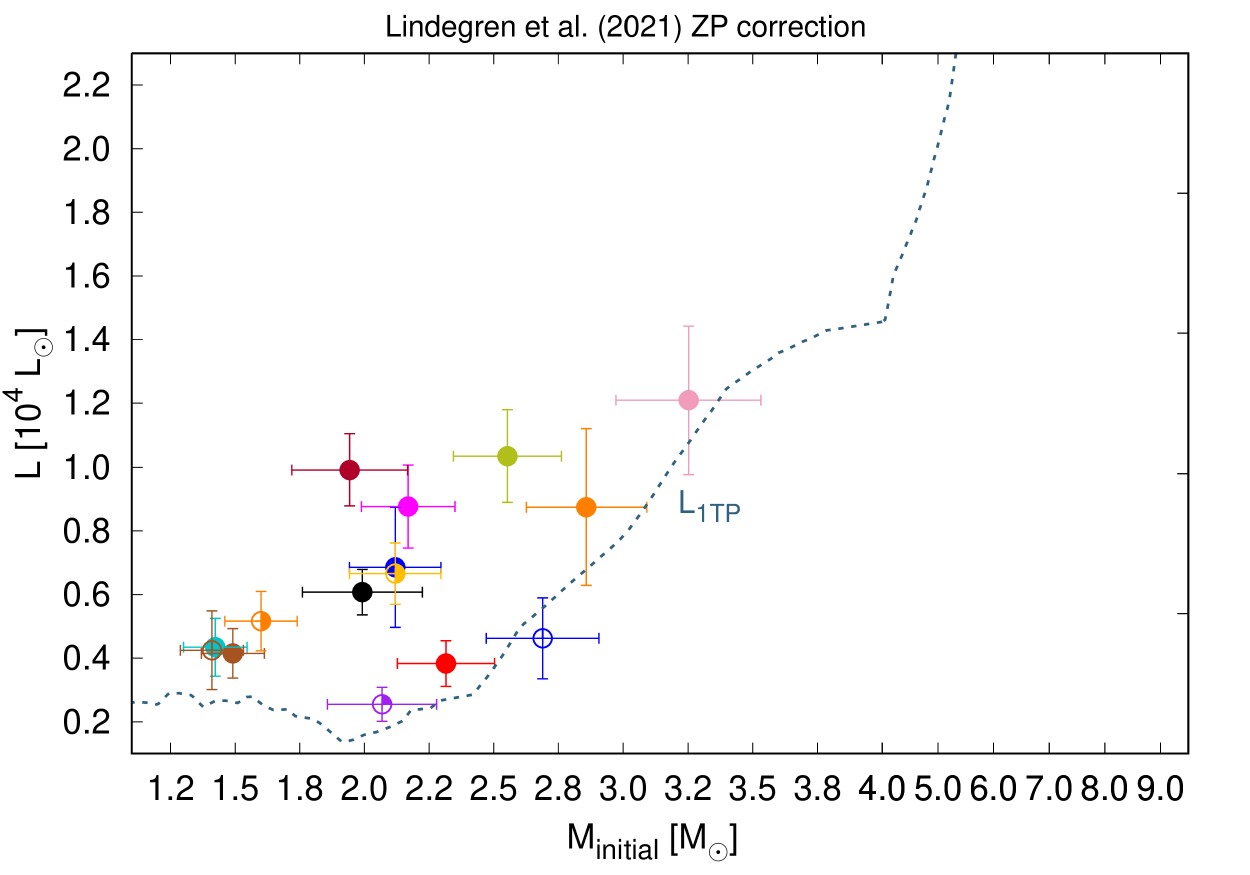}}
\end{minipage}
\begin{minipage}{0.48\textwidth}
\resizebox{\hsize}{!}{\includegraphics{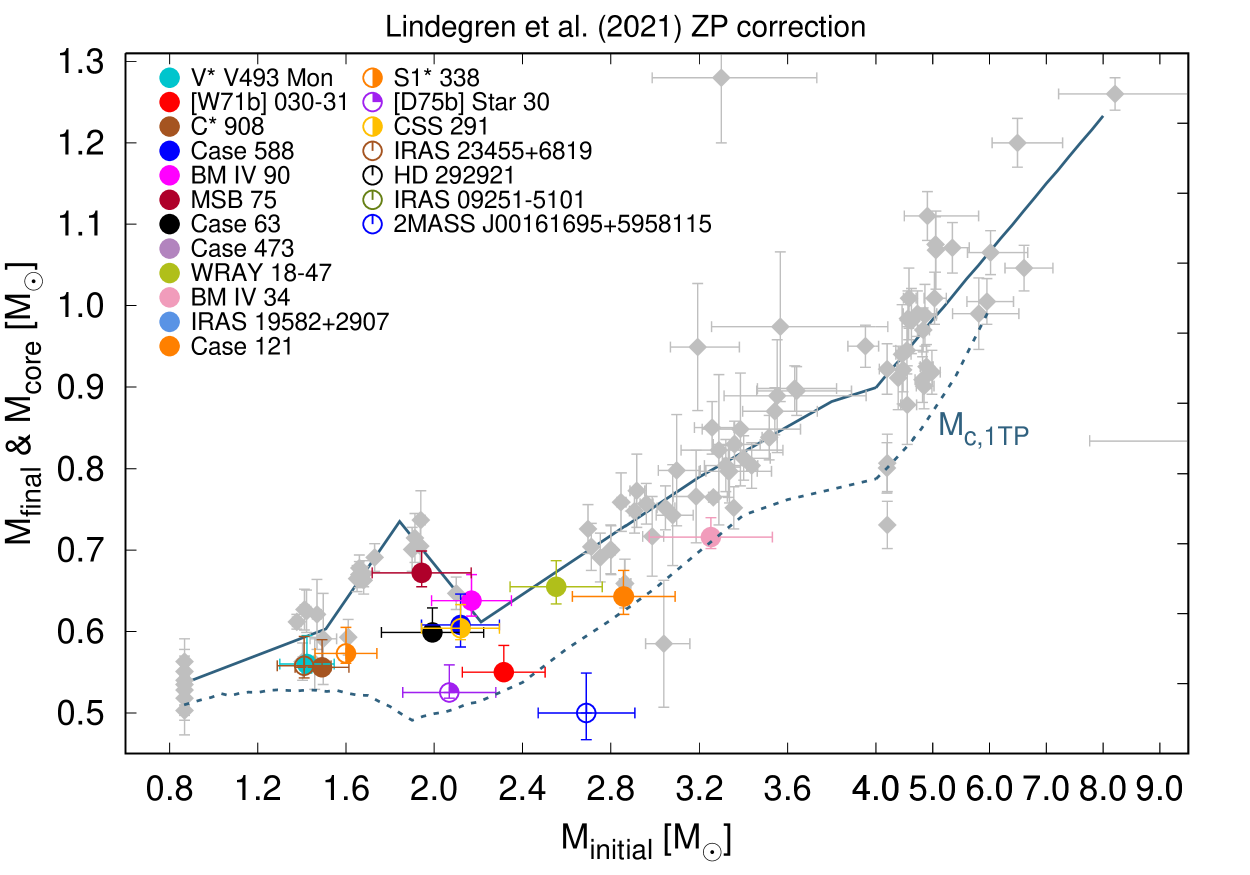}}
\end{minipage}
\begin{minipage}{0.48\textwidth}
\resizebox{\hsize}{!}{\includegraphics{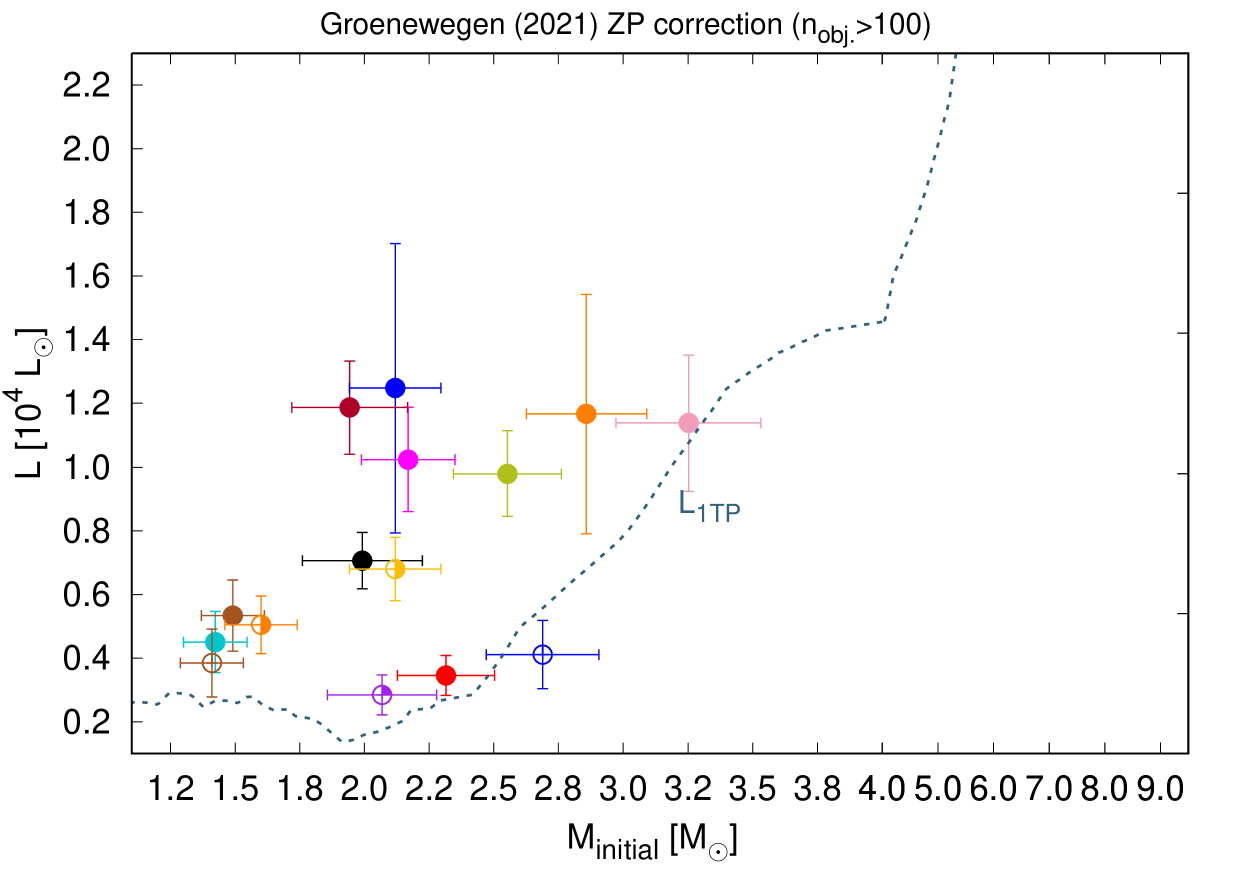}}
\end{minipage}
\begin{minipage}{0.48\textwidth}
\resizebox{\hsize}{!}{\includegraphics{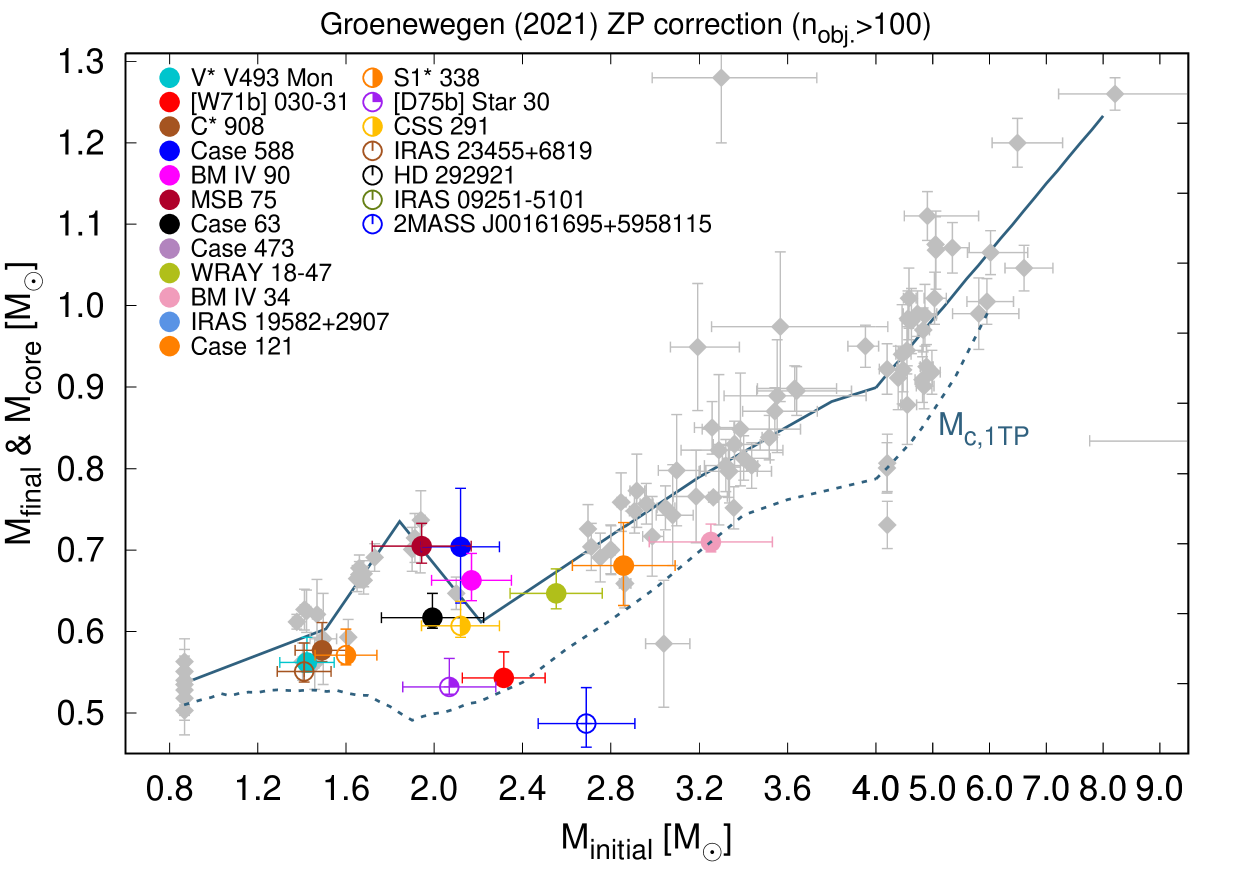}} 
\end{minipage}
\caption{
The same as in Fig.~\ref{fig_lum_ifmr}, but adopting cluster ages and visual extinctions from the work of \citet{Dias_etal_21}. Luminosity  and core mass of all objects are re-derived accordingly.
Note that the stars Case 473, IRAS 19582+2907, HD 292921, and IRAS 09251-5101  are absent, as their host clusters are not included in the catalog.}
\label{fig_lum_ifmr_D21}
\end{figure*}

For the purposes of the discussion that follows, we invite the reader to refer to Tables~\ref{tab_specagb}, \ref{tab_sedfit}, \ref{tab_sedfit:D21}, and Figs.~\ref{fig_lum_ifmr}, \ref{fig_lum_ifmr_D21} which show the luminosity and core mass of the stars under examination.
They contain the results for two cluster age sets \citep{Dias_etal_21, Cantat-Gaudin_etal_20} and three cases of parallax correction (noZP, L21ZP, G21ZP). For simplicity, in the following analysis we will mainly refer to the ages taken from \citet{Dias_etal_21} in conjunction with the noZP case.
Whenever major differences emerge with the other combinations, we will report them.

\subsection{Evolutionary properties of TP-AGB stars in open clusters}
\label{sect_evprop}

\paragraph{Initial masses of carbon stars}\label{para_cstars}
The formation of solar-like metallicity carbon stars is confirmed over an initial-mass range that has the lower extreme,
$M_{\rm i}^{\rm AGB} \simeq 1.5 \,\Msun$, defined by the star C$^\ast$ 908 or [W71b] 030-01 if adopting cluster ages from \citet{Dias_etal_21} or \citep{Cantat-Gaudin_etal_20}.
This value agrees with the minimum mass for carbon stars predicted by TP-AGB models with $Z=0.014$ computed by \citet{Cristallo_etal_11} and \citet{Ventura_etal_18}.
In our sample there is a carbon star, V$^\ast$ V493 Mon,  with a lower initial mass, $M_{\rm i}^{\rm AGB} \approx 1.3-1.4\,\Msun$, but we observe that the host cluster, Trumpler 5, is relatively metal poor \citep[][$\mathrm{[Fe/H] \simeq -0.37}$]{Carrera_etal_19} and therefore the star does not sample the mass range
pertaining to solar metallicity.

As to the upper limit for carbon stars, let us focus on the 3 objects with the most massive progenitors, namely BM IV 34, IRAS 19582+2907 and Case 121 (Case 49 is rejected).

Adopting the age catalog of \citet{Cantat-Gaudin_etal_20},
Case 121, member of the cluster Berkeley 72, is assigned $M_{\rm i}^{\rm AGB} \simeq 7\, \Msun$. This value is very close or even above the maximum limit, $\Mup \approx 6-7\, \Msun$, to develop a degenerate C-O core after helium burning. It might be consistent with a quasi-massive star experiencing the Super-AGB phase once after the completion the carbon burning in the core \citep{Doherty_etal_15,Siess_10}. IRAS 19582+2907, hosted in the cluster FSR 0172, corresponds to a progenitor initial mass $M_{\rm i}^{\rm AGB} \simeq 4.4\, \Msun$, which points towards a standard double-shell TP-AGB phase.

According to stellar evolution theory, the possibility of forming carbon stars at such high masses requires that two conditions are met: the 3DU process is active while the HBB process is weakly efficient or extinguished.
Stellar models \citep[e.g., ][]{Frost_etal_98, Groenewegen_etal_16, Marigo_etal_13} indicate that such configuration can be attained towards the end of evolution if, due to the strong reduction of the envelope mass via stellar winds, the temperature at the base of the convective envelope drops so much that the nuclear reactions of the CNO cycle quench, while some last dredge-up episodes still occur,  enriching the surface with primary carbon, and eventually making the C/O ratio overcome one. This picture is supported by the detection of very bright carbon stars, typically heavily obscured by dusty envelopes produced by intense mass loss \citep{VanLoon_etal_98, Groenewegen_etal_16}.

May this scenario be reasonably applied to IRAS 19582+2907 and Case 121?
We tend to be skeptical about this possibility for the following reasons:
Both stars have a luminosity that places them below the first thermal pulse (Fig.~\ref{fig_lum_ifmr}) for initial masses derived from the age catalog of \citet{Cantat-Gaudin_etal_20}. From a theoretical point of view, this involves a severe difficulty in explaining the surface enrichment in carbon. Furthermore, these stars do not appear dust-enshrouded, so the hypothesis of a late transition to C-rich phase, shortly before ending their evolution, is dropped.
On the other hand, a natural way out of this inconsistency is provided by a reassessment of the age of the parent clusters.
Using \citet{Dias_etal_21} to date Berkeley 72, the initial mass of Case 121 drops to $\simeq 2.9\,\Msun$, so that the values of luminosity and core mass fit well within a standard evolution of TP-AGB, in absence of HBB (see Fig.~\ref{fig_lum_ifmr_D21}). The cluster FSR 0172 is not present in the \citet{Dias_etal_21} catalog. Adopting the age estimate of $\simeq 1.33$ Gyr provided by \citet[][identifier: MWSC 3218]{Kharchenko_etal_13}, IRAS 19582+2907 is assigned an initial mass of $\simeq 1.9\,\Msun$, which brings the star back into the standard TP-AGB framework. These aspects will need to be verified in future studies.

If we reject the ages of \citet{Cantat-Gaudin_etal_20} for Case 121 and IRAS 19582+2907, it turns out that the most massive and brightest carbon star is BM IV 34, member of the cluster Haffner 14, with an initial mass  $M_{\rm i}^{\rm AGB } \simeq 3.3-4.0\,\Msun$. 
It is unclear whether BM IV 34 is a plausible candidate for the occurrence of HBB. TP-AGB models do not agree. In order to check this hypothesis, it would be advisable to carry out spectroscopic measurements to verify a possible increase in the abundance of nitrogen or a lowering of the $^{12}$C/$^{13}$C isotope ratio at the surface. In any case, the SED fitting for BM IV 34  does not indicate the presence of circumstellar dust in significant quantities, nor of large mass-loss rates (see Tables~\ref{tab_sedfit} and \ref{tab_sedfit:D21}).

Overall, the SED fitting results indicate that the luminosities of the 19 C stars are comprised in the range $3\,500-4\,000 \la L/\Lsun \la 11\,000- 16\,000$, depending on the adopted ZP correction for the \gaia\ EDR3 parallaxes.
The corresponding core masses vary in the interval $0.54 \la \Mc/\Msun \la 0.77$.
It is worth recalling that the lower limit of $\Mc$ can actually extend from $\simeq 0.54\ \Msun$ up to $\approx 0.58\,\Msun$, for the effect of the slow luminosity dip thoroughly discussed in Sect.~\ref{ssec_cmlr}.

Adopting the original \gaia\ EDR3  parallaxes (noZP case) the brightest carbon star, BM IV 34, has a core mass $\Mc \simeq 0.77\,\Msun$, while with the L21ZP and G21ZP corrections, the star distance becomes shorter and its bolometric luminosity ($L\approx 11\,000-12\,000 \, \Lsun$) approaches the onset the TP-AGB of a star with $M_{\rm i}^{\rm AGB} \simeq 3.25\, \Msun$, if we use the cluster ages from \citet{Dias_etal_21}.
On the other hand, taking the age catalog of \citet{Cantat-Gaudin_etal_20} the mass of the progenitor increases,  $M_{\rm i}^{\rm AGB} \simeq 3.54\, \Msun$, and its luminosity 
falls just  below the $L_{\rm 1TP}$ with the L21ZP and G21ZP corrections.
In both cases it is hard to explain the existence of a carbon star through the standard 3DU channel in a single star. One may perhaps invoke the mass transfer of carbon-enriched material from a companion star in a binary system, but we tend to disfavor such hypothesis. The SED of BM IV 34 is well reproduced by the spectrum of a single carbon star, and we do not detect  any sign of recent accretion (e.g., infrared excess due to a circumstellar disk).

\paragraph{M and S stars: the onset of the 3DU.} \label{para_sstars} The 7 stars of spectroscopic type M, MS, and S have luminosities in the range $3\,000 \la L/\Lsun \la 9\,500$, and core masses $0.54 \la \Mc/\Msun \la 0.64$ with the noZP case. As predicted by TP-AGB stellar models, M and S stars are in most cases less bright than carbon stars of similar initial mass, consistent with the fact that the carbon enrichment due to the 3DU drives the variation of the spectral type along the M$\rightarrow$S$\rightarrow$C sequence. This does not appear to be entirely the case for the M-type low-mass stars HD 292921 and IRAS 23455+6819 which have a higher average luminosity than the carbon stars V$^\ast$ V493 Mon, [W71b] 030-31, and C$^\ast$ 908.
However, we note that the error bars in L and $M_{\rm i}^{\rm AGB}$ for these stars do not allow us to accurately sort their brightness.

Knowing the luminosity of MS and S stars is important as it can place constraints on the 3DU \citep[see, e.g.,][]{Shetye_etal_21}, in particular on the minimum core mass, $M_{\rm c}^{\rm min}$, for the occurrence of the mixing events.
Solar-metallicity TP-AGB models generally agree that the 3DU is operative at  $M_{\rm i}^{\rm AGB} \approx 2 \Msun$ \citep[e.g.,][]{Ventura_etal_18,Marigo_etal_13,Cristallo_etal_11}, while predictions may disagree at lower masses, especially in identifying the minimum mass for the occurrence of the 3DU
 \citep[see the discussion in][]{Marigo_etal_13,KarakasLugaro_16}.
 According to current TP-AGB models this critical mass is loosely constrained, taking values in the range from $M_{\rm i}^{\rm AGB}\simeq 1.4\,\Msun$  to $M_{\rm i}^{\rm AGB}\simeq 2.0\,\Msun$, mainly depending on the adopted mixing treatment and numerical details.
 
 The new results of this study, based on \gaia\ data, help us to reduce the degree of uncertainty, albeit limited to a narrow range of stellar masses.   
 For example, the MS-type stars [D75b] Star 30 would indicate that for $1.9 \la M_{\rm i}^{\rm AGB}/\Msun \la 2.3$ the onset of the 3DU occurs at $L \simeq 3\,000 \pm 700\, \Lsun $, which corresponds to $0.530\la M_{\rm c}^{\rm min}/\Msun \la 0.575$, taking into account the uncertainties  on distance and flux, together with the effect of the slow luminosity dip.
 
We compared this indication with a few TP-AGB models in the literature that commendably make the relevant quantities available.
The \texttt{FRUITY} model of \citet{Cristallo_etal_11} for $M_{\rm i}^{\rm AGB} = 2.0\,\Msun$ and $Z = 0.014$ agrees well with the observational data, predicting $M_{\rm c}^{\rm min}\simeq 0.56\, \Msun$ with a  quiescent luminosity of about $4\,000 \, \Lsun$. \citet{WeissFerguson_09} find that their TP-AGB model with $M_{\rm i}^{\rm AGB} = 2.0\,\Msun$ and $Z = 0.02$ experiences the first mixing episode much earlier, for $M_{\rm c}^{\rm min} \simeq 0.49 \, \Msun $, but the corresponding luminosity is not indicated. Conversely, a late start of the 3DU seems to characterize the models of \citet{Karakas_14}, who reports $M_{\rm c}^{\rm min} \simeq  0.62\, \Msun$ at $M_{\rm i}^{\rm AGB} = 2 \, \Msun$ and $Z = 0.014$.

The S star CSS 291 with $1.9 \la M_{\rm i}^{\rm AGB}/\Msun \la 2.3$ is attributed a type S4/2, which would correspond to C/O$\,\simeq 0.95$ according to the classification of \citet{Keenan_80}.  If confirmed, this star could probe the luminosity, and therefore the core mass, at the transition from the O-rich regime to the C-rich regime. The brightness of CSS 291 is compatible with the range $6\,500 \la L/\Lsun \la 9\,000 $ which corresponds to a core mass $0.6 \la \Mc/\Msun \la 0.66 $, assuming the noZP case. Somewhat lower values for both $L$ and \Mc\ apply if we take the L21ZP and G21ZP cases.
The \texttt{FRUITY} model for $M_{\rm i}^{\rm AGB} = 2.0\,\Msun$ and $Z = 0.014$ is broadly consistent with the CSS 291 data: just before becoming a carbon star, the model attains a photospheric C/O$\,\simeq 0.94$ when  $L \approx 6\,700\,\Lsun$ and $\Mc \simeq 0.59\,\Msun$.

\paragraph{Pulsation.}
AGB stars exhibit variability due to stellar pulsation in low-order modes \citep[e.g.][]{Wood_15}, possibly non-radial \citep[][and references therein]{Yu_etal_2020}. The generally accepted picture \citep[e.g.][]{LattanzioWood_2004,Wood_15} is that pulsation is dominated by relatively high-order (third or second overtone) modes during the early stages of the LPV phase, when multi-periodicity is common. Overtone modes become gradually stable as the envelope expands \citep{Trabucchi_etal_2019}, until eventually a star pulsates only in the fundamental mode. As the former acquires large amplitude, a star is identified as a Mira variable.

\begin{figure*}[ht!]
    \centering
    \includegraphics[width=\textwidth]{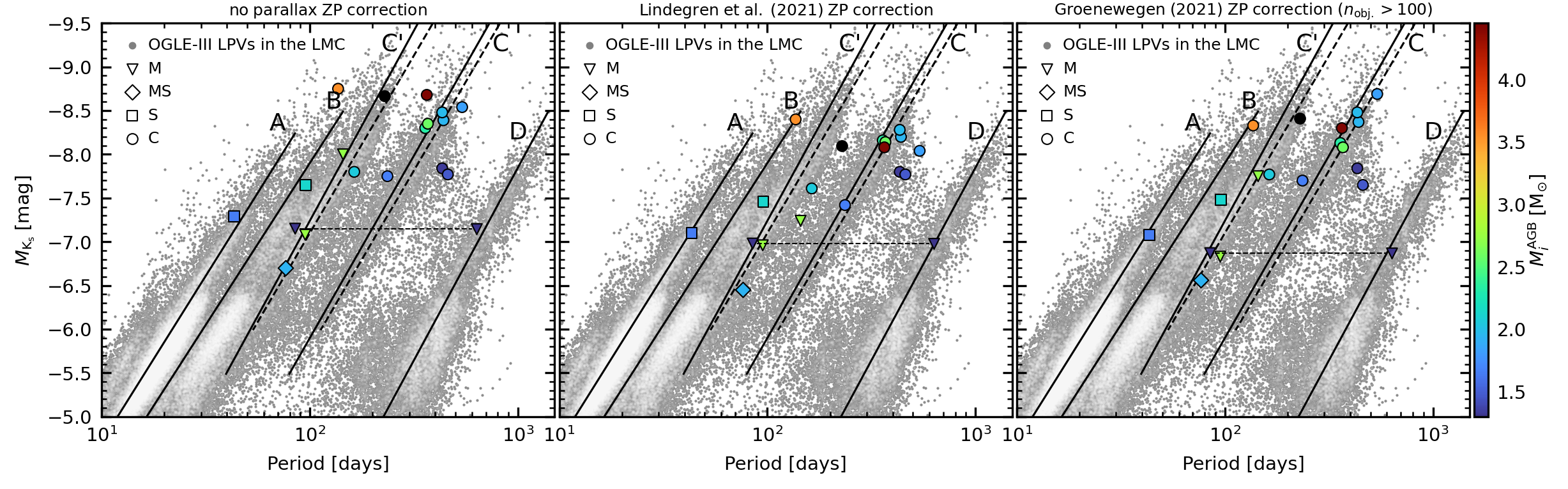}
    \caption{Period-luminosity diagram of the LPVs in open clusters from Table~\ref{tab:obsvar2}, color-coded by initial mass according to \citet{Cantat-Gaudin_etal_20}. The source Case 121 is indicated by the black symbol, as its initial mass according to \citet{Cantat-Gaudin_etal_20} ($\sim7\,\Msun$) is substantially larger compared to all other sources. The spectral type is indicated by symbol shapes (triangles: M-type; diamonds: MS-type; squares: S-type; circles: C-type). For the source IRAS 23455+6918 we show both the pulsation period and the long secondary period, connected by a dashed line. For visual reference, primary periods of LPVs in the LMC from OGLE-III \citep{Soszynski_etal_2009_LMC} are displayed as gray dots in the background. Solid lines represent the best fits to period-luminosity relations A, B, \cprime, C and D derived by \citet{Soszynski_etal_2007} for O-rich LPVs in the LMC, whereas dashed lines correspond to the best fits to C-rich LPVs on sequences \cprime and C. No parallax correction was applied to sources in the left panel, while in the central and right panels, respectively, we have been adopted the \citetalias{Lindegren_etal_21} and \citetalias{Groenewegen21} parallax corrections.}
     \label{fig_pld_minit}
\end{figure*}

\begin{figure*}[ht!]
\centering
\begin{minipage}{0.285\textwidth}
\resizebox{\hsize}{!}{\includegraphics{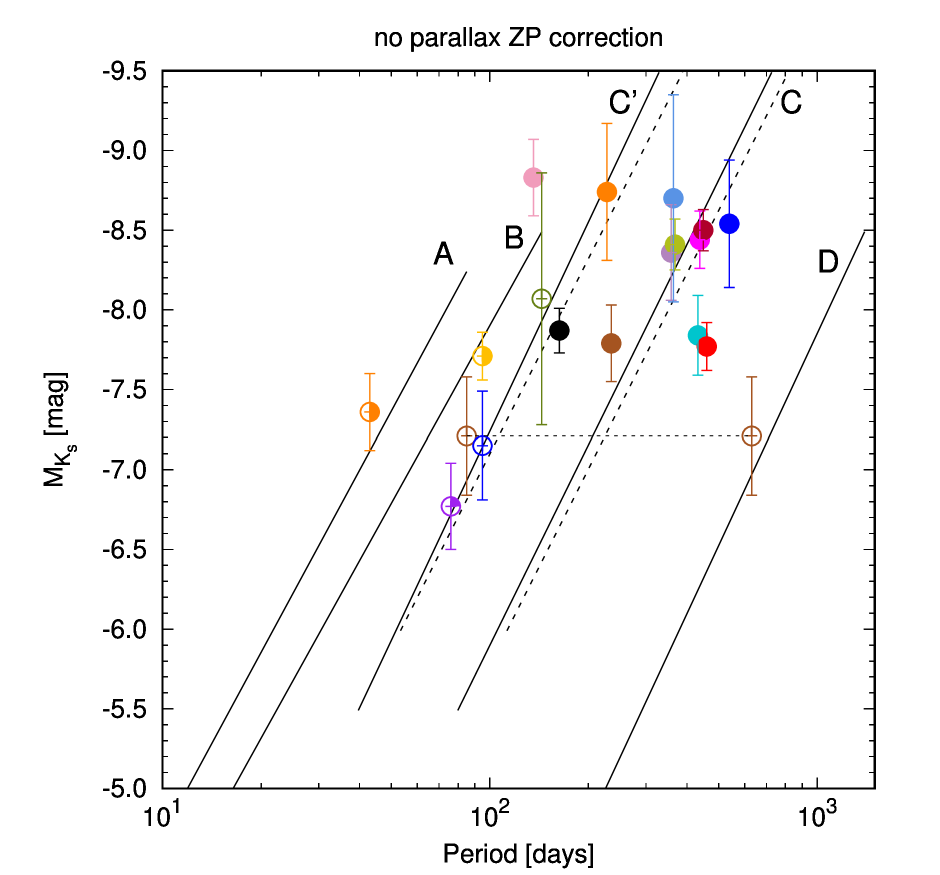}}
\end{minipage}
\begin{minipage}{0.285\textwidth}
\resizebox{\hsize}{!}{\includegraphics{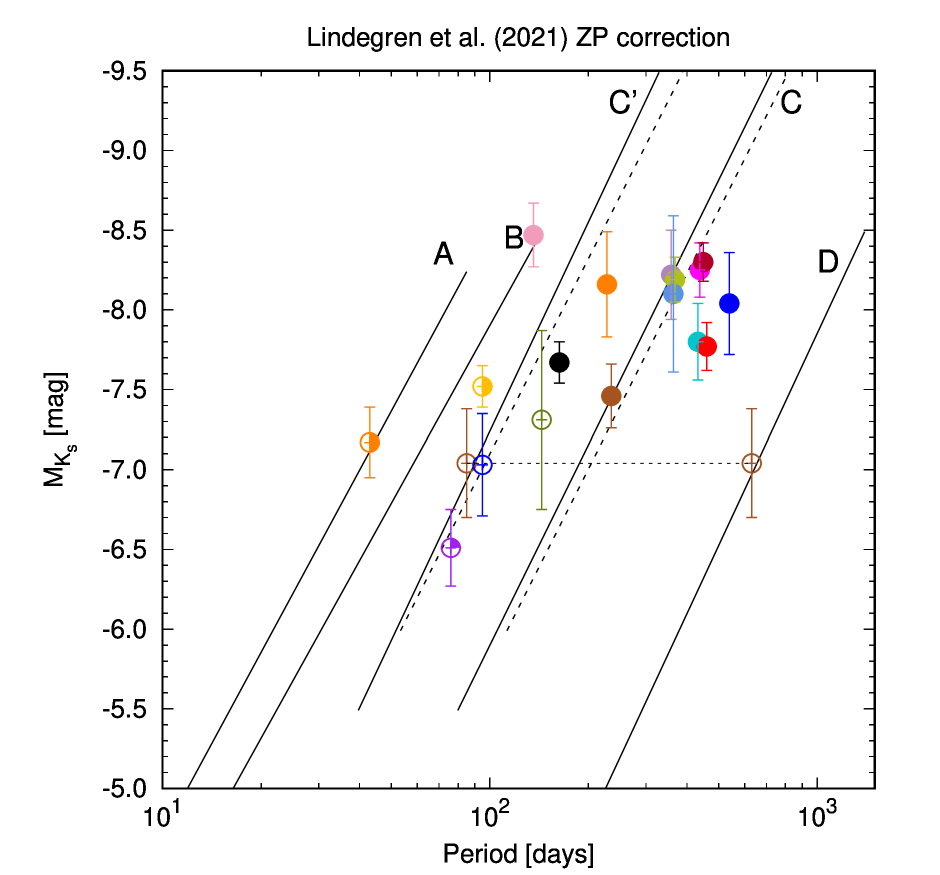}}
\end{minipage}
\begin{minipage}{0.38\textwidth}
\resizebox{\hsize}{!}{\includegraphics{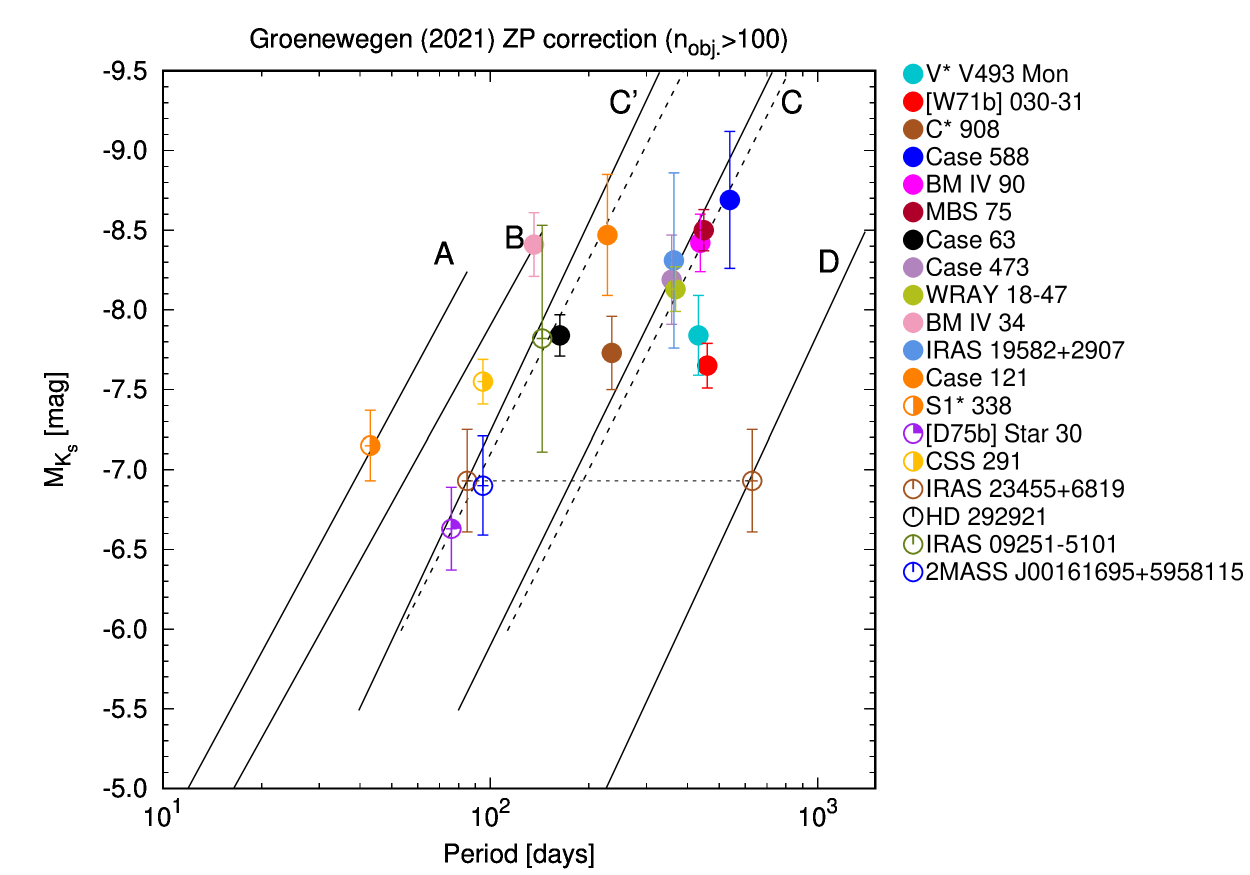}}
\end{minipage}
\caption{The same as in Fig.~\ref{fig_pld_minit},
but with the TP-AGB stars identified by their names, consistently with other figures in the paper.
The absolute magnitude \Mks\ of each star is corrected for interstellar reddening. The results are shown for the three of options of parallax ZP correction, as indicated on the top of each panel.}
\label{fig_lpv}
\end{figure*}

In order to analyze our sample of LPVs within this picture, we examine them in the period-luminosity diagram (PLD) shown in Figure~\ref{fig_pld_minit}. The absolute magnitude $\Mks$ used to construct it includes the contributions of both interstellar and circumstellar extinction, and we consider the three cases of zero-point correction to the \gaia\ EDR3 parallax discussed in Sect.~\ref{ssec_plx}. The same is displayed in Fig.~\ref{fig_lpv}, except there each source is identified by its name as done in Fig.~\ref{fig_lum_ifmr_D21}. The corresponding data are reported in Table~\ref{tab:obsvar2}.

By comparison with the pattern seen in the PLD of LPVs in the LMC as observed by OGLE-III \citep{Soszynski_etal_2007,Soszynski_etal_2009_LMC}, and following the results of \citet[][and references therein]{Trabucchi_etal_17} \citep[see also][]{Trabucchi_etal_2021_SRVs}, we are able to identify the pulsation mode responsible for each period. We identify nine sources with a  period  due to the fundamental mode, eight sources pulsating predominantly in the first overtone mode, and one star (S1$^{\ast}$ 338) whose period is most likely due to pulsation in the second overtone mode. Interestingly, the three different approaches adopted for correcting the parallax zero-point lead to the same mode classification, although a few first overtone mode pulsators are shifted towards the region between sequences \cprime and C, making the identification of their primary period less certain.

Two stars, V$^{\ast}$ V493 Mon and $[$W71b$]$ 030-01, are classified as fundamental mode pulsators, but they are located below the period-luminosity (PL) sequence C compared with other stars with similar periods. This is consistent with the fact that these stars have relatively large mass-loss rates, and the resulting circumstellar extinction makes them appear fainter in the $\ks$ band \citep{ItaMatsunaga_11,Soszynski_etal_2009_LMC,Whitelock_etal_2017}.

The M star IRAS 23455+6819 is the only LPV in our sample whose time series show clear evidence of a long secondary period, which lies very nicely on sequence D regardless of the choice of the parallax zero-point correction. We also point out that, having its primary pulsation period in the area between sequences \cprime and C, this source fits rather well the scenario depicted by \citet{Trabucchi_etal_17}. Indeed, they examined data from the OGLE-III catalog of LPVs in the LMC \citep{Soszynski_etal_2009_LMC} and notice a large fraction of the sources reported to have a period on sequence D also display a period between sequences \cprime and C. They suggested that the large amplitude associated with the long secondary periods on sequence D makes so that it is favored over the true pulsation period in the same star when analyzing the PLD, where normally only one period per star is displayed, thus causing the apparent gap between sequences \cprime and C. Indeed, that the star IRAS 23455+6819 displays a long secondary period whose amplitude is significantly larger than that of the pulsation period in the same star (cf. Fig.~\ref{fig_lcs}).

We note that most of the LPVs in our sample that lie on sequence C, and that we identified as fundamental mode pulsators, have brightness $\Mks\lesssim-8$ mag, consistently with the properties of Miras \citep[see e.g. Fig~1 of][]{BeddingZijlstra_1998}. Moreover, most of these stars appear to have relatively regular light curves, with little evidence of multi-periodicity, also suggestive of Mira-like behavior. Yet, based on their amplitudes, none of this stars would be classified as a Mira. In fact, only the two relatively dusty sources V$^{\ast}$ V493 Mon and $[$W71b$]$ 030-01 have large enough amplitude at visual wavelengths to be possibly identified as Miras, and only the former is classified as such in one of the catalogs we examined (see Table~\ref{tab:obsvar}).

However, it should be noted that the traditional distinction between Miras and semi-regular variables, according to which the former have visual amplitude $\Delta V>2.5$ mag, has been criticized by a number of studies \citep[e.g.][]{Kerschbaum_Hron_1992,Kiss_etal_2000,Lebzelter_Hinkle_2002}. Recently, \citet{Trabucchi_etal_2021_SRVs} have shown that semi-regular variables in the LMC that pulsate only in the fundamental mode follow the same sequence as Miras both in the period-luminosity diagram and in the period-amplitude diagram, suggesting that it would be incorrect to assign them different variability types only because their amplitude is smaller than an arbitrary, although reasonable, threshold.

It is also worth noticing that for most of the carbon stars (except for V$\ast$ V493 Mon and [W71b] 030-31)
the small variability amplitude could be linked with the fact that their C/O is just above unity, which is consistent with the absence of powerful dust-driven winds. Indeed, the spectral absorption features of molecules play a crucial role in determining the large visual amplitudes of Miras and related AGB variables \citep{ReidGoldston_2002}. The low surface temperature corresponding to the point of maximum expansion of the pulsation cycle favors the formation of molecules, which in turn effectively block a large fraction of visual light from escaping the photosphere owing to their high opacity. Towards maximum compression the increased temperature causes these molecules to dissociate, and the star appears much brighter at visual wavelengths. In O-rich stars this effect is mainly associated to metallic oxides, primarily TiO, while in C-stars carbon-bearing molecules such as CN and C$_2$ are the main agents. If C/O is slightly $\ga 1$ almost all oxygen is locked into CO, while a  small excess of carbon, C-O, remains to form other carbon-bearing molecules \citep{Marigo_Aringer_09}.
This could explain the relatively low amplitudes of the sources we have examined.
This picture is supported by observations of Galactic C-rich low-amplitude semi-regular variables \citep{Schoier_Oloffson_01} with measured photospheric C/O ratio \citep{Lambert_etal_86}: the majority of them have $\mathrm{1.04 \la C/O \la 1.3}$.
Putting all the pieces together we may deduce that as long as the carbon excess is small, (1) stars have low-amplitude pulsation, (2) do not form circumstellar dust in significant amount, and (3) do not experience powerful dust-driven outflows.
The three conditions are all met by most of the carbon stars in our sample (Tables~\ref{tab_sedfit}, \ref{tab_sedfit:D21}, \ref{tab:obsvar2}, Fig.~\ref{fig_lpv}), in particular the bright carbon stars MSB 75, BM IV 90 and Case 588. 
As a consequence, a reasonable expectation is that the carbon enrichment in these stars is modest. This point is relevant for the analysis of their core mass and its link to the IFMR of white dwarfs \citep{Marigo_etal_20}.

\begin{table*}
\centering
\scriptsize
\caption{Period, pulsation mode, and absolute $\Mks$ magnitude for the LPVs in our sample. We report magnitudes obtained without correcting the parallaxes zero-point, or by using the corrections of \citetalias{Lindegren_etal_21} and \citetalias{Groenewegen21}.}
\label{tab:obsvar2}
\begin{tabular}{ccccccc}
\hline
\hline
\multicolumn{3}{c}{} & \multicolumn{1}{c}{noZP} & 
\multicolumn{1}{c}{L21ZP} & \multicolumn{1}{c}{G21ZP} \\
\cmidrule(lr){4-6}
\noalign{\smallskip}
Star & $P$ & mode$^{(a)}$ & $\Mks$ & $\Mks$ & $\Mks$ \\
\hline
 & [days] & & [mag] & [mag] & [mag] \\
\hline
V$^{\ast}$ V493 Mon     & 432.4 &  FM & -7.84 & -7.80 & -7.84 \\
$[$W71b$]$ 030-01       & 460.0 &  FM & -7.77 & -7.77 & -7.65 \\
C$^{\ast}$ 908          & 235.9 &  FM & -7.75 & -7.42 & -7.70 \\
Case 588                & 539.0 &  FM & -8.54 & -8.04 & -8.69 \\
BM IV 90                & 438.2 &  FM & -8.39 & -8.20 & -8.37 \\
MSB 75                  & 449.2 &  FM & -8.48 & -8.28 & -8.48 \\
Case 63                 & 163.4 & 1OM & -7.80 & -7.61 & -7.77 \\
Case 473                & 358.4 &  FM & -8.30 & -8.16 & -8.13 \\
WRAY 18-47              & 368.9 &  FM & -8.35 & -8.14 & -8.08 \\
BM IV 34                & 136.7 & 1OM & -8.75 & -8.40 & -8.33 \\
IRAS 19582+2907         & 364.6 &  FM & -8.68 & -8.08 & -8.30 \\
Case 121                & 228.2 & 1OM & -8.67 & -8.10 & -8.41 \\
S1$^{\ast}$ 338         & 43.28 & 2OM & -7.29 & -7.10 & -7.08 \\
$[$D75b$]$ Star 30      & 76.36 & 1OM & -6.70 & -6.45 & -6.56 \\
CSS 291                 & 95.52 & 1OM & -7.65 & -7.46 & -7.48 \\
IRAS 23455+6819         & 85.02 & 1OM & -7.15 & -6.98 & -6.87 \\
HD 292921               &     - &   - & -7.20 & -6.91 & -7.13 \\
IRAS 09251-5101         & 144.4 & 1OM & -8.00 & -7.24 & -7.75 \\
2MASS J00161695+5958115 & 95.01 & 1OM & -7.08 & -6.96 & -6.83 \\
\hline
\noalign{\smallskip}
\end{tabular}
\tablecomments{$^{(a)}$ FM: fundamental mode; 1OM: first overtone mode; 2OM: second overtone mode.}
\end{table*}

\paragraph{Mass loss and dust production.}
In addition to the bolometric luminosity, the SED fitting returns information on the present-day dust mass-loss rate, $\dot M_{\rm d}$, and the optical depth at 10/11.3  $\mu$m for O-/C-rich stars.
All TP-AGB stars analyzed here are optically visible and therefore we do not expect them to be characterized by substantial production of circumstellar dust. In general, in fact, the order of magnitude of $\dot M_{\rm d}$ is very small for most stars, being comprised between $10^{-12}\, \Msun\, {\rm yr}^{-1}$ and $10^{-10}\, \Msun\, {\rm yr}^{-1}$. Correspondingly, the optical depth $\tau_{10}$ or $\tau_{11.3}$ is also low and varies from  $10^{-4}$ and  $10^{-2}$. 

There are 3 stars in the sample that show signs of well-developed stellar winds.
One is the M-type low-mass star IRAS 23455+6819, with $1.2 \la M_{\rm i}^{\rm AGB}/\Msun \la 1.4$. It has $\dot M_{\rm d} \simeq 9\,10^{-10}\,\Msun\, {\rm yr}^{-1}$, that may correspond to a total mass-loss rate approaching $\dot M = \dot M_{\rm d} /\Psi \approx 10^{-6} \,\Msun\, {\rm yr}^{-1}$, assuming a dust-to-gas ratio $\Psi \simeq 10^{-3}$ \citep{Bladh_etal_19_M}.
Interestingly, this star displays a relatively short pulsation period (85 days), that we attribute to the first overtone mode. This suggests that the star is not a Mira, but more likely a semi-regular variable. However, this is also the only source in our sample clearly showing a long secondary period (631 days) on sequence D, which is consistent with relatively strong mass-loss. Indeed, various works \citep[e.g.][]{WoodNicholls_2009,McDonaldTrabucchi_2019} have pointed out the correlation between dust production and the appearance of long secondary periods. More recently, \citet{Soszynski_etal_2021} \citep[see also][]{Wood_etal_1999,Soszynski_2007,SoszynskiUdalski_2014} have put forward strong evidence that long secondary periods are caused by eclipses from a dust cloud dragged probably by a sub-stellar companion.

The other 2 stars with detectable winds are the carbon stars V$^\ast$ V493 Mon and [W71b] 030-31, that have $\dot M_{\rm d} \approx {\rm few}\,10^{-9}\, \Msun\, {\rm yr}^{-1}$ and $\tau_{11.3} \simeq 0.02$. Both stars have low-mass progenitors, $M_{\rm i}^{\rm AGB} \simeq 1.3-1.5\, \Msun$.
Typical values of the dust-to-gas ratio $\Psi$ for Galactic carbon stars vary in the range $2.5\,10^{-3}\la \Psi\la 0.01$  \citep{Groenewegen_etal_98}. Assuming a mean value of $\Psi \approx 5\,10^{-3}$  the total  mass-loss rate, $\dot M$, does not exceed some $10^{-8}\, \Msun\, {\rm yr}^{-1}$ for most stars, while for  V$^\ast$ V493 Mon and [W71b] 030-31 it may reach a few $10^{-7}\, \Msun\, {\rm yr}^{-1}$. We recall that $\dot M \approx\, {\rm 3}\,10^{-7}\, \Msun\, {\rm yr}^{-1}$ is the minimum mass-loss rate for the existence of a  radiation-driven wind triggered by carbonaceous dust grains \citep{Bladh_etal_19_C, Eriksson_etal_14, Mattsson_etal_10}.

Therefore, we can conclude that almost all the C stars in the sample have not yet entered the dust-driven wind regime, while the onset is near or already in progress for V$^\ast$ V493 Mon and [W71b] 030-31.
These predictions agree very well with the observed photometric properties of the carbon stars. Looking at the \gaia-2MASS diagram (Fig.~\ref{fig_lebz}) we see, in fact, that both V$^\ast$ V493 Mon and [W71b] 030-31 are located in the region occupied by the so-called extreme stars, characterized by dusty outflows, and have an intrinsic color $(\jks)_0 \approx 3$.
We note that there is a  third carbon star classified as extreme, Case 49, but it is not considered here as it is rejected from being a member of the cluster NGC 663.
All the others are located in the region where dust-free carbon stars are commonly found; as expected they have  bluer near-infrared colors, typically $1.5 \la (\jks)_0 \la 2$.

\subsection{Core mass: comparison with the IFMR of white dwarfs}
\label{ssec_ifmr}

The following analysis is primarily based on the results for the noZP case;
some comparison with the two zero-point parallax corrections (L21ZP and G21ZP) is done at the end of the section.
Taking an overall look at the luminosity as a function of the initial mass (top-left panels of Fig.~\ref{fig_lum_ifmr} and Fig.~\ref{fig_lum_ifmr_D21}), we note that among the brightest TP-AGB stars in the sample there are the carbon stars  MSB 75 and BM IV 90, members of the clusters NGC 7789 and NGC 2660 ($L\approx 11\,500\,{\rm and}\,10\, 400\, \Lsun $). Their initial masses are estimated to be in the range $1.94 \la M_{\rm i}^{\rm AGB}/\Msun \la 2.17$.
At lower masses, $M_{\rm i}^{\rm AGB} < 1.8\, \Msun$ the luminosity drops to $L\la 6\,000\,\Lsun$, as shown by the stars [W71b] 030-31, C$^\ast$ 908, S1$^\ast$ 338, IRAS 23455+6819, HD 292921. At $M_{\rm i}^{\rm AGB} \simeq 2.1\, \Msun$,  we find two stars, - the S star  CSS 291 and the carbon star Case 63 -, with a lower luminosity, $L\simeq 7\,000-8\,000\,\Lsun$. 
The progenitor mass of carbon star Case 588, $1.8 \la M_{\rm i}^{\rm AGB}/\Msun \la 2.1$,  somewhat  depends on the adopted  cluster age set; independently from that it is one of the most luminous stars, with $L\simeq 9\,000 - 11\,000\,\Lsun$.
Then, for $M_{\rm i}^{\rm AGB} > 2.4\,\Msun$ $L$ increases, reaching a maximum value for the star BM IV 34 with $L\simeq 16\,000\,\Lsun$.

At this point it is interesting to compare the initial-final mass relation of the white dwarfs with the core mass of the TP-AGB stars inferred from their luminosity. We recall that the current \Mc\ sets a lower limit to the final mass (right panels of Fig.~\ref{fig_lum_ifmr} and Fig.~\ref{fig_lum_ifmr_D21}). 
First of all, we note that most sources classified as TP-AGB stars have core masses between the values expected at the first thermal pulse and the final mass of white dwarfs. This represents a remarkable agreement between theory and observations.
Notable exceptions are the  two carbon stars IRAS 19582+2907 and Case 121. Adopting the cluster age estimates from \citet{Cantat-Gaudin_etal_20} their \Mc\
appears located below the expected values at the first thermal pulse (top right panel of Fig.~\ref{fig_lum_ifmr}]).  As we have already discussed in Sect.~\ref{sect_results}, \nameref{para_cstars}, the apparent inconsistency is solved if we assign Case 121 the age from \cite{Dias_etal_21}.
Under this assumption Case 121 is a normal TP-AGB star and the  hypothesis that it could be a Super-AGB star is ruled out. 

Let us focus now on the $1.8 \la M_{\rm i}^{\rm AGB} /\Msun \la 2.1$ mass range. Here, the study of \citet{Marigo_etal_20} identified a kink in the IFMR, which unexpectedly interrupts the commonly assumed monotonic positive correlation between $M_{\rm i}^{\rm AGB}$ and the final mass $M_{\rm f}$. The kink's peak in white dwarf mass of $\simeq 0.70-0.75\,\Msun$ is produced by stars with $M_{\rm i}^{\rm AGB} \simeq 1.80-1.95\, \Msun$, whereas these final masses are typically associated with more massive stellar progenitors, $M_{\rm i}^{\rm AGB} \simeq 3.0-3.5\, \Msun$.

In the framework of single-star evolution, the proposed interpretation links this observational fact to the formation of carbon stars and the modest outflows (mass-loss rate $\dot M < 10^{-7}\, \Msun/{\rm yr}$) they should suffer as long as the carbon excess remains too low to produce dust grains in sufficient amount. 
In other words, the progenitor stars of the IFMR kink would be carbon stars that experienced a shallow 3DU with modest carbon enrichment unable to sustain a powerful dust-driven stellar wind. 
Under these conditions the mass of the carbon-oxygen core can grow more than is generally predicted by stellar models. We refer to the analysis of \citet{Marigo_etal_20} for all the details.

\begin{table*}
\centering
\scriptsize
\begin{threeparttable}
\setlength{\tabcolsep}{4.68pt}
\caption{TP-AGB and carbon stars expected from single-star evolution}
\begin{tabular*}{\textwidth}{@{\extracolsep{\fill}}cccccccccccccccccccc@{}}
\hline\hline
\multicolumn{9}{c}{} &
\multicolumn{3}{c}{NGC 7789 ($N_{\rm G}^{\rm obs}=2275$)} &
\multicolumn{3}{c}{NGC 2660 ($N_{\rm G}^{\rm obs}=376$)}\\
\cmidrule(lr){10-12}\cmidrule(lr){13-15}
\noalign{\smallskip}
\multicolumn{1}{c}{$\log({\rm age/yr})$} &
\multicolumn{1}{c}{$M_{\rm TO}$} & 
\multicolumn{1}{c}{$M_{\rm i}^{\rm AGB}$} & 
\multicolumn{1}{c}{$M_{\rm f}$} & 
\multicolumn{1}{c}{$\mathrm{(C/O)_f}$} & 
\multicolumn{1}{c}{$\tau_{\mathrm{TP{\text -}AGB}}$} &
\multicolumn{1}{c}{$\tau_{\mathrm{C}}$} &
\multicolumn{1}{c}{$N_{\mathrm{TP{\text -}AGB}}^{\rm SSP}$} &
\multicolumn{1}{c}{$N_{\mathrm{C}}^{\rm SSP}$} &
\multicolumn{1}{c}{$N^{\rm SSP}_{\rm G}$} &
\multicolumn{1}{c}{$N_{\rm{TP{\text -}AGB}}^{\rm cl}$} &
\multicolumn{1}{c}{$N_{\rm{C}}^{\rm cl}$} &
\multicolumn{1}{c}{$N^{\rm SSP}_{\rm G}$} &
\multicolumn{1}{c}{$N_{\rm{TP{\text -}AGB}}^{\rm cl}$} &
\multicolumn{1}{c}{$N_{\mathrm{C}}^{\rm cl}$} \\
\noalign{\smallskip}
\hline
\noalign{\smallskip}
\multicolumn{1}{c}{[dex]} & 
\multicolumn{1}{c}{[\Msun]} &
\multicolumn{1}{c}{[\Msun]} &
\multicolumn{1}{c}{[\Msun]} &
\multicolumn{1}{c}{} &
\multicolumn{1}{c}{[Myr]} & 
\multicolumn{1}{c}{[Myr]} & 
\multicolumn{1}{c}{} &
\multicolumn{1}{c}{} &
\multicolumn{1}{c}{} &
\multicolumn{1}{c}{Exp} &
\multicolumn{1}{c}{Exp} &
\multicolumn{1}{c}{} &
\multicolumn{1}{c}{Exp} &
\multicolumn{1}{c}{Exp} \\
\noalign{\smallskip}
\hline
\noalign{\smallskip}
  9.10 &  1.95 &  2.13 &  0.65 &  2.60 &  3.72 &  1.39 &  882 &  402 & 1024958 &  2.0 & 0.9 &  706884 &  0.3 & 0.1 \\
  9.11 &  1.93 &  2.11 &  0.65 &  2.56 &  3.79 &  1.37 & 1774 &  802 & 2043956 &  2.0 & 0.9 & 1408579 &  0.3 & 0.1 \\
  9.12 &  1.92 &  2.10 &  0.65 &  2.52 &  3.86 &  1.35 & 1865 &  800 & 2031698 &  2.1 & 0.9 & 1397565 &  0.3 & 0.1 \\
  9.13 &  1.90 &  2.08 &  0.65 &  2.44 &  3.90 &  1.32 & 2870 & 1143 & 3018628 &  2.2 & 0.9 & 2069930 &  0.3 & 0.1 \\
  9.14 &  1.89 &  2.06 &  0.65 &  2.34 &  3.91 &  1.28 & 2772 & 1050 & 3000056 &  2.1 & 0.8 & 2052403 &  0.3 & 0.1 \\
  9.15 &  1.87 &  2.05 &  0.65 &  2.24 &  3.93 &  1.24 &  875 &  307 &  994128 &  2.0 & 0.7 &  678343 &  0.3 & 0.1 \\
  9.16 &  1.86 &  2.03 &  0.66 &  2.14 &  3.94 &  1.20 &  881 &  289 &  986805 &  2.0 & 0.7 &  672358 &  0.3 & 0.1 \\
  9.17 &  1.84 &  2.01 &  0.66 &  2.02 &  3.96 &  1.17 & 2745 &  826 & 2936153 &  2.1 & 0.6 & 1997876 &  0.3 & 0.1 \\
  9.18 &  1.83 &  2.00 &  0.66 &  1.90 &  3.97 &  1.14 & 1864 &  537 & 1949348 &  2.2 & 0.6 & 1325518 &  0.3 & 0.1 \\
  9.19 &  1.81 &  1.98 &  0.67 &  1.77 &  4.05 &  1.08 & 1912 &  446 & 1919777 &  2.3 & 0.5 & 1298313 &  0.4 & 0.1 \\
  9.20 &  1.80 &  1.96 &  0.69 &  1.64 &  4.16 &  1.01 & 2840 &  652 & 2870420 &  2.3 & 0.5 & 1938252 &  0.4 & 0.1 \\
  9.21 &  1.79 &  1.94 &  0.70 &  1.52 &  4.26 &  0.94 &  928 &  206 &  950643 &  2.2 & 0.5 &  639939 &  0.4 & 0.1 \\
  9.22 &  1.77 &  1.93 &  0.70 &  1.43 &  4.34 &  0.88 &  941 &  211 &  944345 &  2.3 & 0.5 &  633951 &  0.4 & 0.1 \\
  9.23 &  1.76 &  1.91 &  0.71 &  1.35 &  4.43 &  0.81 & 8734 & 1955 & 1848616 & 10.7 & 2.4 & 1231415 &  1.8 & 0.4 \\
  9.24 &  1.74 &  1.83 &  0.75 &  1.32 &  4.23 &  1.06 & 8962 & 2070 & 1798429 & 11.3 & 2.6 & 1182242 &  1.9 & 0.4 \\
  9.25 &  1.73 &  1.81 &  0.74 &  1.32 &  4.01 &  1.06 & 2249 &  668 & 1779231 &  2.9 & 0.9 & 1159680 &  0.5 & 0.1 \\
  9.26 &  1.72 &  1.80 &  0.74 &  1.33 &  3.80 &  1.05 & 2841 &  902 & 2629329 &  2.5 & 0.8 & 1700064 &  0.4 & 0.1 \\
  9.27 &  1.70 &  1.78 &  0.73 &  1.32 &  3.61 &  1.01 & 1761 &  560 & 1744256 &  2.3 & 0.7 & 1125162 &  0.4 & 0.1 \\
  9.28 &  1.69 &  1.76 &  0.71 &  1.28 &  3.40 &  0.88 &  808 &  263 &  859626 &  2.1 & 0.7 &  550197 &  0.4 & 0.1 \\
  9.29 &  1.68 &  1.75 &  0.69 &  1.25 &  3.20 &  0.75 & 1508 &  470 & 1711477 &  2.0 & 0.6 & 1092995 &  0.3 & 0.1 \\
  9.30 &  1.66 &  1.73 &  0.67 &  1.22 &  3.01 &  0.62 & 1341 &  376 & 1695549 &  1.8 & 0.5 & 1077948 &  0.3 & 0.1 \\

\\\noalign{\smallskip}
\hline
\noalign{\smallskip}
\end{tabular*}
\footnotesize{{\bf Notes:} 
Predictions based on 
a simulated SSP with a total mass $M_{\rm SSP}=10^7\,  \Msun$, and solar metallicity,
for several values of the age in the interval $9.1\le \log({\rm age/yr}) \le 9.3$.
For each age the table reports the turn-off mass, $\MTO$, the initial and final mass of the AGB stars, $M_{\rm i}^{\rm AGB}$ and $M_{\rm f}$, the final surface carbon-to-oxygen ratio, (C/O)$_{\rm f}$, the TP-AGB and C-star lifetimes, $\tau_{\rm TP\text{-}AGB}$ and $\tau_{\rm C}$.
For both model and observations, $N_{\rm G}^{\rm SSP}$ and $N_{\rm G}^{\rm obs}$  are the number of stars with apparent magnitude in the interval $12 \le G \le 17$, where \gaia\ observations should be complete.
Cluster's visual extinction $A_V$ is taken from \citet{Cantat-Gaudin_etal_20}: $A_V=0.83$ for NGC 7789 and  $A_V=1.12$ for NGC 2660.
For each cluster we report the expected (Exp) number of TP-AGB stars, $N_{\rm{TP{\text -}AGB}}^{\rm cl}$,  and carbon stars, $N_{\rm C}^{\rm cl}$, computed with Eq.~(\ref{eq_ssp}). TP-AGB models are taken from \citet{Marigo_etal_20}.
}
\label{tab_ssp} 
\end{threeparttable}
\end{table*}

\begin{table*}
\centering
\scriptsize
\begin{threeparttable}
\setlength{\tabcolsep}{4.42pt}
\caption{TP-AGB and carbon stars expected from the blue-straggler channels}
\begin{tabular*}{\textwidth}{@{\extracolsep{\fill}}rcccccccccccccccccccc@{}}
\hline\hline
\multicolumn{12}{c}{} &
\multicolumn{2}{c}{NGC 7789 ($N_{\rm BSS}^{\rm obs}=16$)} &
\multicolumn{2}{c}{NGC 2660 ($N_{\rm BSS}^{\rm obs}=2$)} \\
\cmidrule(lr){13-14}\cmidrule(lr){15-16}
\noalign{\smallskip}
\multicolumn{4}{c}{} &
\multicolumn{8}{c}{MASS TRANSFER ($f_{\rm ch}=0.65$, $f_{\rm m}=0.25$)} &
\multicolumn{4}{c}{} \\
\cmidrule(lr){6-11}
\multicolumn{1}{c}{$q$} & 
\multicolumn{1}{c}{$M_2$} & 
\multicolumn{1}{c}{$\tau_{\rm MS,1}$} &
\multicolumn{1}{c}{$\tau_{\rm MS,2}$} &
\multicolumn{1}{c}{$M_{\rm BSS}$} & 
\multicolumn{1}{c}{$\tau_{\rm MS}$} &
\multicolumn{1}{c}{$\tau_{\mathrm{TP\text{-}AGB}}$} &
\multicolumn{1}{c}{$\tau_{\mathrm{C}}$} &
\multicolumn{1}{c}{$M_{\rm f}$} &  
\multicolumn{1}{c}{${\rm (C/O)_{f}}$} &  
\multicolumn{1}{c}{} &  
\multicolumn{1}{c}{$t_{\rm MS}$} &
\multicolumn{1}{c}{$N_{\mathrm{TP\text{-}AGB}}^{\rm cl}$} &
\multicolumn{1}{c}{$N_{\rm C}^{\rm cl}$} &
\multicolumn{1}{c}{$N_{\mathrm{TP\text{-}AGB}}^{\rm cl}$} &
\multicolumn{1}{c}{$N_{\rm C}^{\rm cl}$} \\
\noalign{\smallskip}
\hline
\noalign{\smallskip}
\multicolumn{1}{c}{} & 
\multicolumn{1}{c}{[\Msun]} &
\multicolumn{1}{c}{[Myr]} & 
\multicolumn{1}{c}{[Myr]} & 
\multicolumn{1}{c}{[\Msun]} & 
\multicolumn{1}{c}{[Myr]} & 
\multicolumn{1}{c}{[Myr]} & 
\multicolumn{1}{c}{[Myr]} & 
\multicolumn{1}{c}{[\Msun]} &
\multicolumn{1}{c}{} & 
\multicolumn{1}{c}{} & 
\multicolumn{1}{c}{[Myr]} & 
\multicolumn{1}{c}{Exp} & 
\multicolumn{1}{c}{Exp} & 
\multicolumn{1}{c}{Exp} & 
\multicolumn{1}{c}{Exp} \\
\noalign{\smallskip}
\hline
\noalign{\smallskip}
 1.0 & 1.80 & 1488 & 1488 & 3.00 &  362 & 1.74 & 0.76 &  0.73 &  2.90 &  & 60 &  7.5E-02 & 3.3E-02 & 9.4E-03 & 4.1E-03\\
 0.7 & 1.26 & 1488 & 4454 & 3.00 &  362 & 1.74 & 0.76 &  0.73 &  2.90 &  & 800 &  5.6E-03 & 2.5E-03 & 7.1E-04 & 3.1E-04\\
\noalign{\smallskip}
\hline 
\multicolumn{4}{c}{} &
\multicolumn{8}{c}{COLLISIONS ($f_{\rm ch}=0.15$, $f_{\rm m}=0.25$)} &
\multicolumn{4}{c}{} \\
\cmidrule(lr){6-11}
\multicolumn{1}{c}{$q$} & 
\multicolumn{1}{c}{$M_2$} & 
\multicolumn{1}{c}{$\tau_{\rm MS,1}$} &
\multicolumn{1}{c}{$\tau_{\rm MS,2}$} &
\multicolumn{1}{c}{$M_{\rm BSS}$} & 
\multicolumn{1}{c}{$\tau_{\rm MS}$} &
\multicolumn{1}{c}{$\tau_{\mathrm{TP\text{-}AGB}}$} &
\multicolumn{1}{c}{$\tau_{\mathrm{C}}$} &
\multicolumn{1}{c}{$M_{\rm f}$} &  
\multicolumn{1}{c}{${\rm (C/O)_{f}}$} &  
\multicolumn{1}{c}{$t_{\rm coll}$} &  
\multicolumn{1}{c}{$t_{\rm MS}$} &
\multicolumn{1}{c}{$N_{\mathrm{TP\text{-}AGB}}^{\rm cl}$} &
\multicolumn{1}{c}{$N_{\rm C}^{\rm cl}$} &
\multicolumn{1}{c}{$N_{\mathrm{TP\text{-}AGB}}^{\rm cl}$} &
\multicolumn{1}{c}{$N_{\rm C}^{\rm cl}$} \\
\noalign{\smallskip}
\hline
\noalign{\smallskip}
\multicolumn{1}{c}{} & 
\multicolumn{1}{c}{[\Msun]} &
\multicolumn{1}{c}{[Myr]} & 
\multicolumn{1}{c}{[Myr]} & 
\multicolumn{1}{c}{[\Msun]} & 
\multicolumn{1}{c}{[Myr]} & 
\multicolumn{1}{c}{[Myr]} & 
\multicolumn{1}{c}{[Myr]} & 
\multicolumn{1}{c}{[\Msun]} &
\multicolumn{1}{c}{} & 
\multicolumn{1}{c}{[Myr]} & 
\multicolumn{1}{c}{[Myr]} & 
\multicolumn{1}{c}{Exp} & 
\multicolumn{1}{c}{Exp} & 
\multicolumn{1}{c}{Exp} & 
\multicolumn{1}{c}{Exp} \\
\noalign{\smallskip}
\hline
\noalign{\smallskip}
 1.0 & 1.80 & 1488 & 1488 & 3.33 &  276 & 1.06 & 0.46 &  0.81 &  2.61 & 400 & 236 & 2.7E-03 & 1.2E-03 & 3.4E-04 & 1.5E-04\\
\multicolumn{10}{c}{} & 800 & 195 & 3.3E-03 & 1.4E-03 & 4.1E-04 & 1.8E-04\\
\multicolumn{10}{c}{} &1200 & 155 & 4.1E-03 & 1.8E-03 & 5.1E-04 & 2.2E-04\\
 0.7 & 1.26 & 1488 & 4454 & 2.84 &  419 & 1.87 & 0.97 &  0.69 &  3.31 & 400 & 373 & 3.0E-03 & 1.6E-03 & 3.8E-04 & 1.9E-04\\
\multicolumn{10}{c}{} & 800 & 327 & 3.4E-03 & 1.8E-03 & 4.3E-04 & 2.2E-04\\
\multicolumn{10}{c}{} &1200 & 282 & 4.0E-03 & 2.1E-03 & 5.0E-04 & 2.6E-04\\

\\\noalign{\smallskip}
\hline
\noalign{\smallskip}
\end{tabular*}
\footnotesize{{\bf Notes:} Predicted numbers of TP-AGB and C stars are obtained with Eq.~(\ref{eq_bss}) for both mass-transfer and collision channels. We assume the primary has a mass $M_1=1.80\,\Msun$ and consider two values of the mass ratio $q$.
We denote with $M_{\rm BSS}$  the blue straggler mass;  $\tau_{\rm MS,1}$, $\tau_{\rm MS,2}$, $\tau_{\rm MS}$ are the main sequence lifetimes of the primary, the secondary and the BSS, all derived from single star evolutionary models; $\tau_{\rm TP-AGB}$ and $\tau_{\rm C}$ denote the duration of the TP-AGB and C-star phases of a single star with $\Mi=M_{\rm BSS}$; 
$M_{\rm f}$ and (C/O)$_{\rm f}$ are the final mass and photospheric carbon-to-oxygen ratio;
$t_{\rm coll}$ is the collision age (only for the collision channel);
$t_{\rm MS}$ is the actual remaining main-sequence lifetime of the BSS. Given the observed number of main-sequence BSS, $N_{\rm BSS}^{\rm obs}$, the last four columns list the expected numbers of TP-AGB and C stars for the two clusters. The TP-AGB models are taken from \citet{Marigo_etal_20}.
}
\label{tab_bss}
\end{threeparttable}
\end{table*}

This scenario now appears strongly supported by the results of this work. Looking at the right panels Figs.~\ref{fig_lum_ifmr} and \ref{fig_lum_ifmr_D21} we see that there are indeed some carbon stars that populates the IFMR kink region. In particular MSB 75 (member of NGC 7789), BM IV 90 (belonging to NGC 2660) and Case 588 (hosted in Dias 2) have current core masses $0.67 \la \Mc/\Msun \la 0.70$. These values are comparable with the final masses that draw the IFMR kink. Another important aspect is that the four white dwarfs that define the kink peak belong to NGC 7789, the same cluster that hosts the brightest carbon star, MSB 75. This coincidence is very relevant as MSB 75 precisely defines the TP-AGB progenitor of the observed white dwarfs.

Furthermore, from the SED fitting of MSB 75, BM IV 90, and Case 588 we derive present-day dust mass-loss rates of some $10^{-10}\, \Msun/{\rm yr}$, which lead to total mass-loss rates of  $\dot M \approx \, {\rm few} \,10^{-8}\, \Msun/{\rm yr}$ assuming a dust-to-gas ratio $\Psi \simeq 200-300$.
These values are well below those that characterize a dust-driven wind.

In conclusion, this study confirms the two main hypotheses of the interpretation proposed by \citet{Marigo_etal_20}: 1) the progenitors of IFMR kink are carbon stars, 2) they are characterized by modest outflows, with mass-loss rates below the dust-driven regime.
Furthermore, a striking agreement is found between the white dwarf masses of the IFMR kink and the current core masses of the progenitor carbon stars, with $0.67 \la \Mc/\Msun\la 0.7$.

\subsection{Bright carbon stars in 1-2-Gyr old open clusters}
\label{ssec_highmc}
Here we discuss possible formation scenarios of the bright carbon stars MSB 75 and  BM IV 90, members of the clusters NGC 7789 and NGC 2660.
As shown in Sect.~\ref{ssec_cmlr} and Table~\ref{tab_sedfit}, their bolometric luminosities indicate current core masses close to $\Mc \simeq 0.68-0.7\, \Msun$, which are unusually higher than those expected from stellar evolutionary models  with initial masses $M_{\rm i}^{\rm AGB}\simeq1.9-2.0\,\Msun$ at solar-like metallicity. At the same time, such values of \Mc\ are consistent with the measurements of white dwarf masses, $\Mf \simeq 0.7-0.74 \, \Msun $, corresponding to progenitors of similar initial mass \citep{Marigo_etal_20}.

Specifically, we will analyze the probability that these observational facts can be explained by two alternative scenarios relating to 1) the evolution of a single star, or 2) an interacting binary system.

\subsubsection{ The single-star channel}
Under this hypothesis, we estimate the predicted number of TP-AGB and C stars in NGC 7789 and NGC 2660 with the aid of the population synthesis technique. The underlying assumption is that a star cluster can be described with a simple stellar population (SSP) of given age and metallicity.

To build the SSPs we first generate stellar isochrones at solar metallicity; they are the same as in \citet{Marigo_etal_17}, except for the TP-AGB phase for which 
we adopt the recent models computed by \citet{Marigo_etal_20}. They successfully reproduce the kink in the IFMR  observed at $\Mi \simeq 1.7-2.0\, \Msun$ where white dwarfs have measured masses up to $\Mf\simeq0.70-0.74 \Msun$.
These TP-AGB models naturally predict carbon stars with high core masses, as a consequence of a mild carbon enrichment and relatively moderate stellar winds.
The new isochrones  are then passed to the  \texttt{TRILEGAL} code \citep{Girardi_etal_05} to generate the SSPs.
The initial mass function is from \citet{Kroupa_02}.

We compute a fine grid of SSPs with ages in the relevant range for the clusters NGC 7789 and NGC 2660:  $9.10\le\mathrm{\log(age/yr)\le 9.30}$, adopting an incremental step of 0.01 dex. An initial total mass of $10^7\, \Msun$ is assumed, which ensures a statistically good representation of the short-lived TP-AGB evolution (a few hundreds of TP-AGB stars are present in each SSP).
For each cluster,
simulated stars are converted to \gaia\ $G$-band photometry, assuming the distance modulus and visual extinction $A_V$ from \citet{Cantat-Gaudin_etal_20}.

The expected number of TP-AGB stars in the cluster is computed with the scaling relation
\begin{equation}
N_{\rm TP\text{-}AGB}^{\rm cl}  \simeq N^{\rm obs}_{G} \times \frac{N_{\rm TP\text{-}AGB}^{\rm SSP}}{N^{\rm SSP}_{G}}\, ,
\label{eq_ssp}
\end{equation}
where $N^{\rm obs}_{G}$ and $N^{\rm SSP}_{G}$ are the observed and simulated number of stars 
within a given apparent G-magnitude range, 
$12 \le G \le 17$, inside which \gaia\ observations should be complete.
For both clusters this magnitude interval comprises a large fraction of  main-sequence stars and core He-burning stars on the red clump. Specifically, we find $N^{\rm obs}_{G}=2275$ for NGC 7789, and $N^{\rm obs}_{G}=248$ for NGC 2660.
The complete results are reported in Table~\ref{tab_ssp}.

\subsubsection{ The blue-straggler channel}

An alternative to  the carbon-star formation hypothesis proposed by \citet{Marigo_etal_20} is that unusually high core masses of 
MSB 75 and BM IV 90 are the result of the TP-AGB evolution of blue straggler stars (BSS) hosted in the parent clusters.
BSS are main-sequence stars that are observed to be bluer and brighter than the main-sequence turnoff in clusters. 
Among  the various BSS formation pathways proposed in literature we mention: mass transfer from a binary companion \citep{McCrea_64},  mergers of close binary systems, and collisions of single stars  \citep{HillsDay_76, Leonard_89}. In principle, all these mechanisms can contribute to the production of BSS: the most recent studies indicate that in open clusters the major fraction ($55\%-75\%$) of BSS  is produced through mass transfer in an interacting binary system, while a smaller role is to be attributed to collisional ($15\%$) or merging processes ($10\%-30\%$) 
\citep[e.g., ][]{LeinerGeller_21, JadhavSubramaniam_21, Gosnell_etal_15,Geller_etal_13, MathieuGeller_09}.

In the following we attempt to estimate the probability that the BSS channel is at work in the two clusters NGC 7789 and NGC 2660, and compare the results 
with the predictions from the single-star evolution scenario.

Using the semi-empirical IFMR and the results of stellar evolution models, we expect that core masses $M_{\rm c} \approx 0.7\, \Msun$ should be produced by BSS of masses $M_{\rm BSS} \approx 3.0\,\Msun$ that evolved through the TP-AGB phase and turned into carbon stars following repeated 3DU episodes. 
Considering that both clusters have a similar turn-off mass ($M_{\rm TO} \sim 1.80\, \Msun$ for $\log({\rm age/yr})\simeq 9.2$), the BSS progenitor should have 
a mass roughly twice the turn-off mass, $M_{\rm BSS} \approx 2 M_{\rm TO}$. This configuration can be reached through an interacting binary system in which the primary and the secondary 
companions have similar initial masses ($M_1 \approx M_2 \approx M_{\rm TO}$), hence a mass ratio not much different from unity ($q=M_1/M_2 \la 1$), provided that no extreme mass loss takes place.

A simple estimate of the expected number of TP-AGB stars progeny from a given BSS formation channel can be obtained with:
\begin{equation}
 N_{\rm TP\text{-}AGB}^{\rm cl}  \simeq f_{\rm ch} \times  f_{\rm m} \times N^{\rm obs}_{\rm BSS} \times \frac{\tau_{\rm TP\text{-}AGB}}{t_{\rm MS}}\, ,
 \label{eq_bss}
\end{equation}
where $N^{\rm obs}(\rm{BSS})$ is the number of observed  BSS in the cluster, $t_{\rm MS}$ and 
$\tau_{\rm TP\text{-}AGB}$ denote the remaining main-sequence lifetime and 
TP-AGB phase duration of the binary product with mass $M_{\rm BSS}$.  
A similar equation can be used to compute the expected number of carbon stars, $N_{\rm C}^{\rm cl}$, by replacing $\tau_{\rm TP\text{-}AGB}$ with the carbon star lifetime, $\tau_{\rm C}$. Stellar lifetimes are obtained from the \texttt{PARSEC-COLIBRI} evolutionary stellar models \citep{Bressan_etal_12, Marigo_etal_20}.
For each cluster the number $N^{\rm obs}_{\rm BSS}$ is taken from the new catalog of BSS in open star clusters \citep{Rain_etal_21}, which is based on \gaia\ DR2 data for astrometric, photometric and membership characterisation.

The multiplicative factor $f_{\rm ch}$ denotes the fractional contribution of a given BSS channel. Following the indications reported at the beginning of the section, we take: $f_{\rm ch}=0.65$ for the mass-transfer channel, $f_{\rm ch}=0.20$ for the merger channel, and $f_{\rm ch}=0.15$ for the collisional channel.
The other factor $f_{\rm m}$ is the probability that BSS are produced with a specified mass, that in our case we set to $M_{\rm BSS} \simeq 3.0\, \Msun$.

 To quantify $f_{\rm m}$ we take advantage of the results of two recent papers \citep{JadhavSubramaniam_21, LeinerGeller_21}, which investigated the population of BSS in several open clusters as a function of age with  \gaia\ DR2.
For clusters with $\mathrm{1\,\la age/Gyr \la 2}$, which encompass the range relevant to our test, both studies derived the semi-empirical mass distribution of BSS.
\citet[][see their figure 3]{LeinerGeller_21} found that, within their sample of 35 open clusters, the difference
between the blue straggler mass and the turnoff mass of the cluster  ($\delta M = M_{\rm BSS}- \MTO$) ranges from -0.2\, \Msun\ to 1.5\, \Msun, with a median of 0.4\, \Msun. Therefore, our test combination ($\MTO \simeq 1.8\,\Msun$ and $M_{\rm BSS}\simeq 2.8-3.0\,\Msun$) gives $\delta M \simeq 1.0-1.2\,\Msun $. This value belongs to the upper bins of the $\delta M$-distribution, and corresponds to an observed frequency of $\simeq 6/35 \approx 17\%$. 

\citet{JadhavSubramaniam_21} introduced the fractional mass excess for BSSs, $M_{\rm e} =
(M_{\rm BSS}-\MTO)/\MTO$ and grouped the clusters in three classes: low-mass excess with $M_{\rm e} < 0.5$, high-mass excess with $0.5 < M_{\rm e} < 1.0$ and extreme-mass excess with $M_{\rm e} > 1$. The 234 BSS hosted in their sample of 77 open clusters with $9.00 \le \log({\rm age/yr}) \le 9.25$ are distributed in the 3 classes with an observed frequency of 47\%, 30\% and 23\%, respectively (see their table 1). Our BSS test  configuration has $M_{\rm e}\simeq 0.56-0.67$, hence it is assigned the high-$M_{\rm e}$ class.

From the above considerations we conclude that the probability of finding BSS with $M_{\rm BSS}\simeq 3\,\Msun$
in NGC 7789 and NGC 2260 is not negligible and may correspond to about $20\%-30\%$.
Accordingly, we set $f_{\rm m} =0.25$ in Eq.(\ref{eq_bss}).

In the following we will analyze the predictions from the  mass-transfer and collision channels, which overall are expected to contribute to $70\%-90\%$ of the BSS populations in open clusters. To evaluate the impact of the merger channel to a first approximation, we can simply consider the complement to unity of the statistical contribution of the other two formation pathways.

\paragraph{Mass-transfer pathway}
Let us start by considering the most probable formation channel, which should be responsible for about $55\%-75\%$ of the BSS population in open clusters. 
Modeling a stable mass-transfer in a binary system is complex and typically involves several free parameters when an analytic approach is adopted \citep[see, for example, the prescriptions of the BSE code developed by ][]{Hurley_etal_02}.

Using the BSE code, \citet{LeinerGeller_21} carried out a systematic investigation of the mass-transfer process with the aid of  synthesis simulations. They pointed out that current binary models tend to underpredict, on average, the number of BSS formed via mass transfer in old open clusters, and concluded that this channel should be more stable than commonly assumed. 
They explored the performance of different prescriptions for $q_{\rm cr}$, that is the critical mass ratio below which mass transfer is found to be stable.
Among the various cases analyzed, one that comes closest to the observed BSS mass distribution (albeit with persisting defects) is the {\emph{L2/L3 Overflow}} model, in which the giant donor exceeds its Roche lobe and mass-transfer flows through the outer Lagrange points ($q_{\rm cr}=1.8$ is assumed).

In order to apply Eq.(\ref{eq_bss}) we need to know the BSS remaining main sequence lifetime, $t_{\rm MS}$.
For the {\emph{L2/L3 Overflow}} model \citet{LeinerGeller_21} computed $t_{\rm MS}$ as a function of the mass ratio $q$
(see their figure 9). 
For our specific application to the clusters NGC 7789 and NGC 2660, it is appropriate to consider the case with $\mathrm{age=1.5\, Gyr}$ (left panel of figure 9).  We see that for $1\ga q \ga 0.7$ the remaining lifetime $t_{\rm MS}$ varies from $\approx 60$ Myr to $\approx 800$ Myr, also depending on the binary period.
We adopt these two extreme values to bracket a wide range of cases.
Furthermore, we assume that the mass-transfer process always leads to the formation of a BSS with $M_{\rm BSS}=3\, \Msun$.

The BSE model accounts in a simple way for the possible rejuvenation of the BSS, if more hydrogen is mixed into the core, which results in a longer $t_{\rm MS}$. The precise amount of  rejuvenation is quite uncertain and recent detailed calculation for the blue straggler binary WOCS 5379 in NGC 188 indicates that 
the effective prolongation of $t_{\rm MS}$ may be larger than predicted by BSE \citep{LeinerGeller_21,Sun_etal_21}. In this respect we note that higher 
$t_{\rm MS}$ values tend to lower the expected star counts.

\paragraph{Collisional pathway.}
Let us now examine the case in which such a BSS is originated through stellar collisions, which should account for $\simeq 15\,\%$ of the BSS population in
open clusters.
Following the results of direct N-body calculations which account for collisions between two main-sequence stars in clusters \citep{Glebbeek_Pols_08, Hurley_etal_05, Hurley_etal_01}, it turns out that the post-main sequence phases of the merger product have similar duration  compared to those of a normal single star with the same initial mass. This implies that in Eq.~(\ref{eq_bss}) the quantity $t_{\rm TP-AGB}$ can be reasonably set equal to the TP-AGB lifetime  of a single star with $\Mi = M_{\rm BSS}$.

Conversely, the main sequence lifetime of the collision product, $t_{\rm MS}$, is shorter than that of a normal single star, $\tau_{\rm MS}$, by an amount that primarily depends on the age of the collision and on how much hydrogen is mixed into the core of the product after the collision \citep{Glebbeek_etal_13,Sills_etal_09,Glebbeek_Pols_08}.

According to \citet{Glebbeek_etal_13} and \citet{Glebbeek_Pols_08}
the  remaining lifetime $t_{\rm MS}$ of the collision product can be expressed as:
\begin{equation}
   t_{\rm MS} = (1-f_{\rm app}) \times \tau_{\rm MS} \, ,
\end{equation}
where  $f_{\rm app}$ is the apparent age of the product, that is the fractional age of a normal main sequence star with the same remaining lifetime $t_{\rm MS}$.

To estimate $f_{\rm app}$ we adopt the analytic recipe proposed by 
\citet{Glebbeek_Pols_08}, which is a function of $M_1$, $q$, the collision time $t_{\rm coll}$, and chemical composition. This relation reproduces fairly well the results of N-body simulations designed for old open clusters. 
As to the collision age, we explore three cases, namely: the collision takes place 1) close to the beginning,  $t_{\rm coll}=400\,{\rm Myr}$, 2) roughly in the middle, $t_{\rm coll}=800\,{\rm Myr}$, and 3) towards the end, $t_{\rm coll}=1200\,{\rm Myr}$ of the MS phase of the primary star. 

In the calculation we take into account the fact that some amount of mass, $\Delta M_{\rm lost}$, is lost through the collision, so that the actual mass of the BSS remnant is
\begin{equation}
M_{\rm BSS} = (M_1+M_2) \times (1 - \phi)\, ,
\end{equation}
where $\phi=0.3\,q/(1+q)^2$ depends on the mass ratio $q$, and $\Delta M_{\rm lost} = \phi \times (M_1+M_2)$.
For the two choices of the mass ratio, $q=1.0, 0.7$, the blue-straggler mass is $M_{\rm BSS}\simeq 3.33,\, {\rm and}\, 2.84\, \Msun$, respectively. 

\subsubsection{Discussion}

Tables~\ref{tab_ssp} and \ref{tab_bss} present the number of 
TP-AGB and carbon stars, $N_{\rm TP-AGB}^{\rm cl}$ and $N_{\rm C}^{\rm cl}$, expected from single-star evolution and the BSS channels.
For both scenarios we also report the predictions of \citet{Marigo_etal_20} regarding the TP-AGB lifetime ($\tau_{\rm TP\text{-}AGB}$), the duration of the carbon star phase ($\tau_{\rm C}$), the final mass ($\Mf$), and the surface carbon-to-oxygen ratio (C/O)$_{\rm f}$ at the end of the AGB evolution.
As to the mass-transfer and collisional pathways, we assume that the primary has a mass $M_1\simeq\MTO=1.8\,\Msun$, and a MS lifetime $\tau_{\rm MS}=1488\, {\rm Myr}$.

In both clusters only one carbon star was identified, which is also the only TP-AGB star detected.
The two clusters have very different population size: considering the candidate members with $p \ge 0.5$,
NGC 7789 contains 2953 stars, NGC 2660 has 376 stars, about a factor of 10 less.
As for NGC 7789, almost over the entire  age interval, the stellar evolution channel predicts $N_{\rm TP\text{-}AGB}^{\rm cl}\approx 2.0 \pm 1.4$, and  $N_{\rm C}^{\rm cl}\approx 0.5 \pm  0.7$ to $\approx 1 \pm  1$. Taking into account the Poisson uncertainty\footnote{It is simply computed as $\sqrt{N_{\rm TP\text{-}AGB}^{\rm cl}}$ and $\sqrt{N_{\rm C}^{\rm cl}}$.}, this is an excellent agreement with the observational data. As for NGC 2660, the predictions drop to $N_{\rm TP\text{-}AGB}^{\rm cl}\approx 0.3-0.4 \pm 0.6$,  and $N_{\rm C}^{\rm cl}\approx 0.1 \pm 0.3$. These numbers are still compatible with the detection of 1 TP-AGB star within the uncertainty interval.

We note that in a narrow age range, $9.23 \le \log({\rm age/yr})\le 9.24$, there is a significant increase in the expected numbers. This fact, known as AGB-boosting,  is related to the abrupt change in the core He-burning lifetime as soon as stellar populations intercept the ages at which red giant branch stars first appear \citep[all details can be found in ][]{Girardi_etal_13}. The boost occurs for $1.74 \la \MTO/\Msun \la 1.76$, and correspondingly for $1.83 \la M_{\rm i}^{\rm AGB}/\Msun \la 1.92$.
From an evolutionary point of view, these specific ages mark the transition between low-mass stars that develop a degenerate core after the main sequence, and intermediate-mass stars that do not undergo electronic degeneracy. From an observational point of view, clusters with ages $\simeq 1.6\,{\rm Gyr}$ show the peculiar morphology of the dual red clump in color-magnitude diagrams \citep{Girardi_etal_00, Girardi_etal_09, Girardi_16}.
Due to their age estimate and elongated red clump, the two clusters are plausible candidates to belong or to be very close to this special class of simple stellar populations. This means that the numbers of TP-AGB stars expected from the single star channel could easily be larger (by factors of a few) than here estimated.

Let us now move to analyze the BSS channels.
For all cases shown in Table~\ref{tab_bss} the predicted $N_{\rm TP\text{-}AGB}^{\rm cl}$ and $N_{\rm C}^{\rm cl}$ are systematically lower than observed, both for the mass-transfer and collision pathways. The expected counts are always of the order of $10^{-2}-10^{-4}$, the higher values applying to NGC 7789.
Given these estimates, we expect that also the merger formation channel, which should statistically account for the remaining $10\%-30\%$ of the BSS, helps negligibly to recover the observed counts.

Among all cases explored, the most favorable one seems to be that of a mass transfer with $q \approx 1$ (binary twins), which yields $N_{\rm TP\text{-}AGB}^{\rm cl} \simeq 0.075$ in NGC 7789.  This number could increase up to $\simeq 0.11$ assuming that all the 16 observed BBS formed via Roche-lobe overflow ($f_{\rm ch}=1$), and up to $\simeq 0.46$ if we hypothesize  that all have the same mass $M_{\rm BSS}\simeq 3\,\Msun$, although this seems rather unrealistic.

From these simple tests, we conclude that the bright carbon stars in NGC 7789 and NGC 2660 can be explained through the TP-AGB evolution of single stars with $1.75 \la M_{\rm i}^{\rm AGB}/\Msun \la 2.0$.
TP-AGB models that include 
a  carbon-dependent mass loss allow the growth of the core mass up to  $\Mc \approx 0.7\,\Msun$, as inferred from the luminosities of the observed carbon stars and measured from the spectra of their white dwarf progeny \citep{Marigo_etal_20}. On the other hand, the BSS channels do not seem to provide an equally convincing alternative to account for the observational data.

\section{Concluding remarks}
\label{sec_conclusion}
This study provides an in-depth analysis of the AGB star population in open clusters, in light of the new \gaia\ data.
We identified 49 AGB candidate stars brighter than the RGB tip. We focused on 19 stars with known spectral types (M, S, C), which should be evolving in the TP-AGB phase.
Their cluster membership was reanalyzed using all the astrometric and kinematic information provided by \gaia\ EDR3, also including zero-point corrections based on recent formulations.
Combining observations with evolutionary and radiative transport models, we characterized each star by assigning distance, spectral energy distribution from the optical to far infrared, initial mass, bolometric luminosity, core mass, circumstellar extinction, mass-loss rate, period and pulsation mode.

Let us briefly summarize the main conclusions:
\begin{itemize}
\item From the bolometric luminosity obtained through the SED fitting  we infer the current core mass  by using TP-AGB models in the literature.
We have paid careful attention to consider  the flash-driven luminosity variations and the evolution of the first pulses occurring below the asymptotic CMLR. These effects are particularly important for low-mass  stars, with $\Mc \la 0.65\, \Msun$.
Luminosity and core mass of almost all stars are well explained by TP-AGB evolutionary models, as they lie between the values predicted at the first thermal pulse and the end  of the AGB phase.
\item
For a few specific cases the results differ depending on the adopted catalog of cluster ages and/or parallax correction. The most striking example is the carbon star Case 121 whose initial mass is about $\simeq 7\,\Msun$ if we use  \citet{Cantat-Gaudin_Anders_20} to date its hosting cluster Berkeley 72, while it drops to $\simeq 3 \,\Msun$ if we use the age catalog of \citet{Dias_etal_21}.
With the former age estimate we face great interpretative difficulties (e.g., explaining how a Super-AGB star can have such a low luminosity,  $L \approx  14\,000- 15\,000 \Lsun$), whereas in the latter case the data for Case 121 is easily explained with the predictions of a  standard TP-AGB phase.
\item The minimum initial mass for carbon star formation at solar-like metallicity should not be higher than $\simeq 1.5\,\Msun$.
The maximum mass should not be lower than $3.0-4.0\,\Msun$, if we exclude that Case 121 has $M_{\rm i}~\simeq 7\,\Msun$ (as discussed in the previous point).
\item
The 3 stars of type MS and S provide information about the onset of the 3DU and the transition to the C-star domain.
\item The 12 carbon stars are all optically visible, and none appear truly dust-enshrouded.
The mass-loss rate for most of them is very low ($\dot M \approx 10^{-8}\,\Msun/{\rm yr}$), below the typical values that characterize a dust-driven wind, except for two carbon stars of low initial mass (V$^{\ast}$ V493 Mon and $[$W71b$]$) which fall in the region of extreme stars in the \gaia-2MASS diagram. For them the estimated mass-loss rate could be of the order of $10^{-7}-10^{-6}\, \Msun/{\rm yr}$.
\item The most massive star in the sample, BM IV 34,  is a carbon star, with $M_{\rm i}^{\rm AGB} \simeq 3.3-4.0\, \Msun$. Excluding Case 121 for the reasons discussed above, we did not find plausible candidates for stars with HBB. No M star brighter than the CMLR is identified with $M_{\rm i}^{\rm AGB} \ga 4\, \Msun$. Among the 4 M stars in the sample the maximum initial mass is $M_{\rm i}^{\rm AGB} \simeq 2.7\,\Msun$.
\item We looked for candidate Super-AGB stars, limiting to the age range $7.38 \la \log{\rm(age/yr)} \la 7.82$, hence $10 \ga \Mi/\Msun \ga 6$. We have identified 10 stars that satisfy the age criterion, but from the SED fitting the conclusion is negative since all luminosities are too low ($L < 40\,000\, \Lsun$).
\item
The photometric variability data we retrieved suggest the stars in the sample are LPVs. The observed periods, in combination with derived absolute magnitudes, are consistent with Mira-like or semi-regular variability. Most of the C-stars appear to be fundamental mode pulsators, while M-, MS- and S-type stars pulsate predominantly in the first overtone mode (consistent with the fact that they are less evolved), except for the S-star S1$^{\ast}$ 338  whose primary period is attributed to pulsation in the second overtone mode. The mode identification for this sample of LPVs is not affected by the choice of the parallax zero-point correction method. The two C-stars for which we derive the largest mass-loss rates (V$^{\ast}$ V493 Mon and $[$W71b$]$ 030-01) lie below the period-luminosity sequence C compared to LPVs with similar periods, in agreement with previous results for LPVs suffering from self-extinction due to circumstellar dust.
\item The comparison of the estimated \Mc\ with the IFMR of the white dwarfs has highlighted a striking fact: the presence of almost dust-free bright carbon stars with $0.65 \la \Mc/\Msun \la 0.70$, and initial masses of $\approx 1.9-2.0\, \Msun$.
Just in the same mass interval a recent study  \citep{Marigo_etal_20} pointed out the existence of a kink in the IFMR, which breaks its increasing monotonicity, with a peak in white dwarf mass of $\simeq  0.70-0.74\, \Msun$.

Therefore, the new findings of this study not only support the existence of the IFMR  kink, but also the underlying interpretative hypotheses: the progenitors are 1) carbon stars that 2) experienced modest outflows for a significant fraction of their C-rich phase, 3) with inefficient dust production. 
In fact, the carbon stars MSB 75 and BM IV 90 ($L \approx 10\,000 - 13\,000\,\Lsun$), have an estimated mass-loss rate of $\approx {\rm few}\, 10^{-8} \,\Msun/{\rm yr}$, while their variability is characterized by  low-amplitude pulsation. 
\citet{Marigo_etal_20} advanced the hypothesis that these stars are poorly enriched in carbon (following a shallow 3DU) and therefore dust cannot form in sufficient quantities to trigger a powerful wind \citep{Bladh_etal_19_C, Mattsson_etal_10}. Although we do not have photospheric C/O measurements for MSB 75 and BM IV 90,  their overall properties seem to fit very well within the suggested picture.
\item The above results are particularly intriguing as the open clusters NGC 7789 and NGC 2660 have ages $\approx 1.3-1.6\,{\rm Gyr}$ and show signs of dual clump morphology \citep{Girardi_etal_00, Girardi_16}. Therefore, their carbon star progenitors are expected to be close to the initial mass limit, $M_{\rm HeF}$, at the transition between low-mass stars, that develop degenerate He-cores after the main sequence, and intermediate-mass stars that avoid electron degeneracy.
\item Finally, to complete our analysis, we compare two possible pathways for the formation of carbon stars with $\Mc\simeq 0.7\,\Msun$ belonging to intermediate-age $1.3-1.6$ Gyr old clusters.
Our  calculations suggest that, while the evolution of single stars provides a consistent interpretation of the observed star counts, the
blue-straggler channel appears rather unlikely.
\end{itemize}

\appendix
\section{Cluster membership revisited with \gaia\ EDR3}
\label{sec_app}
Here we include the full table with the results of the new analysis, based on \gaia\ EDR3, to assess the cluster membership of the TP-AGB stars of known spectroscopic type.
\begin{table}
\centering
\begin{threeparttable}
\tiny
\caption{\gaia\ EDR3 astrometric parameters of the  C, S, and M stars and their candidate parent clusters. Cluster membership assessment is reported according to the original parallaxes and two cases of zero-point corrections, as indicated.}
\label{tab_stars}

\begin{tabular*}{\textwidth}{@{\extracolsep{\fill}}rcccccccccccc@{}}
\noalign{\smallskip}
\hline\hline
\

star          &  \multirow{2}{*}{$\mu_{\alpha}$ [mas/r]} &  \multirow{2}{*}{$\mu_{\delta}$ [mas/yr]}   & \multicolumn{3}{c}{$\pi_{\rm t}$ [mas]}    & \multicolumn{2}{c}{ZP [mas]}   & \multicolumn{3}{c}{membership}   & \multirow{2}{*}{type} \\ \cline{4-6} \cline{7-8} \cline{9-11}  
{\it cluster} &                                          &                                             &         noZP      & L21 &  G21             &     L21         &     G21        &    noZP      & L21 &  G21        &                       \\    
\hline\hline
                    V* V493 Mon &                  $-0.48\pm0.04$ &                  $+0.40\pm0.03$ &                  $+0.37\pm0.04$ &                  $+0.38\pm0.04$ &                  $+0.37\pm0.04$ &                        $-0.007$ &                        $-0.000$ & \multirow{3}{*}{\parbox{0.6cm}{\centering $\subseteq$ $68$\% C.L.}} & \multirow{3}{*}{\parbox{0.6cm}{\centering $\subseteq$ $68$\% C.L.}} & \multirow{3}{*}{\parbox{0.6cm}{\centering $\subseteq$ $68$\% C.L.}} &                               C\\
 \multirow{2}{*}{\it Trumpler 5} &             {\it [-0.76;-0.48]} &              {\it [+0.15;0.40]} &              {\it [+0.22;0.38]} &              {\it [+0.26;0.41]} &              {\it [+0.23;0.39]} &                                 &                                 &                                 &                                 &                                 &                                \\
                                &             {\it [-1.07;-0.04]} &              {\it [-0.24;0.79]} &              {\it [+0.06;0.54]} &              {\it [+0.09;0.56]} &              {\it [+0.07;0.54]} &                                 &                                 &                                 &                                 &                                 &                                \\
\hline
                  [W71b] 030-01 &                  $-4.65\pm0.03$ &                  $+6.78\pm0.04$ &                  $+0.56\pm0.03$ &                  $+0.56\pm0.03$ &                  $+0.59\pm0.03$ &                        $-0.001$ &                        $-0.030$ & \multirow{3}{*}{\parbox{0.6cm}{\centering $\subseteq$ $99$\% C.L.}} & \multirow{3}{*}{\parbox{0.6cm}{\centering $\subseteq$ $99$\% C.L.}} & \multirow{3}{*}{\parbox{0.6cm}{\centering $\subseteq$ $99$\% C.L.}} &                               C\\
   \multirow{2}{*}{\it Pismis 3} &             {\it [-4.91;-4.66]} &              {\it [+6.57;6.83]} &              {\it [+0.39;0.50]} &              {\it [+0.42;0.53]} &              {\it [+0.43;0.54]} &                                 &                                 &                                 &                                 &                                 &                                \\
                                &             {\it [-5.15;-4.39]} &              {\it [+6.30;7.03]} &              {\it [+0.27;0.61]} &              {\it [+0.30;0.64]} &              {\it [+0.31;0.65]} &                                 &                                 &                                 &                                 &                                 &                                \\
\hline
                         C* 908 &                  $-1.93\pm0.02$ &                  $+2.15\pm0.02$ &                  $+0.26\pm0.02$ &                  $+0.30\pm0.02$ &                  $+0.27\pm0.02$ &                        $-0.043$ &                        $-0.007$ & \multirow{3}{*}{\parbox{0.6cm}{\centering $\subseteq$ $99$\% C.L.}} & \multirow{3}{*}{\parbox{0.6cm}{\centering $\subseteq$ $99$\% C.L.}} & \multirow{3}{*}{\parbox{0.6cm}{\centering $\subseteq$ $99$\% C.L.}} &                               C\\
\multirow{2}{*}{\it Ruprecht 37} &             {\it [-1.75;-1.61]} &              {\it [+2.37;2.48]} &              {\it [+0.12;0.24]} &              {\it [+0.16;0.27]} &              {\it [+0.14;0.25]} &                                 &                                 &                                 &                                 &                                 &                                \\
                                &             {\it [-1.92;-1.41]} &              {\it [+2.13;2.68]} &              {\it [-0.00;0.32]} &              {\it [+0.03;0.35]} &              {\it [+0.01;0.33]} &                                 &                                 &                                 &                                 &                                 &                                \\
\hline
                       Case 588 &                  $-0.98\pm0.04$ &                  $+1.28\pm0.03$ &                  $+0.21\pm0.03$ &                  $+0.27\pm0.03$ &                  $+0.20\pm0.03$ &                        $-0.056$ &                         $0.014$ & \multirow{3}{*}{\parbox{0.6cm}{\centering $\subseteq$ $99$\% C.L.}} & \multirow{3}{*}{\parbox{0.6cm}{\centering $\subseteq$ $99$\% C.L.}} & \multirow{3}{*}{\parbox{0.6cm}{\centering $\subseteq$ $99$\% C.L.}} &                               C\\
     \multirow{2}{*}{\it Dias 2} &             {\it [-0.89;-0.68]} &              {\it [+1.16;1.34]} &              {\it [+0.15;0.28]} &              {\it [+0.18;0.32]} &              {\it [+0.14;0.28]} &                                 &                                 &                                 &                                 &                                 &                                \\
                                &             {\it [-1.37;-0.42]} &              {\it [+1.06;1.81]} &              {\it [-0.02;0.41]} &              {\it [+0.01;0.44]} &              {\it [-0.03;0.39]} &                                 &                                 &                                 &                                 &                                 &                                \\
\hline
                       BM IV 90 &                  $-2.77\pm0.03$ &                  $+5.30\pm0.03$ &                  $+0.35\pm0.03$ &                  $+0.38\pm0.03$ &                  $+0.35\pm0.03$ &                        $-0.032$ &                        $-0.004$ & \multirow{3}{*}{\parbox{0.6cm}{\centering $\subseteq$ $68$\% C.L.}} & \multirow{3}{*}{\parbox{0.6cm}{\centering $\subseteq$ $68$\% C.L.}} & \multirow{3}{*}{\parbox{0.6cm}{\centering $\subseteq$ $68$\% C.L.}} &                               C\\
   \multirow{2}{*}{\it NGC 2660} &             {\it [-2.83;-2.65]} &              {\it [+5.12;5.29]} &              {\it [+0.29;0.38]} &              {\it [+0.32;0.41]} &              {\it [+0.30;0.39]} &                                 &                                 &                                 &                                 &                                 &                                \\
                                &             {\it [-2.99;-2.46]} &              {\it [+4.87;5.55]} &              {\it [+0.18;0.50]} &              {\it [+0.20;0.53]} &              {\it [+0.18;0.51]} &                                 &                                 &                                 &                                 &                                 &                                \\
\hline
                         MSB 75 &                  $-1.07\pm0.02$ &                  $-2.20\pm0.02$ &                  $+0.49\pm0.03$ &                  $+0.53\pm0.03$ &                  $+0.49\pm0.03$ &                        $-0.047$ &                        $-0.001$ & \multirow{3}{*}{\parbox{0.6cm}{\centering $\subseteq$ $99$\% C.L.}} & \multirow{3}{*}{\parbox{0.6cm}{\centering $\subseteq$ $99$\% C.L.}} & \multirow{3}{*}{\parbox{0.6cm}{\centering $\subseteq$ $99$\% C.L.}} &                               C\\
   \multirow{2}{*}{\it NGC 7789} &             {\it [-1.04;-0.80]} &             {\it [-2.08;-1.83]} &              {\it [+0.44;0.52]} &              {\it [+0.47;0.54]} &              {\it [+0.45;0.53]} &                                 &                                 &                                 &                                 &                                 &                                \\
                                &             {\it [-1.29;-0.53]} &             {\it [-2.35;-1.59]} &              {\it [+0.32;0.63]} &              {\it [+0.35;0.65]} &              {\it [+0.33;0.64]} &                                 &                                 &                                 &                                 &                                 &                                \\
\hline
                        Case 63 &                  $+1.54\pm0.03$ &                  $+0.08\pm0.03$ &                  $+0.57\pm0.03$ &                  $+0.63\pm0.03$ &                  $+0.58\pm0.03$ &                        $-0.054$ &                        $-0.009$ & \multirow{3}{*}{\parbox{0.6cm}{\centering $\subseteq$ $99$\% C.L.}} & \multirow{3}{*}{\parbox{0.6cm}{\centering $\subseteq$ $99$\% C.L.}} & \multirow{3}{*}{\parbox{0.6cm}{\centering $\subseteq$ $99$\% C.L.}} &                               C\\
 \multirow{2}{*}{\it Berkeley 9} &              {\it [+1.41;1.61]} &              {\it [-0.10;0.10]} &              {\it [+0.49;0.61]} &              {\it [+0.52;0.65]} &              {\it [+0.50;0.63]} &                                 &                                 &                                 &                                 &                                 &                                \\
                                &              {\it [+1.07;1.85]} &              {\it [-0.37;0.73]} &              {\it [+0.36;0.76]} &              {\it [+0.40;0.79]} &              {\it [+0.37;0.77]} &                                 &                                 &                                 &                                 &                                 &                                \\
\hline
                       Case 473 &                  $-3.90\pm0.04$ &                  $-5.78\pm0.04$ &                  $+0.27\pm0.03$ &                  $+0.29\pm0.03$ &                  $+0.30\pm0.03$ &                        $-0.019$ &                        $-0.023$ & \multirow{3}{*}{\parbox{0.6cm}{\centering $\subseteq$ $68$\% C.L.}} & \multirow{3}{*}{\parbox{0.6cm}{\centering $\subseteq$ $68$\% C.L.}} & \multirow{3}{*}{\parbox{0.6cm}{\centering $\subseteq$ $68$\% C.L.}} &                               C\\
\multirow{2}{*}{\it Berkeley 53} &             {\it [-4.00;-3.69]} &             {\it [-5.81;-5.55]} &              {\it [+0.17;0.33]} &              {\it [+0.22;0.37]} &              {\it [+0.19;0.34]} &                                 &                                 &                                 &                                 &                                 &                                \\
                                &             {\it [-4.51;-3.23]} &             {\it [-6.31;-5.02]} &              {\it [+0.01;0.53]} &              {\it [+0.06;0.56]} &              {\it [+0.03;0.54]} &                                 &                                 &                                 &                                 &                                 &                                \\
\hline
                     Wray 18-47 &                  $-3.26\pm0.02$ &                  $+5.11\pm0.02$ &                  $+0.36\pm0.02$ &                  $+0.40\pm0.02$ &                  $+0.41\pm0.02$ &                        $-0.037$ &                        $-0.048$ & \multirow{3}{*}{\parbox{0.6cm}{\centering $\subseteq$ $99$\% C.L.}} & \multirow{3}{*}{\parbox{0.6cm}{\centering $\subseteq$ $99$\% C.L.}} & \multirow{3}{*}{\parbox{0.6cm}{\centering $\subseteq$ $99$\% C.L.}} &                               C\\
   \multirow{2}{*}{\it NGC 2533} &             {\it [-3.21;-3.15]} &              {\it [+5.02;5.11]} &              {\it [+0.33;0.37]} &              {\it [+0.36;0.41]} &              {\it [+0.38;0.43]} &                                 &                                 &                                 &                                 &                                 &                                \\
                                &             {\it [-3.31;-2.96]} &              {\it [+4.94;5.24]} &              {\it [+0.26;0.44]} &              {\it [+0.29;0.47]} &              {\it [+0.31;0.50]} &                                 &                                 &                                 &                                 &                                 &                                \\
\hline
                       BM IV 34 &                  $-1.77\pm0.02$ &                  $+1.68\pm0.02$ &                  $+0.22\pm0.02$ &                  $+0.26\pm0.02$ &                  $+0.27\pm0.02$ &                        $-0.039$ &                        $-0.047$ & \multirow{3}{*}{\parbox{0.6cm}{\centering $\subseteq$ $68$\% C.L.}} & \multirow{3}{*}{\parbox{0.6cm}{\centering $\subseteq$ $68$\% C.L.}} & \multirow{3}{*}{\parbox{0.6cm}{\centering $\subseteq$ $68$\% C.L.}} &                               C\\
 \multirow{2}{*}{\it Haffner 14} &             {\it [-1.87;-1.77]} &              {\it [+1.68;1.79]} &              {\it [+0.21;0.28]} &              {\it [+0.25;0.31]} &              {\it [+0.27;0.34]} &                                 &                                 &                                 &                                 &                                 &                                \\
                                &             {\it [-2.04;-1.60]} &              {\it [+1.49;1.96]} &              {\it [+0.12;0.40]} &              {\it [+0.15;0.43]} &              {\it [+0.17;0.46]} &                                 &                                 &                                 &                                 &                                 &                                \\
\hline
                IRAS 19582+2907 &                  $-2.28\pm0.04$ &                  $-5.74\pm0.05$ &                  $+0.20\pm0.05$ &                  $+0.27\pm0.05$ &                  $+0.24\pm0.05$ &                        $-0.065$ &                        $-0.039$ & \multirow{3}{*}{\parbox{0.6cm}{\centering $\subseteq$ $99$\% C.L.}} & \multirow{3}{*}{\parbox{0.6cm}{\centering $\subseteq$ $99$\% C.L.}} & \multirow{3}{*}{\parbox{0.6cm}{\centering $\subseteq$ $99$\% C.L.}} &                               C\\
    \multirow{2}{*}{\it FSR 172} &             {\it [-2.61;-2.49]} &             {\it [-6.05;-5.86]} &              {\it [+0.26;0.33]} &              {\it [+0.30;0.37]} &              {\it [+0.29;0.37]} &                                 &                                 &                                 &                                 &                                 &                                \\
                                &             {\it [-3.01;-2.28]} &             {\it [-6.12;-5.43]} &              {\it [+0.10;0.40]} &              {\it [+0.16;0.43]} &              {\it [+0.13;0.43]} &                                 &                                 &                                 &                                 &                                 &                                \\
\hline
                       Case 121 &                  $+0.61\pm0.04$ &                  $-0.10\pm0.03$ &                  $+0.20\pm0.03$ &                  $+0.26\pm0.03$ &                  $+0.22\pm0.03$ &                        $-0.060$ &                        $-0.026$ & \multirow{3}{*}{\parbox{0.6cm}{\centering $\subseteq$ $99$\% C.L.}} & \multirow{3}{*}{\parbox{0.6cm}{\centering $\subseteq$ $99$\% C.L.}} & \multirow{3}{*}{\parbox{0.6cm}{\centering $\subseteq$ $99$\% C.L.}} &                               C\\
\multirow{2}{*}{\it Berkeley 72} &              {\it [+0.67;0.88]} &             {\it [-0.35;-0.13]} &              {\it [+0.07;0.22]} &              {\it [+0.11;0.25]} &              {\it [+0.10;0.25]} &                                 &                                 &                                 &                                 &                                 &                                \\
                                &              {\it [+0.30;1.26]} &              {\it [-0.65;0.24]} &              {\it [+0.01;0.48]} &              {\it [+0.04;0.51]} &              {\it [+0.04;0.51]} &                                 &                                 &                                 &                                 &                                 &                                \\
\hline
                      NIKC 3-81 &                  $+1.31\pm0.02$ &                  $+0.43\pm0.02$ &                  $+0.25\pm0.02$ &                  $+0.26\pm0.02$ &                  $+0.26\pm0.02$ &                        $-0.006$ &                        $-0.007$ & \multirow{3}{*}{\parbox{0.6cm}{\centering $\subseteq$ $68$\% C.L.}} & \multirow{3}{*}{\parbox{0.6cm}{\centering $\subseteq$ $68$\% C.L.}} & \multirow{3}{*}{\parbox{0.6cm}{\centering $\subseteq$ $68$\% C.L.}} &                               C\\
\multirow{2}{*}{\it Berkeley 14} &              {\it [+1.29;1.53]} &              {\it [+0.29;0.51]} &              {\it [+0.12;0.27]} &              {\it [+0.15;0.30]} &              {\it [+0.13;0.29]} &                                 &                                 &                                 &                                 &                                 &                                \\
                                &              {\it [+0.90;1.80]} &              {\it [+0.02;0.79]} &              {\it [-0.05;0.40]} &              {\it [-0.02;0.43]} &              {\it [-0.03;0.41]} &                                 &                                 &                                 &                                 &                                 &                                \\
\hline
                        Case 49 &                  $-1.40\pm0.02$ &                  $-0.38\pm0.02$ &                  $+0.43\pm0.03$ &                  $+0.44\pm0.03$ &                  $+0.45\pm0.03$ &                        $-0.010$ &                        $-0.018$ & \multirow{3}{*}{\parbox{0.6cm}{\centering rejected}} & \multirow{3}{*}{\parbox{0.6cm}{\centering rejected}} & \multirow{3}{*}{\parbox{0.6cm}{\centering rejected}} &                               C\\
    \multirow{2}{*}{\it NGC 663} &             {\it [-1.20;-1.07]} &             {\it [-0.40;-0.25]} &              {\it [+0.31;0.38]} &              {\it [+0.35;0.40]} &              {\it [+0.34;0.40]} &                                 &                                 &                                 &                                 &                                 &                                \\
                                &             {\it [-1.32;-0.95]} &             {\it [-0.55;-0.08]} &              {\it [+0.24;0.44]} &              {\it [+0.28;0.47]} &              {\it [+0.27;0.47]} &                                 &                                 &                                 &                                 &                                 &                                \\
\hline
                        S1* 338 &                  $-4.15\pm0.02$ &                  $+3.19\pm0.02$ &                  $+0.21\pm0.02$ &                  $+0.23\pm0.02$ &                  $+0.23\pm0.02$ &                        $-0.019$ &                        $-0.022$ & \multirow{3}{*}{\parbox{0.6cm}{\centering $\subseteq$ $99$\% C.L.}} & \multirow{3}{*}{\parbox{0.6cm}{\centering $\subseteq$ $99$\% C.L.}} & \multirow{3}{*}{\parbox{0.6cm}{\centering $\subseteq$ $99$\% C.L.}} &                               S\\
      \multirow{2}{*}{\it BH 55} &             {\it [-4.03;-3.85]} &              {\it [+3.09;3.23]} &              {\it [+0.16;0.27]} &              {\it [+0.19;0.30]} &              {\it [+0.18;0.28]} &                                 &                                 &                                 &                                 &                                 &                                \\
                                &             {\it [-4.20;-3.63]} &              {\it [+2.97;3.34]} &              {\it [+0.02;0.36]} &              {\it [+0.05;0.39]} &              {\it [+0.04;0.38]} &                                 &                                 &                                 &                                 &                                 &                                \\
\hline
                 [D75b] Star 30 &                  $+0.64\pm0.04$ &                  $-0.20\pm0.03$ &                  $+0.29\pm0.03$ &                  $+0.33\pm0.03$ &                  $+0.31\pm0.03$ &                        $-0.037$ &                        $-0.019$ & \multirow{3}{*}{\parbox{0.6cm}{\centering $\subseteq$ $99$\% C.L.}} & \multirow{3}{*}{\parbox{0.6cm}{\centering $\subseteq$ $99$\% C.L.}} & \multirow{3}{*}{\parbox{0.6cm}{\centering $\subseteq$ $99$\% C.L.}} &                              MS\\
   \multirow{2}{*}{\it NGC 1798} &              {\it [+0.71;0.90]} &             {\it [-0.44;-0.28]} &              {\it [+0.15;0.25]} &              {\it [+0.19;0.28]} &              {\it [+0.17;0.27]} &                                 &                                 &                                 &                                 &                                 &                                \\
                                &              {\it [+0.44;1.19]} &              {\it [-0.75;0.08]} &              {\it [+0.07;0.35]} &              {\it [+0.09;0.38]} &              {\it [+0.08;0.37]} &                                 &                                 &                                 &                                 &                                 &                                \\
\hline
                        CSS 291 &                  $-2.61\pm0.02$ &                  $+3.72\pm0.02$ &                  $+0.42\pm0.02$ &                  $+0.46\pm0.02$ &                  $+0.45\pm0.02$ &                        $-0.039$ &                        $-0.034$ & \multirow{3}{*}{\parbox{0.6cm}{\centering $\subseteq$ $99$\% C.L.}} & \multirow{3}{*}{\parbox{0.6cm}{\centering $\subseteq$ $99$\% C.L.}} & \multirow{3}{*}{\parbox{0.6cm}{\centering $\subseteq$ $99$\% C.L.}} &                               S\\
 \multirow{2}{*}{\it Tombaugh 1} &             {\it [-2.63;-2.49]} &              {\it [+3.76;3.91]} &              {\it [+0.35;0.42]} &              {\it [+0.39;0.45]} &              {\it [+0.39;0.46]} &                                 &                                 &                                 &                                 &                                 &                                \\
                                &             {\it [-2.83;-2.27]} &              {\it [+3.64;4.03]} &              {\it [+0.28;0.50]} &              {\it [+0.31;0.53]} &              {\it [+0.31;0.54]} &                                 &                                 &                                 &                                 &                                 &                                \\
\hline
                IRAS 23455+6819 &                  $-3.37\pm0.05$ &                  $-0.70\pm0.05$ &                  $+0.33\pm0.05$ &                  $+0.36\pm0.05$ &                  $+0.37\pm0.05$ &                        $-0.027$ &                        $-0.045$ & \multirow{3}{*}{\parbox{0.6cm}{\centering $\subseteq$ $68$\% C.L.}} & \multirow{3}{*}{\parbox{0.6cm}{\centering $\subseteq$ $68$\% C.L.}} & \multirow{3}{*}{\parbox{0.6cm}{\centering $\subseteq$ $68$\% C.L.}} &                               M\\
    \multirow{2}{*}{\it King 11} &             {\it [-3.49;-3.30]} &             {\it [-0.76;-0.56]} &              {\it [+0.24;0.36]} &              {\it [+0.27;0.40]} &              {\it [+0.27;0.40]} &                                 &                                 &                                 &                                 &                                 &                                \\
                                &             {\it [-3.90;-3.00]} &             {\it [-0.97;-0.33]} &              {\it [+0.08;0.58]} &              {\it [+0.11;0.61]} &              {\it [+0.11;0.60]} &                                 &                                 &                                 &                                 &                                 &                                \\
\hline
                      HD 292921 &                  $-1.17\pm0.02$ &                  $+0.36\pm0.02$ &                  $+0.25\pm0.02$ &                  $+0.28\pm0.02$ &                  $+0.25\pm0.02$ &                        $-0.035$ &                        $-0.008$ & \multirow{3}{*}{\parbox{0.6cm}{\centering $\subseteq$ $99$\% C.L.}} & \multirow{3}{*}{\parbox{0.6cm}{\centering $\subseteq$ $99$\% C.L.}} & \multirow{3}{*}{\parbox{0.6cm}{\centering $\subseteq$ $99$\% C.L.}} &                               M\\
\multirow{2}{*}{\it Berkeley 34} &             {\it [-1.44;-1.27]} &              {\it [+0.12;0.29]} &              {\it [+0.08;0.24]} &              {\it [+0.12;0.27]} &              {\it [+0.09;0.24]} &                                 &                                 &                                 &                                 &                                 &                                \\
                                &             {\it [-1.78;-0.91]} &              {\it [+0.06;0.38]} &              {\it [-0.01;0.34]} &              {\it [+0.01;0.37]} &              {\it [-0.00;0.35]} &                                 &                                 &                                 &                                 &                                 &                                \\
\hline
                IRAS 09251-5101 &                  $-3.79\pm0.05$ &                  $+2.90\pm0.05$ &                  $+0.14\pm0.05$ &                  $+0.20\pm0.05$ &                  $+0.16\pm0.05$ &                        $-0.060$ &                        $-0.018$ & \multirow{3}{*}{\parbox{0.6cm}{\centering $\subseteq$ $68$\% C.L.}} & \multirow{3}{*}{\parbox{0.6cm}{\centering $\subseteq$ $68$\% C.L.}} & \multirow{3}{*}{\parbox{0.6cm}{\centering $\subseteq$ $68$\% C.L.}} &                               M\\
      \multirow{2}{*}{\it BH 67} &             {\it [-3.74;-3.54]} &              {\it [+2.72;2.92]} &              {\it [+0.04;0.17]} &              {\it [+0.07;0.20]} &              {\it [+0.04;0.17]} &                                 &                                 &                                 &                                 &                                 &                                \\
                                &             {\it [-4.19;-3.18]} &              {\it [+2.42;3.01]} &              {\it [-0.03;0.28]} &              {\it [+0.00;0.31]} &              {\it [-0.02;0.29]} &                                 &                                 &                                 &                                 &                                 &                                \\
\hline
2MASS        J00161695+5958115 &                  $-1.79\pm0.03$ &                  $-1.56\pm0.03$ &                  $+0.23\pm0.03$ &                  $+0.24\pm0.03$ &                  $+0.25\pm0.03$ &                        $-0.013$ &                        $-0.028$ & \multirow{3}{*}{\parbox{0.6cm}{\centering $\subseteq$ $99$\% C.L.}} & \multirow{3}{*}{\parbox{0.6cm}{\centering $\subseteq$ $99$\% C.L.}} & \multirow{3}{*}{\parbox{0.6cm}{\centering $\subseteq$ $99$\% C.L.}} &                               M\\
\multirow{2}{*}{\it Juchert Saloran 1} &             {\it [-1.73;-1.55]} &             {\it [-1.42;-1.21]} &              {\it [+0.13;0.26]} &              {\it [+0.16;0.29]} &              {\it [+0.16;0.29]} &                                 &                                 &                                 &                                 &                                 &                                \\
                                &             {\it [-2.17;-1.33]} &             {\it [-1.73;-1.07]} &              {\it [-0.06;0.38]} &              {\it [-0.03;0.41]} &              {\it [-0.03;0.41]} &                                 &                                 &                                 &                                 &                                 &                                \\
\hline
\end{tabular*}
\medskip
\footnotesize{{\bf Notes:} 
For each star, moving leftward  we report: designation,  $\mu_{\alpha},\, \mu_{\delta}$, $\pi_{\rm t}$ (for noZP, L21ZP and G21ZP cases) and their uncertainty, the parallax corrections according to \citet{Lindegren_etal_21} and \citet{Groenewegen21},  the membership assessments, and the spectral type.
For each cluster, in correspondence to each parameter the two intervals indicate the extremes of the  confidence regions of 68\% and 99\%, respectively.
}
\end{threeparttable}
\end{table}

\bibliography{AGBstars_oc.bib}{}

\begin{thebibliography}{}
\expandafter\ifx\csname natexlab\endcsname\relax\def\natexlab#1{#1}\fi
\providecommand{\url}[1]{\href{#1}{#1}}
\providecommand{\dodoi}[1]{doi:~\href{http://doi.org/#1}{\nolinkurl{#1}}}
\providecommand{\doeprint}[1]{\href{http://ascl.net/#1}{\nolinkurl{http://ascl.net/#1}}}
\providecommand{\doarXiv}[1]{\href{https://arxiv.org/abs/#1}{\nolinkurl{https://arxiv.org/abs/#1}}}

\bibitem[{{Abia} {et~al.}(2020){Abia}, {de Laverny}, {Cristallo}, {Kordopatis},
  \& {Straniero}}]{Abia_etal_20}
{Abia}, C., {de Laverny}, P., {Cristallo}, S., {Kordopatis}, G., \&
  {Straniero}, O. 2020, \aap, 633, A135, \dodoi{10.1051/0004-6361/201936831}

\bibitem[{Addari(2020)}]{Addari_2020}
Addari, F. 2020, Master's thesis, University of Padova, Italy

\bibitem[{{Alksnis} {et~al.}(2001){Alksnis}, {Balklavs}, {Dzervitis},
  {Eglitis}, {Paupers}, \& {Pundure}}]{Alksnis_etal_01}
{Alksnis}, A., {Balklavs}, A., {Dzervitis}, U., {et~al.} 2001, Baltic
  Astronomy, 10, 1, \dodoi{10.1515/astro-2001-1-202}

\bibitem[{{Althaus} {et~al.}(2021){Althaus}, {Gil-Pons}, {C{\'o}rsico}, {Miller
  Bertolami}, {De Ger{\'o}nimo}, {Camisassa}, {Torres}, {Gutierrez}, \&
  {Rebassa-Mansergas}}]{Althaus_etal_21}
{Althaus}, L.~G., {Gil-Pons}, P., {C{\'o}rsico}, A.~H., {et~al.} 2021, \aap,
  646, A30, \dodoi{10.1051/0004-6361/202038930}

\bibitem[{{Aringer} {et~al.}(2009){Aringer}, {Girardi}, {Nowotny}, {Marigo}, \&
  {Lederer}}]{Aringer_etal_09}
{Aringer}, B., {Girardi}, L., {Nowotny}, W., {Marigo}, P., \& {Lederer}, M.~T.
  2009, \aap, 503, 913, \dodoi{10.1051/0004-6361/200911703}

\bibitem[{{Asplund} {et~al.}(2009){Asplund}, {Grevesse}, {Sauval}, \&
  {Scott}}]{AGSS09}
{Asplund}, M., {Grevesse}, N., {Sauval}, A.~J., \& {Scott}, P. 2009, \araa, 47,
  481, \dodoi{10.1146/annurev.astro.46.060407.145222}

\bibitem[{{Barnett} {et~al.}(2021){Barnett}, {Williams}, {B{\'e}dard}, \&
  {Bolte}}]{Barnett_etal_21}
{Barnett}, J.~W., {Williams}, K.~A., {B{\'e}dard}, A., \& {Bolte}, M. 2021,
  arXiv e-prints, arXiv:2107.06373.
\newblock \doarXiv{2107.06373}

\bibitem[{{Bayo} {et~al.}(2008){Bayo}, {Rodrigo}, {Barrado Y Navascu{\'e}s},
  {Solano}, {Guti{\'e}rrez}, {Morales-Calder{\'o}n}, \& {Allard}}]{vosa08}
{Bayo}, A., {Rodrigo}, C., {Barrado Y Navascu{\'e}s}, D., {et~al.} 2008, \aap,
  492, 277, \dodoi{10.1051/0004-6361:200810395}

\bibitem[{{Bedding} \& {Zijlstra}(1998)}]{BeddingZijlstra_1998}
{Bedding}, T.~R., \& {Zijlstra}, A.~A. 1998, \apjl, 506, L47,
  \dodoi{10.1086/311632}

\bibitem[{{Beichman} et al.(1988)}]{IRAS_88}
{Beichman} et al. 1988, {Infrared Astronomical Satellite (IRAS) Catalogs and
  Atlases.Volume 1: Explanatory Supplement.}, Vol.~1

\bibitem[{{Bellm} {et~al.}(2019){Bellm}, {Kulkarni}, {Graham}, {Dekany},
  {Smith}, {Riddle}, {Masci}, {Helou}, {Prince}, {Adams}, {Barbarino},
  {Barlow}, {Bauer}, {Beck}, {Belicki}, {Biswas}, {Blagorodnova}, {Bodewits},
  {Bolin}, {Brinnel}, {Brooke}, {Bue}, {Bulla}, {Burruss}, {Cenko}, {Chang},
  {Connolly}, {Coughlin}, {Cromer}, {Cunningham}, {De}, {Delacroix}, {Desai},
  {Duev}, {Eadie}, {Farnham}, {Feeney}, {Feindt}, {Flynn}, {Franckowiak},
  {Frederick}, {Fremling}, {Gal-Yam}, {Gezari}, {Giomi}, {Goldstein},
  {Golkhou}, {Goobar}, {Groom}, {Hacopians}, {Hale}, {Henning}, {Ho}, {Hover},
  {Howell}, {Hung}, {Huppenkothen}, {Imel}, {Ip}, {Ivezi{\'c}}, {Jackson},
  {Jones}, {Juric}, {Kasliwal}, {Kaspi}, {Kaye}, {Kelley}, {Kowalski},
  {Kramer}, {Kupfer}, {Landry}, {Laher}, {Lee}, {Lin}, {Lin}, {Lunnan},
  {Giomi}, {Mahabal}, {Mao}, {Miller}, {Monkewitz}, {Murphy}, {Ngeow},
  {Nordin}, {Nugent}, {Ofek}, {Patterson}, {Penprase}, {Porter}, {Rauch},
  {Rebbapragada}, {Reiley}, {Rigault}, {Rodriguez}, {van Roestel}, {Rusholme},
  {van Santen}, {Schulze}, {Shupe}, {Singer}, {Soumagnac}, {Stein}, {Surace},
  {Sollerman}, {Szkody}, {Taddia}, {Terek}, {Van Sistine}, {van Velzen},
  {Vestrand}, {Walters}, {Ward}, {Ye}, {Yu}, {Yan}, \&
  {Zolkower}}]{Bellm_etal_2019}
{Bellm}, E.~C., {Kulkarni}, S.~R., {Graham}, M.~J., {et~al.} 2019, \pasp, 131,
  018002, \dodoi{10.1088/1538-3873/aaecbe}

\bibitem[{{Bhardwaj} {et~al.}(2021){Bhardwaj}, {Rejkuba}, {de Grijs}, {Yang},
  {Herczeg}, {Marconi}, {Singh}, {Kanbur}, \& {Ngeow}}]{Bhardwaj_etal_21}
{Bhardwaj}, A., {Rejkuba}, M., {de Grijs}, R., {et~al.} 2021, \apj, 909, 200,
  \dodoi{10.3847/1538-4357/abdf48}

\bibitem[{{Bladh} {et~al.}(2019{\natexlab{a}}){Bladh}, {Eriksson}, {Marigo},
  {Liljegren}, \& {Aringer}}]{Bladh_etal_19_C}
{Bladh}, S., {Eriksson}, K., {Marigo}, P., {Liljegren}, S., \& {Aringer}, B.
  2019{\natexlab{a}}, \aap, 623, A119, \dodoi{10.1051/0004-6361/201834778}

\bibitem[{{Bladh} {et~al.}(2019{\natexlab{b}}){Bladh}, {Liljegren},
  {H{\"o}fner}, {Aringer}, \& {Marigo}}]{Bladh_etal_19_M}
{Bladh}, S., {Liljegren}, S., {H{\"o}fner}, S., {Aringer}, B., \& {Marigo}, P.
  2019{\natexlab{b}}, \aap, 626, A100, \dodoi{10.1051/0004-6361/201935366}

\bibitem[{{Bloecker} \& {Schoenberner}(1991)}]{Blocker_91}
{Bloecker}, T., \& {Schoenberner}, D. 1991, \aap, 244, L43

\bibitem[{{Blum} {et~al.}(2006){Blum}, {Mould}, {Olsen}, {Frogel}, {Werner},
  {Meixner}, {Markwick-Kemper}, {Indebetouw}, {Whitney}, {Meade}, {Babler},
  {Churchwell}, {Gordon}, {Engelbracht}, {For}, {Misselt}, {Vijh}, {Leitherer},
  {Volk}, {Points}, {Reach}, {Hora}, {Bernard}, {Boulanger}, {Bracker},
  {Cohen}, {Fukui}, {Gallagher}, {Gorjian}, {Harris}, {Kelly}, {Kawamura},
  {Latter}, {Madden}, {Mizuno}, {Mizuno}, {Nota}, {Oey}, {Onishi}, {Paladini},
  {Panagia}, {Perez-Gonzalez}, {Shibai}, {Sato}, {Smith}, {Staveley-Smith},
  {Tielens}, {Ueta}, {Van Dyk}, \& {Zaritsky}}]{Blum_etal_06}
{Blum}, R.~D., {Mould}, J.~R., {Olsen}, K.~A., {et~al.} 2006, \aj, 132, 2034,
  \dodoi{10.1086/508227}

\bibitem[{{Boothroyd} \& {Sackmann}(1988)}]{BoothroydSackmann_88}
{Boothroyd}, A.~I., \& {Sackmann}, I.~J. 1988, \apj, 328, 641,
  \dodoi{10.1086/166322}

\bibitem[{{Boothroyd} \& {Sackmann}(1992)}]{BoothroydSackmann_91}
---. 1992, \apjl, 393, L21, \dodoi{10.1086/186441}

\bibitem[{{Bossini} {et~al.}(2019){Bossini}, {Vallenari}, {Bragaglia},
  {Cantat-Gaudin}, {Sordo}, {Balaguer-N{\'u}{\~n}ez}, {Jordi}, {Moitinho},
  {Soubiran}, {Casamiquela}, {Carrera}, \& {Heiter}}]{Bossini_etal_19}
{Bossini}, D., {Vallenari}, A., {Bragaglia}, A., {et~al.} 2019, \aap, 623,
  A108, \dodoi{10.1051/0004-6361/201834693}

\bibitem[{{Boyer} {et~al.}(2008){Boyer}, {McDonald}, {van Loon}, {Woodward},
  {Gehrz}, {Evans}, \& {Dupree}}]{Boyer_etal_08}
{Boyer}, M.~L., {McDonald}, I., {van Loon}, J.~T., {et~al.} 2008, \aj, 135,
  1395, \dodoi{10.1088/0004-6256/135/4/1395}

\bibitem[{{Boyer} {et~al.}(2011){Boyer}, {Srinivasan}, {van Loon}, {McDonald},
  {Meixner}, {Zaritsky}, {Gordon}, {Kemper}, {Babler}, {Block}, {Bracker},
  {Engelbracht}, {Hora}, {Indebetouw}, {Meade}, {Misselt}, {Robitaille},
  {Sewi{\l}o}, {Shiao}, \& {Whitney}}]{Boyer_etal_11}
{Boyer}, M.~L., {Srinivasan}, S., {van Loon}, J.~T., {et~al.} 2011, \aj, 142,
  103, \dodoi{10.1088/0004-6256/142/4/103}

\bibitem[{{Boyer} {et~al.}(2013){Boyer}, {Girardi}, {Marigo}, {Williams},
  {Aringer}, {Nowotny}, {Rosenfield}, {Dorman}, {Guhathakurta}, {Dalcanton},
  {Melbourne}, {Olsen}, \& {Weisz}}]{Boyer_etal_13}
{Boyer}, M.~L., {Girardi}, L., {Marigo}, P., {et~al.} 2013, \apj, 774, 83,
  \dodoi{10.1088/0004-637X/774/1/83}

\bibitem[{{Bressan} {et~al.}(2012){Bressan}, {Marigo}, {Girardi}, {Salasnich},
  {Dal Cero}, {Rubele}, \& {Nanni}}]{Bressan_etal_12}
{Bressan}, A., {Marigo}, P., {Girardi}, L., {et~al.} 2012, \mnras, 427, 127,
  \dodoi{10.1111/j.1365-2966.2012.21948.x}

\bibitem[{{Bruzual}(2007)}]{Bruzual_07}
{Bruzual}, A.~G. 2007, in Stellar Populations as Building Blocks of Galaxies,
  ed. A.~{Vazdekis} \& R.~{Peletier}, Vol. 241, 125--132,
  \dodoi{10.1017/S1743921307007624}

\bibitem[{{Caffau} {et~al.}(2011){Caffau}, {Ludwig}, {Steffen}, {Freytag}, \&
  {Bonifacio}}]{Caffau_etal_11}
{Caffau}, E., {Ludwig}, H.~G., {Steffen}, M., {Freytag}, B., \& {Bonifacio}, P.
  2011, \solphys, 268, 255, \dodoi{10.1007/s11207-010-9541-4}

\bibitem[{{Cantat-Gaudin} \& {Anders}(2020)}]{Cantat-Gaudin_Anders_20}
{Cantat-Gaudin}, T., \& {Anders}, F. 2020, \aap, 633, A99,
  \dodoi{10.1051/0004-6361/201936691}

\bibitem[{{Cantat-Gaudin} {et~al.}(2018){Cantat-Gaudin}, {Jordi}, {Vallenari},
  {Bragaglia}, {Balaguer-N{\'u}{\~n}ez}, {Soubiran}, {Bossini}, {Moitinho},
  {Castro-Ginard}, {Krone-Martins}, {Casamiquela}, {Sordo}, \&
  {Carrera}}]{Cantat-Gaudin_etal_18}
{Cantat-Gaudin}, T., {Jordi}, C., {Vallenari}, A., {et~al.} 2018, \aap, 618,
  A93, \dodoi{10.1051/0004-6361/201833476}

\bibitem[{{Cantat-Gaudin} {et~al.}(2020){Cantat-Gaudin}, {Anders},
  {Castro-Ginard}, {Jordi}, {Romero-G{\'o}mez}, {Soubiran}, {Casamiquela},
  {Tarricq}, {Moitinho}, {Vallenari}, {Bragaglia}, {Krone-Martins}, \&
  {Kounkel}}]{Cantat-Gaudin_etal_20}
{Cantat-Gaudin}, T., {Anders}, F., {Castro-Ginard}, A., {et~al.} 2020, \aap,
  640, A1, \dodoi{10.1051/0004-6361/202038192}

\bibitem[{{Cardelli} {et~al.}(1989){Cardelli}, {Clayton}, \&
  {Mathis}}]{Cardelli_etal_89}
{Cardelli}, J.~A., {Clayton}, G.~C., \& {Mathis}, J.~S. 1989, \apj, 345, 245,
  \dodoi{10.1086/167900}

\bibitem[{{Carraro} \& {Ortolani}(1994)}]{CarraroOrtolani_94}
{Carraro}, G., \& {Ortolani}, S. 1994, \aap, 291, 106

\bibitem[{{Carrera} {et~al.}(2019){Carrera}, {Bragaglia}, {Cantat-Gaudin},
  {Vallenari}, {Balaguer-N{\'u}{\~n}ez}, {Bossini}, {Casamiquela}, {Jordi},
  {Sordo}, \& {Soubiran}}]{Carrera_etal_19}
{Carrera}, R., {Bragaglia}, A., {Cantat-Gaudin}, T., {et~al.} 2019, \aap, 623,
  A80, \dodoi{10.1051/0004-6361/201834546}

\bibitem[{{Castro-Ginard} {et~al.}(2019){Castro-Ginard}, {Jordi}, {Luri},
  {Cantat-Gaudin}, \& {Balaguer-N{\'u}{\~n}ez}}]{Castro-Ginard_etal_19}
{Castro-Ginard}, A., {Jordi}, C., {Luri}, X., {Cantat-Gaudin}, T., \&
  {Balaguer-N{\'u}{\~n}ez}, L. 2019, \aap, 627, A35,
  \dodoi{10.1051/0004-6361/201935531}

\bibitem[{{Castro-Ginard} {et~al.}(2020){Castro-Ginard}, {Jordi}, {Luri},
  {{\'A}lvarez Cid-Fuentes}, {Casamiquela}, {Anders}, {Cantat-Gaudin},
  {Mongui{\'o}}, {Balaguer-N{\'u}{\~n}ez}, {Sol{\`a}}, \&
  {Badia}}]{Castro-Ginard_etal_20}
{Castro-Ginard}, A., {Jordi}, C., {Luri}, X., {et~al.} 2020, \aap, 635, A45,
  \dodoi{10.1051/0004-6361/201937386}

\bibitem[{{Catchpole} \& {Feast}(1973)}]{CatchpoleFeast_73}
{Catchpole}, R.~M., \& {Feast}, M.~W. 1973, \mnras, 164, 11P,
  \dodoi{10.1093/mnras/164.1.11P}

\bibitem[{{Chen} {et~al.}(2020){Chen}, {Wang}, {Deng}, {de Grijs}, {Yang}, \&
  {Tian}}]{Chen_etal_2020_ZTF}
{Chen}, X., {Wang}, S., {Deng}, L., {et~al.} 2020, \apjs, 249, 18,
  \dodoi{10.3847/1538-4365/ab9cae}

\bibitem[{{Chen} {et~al.}(2015){Chen}, {Bressan}, {Girardi}, {Marigo}, {Kong},
  \& {Lanza}}]{Chen_etal_15}
{Chen}, Y., {Bressan}, A., {Girardi}, L., {et~al.} 2015, \mnras, 452, 1068,
  \dodoi{10.1093/mnras/stv1281}

\bibitem[{{Chiavassa} {et~al.}(2018){Chiavassa}, {Freytag}, \&
  {Schultheis}}]{Chiavassa_etal_18}
{Chiavassa}, A., {Freytag}, B., \& {Schultheis}, M. 2018, \aap, 617, L1,
  \dodoi{10.1051/0004-6361/201833844}

\bibitem[{{Cioni} {et~al.}(2003){Cioni}, {Blommaert}, {Groenewegen}, {Habing},
  {Hron}, {Kerschbaum}, {Loup}, {Omont}, {van Loon}, {Whitelock}, \&
  {Zijlstra}}]{Cioni_etal_2003}
{Cioni}, M. R.~L., {Blommaert}, J.~A.~D.~L., {Groenewegen}, M.~A.~T., {et~al.}
  2003, \aap, 406, 51, \dodoi{10.1051/0004-6361:20030707}

\bibitem[{{Costa} {et~al.}(2021){Costa}, {Bressan}, {Mapelli}, {Marigo},
  {Iorio}, \& {Spera}}]{Costa_etal_21}
{Costa}, G., {Bressan}, A., {Mapelli}, M., {et~al.} 2021, \mnras, 501, 4514,
  \dodoi{10.1093/mnras/staa3916}

\bibitem[{{Cristallo} {et~al.}(2011){Cristallo}, {Piersanti}, {Straniero},
  {Gallino}, {Dom{\'\i}nguez}, {Abia}, {Di Rico}, {Quintini}, \&
  {Bisterzo}}]{Cristallo_etal_11}
{Cristallo}, S., {Piersanti}, L., {Straniero}, O., {et~al.} 2011, \apjs, 197,
  17, \dodoi{10.1088/0067-0049/197/2/17}

\bibitem[{{Cummings} {et~al.}(2018){Cummings}, {Kalirai}, {Tremblay},
  {Ramirez-Ruiz}, \& {Choi}}]{Cummings_etal_18}
{Cummings}, J.~D., {Kalirai}, J.~S., {Tremblay}, P.~E., {Ramirez-Ruiz}, E., \&
  {Choi}, J. 2018, \apj, 866, 21, \dodoi{10.3847/1538-4357/aadfd6}

\bibitem[{{Cutri} {et~al.}(2014){Cutri}, {1}, {2}, {1}, {2}, {1}, {2}, \&
  {2}}]{wise14}
{Cutri}, R.~M., {1}, a., {2}, a., {et~al.} 2014, VizieR Online Data Catalog,
  II/328

\bibitem[{{Cutri} {et~al.}(2003){Cutri}, {Skrutskie}, {van Dyk}, {Beichman},
  {Carpenter}, {Chester}, {Cambresy}, {Evans}, {Fowler}, {Gizis}, {Howard},
  {Huchra}, {Jarrett}, {Kopan}, {Kirkpatrick}, {Light}, {Marsh}, {McCallon},
  {Schneider}, {Stiening}, {Sykes}, {Weinberg}, {Wheaton}, {Wheelock}, \&
  {Zacarias}}]{Cutri_etal_03}
{Cutri}, R.~M., {Skrutskie}, M.~F., {van Dyk}, S., {et~al.} 2003, VizieR Online
  Data Catalog, II/246

\bibitem[{{Dalcanton} {et~al.}(2009){Dalcanton}, {Williams}, {Seth}, {Dolphin},
  {Holtzman}, {Rosema}, {Skillman}, {Cole}, {Girardi}, {Gogarten},
  {Karachentsev}, {Olsen}, {Weisz}, {Christensen}, {Freeman}, {Gilbert},
  {Gallart}, {Harris}, {Hodge}, {de Jong}, {Karachentseva}, {Mateo}, {Stetson},
  {Tavarez}, {Zaritsky}, {Governato}, \& {Quinn}}]{Dalcanton_etal_09}
{Dalcanton}, J.~J., {Williams}, B.~F., {Seth}, A.~C., {et~al.} 2009, \apjs,
  183, 67, \dodoi{10.1088/0067-0049/183/1/67}

\bibitem[{{Dalcanton} {et~al.}(2012){Dalcanton}, {Williams}, {Lang}, {Lauer},
  {Kalirai}, {Seth}, {Dolphin}, {Rosenfield}, {Weisz}, {Bell}, {Bianchi},
  {Boyer}, {Caldwell}, {Dong}, {Dorman}, {Gilbert}, {Girardi}, {Gogarten},
  {Gordon}, {Guhathakurta}, {Hodge}, {Holtzman}, {Johnson}, {Larsen}, {Lewis},
  {Melbourne}, {Olsen}, {Rix}, {Rosema}, {Saha}, {Sarajedini}, {Skillman}, \&
  {Stanek}}]{Dalcanton_etal_12}
{Dalcanton}, J.~J., {Williams}, B.~F., {Lang}, D., {et~al.} 2012, \apjs, 200,
  18, \dodoi{10.1088/0067-0049/200/2/18}

\bibitem[{{Dell'Agli} {et~al.}(2015){Dell'Agli}, {Ventura}, {Schneider}, {Di
  Criscienzo}, {Garc{\'\i}a-Hern{\'a}ndez}, {Rossi}, \&
  {Brocato}}]{DellAgli_etal_15}
{Dell'Agli}, F., {Ventura}, P., {Schneider}, R., {et~al.} 2015, \mnras, 447,
  2992, \dodoi{10.1093/mnras/stu2559}

\bibitem[{{Dias} {et~al.}(2021){Dias}, {Monteiro}, {Moitinho}, {L{\'e}pine},
  {Carraro}, {Paunzen}, {Alessi}, \& {Villela}}]{Dias_etal_21}
{Dias}, W.~S., {Monteiro}, H., {Moitinho}, A., {et~al.} 2021, \mnras, 504, 356,
  \dodoi{10.1093/mnras/stab770}

\bibitem[{{Doherty} {et~al.}(2015){Doherty}, {Gil-Pons}, {Siess}, {Lattanzio},
  \& {Lau}}]{Doherty_etal_15}
{Doherty}, C.~L., {Gil-Pons}, P., {Siess}, L., {Lattanzio}, J.~C., \& {Lau}, H.
  H.~B. 2015, \mnras, 446, 2599, \dodoi{10.1093/mnras/stu2180}

\bibitem[{{Egan} {et~al.}(2003){Egan}, {Price}, {Kraemer}, {Mizuno}, {Carey},
  {Wright}, {Engelke}, {Cohen}, \& {Gugliotti}}]{Egan_etal_03}
{Egan}, M.~P., {Price}, S.~D., {Kraemer}, K.~E., {et~al.} 2003, VizieR Online
  Data Catalog, V/114

\bibitem[{{Eggen} \& {Iben}(1991)}]{EggenIben_91}
{Eggen}, O.~J., \& {Iben}, Icko, J. 1991, \aj, 101, 1377,
  \dodoi{10.1086/115773}

\bibitem[{{Eggleton}(1967)}]{Eggleton_67}
{Eggleton}, P.~P. 1967, \mnras, 135, 243, \dodoi{10.1093/mnras/135.3.243}

\bibitem[{{El-Badry} {et~al.}(2018){El-Badry}, {Rix}, \&
  {Weisz}}]{El-Badry_etal_18}
{El-Badry}, K., {Rix}, H.-W., \& {Weisz}, D.~R. 2018, \apjl, 860, L17,
  \dodoi{10.3847/2041-8213/aaca9c}

\bibitem[{{Eriksson} {et~al.}(2014){Eriksson}, {Nowotny}, {H{\"o}fner},
  {Aringer}, \& {Wachter}}]{Eriksson_etal_14}
{Eriksson}, K., {Nowotny}, W., {H{\"o}fner}, S., {Aringer}, B., \& {Wachter},
  A. 2014, \aap, 566, A95, \dodoi{10.1051/0004-6361/201323241}

\bibitem[{{Feast} {et~al.}(1989){Feast}, {Glass}, {Whitelock}, \&
  {Catchpole}}]{Feast_etal_1989}
{Feast}, M.~W., {Glass}, I.~S., {Whitelock}, P.~A., \& {Catchpole}, R.~M. 1989,
  \mnras, 241, 375, \dodoi{10.1093/mnras/241.3.375}

\bibitem[{{Ferrarotti} \& {Gail}(2006)}]{FerrarottiGail_06}
{Ferrarotti}, A.~S., \& {Gail}, H.~P. 2006, \aap, 447, 553,
  \dodoi{10.1051/0004-6361:20041198}

\bibitem[{{Ferreira} {et~al.}(2020){Ferreira}, {Corradi}, {Maia}, {Angelo}, \&
  {Santos}}]{Ferreira_etal_20}
{Ferreira}, F.~A., {Corradi}, W.~J.~B., {Maia}, F.~F.~S., {Angelo}, M.~S., \&
  {Santos}, J.~F.~C., J. 2020, \mnras, 496, 2021,
  \dodoi{10.1093/mnras/staa1684}

\bibitem[{{Fragkou} {et~al.}(2019){Fragkou}, {Parker}, {Zijlstra}, {Shaw}, \&
  {Lykou}}]{Fragkou_etal_19}
{Fragkou}, V., {Parker}, Q.~A., {Zijlstra}, A., {Shaw}, R., \& {Lykou}, F.
  2019, \mnras, 484, 3078, \dodoi{10.1093/mnras/stz108}

\bibitem[{{Freedman} {et~al.}(2020){Freedman}, {Madore}, {Hoyt}, {Jang},
  {Beaton}, {Lee}, {Monson}, {Neeley}, \& {Rich}}]{Freedman_etal_20}
{Freedman}, W.~L., {Madore}, B.~F., {Hoyt}, T., {et~al.} 2020, \apj, 891, 57,
  \dodoi{10.3847/1538-4357/ab7339}

\bibitem[{{Frogel} {et~al.}(1990){Frogel}, {Mould}, \&
  {Blanco}}]{FrogelMouldBlanco_90}
{Frogel}, J.~A., {Mould}, J., \& {Blanco}, V.~M. 1990, \apj, 352, 96,
  \dodoi{10.1086/168518}

\bibitem[{{Frost} {et~al.}(1998){Frost}, {Cannon}, {Lattanzio}, {Wood}, \&
  {Forestini}}]{Frost_etal_98}
{Frost}, C.~A., {Cannon}, R.~C., {Lattanzio}, J.~C., {Wood}, P.~R., \&
  {Forestini}, M. 1998, \aap, 332, L17.
\newblock \doarXiv{astro-ph/9710054}

\bibitem[{{Frost} \& {Lattanzio}(1996)}]{FrostLattanzio_96}
{Frost}, C.~A., \& {Lattanzio}, J.~C. 1996, \apj, 473, 383,
  \dodoi{10.1086/178152}

\bibitem[{{Gaia Collaboration} {et~al.}(2018){Gaia Collaboration}, {Brown},
  {Vallenari}, {Prusti}, {de Bruijne}, {Babusiaux}, {Bailer-Jones}, {Biermann},
  {Evans}, {Eyer}, {Jansen}, {Jordi}, {Klioner}, {Lammers}, {Lindegren},
  {Luri}, {Mignard}, {Panem}, {Pourbaix}, {Randich}, {Sartoretti}, {Siddiqui},
  {Soubiran}, {van Leeuwen}, {Walton}, {Arenou}, {Bastian}, {Cropper},
  {Drimmel}, {Katz}, {Lattanzi}, {Bakker}, {Cacciari}, {Casta{\~n}eda},
  {Chaoul}, {Cheek}, {De Angeli}, {Fabricius}, {Guerra}, {Holl}, {Masana},
  {Messineo}, {Mowlavi}, {Nienartowicz}, {Panuzzo}, {Portell}, {Riello},
  {Seabroke}, {Tanga}, {Th{\'e}venin}, {Gracia-Abril}, {Comoretto},
  {Garcia-Reinaldos}, {Teyssier}, {Altmann}, {Andrae}, {Audard},
  {Bellas-Velidis}, {Benson}, {Berthier}, {Blomme}, {Burgess}, {Busso},
  {Carry}, {Cellino}, {Clementini}, {Clotet}, {Creevey}, {Davidson}, {De
  Ridder}, {Delchambre}, {Dell'Oro}, {Ducourant},
  {Fern{\'a}ndez-Hern{\'a}ndez}, {Fouesneau}, {Fr{\'e}mat}, {Galluccio},
  {Garc{\'\i}a-Torres}, {Gonz{\'a}lez-N{\'u}{\~n}ez}, {Gonz{\'a}lez-Vidal},
  {Gosset}, {Guy}, {Halbwachs}, {Hambly}, {Harrison}, {Hern{\'a}ndez},
  {Hestroffer}, {Hodgkin}, {Hutton}, {Jasniewicz}, {Jean-Antoine-Piccolo},
  {Jordan}, {Korn}, {Krone-Martins}, {Lanzafame}, {Lebzelter}, {L{\"o}ffler},
  {Manteiga}, {Marrese}, {Mart{\'\i}n-Fleitas}, {Moitinho}, {Mora}, {Muinonen},
  {Osinde}, {Pancino}, {Pauwels}, {Petit}, {Recio-Blanco}, {Richards},
  {Rimoldini}, {Robin}, {Sarro}, {Siopis}, {Smith}, {Sozzetti}, {S{\"u}veges},
  {Torra}, {van Reeven}, {Abbas}, {Abreu Aramburu}, {Accart}, {Aerts},
  {Altavilla}, {{\'A}lvarez}, {Alvarez}, {Alves}, {Anderson}, {Andrei},
  {Anglada Varela}, {Antiche}, {Antoja}, {Arcay}, {Astraatmadja}, {Bach},
  {Baker}, {Balaguer-N{\'u}{\~n}ez}, {Balm}, {Barache}, {Barata}, {Barbato},
  {Barblan}, {Barklem}, {Barrado}, {Barros}, {Barstow}, {Bartholom{\'e}
  Mu{\~n}oz}, {Bassilana}, {Becciani}, {Bellazzini}, {Berihuete}, {Bertone},
  {Bianchi}, {Bienaym{\'e}}, {Blanco-Cuaresma}, {Boch}, {Boeche}, {Bombrun},
  {Borrachero}, {Bossini}, {Bouquillon}, {Bourda}, {Bragaglia}, {Bramante},
  {Breddels}, {Bressan}, {Brouillet}, {Br{\"u}semeister}, {Brugaletta},
  {Bucciarelli}, {Burlacu}, {Busonero}, {Butkevich}, {Buzzi}, {Caffau},
  {Cancelliere}, {Cannizzaro}, {Cantat-Gaudin}, {Carballo}, {Carlucci},
  {Carrasco}, {Casamiquela}, {Castellani}, {Castro-Ginard}, {Charlot},
  {Chemin}, {Chiavassa}, {Cocozza}, {Costigan}, {Cowell}, {Crifo}, {Crosta},
  {Crowley}, {Cuypers}, {Dafonte}, {Damerdji}, {Dapergolas}, {David}, {David},
  {de Laverny}, {De Luise}, {De March}, {de Martino}, {de Souza}, {de Torres},
  {Debosscher}, {del Pozo}, {Delbo}, {Delgado}, {Delgado}, {Di Matteo},
  {Diakite}, {Diener}, {Distefano}, {Dolding}, {Drazinos}, {Dur{\'a}n},
  {Edvardsson}, {Enke}, {Eriksson}, {Esquej}, {Eynard Bontemps}, {Fabre},
  {Fabrizio}, {Faigler}, {Falc{\~a}o}, {Farr{\`a}s Casas}, {Federici},
  {Fedorets}, {Fernique}, {Figueras}, {Filippi}, {Findeisen}, {Fonti},
  {Fraile}, {Fraser}, {Fr{\'e}zouls}, {Gai}, {Galleti}, {Garabato},
  {Garc{\'\i}a-Sedano}, {Garofalo}, {Garralda}, {Gavel}, {Gavras}, {Gerssen},
  {Geyer}, {Giacobbe}, {Gilmore}, {Girona}, {Giuffrida}, {Glass}, {Gomes},
  {Granvik}, {Gueguen}, {Guerrier}, {Guiraud}, {Guti{\'e}rrez-S{\'a}nchez},
  {Haigron}, {Hatzidimitriou}, {Hauser}, {Haywood}, {Heiter}, {Helmi}, {Heu},
  {Hilger}, {Hobbs}, {Hofmann}, {Holland}, {Huckle}, {Hypki}, {Icardi},
  {Jan{\ss}en}, {Jevardat de Fombelle}, {Jonker}, {Juh{\'a}sz}, {Julbe},
  {Karampelas}, {Kewley}, {Klar}, {Kochoska}, {Kohley}, {Kolenberg},
  {Kontizas}, {Kontizas}, {Koposov}, {Kordopatis}, {Kostrzewa-Rutkowska},
  {Koubsky}, {Lambert}, {Lanza}, {Lasne}, {Lavigne}, {Le Fustec}, {Le
  Poncin-Lafitte}, {Lebreton}, {Leccia}, {Leclerc}, {Lecoeur-Taibi},
  {Lenhardt}, {Leroux}, {Liao}, {Licata}, {Lindstr{\o}m}, {Lister}, {Livanou},
  {Lobel}, {L{\'o}pez}, {Managau}, {Mann}, {Mantelet}, {Marchal}, {Marchant},
  {Marconi}, {Marinoni}, {Marschalk{\'o}}, {Marshall}, {Martino}, {Marton},
  {Mary}, {Massari}, {Matijevi{\v{c}}}, {Mazeh}, {McMillan}, {Messina},
  {Michalik}, {Millar}, {Molina}, {Molinaro}, {Moln{\'a}r}, {Montegriffo},
  {Mor}, {Morbidelli}, {Morel}, {Morris}, {Mulone}, {Muraveva}, {Musella},
  {Nelemans}, {Nicastro}, {Noval}, {O'Mullane}, {Ord{\'e}novic},
  {Ord{\'o}{\~n}ez-Blanco}, {Osborne}, {Pagani}, {Pagano}, {Pailler},
  {Palacin}, {Palaversa}, {Panahi}, {Pawlak}, {Piersimoni}, {Pineau}, {Plachy},
  {Plum}, {Poggio}, {Poujoulet}, {Pr{\v{s}}a}, {Pulone}, {Racero}, {Ragaini},
  {Rambaux}, {Ramos-Lerate}, {Regibo}, {Reyl{\'e}}, {Riclet}, {Ripepi}, {Riva},
  {Rivard}, {Rixon}, {Roegiers}, {Roelens}, {Romero-G{\'o}mez}, {Rowell},
  {Royer}, {Ruiz-Dern}, {Sadowski}, {Sagrist{\`a} Sell{\'e}s}, {Sahlmann},
  {Salgado}, {Salguero}, {Sanna}, {Santana-Ros}, {Sarasso}, {Savietto},
  {Schultheis}, {Sciacca}, {Segol}, {Segovia}, {S{\'e}gransan}, {Shih},
  {Siltala}, {Silva}, {Smart}, {Smith}, {Solano}, {Solitro}, {Sordo}, {Soria
  Nieto}, {Souchay}, {Spagna}, {Spoto}, {Stampa}, {Steele},
  {Steidelm{\"u}ller}, {Stephenson}, {Stoev}, {Suess}, {Surdej}, {Szabados},
  {Szegedi-Elek}, {Tapiador}, {Taris}, {Tauran}, {Taylor}, {Teixeira},
  {Terrett}, {Teyssand ier}, {Thuillot}, {Titarenko}, {Torra Clotet}, {Turon},
  {Ulla}, {Utrilla}, {Uzzi}, {Vaillant}, {Valentini}, {Valette}, {van Elteren},
  {Van Hemelryck}, {van Leeuwen}, {Vaschetto}, {Vecchiato}, {Veljanoski},
  {Viala}, {Vicente}, {Vogt}, {von Essen}, {Voss}, {Votruba}, {Voutsinas},
  {Walmsley}, {Weiler}, {Wertz}, {Wevers}, {Wyrzykowski}, {Yoldas},
  {{\v{Z}}erjal}, {Ziaeepour}, {Zorec}, {Zschocke}, {Zucker}, {Zurbach}, \&
  {Zwitter}}]{Gaia_DR2_18}
{Gaia Collaboration}, {Brown}, A.~G.~A., {Vallenari}, A., {et~al.} 2018, \aap,
  616, A1, \dodoi{10.1051/0004-6361/201833051}

\bibitem[{{Gaia Collaboration} {et~al.}(2021){Gaia Collaboration}, {Brown},
  {Vallenari}, {Prusti}, {de Bruijne}, {Babusiaux}, {Biermann}, {Creevey},
  {Evans}, {Eyer}, {Hutton}, {Jansen}, {Jordi}, {Klioner}, {Lammers},
  {Lindegren}, {Luri}, {Mignard}, {Panem}, {Pourbaix}, {Randich}, {Sartoretti},
  {Soubiran}, {Walton}, {Arenou}, {Bailer-Jones}, {Bastian}, {Cropper},
  {Drimmel}, {Katz}, {Lattanzi}, {van Leeuwen}, {Bakker}, {Cacciari},
  {Casta{\~n}eda}, {De Angeli}, {Ducourant}, {Fabricius}, {Fouesneau},
  {Fr{\'e}mat}, {Guerra}, {Guerrier}, {Guiraud}, {Jean-Antoine Piccolo},
  {Masana}, {Messineo}, {Mowlavi}, {Nicolas}, {Nienartowicz}, {Pailler},
  {Panuzzo}, {Riclet}, {Roux}, {Seabroke}, {Sordo}, {Tanga}, {Th{\'e}venin},
  {Gracia-Abril}, {Portell}, {Teyssier}, {Altmann}, {Andrae}, {Bellas-Velidis},
  {Benson}, {Berthier}, {Blomme}, {Brugaletta}, {Burgess}, {Busso}, {Carry},
  {Cellino}, {Cheek}, {Clementini}, {Damerdji}, {Davidson}, {Delchambre},
  {Dell'Oro}, {Fern{\'a}ndez-Hern{\'a}ndez}, {Galluccio}, {Garc{\'\i}a-Lario},
  {Garcia-Reinaldos}, {Gonz{\'a}lez-N{\'u}{\~n}ez}, {Gosset}, {Haigron},
  {Halbwachs}, {Hambly}, {Harrison}, {Hatzidimitriou}, {Heiter},
  {Hern{\'a}ndez}, {Hestroffer}, {Hodgkin}, {Holl}, {Jan{\ss}en}, {Jevardat de
  Fombelle}, {Jordan}, {Krone-Martins}, {Lanzafame}, {L{\"o}ffler}, {Lorca},
  {Manteiga}, {Marchal}, {Marrese}, {Moitinho}, {Mora}, {Muinonen}, {Osborne},
  {Pancino}, {Pauwels}, {Petit}, {Recio-Blanco}, {Richards}, {Riello},
  {Rimoldini}, {Robin}, {Roegiers}, {Rybizki}, {Sarro}, {Siopis}, {Smith},
  {Sozzetti}, {Ulla}, {Utrilla}, {van Leeuwen}, {van Reeven}, {Abbas}, {Abreu
  Aramburu}, {Accart}, {Aerts}, {Aguado}, {Ajaj}, {Altavilla}, {{\'A}lvarez},
  {{\'A}lvarez Cid-Fuentes}, {Alves}, {Anderson}, {Anglada Varela}, {Antoja},
  {Audard}, {Baines}, {Baker}, {Balaguer-N{\'u}{\~n}ez}, {Balbinot}, {Balog},
  {Barache}, {Barbato}, {Barros}, {Barstow}, {Bartolom{\'e}}, {Bassilana},
  {Bauchet}, {Baudesson-Stella}, {Becciani}, {Bellazzini}, {Bernet}, {Bertone},
  {Bianchi}, {Blanco-Cuaresma}, {Boch}, {Bombrun}, {Bossini}, {Bouquillon},
  {Bragaglia}, {Bramante}, {Breedt}, {Bressan}, {Brouillet}, {Bucciarelli},
  {Burlacu}, {Busonero}, {Butkevich}, {Buzzi}, {Caffau}, {Cancelliere},
  {C{\'a}novas}, {Cantat-Gaudin}, {Carballo}, {Carlucci}, {Carnerero},
  {Carrasco}, {Casamiquela}, {Castellani}, {Castro-Ginard}, {Castro Sampol},
  {Chaoul}, {Charlot}, {Chemin}, {Chiavassa}, {Cioni}, {Comoretto}, {Cooper},
  {Cornez}, {Cowell}, {Crifo}, {Crosta}, {Crowley}, {Dafonte}, {Dapergolas},
  {David}, {David}, {de Laverny}, {De Luise}, {De March}, {De Ridder}, {de
  Souza}, {de Teodoro}, {de Torres}, {del Peloso}, {del Pozo}, {Delbo},
  {Delgado}, {Delgado}, {Delisle}, {Di Matteo}, {Diakite}, {Diener},
  {Distefano}, {Dolding}, {Eappachen}, {Edvardsson}, {Enke}, {Esquej}, {Fabre},
  {Fabrizio}, {Faigler}, {Fedorets}, {Fernique}, {Fienga}, {Figueras},
  {Fouron}, {Fragkoudi}, {Fraile}, {Franke}, {Gai}, {Garabato},
  {Garcia-Gutierrez}, {Garc{\'\i}a-Torres}, {Garofalo}, {Gavras}, {Gerlach},
  {Geyer}, {Giacobbe}, {Gilmore}, {Girona}, {Giuffrida}, {Gomel}, {Gomez},
  {Gonzalez-Santamaria}, {Gonz{\'a}lez-Vidal}, {Granvik},
  {Guti{\'e}rrez-S{\'a}nchez}, {Guy}, {Hauser}, {Haywood}, {Helmi}, {Hidalgo},
  {Hilger}, {H{\l}adczuk}, {Hobbs}, {Holland}, {Huckle}, {Jasniewicz},
  {Jonker}, {Juaristi Campillo}, {Julbe}, {Karbevska}, {Kervella}, {Khanna},
  {Kochoska}, {Kontizas}, {Kordopatis}, {Korn}, {Kostrzewa-Rutkowska},
  {Kruszy{\'n}ska}, {Lambert}, {Lanza}, {Lasne}, {Le Campion}, {Le Fustec},
  {Lebreton}, {Lebzelter}, {Leccia}, {Leclerc}, {Lecoeur-Taibi}, {Liao},
  {Licata}, {Lindstr{\o}m}, {Lister}, {Livanou}, {Lobel}, {Madrero Pardo},
  {Managau}, {Mann}, {Marchant}, {Marconi}, {Marcos Santos}, {Marinoni},
  {Marocco}, {Marshall}, {Martin Polo}, {Mart{\'\i}n-Fleitas}, {Masip},
  {Massari}, {Mastrobuono-Battisti}, {Mazeh}, {McMillan}, {Messina},
  {Michalik}, {Millar}, {Mints}, {Molina}, {Molinaro}, {Moln{\'a}r},
  {Montegriffo}, {Mor}, {Morbidelli}, {Morel}, {Morris}, {Mulone}, {Munoz},
  {Muraveva}, {Murphy}, {Musella}, {Noval}, {Ord{\'e}novic}, {Orr{\`u}},
  {Osinde}, {Pagani}, {Pagano}, {Palaversa}, {Palicio}, {Panahi}, {Pawlak},
  {Pe{\~n}alosa Esteller}, {Penttil{\"a}}, {Piersimoni}, {Pineau}, {Plachy},
  {Plum}, {Poggio}, {Poretti}, {Poujoulet}, {Pr{\v{s}}a}, {Pulone}, {Racero},
  {Ragaini}, {Rainer}, {Raiteri}, {Rambaux}, {Ramos}, {Ramos-Lerate}, {Re
  Fiorentin}, {Regibo}, {Reyl{\'e}}, {Ripepi}, {Riva}, {Rixon}, {Robichon},
  {Robin}, {Roelens}, {Rohrbasser}, {Romero-G{\'o}mez}, {Rowell}, {Royer},
  {Rybicki}, {Sadowski}, {Sagrist{\`a} Sell{\'e}s}, {Sahlmann}, {Salgado},
  {Salguero}, {Samaras}, {Sanchez Gimenez}, {Sanna}, {Santove{\~n}a},
  {Sarasso}, {Schultheis}, {Sciacca}, {Segol}, {Segovia}, {S{\'e}gransan},
  {Semeux}, {Shahaf}, {Siddiqui}, {Siebert}, {Siltala}, {Slezak}, {Smart},
  {Solano}, {Solitro}, {Souami}, {Souchay}, {Spagna}, {Spoto}, {Steele},
  {Steidelm{\"u}ller}, {Stephenson}, {S{\"u}veges}, {Szabados}, {Szegedi-Elek},
  {Taris}, {Tauran}, {Taylor}, {Teixeira}, {Thuillot}, {Tonello}, {Torra},
  {Torra}, {Turon}, {Unger}, {Vaillant}, {van Dillen}, {Vanel}, {Vecchiato},
  {Viala}, {Vicente}, {Voutsinas}, {Weiler}, {Wevers}, {Wyrzykowski}, {Yoldas},
  {Yvard}, {Zhao}, {Zorec}, {Zucker}, {Zurbach}, \& {Zwitter}}]{Gaia_EDR3_21}
---. 2021, \aap, 649, A1, \dodoi{10.1051/0004-6361/202039657}

\bibitem[{{Gaustad} \& {Conti}(1971)}]{GaustadConti_71}
{Gaustad}, J.~E., \& {Conti}, P.~S. 1971, \pasp, 83, 351,
  \dodoi{10.1086/129135}

\bibitem[{{Geller} {et~al.}(2013){Geller}, {Hurley}, \&
  {Mathieu}}]{Geller_etal_13}
{Geller}, A.~M., {Hurley}, J.~R., \& {Mathieu}, R.~D. 2013, \aj, 145, 8,
  \dodoi{10.1088/0004-6256/145/1/8}

\bibitem[{{Girardi}(2016)}]{Girardi_16}
{Girardi}, L. 2016, \araa, 54, 95, \dodoi{10.1146/annurev-astro-081915-023354}

\bibitem[{{Girardi} {et~al.}(2005){Girardi}, {Groenewegen}, {Hatziminaoglou},
  \& {da Costa}}]{Girardi_etal_05}
{Girardi}, L., {Groenewegen}, M.~A.~T., {Hatziminaoglou}, E., \& {da Costa}, L.
  2005, \aap, 436, 895, \dodoi{10.1051/0004-6361:20042352}

\bibitem[{{Girardi} \& {Marigo}(2007)}]{GirardiMarigo_07}
{Girardi}, L., \& {Marigo}, P. 2007, \aap, 462, 237,
  \dodoi{10.1051/0004-6361:20065249}

\bibitem[{{Girardi} {et~al.}(2013){Girardi}, {Marigo}, {Bressan}, \&
  {Rosenfield}}]{Girardi_etal_13}
{Girardi}, L., {Marigo}, P., {Bressan}, A., \& {Rosenfield}, P. 2013, \apj,
  777, 142, \dodoi{10.1088/0004-637X/777/2/142}

\bibitem[{{Girardi} {et~al.}(2000){Girardi}, {Mermilliod}, \&
  {Carraro}}]{Girardi_etal_00}
{Girardi}, L., {Mermilliod}, J.~C., \& {Carraro}, G. 2000, \aap, 354, 892.
\newblock \doarXiv{astro-ph/0001068}

\bibitem[{{Girardi} {et~al.}(2009){Girardi}, {Rubele}, \&
  {Kerber}}]{Girardi_etal_09}
{Girardi}, L., {Rubele}, S., \& {Kerber}, L. 2009, \mnras, 394, L74,
  \dodoi{10.1111/j.1745-3933.2008.00614.x}

\bibitem[{{Girardi} {et~al.}(2010){Girardi}, {Williams}, {Gilbert},
  {Rosenfield}, {Dalcanton}, {Marigo}, {Boyer}, {Dolphin}, {Weisz},
  {Melbourne}, {Olsen}, {Seth}, \& {Skillman}}]{Girardi_etal_10}
{Girardi}, L., {Williams}, B.~F., {Gilbert}, K.~M., {et~al.} 2010, \apj, 724,
  1030, \dodoi{10.1088/0004-637X/724/2/1030}

\bibitem[{{Girardi} {et~al.}(2020){Girardi}, {Boyer}, {Johnson}, {Dalcanton},
  {Rosenfield}, {Seth}, {Skillman}, {Weisz}, {Williams}, {Bhattacharya},
  {Bressan}, {Caldwell}, {Chen}, {Dolphin}, {Fouesneau}, {Goldman},
  {Guhathakurta}, {Marigo}, {Mukherjee}, {Pastorelli}, {Quirk}, {Soraisam}, \&
  {Trabucchi}}]{Girardi_etal_20}
{Girardi}, L., {Boyer}, M.~L., {Johnson}, L.~C., {et~al.} 2020, \apj, 901, 19,
  \dodoi{10.3847/1538-4357/abad3a}

\bibitem[{{Glebbeek} {et~al.}(2013){Glebbeek}, {Gaburov}, {Portegies Zwart}, \&
  {Pols}}]{Glebbeek_etal_13}
{Glebbeek}, E., {Gaburov}, E., {Portegies Zwart}, S., \& {Pols}, O.~R. 2013,
  \mnras, 434, 3497, \dodoi{10.1093/mnras/stt1268}

\bibitem[{{Glebbeek} \& {Pols}(2008)}]{Glebbeek_Pols_08}
{Glebbeek}, E., \& {Pols}, O.~R. 2008, \aap, 488, 1017,
  \dodoi{10.1051/0004-6361:200809931}

\bibitem[{{Goldman} {et~al.}(2019){Goldman}, {Boyer}, {McQuinn}, {Whitelock},
  {McDonald}, {van Loon}, {Skillman}, {Gehrz}, {Javadi}, {Sloan}, {Jones},
  {Groenewegen}, \& {Menzies}}]{Goldman_etal_2019}
{Goldman}, S.~R., {Boyer}, M.~L., {McQuinn}, K.~B.~W., {et~al.} 2019, \apj,
  877, 49, \dodoi{10.3847/1538-4357/ab0965}

\bibitem[{{Gosnell} {et~al.}(2015){Gosnell}, {Mathieu}, {Geller}, {Sills},
  {Leigh}, \& {Knigge}}]{Gosnell_etal_15}
{Gosnell}, N.~M., {Mathieu}, R.~D., {Geller}, A.~M., {et~al.} 2015, \apj, 814,
  163, \dodoi{10.1088/0004-637X/814/2/163}

\bibitem[{{Goswami} {et~al.}(2021){Goswami}, {Slemer}, {Marigo}, {Bressan},
  {Silva}, {Spera}, {Boco}, {Grisoni}, {Pantoni}, \& {Lapi}}]{Goswami_etal_21}
{Goswami}, S., {Slemer}, A., {Marigo}, P., {et~al.} 2021, \aap, 650, A203,
  \dodoi{10.1051/0004-6361/202039842}

\bibitem[{{Groenewegen}(2021)}]{Groenewegen21}
{Groenewegen}, M. 2021, arXiv e-prints, arXiv:2106.08128.
\newblock \doarXiv{2106.08128}

\bibitem[{{Groenewegen} \& {Sloan}(2018)}]{GroenewegenSloan_18}
{Groenewegen}, M.~A.~T., \& {Sloan}, G.~C. 2018, \aap, 609, A114,
  \dodoi{10.1051/0004-6361/201731089}

\bibitem[{{Groenewegen} {et~al.}(2009){Groenewegen}, {Sloan}, {Soszy{\'n}ski},
  \& {Petersen}}]{Groenewegen_etal_09}
{Groenewegen}, M.~A.~T., {Sloan}, G.~C., {Soszy{\'n}ski}, I., \& {Petersen},
  E.~A. 2009, \aap, 506, 1277, \dodoi{10.1051/0004-6361/200912678}

\bibitem[{{Groenewegen} {et~al.}(1995){Groenewegen}, {van den Hoek}, \& {de
  Jong}}]{Groenewegen_etal_95}
{Groenewegen}, M.~A.~T., {van den Hoek}, L.~B., \& {de Jong}, T. 1995, \aap,
  293, 381

\bibitem[{{Groenewegen} {et~al.}(1998){Groenewegen}, {Whitelock}, {Smith}, \&
  {Kerschbaum}}]{Groenewegen_etal_98}
{Groenewegen}, M.~A.~T., {Whitelock}, P.~A., {Smith}, C.~H., \& {Kerschbaum},
  F. 1998, \mnras, 293, 18, \dodoi{10.1046/j.1365-8711.1998.01113.x}

\bibitem[{{Groenewegen} {et~al.}(2016){Groenewegen}, {Vlemmings}, {Marigo},
  {Sloan}, {Decin}, {Feast}, {Goldman}, {Justtanont}, {Kerschbaum}, {Matsuura},
  {McDonald}, {Olofsson}, {Sahai}, {van Loon}, {Wood}, {Zijlstra},
  {Bernard-Salas}, {Boyer}, {Guzman-Ramirez}, {Jones}, {Lagadec}, {Meixner},
  {Rawlings}, \& {Srinivasan}}]{Groenewegen_etal_16}
{Groenewegen}, M.~A.~T., {Vlemmings}, W.~H.~T., {Marigo}, P., {et~al.} 2016,
  \aap, 596, A50, \dodoi{10.1051/0004-6361/201629590}

\bibitem[{{Gullieuszik} {et~al.}(2012){Gullieuszik}, {Groenewegen}, {Cioni},
  {de Grijs}, {van Loon}, {Girardi}, {Ivanov}, {Oliveira}, {Emerson}, \&
  {Guandalini}}]{GullieuszikGroenewegen_12}
{Gullieuszik}, M., {Groenewegen}, M.~A.~T., {Cioni}, M. R.~L., {et~al.} 2012,
  \aap, 537, A105, \dodoi{10.1051/0004-6361/201117493}

\bibitem[{{Hartwick} \& {Hesser}(1971)}]{HartwickHesser_71}
{Hartwick}, F.~D.~A., \& {Hesser}, J.~E. 1971, \pasp, 83, 53,
  \dodoi{10.1086/129067}

\bibitem[{{Hartwick} \& {Hesser}(1973)}]{HartwickHesser_73}
---. 1973, \apj, 183, 883, \dodoi{10.1086/152275}

\bibitem[{{Herwig} {et~al.}(1998){Herwig}, {Schoenberner}, \&
  {Bloecker}}]{Herwig_etal_98}
{Herwig}, F., {Schoenberner}, D., \& {Bloecker}, T. 1998, \aap, 340, L43.
\newblock \doarXiv{astro-ph/9811076}

\bibitem[{{Hills} \& {Day}(1976)}]{HillsDay_76}
{Hills}, J.~G., \& {Day}, C.~A. 1976, \aplett, 17, 87

\bibitem[{{Holl} {et~al.}(2018){Holl}, {Audard}, {Nienartowicz}, {Jevardat de
  Fombelle}, {Marchal}, {Mowlavi}, {Clementini}, {De Ridder}, {Evans}, {Guy},
  {Lanzafame}, {Lebzelter}, {Rimoldini}, {Roelens}, {Zucker}, {Distefano},
  {Garofalo}, {Lecoeur-Ta{\"\i}bi}, {Lopez}, {Molinaro}, {Muraveva}, {Panahi},
  {Regibo}, {Ripepi}, {Sarro}, {Aerts}, {Anderson}, {Charnas}, {Barblan},
  {Blanco-Cuaresma}, {Busso}, {Cuypers}, {De Angeli}, {Glass}, {Grenon},
  {Juh{\'a}sz}, {Kochoska}, {Koubsky}, {Lanza}, {Leccia}, {Lorenz}, {Marconi},
  {Marschalk{\'o}}, {Mazeh}, {Messina}, {Mignard}, {Moitinho}, {Moln{\'a}r},
  {Morgenthaler}, {Musella}, {Ordenovic}, {Ord{\'o}{\~n}ez}, {Pagano},
  {Palaversa}, {Pawlak}, {Plachy}, {Pr{\v{s}}a}, {Riello}, {S{\"u}veges},
  {Szabados}, {Szegedi-Elek}, {Votruba}, \& {Eyer}}]{Holl_etal_2018}
{Holl}, B., {Audard}, M., {Nienartowicz}, K., {et~al.} 2018, \aap, 618, A30,
  \dodoi{10.1051/0004-6361/201832892}

\bibitem[{{Huang} {et~al.}(2021){Huang}, {Yuan}, {Beers}, \&
  {Zhang}}]{Huang_etal21}
{Huang}, Y., {Yuan}, H., {Beers}, T.~C., \& {Zhang}, H. 2021, \apjl, 910, L5,
  \dodoi{10.3847/2041-8213/abe69a}

\bibitem[{{Humphreys} \& {Davidson}(1979)}]{HDlimit}
{Humphreys}, R.~M., \& {Davidson}, K. 1979, \apj, 232, 409,
  \dodoi{10.1086/157301}

\bibitem[{{Hurley} {et~al.}(2005){Hurley}, {Pols}, {Aarseth}, \&
  {Tout}}]{Hurley_etal_05}
{Hurley}, J.~R., {Pols}, O.~R., {Aarseth}, S.~J., \& {Tout}, C.~A. 2005,
  \mnras, 363, 293, \dodoi{10.1111/j.1365-2966.2005.09448.x}

\bibitem[{{Hurley} {et~al.}(2001){Hurley}, {Tout}, {Aarseth}, \&
  {Pols}}]{Hurley_etal_01}
{Hurley}, J.~R., {Tout}, C.~A., {Aarseth}, S.~J., \& {Pols}, O.~R. 2001,
  \mnras, 323, 630, \dodoi{10.1046/j.1365-8711.2001.04220.x}

\bibitem[{{Hurley} {et~al.}(2002){Hurley}, {Tout}, \& {Pols}}]{Hurley_etal_02}
{Hurley}, J.~R., {Tout}, C.~A., \& {Pols}, O.~R. 2002, \mnras, 329, 897,
  \dodoi{10.1046/j.1365-8711.2002.05038.x}

\bibitem[{{Ishihara} {et~al.}(2010){Ishihara}, {Onaka}, {Kataza}, {Salama},
  {Alfageme}, {Cassatella}, {Cox}, {Garc{\'\i}a-Lario}, {Stephenson}, {Cohen},
  {Fujishiro}, {Fujiwara}, {Hasegawa}, {Ita}, {Kim}, {Matsuhara}, {Murakami},
  {M{\"u}ller}, {Nakagawa}, {Ohyama}, {Oyabu}, {Pyo}, {Sakon}, {Shibai},
  {Takita}, {Tanab{\'e}}, {Uemizu}, {Ueno}, {Usui}, {Wada}, {Watarai},
  {Yamamura}, \& {Yamauchi}}]{Tshihara_etal_10}
{Ishihara}, D., {Onaka}, T., {Kataza}, H., {et~al.} 2010, \aap, 514, A1,
  \dodoi{10.1051/0004-6361/200913811}

\bibitem[{{Ita} \& {Matsunaga}(2011)}]{ItaMatsunaga_11}
{Ita}, Y., \& {Matsunaga}, N. 2011, \mnras, 412, 2345,
  \dodoi{10.1111/j.1365-2966.2010.18056.x}

\bibitem[{{Ita} {et~al.}(2018){Ita}, {Matsunaga}, {Tanab{\'e}}, {Nakada},
  {Kato}, {Nagayama}, {Nagashima}, {Kurita}, {Nakajima}, {Whitelock},
  {Menzies}, {Feast}, {Nagata}, {Tamura}, \& {Nakaya}}]{Ita_etal_2018}
{Ita}, Y., {Matsunaga}, N., {Tanab{\'e}}, T., {et~al.} 2018, \mnras, 481, 4206,
  \dodoi{10.1093/mnras/sty2539}

\bibitem[{{Ita} {et~al.}(2021){Ita}, {Menzies}, {Whitelock}, {Matsunaga},
  {Takayama}, {Nakada}, {Tanab{\'e}}, {Feast}, \& {Nagayama}}]{Ita_etal_2021}
{Ita}, Y., {Menzies}, J.~W., {Whitelock}, P.~A., {et~al.} 2021, \mnras, 500,
  82, \dodoi{10.1093/mnras/staa3251}

\bibitem[{{Iwanek} {et~al.}(2021){Iwanek}, {Koz{\l}owski}, {Gromadzki},
  {Soszy{\'n}ski}, {Wrona}, {Skowron}, {Ratajczak}, {Udalski}, {Szyma{\'n}ski},
  {Pietrukowicz}, {Ulaczyk}, {Poleski}, {Mr{\'o}z}, {Skowron}, \&
  {Rybicki}}]{Iwanek_etal_2021}
{Iwanek}, P., {Koz{\l}owski}, S., {Gromadzki}, M., {et~al.} 2021, arXiv
  e-prints, arXiv:2107.03397.
\newblock \doarXiv{2107.03397}

\bibitem[{{Jadhav} \& {Subramaniam}(2021)}]{JadhavSubramaniam_21}
{Jadhav}, V.~V., \& {Subramaniam}, A. 2021, \mnras, 507, 1699,
  \dodoi{10.1093/mnras/stab2264}

\bibitem[{{Jaschek} \& {Keenan}(1985)}]{Jorgensen_85}
{Jaschek}, M., \& {Keenan}, P.~C., eds. 1985, {Carbon stars and S stars near
  open clusters -a statistical approach.}, Vol. 114, 181,
  \dodoi{10.1007/978-94-009-5325-3\_22}

\bibitem[{{Jayasinghe} {et~al.}(2019){Jayasinghe}, {Stanek}, {Kochanek},
  {Shappee}, {Holoien}, {Thompson}, {Prieto}, {Dong}, {Pawlak}, {Pejcha},
  {Shields}, {Pojmanski}, {Otero}, {Hurst}, {Britt}, \&
  {Will}}]{Jayasinghe_etal_2019}
{Jayasinghe}, T., {Stanek}, K.~Z., {Kochanek}, C.~S., {et~al.} 2019, \mnras,
  485, 961, \dodoi{10.1093/mnras/stz444}

\bibitem[{{Jeffery} {et~al.}(2016){Jeffery}, {von Hippel}, {van Dyk},
  {Stenning}, {Robinson}, {Stein}, \& {Jefferys}}]{Jeffery_etal_16}
{Jeffery}, E.~J., {von Hippel}, T., {van Dyk}, D.~A., {et~al.} 2016, \apj, 828,
  79, \dodoi{10.3847/0004-637X/828/2/79}

\bibitem[{{Jorgensen} \& {Westerlund}(1988)}]{JorgensenWesterlund_88}
{Jorgensen}, U.~G., \& {Westerlund}, B.~E. 1988, \aaps, 72, 193

\bibitem[{{Kalinowski} {et~al.}(1974){Kalinowski}, {Burkhead}, \&
  {Honeycutt}}]{Kalinowsli_etal_74}
{Kalinowski}, J.~K., {Burkhead}, M.~S., \& {Honeycutt}, R.~K. 1974, \apjl, 193,
  L77, \dodoi{10.1086/181636}

\bibitem[{{Kalirai} {et~al.}(2014){Kalirai}, {Marigo}, \&
  {Tremblay}}]{Kalirai_etal_14}
{Kalirai}, J.~S., {Marigo}, P., \& {Tremblay}, P.-E. 2014, \apj, 782, 17,
  \dodoi{10.1088/0004-637X/782/1/17}

\bibitem[{{Kamath} {et~al.}(2012){Kamath}, {Karakas}, \&
  {Wood}}]{Kamath_etal_12}
{Kamath}, D., {Karakas}, A.~I., \& {Wood}, P.~R. 2012, \apj, 746, 20,
  \dodoi{10.1088/0004-637X/746/1/20}

\bibitem[{{Kamath} {et~al.}(2010){Kamath}, {Wood}, {Soszy{\'n}ski}, \&
  {Lebzelter}}]{Kamath_etal_10}
{Kamath}, D., {Wood}, P.~R., {Soszy{\'n}ski}, I., \& {Lebzelter}, T. 2010,
  \mnras, 408, 522, \dodoi{10.1111/j.1365-2966.2010.17137.x}

\bibitem[{{Karakas}(2014)}]{Karakas_14}
{Karakas}, A.~I. 2014, \mnras, 445, 347, \dodoi{10.1093/mnras/stu1727}

\bibitem[{{Karakas} {et~al.}(2002){Karakas}, {Lattanzio}, \&
  {Pols}}]{KarakasLattanzio_02}
{Karakas}, A.~I., {Lattanzio}, J.~C., \& {Pols}, O.~R. 2002, \pasa, 19, 515,
  \dodoi{10.1071/AS02013}

\bibitem[{{Karakas} \& {Lugaro}(2016)}]{KarakasLugaro_16}
{Karakas}, A.~I., \& {Lugaro}, M. 2016, \apj, 825, 26,
  \dodoi{10.3847/0004-637X/825/1/26}

\bibitem[{{Keenan} \& {Boeshaar}(1980)}]{Keenan_80}
{Keenan}, P.~C., \& {Boeshaar}, P.~C. 1980, \apjs, 43, 379,
  \dodoi{10.1086/190673}

\bibitem[{{Kerschbaum} \& {Hron}(1992)}]{Kerschbaum_Hron_1992}
{Kerschbaum}, F., \& {Hron}, J. 1992, \aap, 263, 97

\bibitem[{{Kharchenko} {et~al.}(2013){Kharchenko}, {Piskunov}, {Schilbach},
  {R{\"o}ser}, \& {Scholz}}]{Kharchenko_etal_13}
{Kharchenko}, N.~V., {Piskunov}, A.~E., {Schilbach}, E., {R{\"o}ser}, S., \&
  {Scholz}, R.~D. 2013, \aap, 558, A53, \dodoi{10.1051/0004-6361/201322302}

\bibitem[{{Kiss} {et~al.}(2000){Kiss}, {Szatm{\'a}ry}, {Szab{\'o}}, \&
  {Mattei}}]{Kiss_etal_2000}
{Kiss}, L.~L., {Szatm{\'a}ry}, K., {Szab{\'o}}, G., \& {Mattei}, J.~A. 2000,
  \aaps, 145, 283, \dodoi{10.1051/aas:2000353}

\bibitem[{{Krone-Martins} \& {Moitinho}(2014)}]{Krone_etal_14}
{Krone-Martins}, A., \& {Moitinho}, A. 2014, \aap, 561, A57,
  \dodoi{10.1051/0004-6361/201321143}

\bibitem[{{Kroupa}(2002)}]{Kroupa_02}
{Kroupa}, P. 2002, Science, 295, 82, \dodoi{10.1126/science.1067524}

\bibitem[{{Ku{\v{c}}inskas} {et~al.}(2005){Ku{\v{c}}inskas}, {Hauschildt},
  {Ludwig}, {Brott}, {Vansevi{\v{c}}ius}, {Lindegren}, {Tanab{\'e}}, \&
  {Allard}}]{Kucinskas_etal_05}
{Ku{\v{c}}inskas}, A., {Hauschildt}, P.~H., {Ludwig}, H.~G., {et~al.} 2005,
  \aap, 442, 281, \dodoi{10.1051/0004-6361:20053028}

\bibitem[{{Lambert} {et~al.}(1986){Lambert}, {Gustafsson}, {Eriksson}, \&
  {Hinkle}}]{Lambert_etal_86}
{Lambert}, D.~L., {Gustafsson}, B., {Eriksson}, K., \& {Hinkle}, K.~H. 1986,
  \apjs, 62, 373, \dodoi{10.1086/191145}

\bibitem[{{Lattanzio} \& {Wood}(2004)}]{LattanzioWood_2004}
{Lattanzio}, J.~C., \& {Wood}, P.~R. 2004, {Evolution, Nucleosynthesis, and
  Pulsation of AGB Stars}, 23--104, \dodoi{10.1007/978-1-4757-3876-6\_2}

\bibitem[{{Lebzelter} \& {Hinkle}(2002)}]{Lebzelter_Hinkle_2002}
{Lebzelter}, T., \& {Hinkle}, K.~H. 2002, \aap, 393, 563,
  \dodoi{10.1051/0004-6361:20021085}

\bibitem[{{Lebzelter} {et~al.}(2008){Lebzelter}, {Lederer}, {Cristallo},
  {Hinkle}, {Straniero}, \& {Aringer}}]{Lebzelter_etal_08}
{Lebzelter}, T., {Lederer}, M.~T., {Cristallo}, S., {et~al.} 2008, \aap, 486,
  511, \dodoi{10.1051/0004-6361:200809363}

\bibitem[{{Lebzelter} {et~al.}(2018){Lebzelter}, {Mowlavi}, {Marigo},
  {Pastorelli}, {Trabucchi}, {Wood}, \&
  {Lecoeur-Ta{\"i}bi}}]{Lebzelter_etal_18}
{Lebzelter}, T., {Mowlavi}, N., {Marigo}, P., {et~al.} 2018, \aap, 616, L13,
  \dodoi{10.1051/0004-6361/201833615}

\bibitem[{{Lebzelter} {et~al.}(2014){Lebzelter}, {Nowotny}, {Hinkle},
  {H{\"o}fner}, \& {Aringer}}]{Lebzelter_etal_14}
{Lebzelter}, T., {Nowotny}, W., {Hinkle}, K.~H., {H{\"o}fner}, S., \&
  {Aringer}, B. 2014, \aap, 567, A143, \dodoi{10.1051/0004-6361/201424078}

\bibitem[{{Lebzelter} {et~al.}(2019){Lebzelter}, {Trabucchi}, {Mowlavi},
  {Wood}, {Marigo}, {Pastorelli}, \& {Lecoeur-Ta{\"\i}bi}}]{Lebzelter_etal_19}
{Lebzelter}, T., {Trabucchi}, M., {Mowlavi}, N., {et~al.} 2019, \aap, 631, A24,
  \dodoi{10.1051/0004-6361/201936395}

\bibitem[{{Lebzelter} \& {Wood}(2005)}]{LebzelterWood_05}
{Lebzelter}, T., \& {Wood}, P.~R. 2005, \aap, 441, 1117,
  \dodoi{10.1051/0004-6361:20053464}

\bibitem[{{Leiner} \& {Geller}(2021)}]{LeinerGeller_21}
{Leiner}, E.~M., \& {Geller}, A. 2021, \apj, 908, 229,
  \dodoi{10.3847/1538-4357/abd7e9}

\bibitem[{{Leonard}(1989)}]{Leonard_89}
{Leonard}, P. J.~T. 1989, \aj, 98, 217, \dodoi{10.1086/115138}

\bibitem[{{Lindegren} {et~al.}(2018){Lindegren}, {Hern{\'a}ndez}, {Bombrun},
  {Klioner}, {Bastian}, {Ramos-Lerate}, {de Torres}, {Steidelm{\"u}ller},
  {Stephenson}, {Hobbs}, {Lammers}, {Biermann}, {Geyer}, {Hilger}, {Michalik},
  {Stampa}, {McMillan}, {Casta{\~n}eda}, {Clotet}, {Comoretto}, {Davidson},
  {Fabricius}, {Gracia}, {Hambly}, {Hutton}, {Mora}, {Portell}, {van Leeuwen},
  {Abbas}, {Abreu}, {Altmann}, {Andrei}, {Anglada}, {Balaguer-N{\'u}{\~n}ez},
  {Barache}, {Becciani}, {Bertone}, {Bianchi}, {Bouquillon}, {Bourda},
  {Br{\"u}semeister}, {Bucciarelli}, {Busonero}, {Buzzi}, {Cancelliere},
  {Carlucci}, {Charlot}, {Cheek}, {Crosta}, {Crowley}, {de Bruijne}, {de
  Felice}, {Drimmel}, {Esquej}, {Fienga}, {Fraile}, {Gai}, {Garralda},
  {Gonz{\'a}lez-Vidal}, {Guerra}, {Hauser}, {Hofmann}, {Holl}, {Jordan},
  {Lattanzi}, {Lenhardt}, {Liao}, {Licata}, {Lister}, {L{\"o}ffler},
  {Marchant}, {Martin-Fleitas}, {Messineo}, {Mignard}, {Morbidelli}, {Poggio},
  {Riva}, {Rowell}, {Salguero}, {Sarasso}, {Sciacca}, {Siddiqui}, {Smart},
  {Spagna}, {Steele}, {Taris}, {Torra}, {van Elteren}, {van Reeven}, \&
  {Vecchiato}}]{Lindegren_etal_18}
{Lindegren}, L., {Hern{\'a}ndez}, J., {Bombrun}, A., {et~al.} 2018, \aap, 616,
  A2, \dodoi{10.1051/0004-6361/201832727}

\bibitem[{{Lindegren} {et~al.}(2021{\natexlab{a}}){Lindegren}, {Bastian},
  {Biermann}, {Bombrun}, {de Torres}, {Gerlach}, {Geyer}, {Hern{\'a}ndez},
  {Hilger}, {Hobbs}, {Klioner}, {Lammers}, {McMillan}, {Ramos-Lerate},
  {Steidelm{\"u}ller}, {Stephenson}, \& {van Leeuwen}}]{Lindegren_etal_21}
{Lindegren}, L., {Bastian}, U., {Biermann}, M., {et~al.} 2021{\natexlab{a}},
  \aap, 649, A4, \dodoi{10.1051/0004-6361/202039653}

\bibitem[{{Lindegren} {et~al.}(2021{\natexlab{b}}){Lindegren}, {Klioner},
  {Hern{\'a}ndez}, {Bombrun}, {Ramos-Lerate}, {Steidelm{\"u}ller}, {Bastian},
  {Biermann}, {de Torres}, {Gerlach}, {Geyer}, {Hilger}, {Hobbs}, {Lammers},
  {McMillan}, {Stephenson}, {Casta{\~n}eda}, {Davidson}, {Fabricius},
  {Gracia-Abril}, {Portell}, {Rowell}, {Teyssier}, {Torra}, {Bartolom{\'e}},
  {Clotet}, {Garralda}, {Gonz{\'a}lez-Vidal}, {Torra}, {Abbas}, {Altmann},
  {Anglada Varela}, {Balaguer-N{\'u}{\~n}ez}, {Balog}, {Barache}, {Becciani},
  {Bernet}, {Bertone}, {Bianchi}, {Bouquillon}, {Brown}, {Bucciarelli},
  {Busonero}, {Butkevich}, {Buzzi}, {Cancelliere}, {Carlucci}, {Charlot},
  {Cioni}, {Crosta}, {Crowley}, {del Peloso}, {del Pozo}, {Drimmel}, {Esquej},
  {Fienga}, {Fraile}, {Gai}, {Garcia-Reinaldos}, {Guerra}, {Hambly}, {Hauser},
  {Jan{\ss}en}, {Jordan}, {Kostrzewa-Rutkowska}, {Lattanzi}, {Liao}, {Licata},
  {Lister}, {L{\"o}ffler}, {Marchant}, {Masip}, {Mignard}, {Mints}, {Molina},
  {Mora}, {Morbidelli}, {Murphy}, {Pagani}, {Panuzzo}, {Pe{\~n}alosa Esteller},
  {Poggio}, {Re Fiorentin}, {Riva}, {Sagrist{\`a} Sell{\'e}s}, {Sanchez
  Gimenez}, {Sarasso}, {Sciacca}, {Siddiqui}, {Smart}, {Souami}, {Spagna},
  {Steele}, {Taris}, {Utrilla}, {van Reeven}, \&
  {Vecchiato}}]{Lindegren_etal_21a}
{Lindegren}, L., {Klioner}, S.~A., {Hern{\'a}ndez}, J., {et~al.}
  2021{\natexlab{b}}, \aap, 649, A2, \dodoi{10.1051/0004-6361/202039709}

\bibitem[{{Liu} \& {Pang}(2019)}]{LiuPang_19}
{Liu}, L., \& {Pang}, X. 2019, \apjs, 245, 32, \dodoi{10.3847/1538-4365/ab530a}

\bibitem[{{Madore}(1982)}]{Madore_1982}
{Madore}, B.~F. 1982, \apj, 253, 575, \dodoi{10.1086/159659}

\bibitem[{{Maraston}(2005)}]{Maraston05}
{Maraston}, C. 2005, \mnras, 362, 799, \dodoi{10.1111/j.1365-2966.2005.09270.x}

\bibitem[{{Maraston} {et~al.}(2006){Maraston}, {Daddi}, {Renzini}, {Cimatti},
  {Dickinson}, {Papovich}, {Pasquali}, \& {Pirzkal}}]{Maraston_etal_06}
{Maraston}, C., {Daddi}, E., {Renzini}, A., {et~al.} 2006, \apj, 652, 85,
  \dodoi{10.1086/508143}

\bibitem[{{Marigo}(2001)}]{Marigo_01}
{Marigo}, P. 2001, \aap, 370, 194, \dodoi{10.1051/0004-6361:20000247}

\bibitem[{{Marigo}(2015)}]{Marigo_15}
{Marigo}, P. 2015, in Astronomical Society of the Pacific Conference Series,
  Vol. 497, Why Galaxies Care about AGB Stars III: A Closer Look in Space and
  Time, ed. F.~{Kerschbaum}, R.~F. {Wing}, \& J.~{Hron}, 229.
\newblock \doarXiv{1411.3126}

\bibitem[{{Marigo} \& {Aringer}(2009)}]{Marigo_Aringer_09}
{Marigo}, P., \& {Aringer}, B. 2009, \aap, 508, 1539,
  \dodoi{10.1051/0004-6361/200912598}

\bibitem[{{Marigo} {et~al.}(2013){Marigo}, {Bressan}, {Nanni}, {Girardi}, \&
  {Pumo}}]{Marigo_etal_13}
{Marigo}, P., {Bressan}, A., {Nanni}, A., {Girardi}, L., \& {Pumo}, M.~L. 2013,
  \mnras, 434, 488, \dodoi{10.1093/mnras/stt1034}

\bibitem[{{Marigo} \& {Girardi}(2007)}]{Marigo_etal_07}
{Marigo}, P., \& {Girardi}, L. 2007, \aap, 469, 239,
  \dodoi{10.1051/0004-6361:20066772}

\bibitem[{{Marigo} {et~al.}(1999){Marigo}, {Girardi}, \&
  {Bressan}}]{Marigo_etal_99}
{Marigo}, P., {Girardi}, L., \& {Bressan}, A. 1999, \aap, 344, 123.
\newblock \doarXiv{astro-ph/9901235}

\bibitem[{{Marigo} {et~al.}(2017){Marigo}, {Girardi}, {Bressan}, {Rosenfield},
  {Aringer}, {Chen}, {Dussin}, {Nanni}, {Pastorelli}, {Rodrigues}, {Trabucchi},
  {Bladh}, {Dalcanton}, {Groenewegen}, {Montalb{\'a}n}, \&
  {Wood}}]{Marigo_etal_17}
{Marigo}, P., {Girardi}, L., {Bressan}, A., {et~al.} 2017, \apj, 835, 77,
  \dodoi{10.3847/1538-4357/835/1/77}

\bibitem[{{Marigo} {et~al.}(2020){Marigo}, {Cummings}, {Curtis}, {Kalirai},
  {Chen}, {Tremblay}, {Ramirez-Ruiz}, {Bergeron}, {Bladh}, {Bressan},
  {Girardi}, {Pastorelli}, {Trabucchi}, {Cheng}, {Aringer}, \&
  {Tio}}]{Marigo_etal_20}
{Marigo}, P., {Cummings}, J.~D., {Curtis}, J.~L., {et~al.} 2020, Nature
  Astronomy, 4, 1102, \dodoi{10.1038/s41550-020-1132-1}

\bibitem[{{Masci} {et~al.}(2019){Masci}, {Laher}, {Rusholme}, {Shupe}, {Groom},
  {Surace}, {Jackson}, {Monkewitz}, {Beck}, {Flynn}, {Terek}, {Landry},
  {Hacopians}, {Desai}, {Howell}, {Brooke}, {Imel}, {Wachter}, {Ye}, {Lin},
  {Cenko}, {Cunningham}, {Rebbapragada}, {Bue}, {Miller}, {Mahabal}, {Bellm},
  {Patterson}, {Juri{\'c}}, {Golkhou}, {Ofek}, {Walters}, {Graham}, {Kasliwal},
  {Dekany}, {Kupfer}, {Burdge}, {Cannella}, {Barlow}, {Van Sistine}, {Giomi},
  {Fremling}, {Blagorodnova}, {Levitan}, {Riddle}, {Smith}, {Helou}, {Prince},
  \& {Kulkarni}}]{Masci_etal_2019_ZTFdata}
{Masci}, F.~J., {Laher}, R.~R., {Rusholme}, B., {et~al.} 2019, \pasp, 131,
  018003, \dodoi{10.1088/1538-3873/aae8ac}

\bibitem[{{Mathieu} \& {Geller}(2009)}]{MathieuGeller_09}
{Mathieu}, R.~D., \& {Geller}, A.~M. 2009, \nat, 462, 1032,
  \dodoi{10.1038/nature08568}

\bibitem[{{Mattsson} {et~al.}(2010){Mattsson}, {Wahlin}, \&
  {H{\"o}fner}}]{Mattsson_etal_10}
{Mattsson}, L., {Wahlin}, R., \& {H{\"o}fner}, S. 2010, \aap, 509, A14,
  \dodoi{10.1051/0004-6361/200912084}

\bibitem[{{McCrea}(1964)}]{McCrea_64}
{McCrea}, W.~H. 1964, \mnras, 128, 147, \dodoi{10.1093/mnras/128.2.147}

\bibitem[{{McDonald} {et~al.}(2011{\natexlab{a}}){McDonald}, {Boyer}, {van
  Loon}, \& {Zijlstra}}]{McDonald_etal_11a}
{McDonald}, I., {Boyer}, M.~L., {van Loon}, J.~T., \& {Zijlstra}, A.~A.
  2011{\natexlab{a}}, \apj, 730, 71, \dodoi{10.1088/0004-637X/730/2/71}

\bibitem[{{McDonald} \& {Trabucchi}(2019)}]{McDonaldTrabucchi_2019}
{McDonald}, I., \& {Trabucchi}, M. 2019, \mnras, 484, 4678,
  \dodoi{10.1093/mnras/stz324}

\bibitem[{{McDonald} {et~al.}(2009){McDonald}, {van Loon}, {Decin}, {Boyer},
  {Dupree}, {Evans}, {Gehrz}, \& {Woodward}}]{McDonald_etal_09}
{McDonald}, I., {van Loon}, J.~T., {Decin}, L., {et~al.} 2009, \mnras, 394,
  831, \dodoi{10.1111/j.1365-2966.2008.14370.x}

\bibitem[{{McDonald} {et~al.}(2011{\natexlab{b}}){McDonald}, {van Loon},
  {Sloan}, {Dupree}, {Zijlstra}, {Boyer}, {Gehrz}, {Evans}, {Woodward}, \&
  {Johnson}}]{McDonald_etal_11b}
{McDonald}, I., {van Loon}, J.~T., {Sloan}, G.~C., {et~al.} 2011{\natexlab{b}},
  \mnras, 417, 20, \dodoi{10.1111/j.1365-2966.2011.18963.x}

\bibitem[{{Meixner} {et~al.}(2006){Meixner}, {Gordon}, {Indebetouw}, {Hora},
  {Whitney}, {Blum}, {Reach}, {Bernard}, {Meade}, {Babler}, {Engelbracht},
  {For}, {Misselt}, {Vijh}, {Leitherer}, {Cohen}, {Churchwell}, {Boulanger},
  {Frogel}, {Fukui}, {Gallagher}, {Gorjian}, {Harris}, {Kelly}, {Kawamura},
  {Kim}, {Latter}, {Madden}, {Markwick-Kemper}, {Mizuno}, {Mizuno}, {Mould},
  {Nota}, {Oey}, {Olsen}, {Onishi}, {Paladini}, {Panagia}, {Perez-Gonzalez},
  {Shibai}, {Sato}, {Smith}, {Staveley-Smith}, {Tielens}, {Ueta}, {van Dyk},
  {Volk}, {Werner}, \& {Zaritsky}}]{Meixner_etal_06}
{Meixner}, M., {Gordon}, K.~D., {Indebetouw}, R., {et~al.} 2006, \aj, 132,
  2268, \dodoi{10.1086/508185}

\bibitem[{{Momany} {et~al.}(2012){Momany}, {Saviane}, {Smette}, {Bayo},
  {Girardi}, {Marconi}, {Milone}, \& {Bressan}}]{Momany_etal_12}
{Momany}, Y., {Saviane}, I., {Smette}, A., {et~al.} 2012, \aap, 537, A2,
  \dodoi{10.1051/0004-6361/201117223}

\bibitem[{{Monteiro} {et~al.}(2020){Monteiro}, {Dias}, {Moitinho},
  {Cantat-Gaudin}, {L{\'e}pine}, {Carraro}, \& {Paunzen}}]{Monteiro_etal_20}
{Monteiro}, H., {Dias}, W.~S., {Moitinho}, A., {et~al.} 2020, \mnras, 499,
  1874, \dodoi{10.1093/mnras/staa2983}

\bibitem[{{Mowlavi} {et~al.}(2018){Mowlavi}, {Lecoeur-Ta{\"\i}bi}, {Lebzelter},
  {Rimoldini}, {Lorenz}, {Audard}, {De Ridder}, {Eyer}, {Guy}, {Holl},
  {Jevardat de Fombelle}, {Marchal}, {Nienartowicz}, {Regibo}, {Roelens}, \&
  {Sarro}}]{Mowlavi_etal_18}
{Mowlavi}, N., {Lecoeur-Ta{\"\i}bi}, I., {Lebzelter}, T., {et~al.} 2018, \aap,
  618, A58, \dodoi{10.1051/0004-6361/201833366}

\bibitem[{{Nanni} {et~al.}(2014){Nanni}, {Bressan}, {Marigo}, \&
  {Girardi}}]{Nanni_etal_14}
{Nanni}, A., {Bressan}, A., {Marigo}, P., \& {Girardi}, L. 2014, \mnras, 438,
  2328, \dodoi{10.1093/mnras/stt2348}

\bibitem[{{Nanni} {et~al.}(2019){Nanni}, {Groenewegen}, {Aringer}, {Rubele},
  {Bressan}, {van Loon}, {Goldman}, \& {Boyer}}]{Nanni_etal_19}
{Nanni}, A., {Groenewegen}, M. A.~T., {Aringer}, B., {et~al.} 2019, \mnras,
  487, 502, \dodoi{10.1093/mnras/stz1255}

\bibitem[{{Nanni} {et~al.}(2018){Nanni}, {Marigo}, {Girardi}, {Rubele},
  {Bressan}, {Groenewegen}, {Pastorelli}, \& {Aringer}}]{Nanni_etal_18}
{Nanni}, A., {Marigo}, P., {Girardi}, L., {et~al.} 2018, \mnras, 473, 5492,
  \dodoi{10.1093/mnras/stx2641}

\bibitem[{{No{\"e}l} {et~al.}(2013){No{\"e}l}, {Greggio}, {Renzini}, {Carollo},
  \& {Maraston}}]{Noel_etal_13}
{No{\"e}l}, N.~E.~D., {Greggio}, L., {Renzini}, A., {Carollo}, C.~M., \&
  {Maraston}, C. 2013, \apj, 772, 58, \dodoi{10.1088/0004-637X/772/1/58}

\bibitem[{{O'Donnell}(1994)}]{ODonnell_James_94}
{O'Donnell}, J.~E. 1994, \apj, 422, 158, \dodoi{10.1086/173713}

\bibitem[{{Paczy{\'n}ski}(1970)}]{Paczynski_70}
{Paczy{\'n}ski}, B. 1970, \actaa, 20, 47

\bibitem[{{Pal} \& {Worthey}(2021)}]{Pal_Worthey_21}
{Pal}, T., \& {Worthey}, G. 2021, \mnras, 506, 3669,
  \dodoi{10.1093/mnras/stab1967}

\bibitem[{{Pastorelli} {et~al.}(2019){Pastorelli}, {Marigo}, {Girardi}, {Chen},
  {Rubele}, {Trabucchi}, {Aringer}, {Bladh}, {Bressan}, {Montalb{\'a}n},
  {Boyer}, {Dalcanton}, {Eriksson}, {Groenewegen}, {H{\"o}fner}, {Lebzelter},
  {Nanni}, {Rosenfield}, {Wood}, \& {Cioni}}]{Pastorelli_etal_19}
{Pastorelli}, G., {Marigo}, P., {Girardi}, L., {et~al.} 2019, \mnras, 485,
  5666, \dodoi{10.1093/mnras/stz725}

\bibitem[{{Pastorelli} {et~al.}(2020){Pastorelli}, {Marigo}, {Girardi},
  {Aringer}, {Chen}, {Rubele}, {Trabucchi}, {Bladh}, {Boyer}, {Bressan},
  {Dalcanton}, {Groenewegen}, {Lebzelter}, {Mowlavi}, {Chubb}, {Cioni}, {de
  Grijs}, {Ivanov}, {Nanni}, {van Loon}, \& {Zaggia}}]{Pastorelli_etal_20}
---. 2020, \mnras, 498, 3283, \dodoi{10.1093/mnras/staa2565}

\bibitem[{{Paxton} {et~al.}(2011){Paxton}, {Bildsten}, {Dotter}, {Herwig},
  {Lesaffre}, \& {Timmes}}]{Paxton_etal_11}
{Paxton}, B., {Bildsten}, L., {Dotter}, A., {et~al.} 2011, \apjs, 192, 3,
  \dodoi{10.1088/0067-0049/192/1/3}

\bibitem[{{Paxton} {et~al.}(2013){Paxton}, {Cantiello}, {Arras}, {Bildsten},
  {Brown}, {Dotter}, {Mankovich}, {Montgomery}, {Stello}, {Timmes}, \&
  {Townsend}}]{mesa13}
{Paxton}, B., {Cantiello}, M., {Arras}, P., {et~al.} 2013, \apjs, 208, 4,
  \dodoi{10.1088/0067-0049/208/1/4}

\bibitem[{{Paxton} {et~al.}(2015){Paxton}, {Marchant}, {Schwab}, {Bauer},
  {Bildsten}, {Cantiello}, {Dessart}, {Farmer}, {Hu}, {Langer}, {Townsend},
  {Townsley}, \& {Timmes}}]{mesa15}
{Paxton}, B., {Marchant}, P., {Schwab}, J., {et~al.} 2015, \apjs, 220, 15,
  \dodoi{10.1088/0067-0049/220/1/15}

\bibitem[{{Paxton} {et~al.}(2018){Paxton}, {Schwab}, {Bauer}, {Bildsten},
  {Blinnikov}, {Duffell}, {Farmer}, {Goldberg}, {Marchant}, {Sorokina},
  {Thoul}, {Townsend}, \& {Timmes}}]{mesa18}
{Paxton}, B., {Schwab}, J., {Bauer}, E.~B., {et~al.} 2018, \apjs, 234, 34,
  \dodoi{10.3847/1538-4365/aaa5a8}

\bibitem[{{Pessev} {et~al.}(2008){Pessev}, {Goudfrooij}, {Puzia}, \&
  {Chandar}}]{Pessev_etal_08}
{Pessev}, P.~M., {Goudfrooij}, P., {Puzia}, T.~H., \& {Chandar}, R. 2008,
  \mnras, 385, 1535, \dodoi{10.1111/j.1365-2966.2008.12935.x}

\bibitem[{{Piatti} {et~al.}(2019){Piatti}, {Angelo}, \&
  {Dias}}]{Piatti_etal_19}
{Piatti}, A.~E., {Angelo}, M.~S., \& {Dias}, W.~S. 2019, \mnras, 488, 4648,
  \dodoi{10.1093/mnras/stz2050}

\bibitem[{{Platais} {et~al.}(2003){Platais}, {Pourbaix}, {Jorissen}, {Makarov},
  {Berdnikov}, {Samus}, {Lloyd Evans}, {Lebzelter}, \&
  {Sperauskas}}]{Platais_etal_03}
{Platais}, I., {Pourbaix}, D., {Jorissen}, A., {et~al.} 2003, \aap, 397, 997,
  \dodoi{10.1051/0004-6361:20021589}

\bibitem[{{Rain} {et~al.}(2021){Rain}, {Ahumada}, \& {Carraro}}]{Rain_etal_21}
{Rain}, M.~J., {Ahumada}, J., \& {Carraro}, G. 2021, arXiv e-prints,
  arXiv:2103.06004.
\newblock \doarXiv{2103.06004}

\bibitem[{{Reid} \& {Goldston}(2002)}]{ReidGoldston_2002}
{Reid}, M.~J., \& {Goldston}, J.~E. 2002, \apj, 568, 931,
  \dodoi{10.1086/338947}

\bibitem[{{Riebel} {et~al.}(2015){Riebel}, {Boyer}, {Srinivasan}, {Whitelock},
  {Meixner}, {Babler}, {Feast}, {Groenewegen}, {Ita}, {Meade}, {Shiao}, \&
  {Whitney}}]{Riebel_etal_2015}
{Riebel}, D., {Boyer}, M.~L., {Srinivasan}, S., {et~al.} 2015, \apj, 807, 1,
  \dodoi{10.1088/0004-637X/807/1/1}

\bibitem[{{Riess} {et~al.}(2021){Riess}, {Casertano}, {Yuan}, {Bowers},
  {Macri}, {Zinn}, \& {Scolnic}}]{Riess_etal21}
{Riess}, A.~G., {Casertano}, S., {Yuan}, W., {et~al.} 2021, \apjl, 908, L6,
  \dodoi{10.3847/2041-8213/abdbaf}

\bibitem[{{Rosenfield} {et~al.}(2014){Rosenfield}, {Marigo}, {Girardi},
  {Dalcanton}, {Bressan}, {Gullieuszik}, {Weisz}, {Williams}, {Dolphin}, \&
  {Aringer}}]{Rosenfield_etal_14}
{Rosenfield}, P., {Marigo}, P., {Girardi}, L., {et~al.} 2014, \apj, 790, 22,
  \dodoi{10.1088/0004-637X/790/1/22}

\bibitem[{{Salaris} {et~al.}(2009){Salaris}, {Serenelli}, {Weiss}, \& {Miller
  Bertolami}}]{Salaris_etal_09}
{Salaris}, M., {Serenelli}, A., {Weiss}, A., \& {Miller Bertolami}, M. 2009,
  \apj, 692, 1013, \dodoi{10.1088/0004-637X/692/2/1013}

\bibitem[{{Salaris} {et~al.}(2004){Salaris}, {Weiss}, \&
  {Percival}}]{Salaris_etal_04}
{Salaris}, M., {Weiss}, A., \& {Percival}, S.~M. 2004, \aap, 414, 163,
  \dodoi{10.1051/0004-6361:20031578}

\bibitem[{{Samus'} {et~al.}(2017){Samus'}, {Kazarovets}, {Durlevich},
  {Kireeva}, \& {Pastukhova}}]{Samus_etal_17}
{Samus'}, N.~N., {Kazarovets}, E.~V., {Durlevich}, O.~V., {Kireeva}, N.~N., \&
  {Pastukhova}, E.~N. 2017, Astronomy Reports, 61, 80,
  \dodoi{10.1134/S1063772917010085}

\bibitem[{{Sargent} {et~al.}(2011){Sargent}, {Srinivasan}, \&
  {Meixner}}]{Sargent_etal_11}
{Sargent}, B.~A., {Srinivasan}, S., \& {Meixner}, M. 2011, \apj, 728, 93,
  \dodoi{10.1088/0004-637X/728/2/93}

\bibitem[{{Sch{\"o}ier} \& {Olofsson}(2001)}]{Schoier_Oloffson_01}
{Sch{\"o}ier}, F.~L., \& {Olofsson}, H. 2001, \aap, 368, 969,
  \dodoi{10.1051/0004-6361:20010072}

\bibitem[{{Shappee} {et~al.}(2014){Shappee}, {Prieto}, {Stanek}, {Kochanek},
  {Holoien}, {Jencson}, {Basu}, {Beacom}, {Szczygiel}, {Pojmanski},
  {Brimacombe}, {Dubberley}, {Elphick}, {Foale}, {Hawkins}, {Mullins},
  {Rosing}, {Ross}, \& {Walker}}]{Shappee_etal_2014}
{Shappee}, B., {Prieto}, J., {Stanek}, K.~Z., {et~al.} 2014, in American
  Astronomical Society Meeting Abstracts, Vol. 223, American Astronomical
  Society Meeting Abstracts \#223, 236.03

\bibitem[{{Shetye} {et~al.}(2021){Shetye}, {Van Eck}, {Jorissen}, {Goriely},
  {Siess}, {Van Winckel}, {Plez}, {Godefroid}, \&
  {Wallerstein}}]{Shetye_etal_21}
{Shetye}, S., {Van Eck}, S., {Jorissen}, A., {et~al.} 2021, \aap, 650, A118,
  \dodoi{10.1051/0004-6361/202040207}

\bibitem[{{Siegel} {et~al.}(2019){Siegel}, {LaPorte}, {Porterfield}, {Hagen},
  \& {Gronwall}}]{Siegel_etal_19}
{Siegel}, M.~H., {LaPorte}, S.~J., {Porterfield}, B.~L., {Hagen}, L. M.~Z., \&
  {Gronwall}, C.~A. 2019, \aj, 158, 35, \dodoi{10.3847/1538-3881/ab21e1}

\bibitem[{{Siess}(2010)}]{Siess_10}
{Siess}, L. 2010, \aap, 512, A10, \dodoi{10.1051/0004-6361/200913556}

\bibitem[{{Sills} {et~al.}(2009){Sills}, {Karakas}, \&
  {Lattanzio}}]{Sills_etal_09}
{Sills}, A., {Karakas}, A., \& {Lattanzio}, J. 2009, \apj, 692, 1411,
  \dodoi{10.1088/0004-637X/692/2/1411}

\bibitem[{{Slemer} {et~al.}(2017){Slemer}, {Marigo}, {Piatti}, {Aliotta},
  {Bemmerer}, {Best}, {Boeltzig}, {Bressan}, {Broggini}, {Bruno}, {Caciolli},
  {Cavanna}, {Ciani}, {Corvisiero}, {Davinson}, {Depalo}, {Di Leva}, {Elekes},
  {Ferraro}, {Formicola}, {F{\"u}l{\"o}p}, {Gervino}, {Guglielmetti},
  {Gustavino}, {Gy{\"u}rky}, {Imbriani}, {Junker}, {Menegazzo}, {Mossa},
  {Pantaleo}, {Prati}, {Straniero}, {Sz{\"u}cs}, {Tak{\'a}cs}, \&
  {Trezzi}}]{Slemer_etal_17}
{Slemer}, A., {Marigo}, P., {Piatti}, D., {et~al.} 2017, \mnras, 465, 4817,
  \dodoi{10.1093/mnras/stw3029}

\bibitem[{{Soszy{\'n}ski}(2007)}]{Soszynski_2007}
{Soszy{\'n}ski}, I. 2007, \apj, 660, 1486, \dodoi{10.1086/513012}

\bibitem[{{Soszy{\'n}ski} \& {Udalski}(2014)}]{SoszynskiUdalski_2014}
{Soszy{\'n}ski}, I., \& {Udalski}, A. 2014, \apj, 788, 13,
  \dodoi{10.1088/0004-637X/788/1/13}

\bibitem[{{Soszy{\'n}ski} {et~al.}(2013){Soszy{\'n}ski}, {Wood}, \&
  {Udalski}}]{Soszynski_etal_2013}
{Soszy{\'n}ski}, I., {Wood}, P.~R., \& {Udalski}, A. 2013, \apj, 779, 167,
  \dodoi{10.1088/0004-637X/779/2/167}

\bibitem[{{Soszynski} {et~al.}(2005){Soszynski}, {Udalski}, {Kubiak},
  {Szymanski}, {Pietrzynski}, {Zebrun}, {Szewczyk}, {Wyrzykowski}, \&
  {Ulaczyk}}]{Soszynski_etal_2005}
{Soszynski}, I., {Udalski}, A., {Kubiak}, M., {et~al.} 2005, \actaa, 55, 331.
\newblock \doarXiv{astro-ph/0512578}

\bibitem[{{Soszy{\'n}ski} {et~al.}(2007){Soszy{\'n}ski}, {Dziembowski},
  {Udalski}, {Kubiak}, {Szyma{\'n}ski}, {Pietrzy{\'n}ski}, {Wyrzykowski},
  {Szewczyk}, \& {Ulaczyk}}]{Soszynski_etal_2007}
{Soszy{\'n}ski}, I., {Dziembowski}, W.~A., {Udalski}, A., {et~al.} 2007,
  \actaa, 57, 201

\bibitem[{{Soszy{\'n}ski} {et~al.}(2009){Soszy{\'n}ski}, {Udalski},
  {Szyma{\'n}ski}, {Kubiak}, {Pietrzy{\'n}ski}, {Wyrzykowski}, {Szewczyk},
  {Ulaczyk}, \& {Poleski}}]{Soszynski_etal_2009_LMC}
{Soszy{\'n}ski}, I., {Udalski}, A., {Szyma{\'n}ski}, M.~K., {et~al.} 2009,
  \actaa, 59, 239.
\newblock \doarXiv{0910.1354}

\bibitem[{{Soszy{\'n}ski} {et~al.}(2011){Soszy{\'n}ski}, {Udalski},
  {Szyma{\'n}ski}, {Kubiak}, {Pietrzy{\'n}ski}, {Wyrzykowski}, {Ulaczyk},
  {Poleski}, {Koz{\l}owski}, \& {Pietrukowicz}}]{Soszynski_etal_2011_SMC}
---. 2011, \actaa, 61, 217.
\newblock \doarXiv{1109.1143}

\bibitem[{{Soszy{\'n}ski} {et~al.}(2021){Soszy{\'n}ski}, {Olechowska},
  {Ratajczak}, {Iwanek}, {Skowron}, {Mr{\'o}z}, {Pietrukowicz}, {Udalski},
  {Szyma{\'n}ski}, {Skowron}, {Gromadzki}, {Poleski}, {Koz{\l}owski}, {Wrona},
  {Ulaczyk}, \& {Rybicki}}]{Soszynski_etal_2021}
{Soszy{\'n}ski}, I., {Olechowska}, A., {Ratajczak}, M., {et~al.} 2021, \apjl,
  911, L22, \dodoi{10.3847/2041-8213/abf3c9}

\bibitem[{{Srinivasan} {et~al.}(2011){Srinivasan}, {Sargent}, \&
  {Meixner}}]{Srinivasan_etal_11}
{Srinivasan}, S., {Sargent}, B.~A., \& {Meixner}, M. 2011, \aap, 532, A54,
  \dodoi{10.1051/0004-6361/201117033}

\bibitem[{{Stassun} \& {Torres}(2021)}]{Stassun_etal_21}
{Stassun}, K.~G., \& {Torres}, G. 2021, \apjl, 907, L33,
  \dodoi{10.3847/2041-8213/abdaad}

\bibitem[{{Su{\'a}rez} {et~al.}(2006){Su{\'a}rez}, {Garc{\'\i}a-Lario},
  {Manchado}, {Manteiga}, {Ulla}, \& {Pottasch}}]{Suarez_etal_06}
{Su{\'a}rez}, O., {Garc{\'\i}a-Lario}, P., {Manchado}, A., {et~al.} 2006, \aap,
  458, 173, \dodoi{10.1051/0004-6361:20054108}

\bibitem[{{Sun} {et~al.}(2021){Sun}, {Mathieu}, {Leiner}, \&
  {Townsend}}]{Sun_etal_21}
{Sun}, M., {Mathieu}, R.~D., {Leiner}, E.~M., \& {Townsend}, R.~H.~D. 2021,
  \apj, 908, 7, \dodoi{10.3847/1538-4357/abd402}

\bibitem[{{Tadross}(2009)}]{Tadross_09}
{Tadross}, A.~L. 2009, \na, 14, 200, \dodoi{10.1016/j.newast.2008.08.004}

\bibitem[{{Trabucchi} {et~al.}(2021{\natexlab{a}}){Trabucchi}, {Mowlavi}, \&
  {Lebzelter}}]{Trabucchi_etal_2021_SRVs}
{Trabucchi}, M., {Mowlavi}, N., \& {Lebzelter}, T. 2021{\natexlab{a}}, arXiv
  e-prints, arXiv:2109.04293.
\newblock \doarXiv{2109.04293}

\bibitem[{{Trabucchi} {et~al.}(2017){Trabucchi}, {Wood}, {Montalb{\'a}n},
  {Marigo}, {Pastorelli}, \& {Girardi}}]{Trabucchi_etal_17}
{Trabucchi}, M., {Wood}, P.~R., {Montalb{\'a}n}, J., {et~al.} 2017, \apj, 847,
  139, \dodoi{10.3847/1538-4357/aa8998}

\bibitem[{{Trabucchi} {et~al.}(2019){Trabucchi}, {Wood}, {Montalb{\'a}n},
  {Marigo}, {Pastorelli}, \& {Girardi}}]{Trabucchi_etal_2019}
---. 2019, \mnras, 482, 929, \dodoi{10.1093/mnras/sty2745}

\bibitem[{{Trabucchi} {et~al.}(2021{\natexlab{b}}){Trabucchi}, {Wood},
  {Mowlavi}, {Pastorelli}, {Marigo}, {Girardi}, \&
  {Lebzelter}}]{Trabucchi_etal_2021}
{Trabucchi}, M., {Wood}, P.~R., {Mowlavi}, N., {et~al.} 2021{\natexlab{b}},
  \mnras, 500, 1575, \dodoi{10.1093/mnras/staa3356}

\bibitem[{{Tuchman} {et~al.}(1983){Tuchman}, {Glasner}, \&
  {Barkat}}]{Tuchman_etal_83}
{Tuchman}, Y., {Glasner}, A., \& {Barkat}, Z. 1983, \apj, 268, 356,
  \dodoi{10.1086/160958}

\bibitem[{{van Loon} {et~al.}(1998){van Loon}, {Zijlstra}, {Whitelock}, {te
  Lintel Hekkert}, {Chapman}, {Loup}, {Groenewegen}, {Waters}, \&
  {Trams}}]{VanLoon_etal_98}
{van Loon}, J.~T., {Zijlstra}, A.~A., {Whitelock}, P.~A., {et~al.} 1998, \aap,
  329, 169.
\newblock \doarXiv{astro-ph/9709119}

\bibitem[{{VanderPlas}(2018)}]{vanderplas_2018}
{VanderPlas}, J.~T. 2018, \apjs, 236, 16, \dodoi{10.3847/1538-4365/aab766}

\bibitem[{{Vassiliadis} \& {Wood}(1993)}]{VassiliadisWood_93}
{Vassiliadis}, E., \& {Wood}, P.~R. 1993, \apj, 413, 641,
  \dodoi{10.1086/173033}

\bibitem[{{Ventura} \& {D'Antona}(2005)}]{VenturaDantona_05}
{Ventura}, P., \& {D'Antona}, F. 2005, \aap, 431, 279,
  \dodoi{10.1051/0004-6361:20041917}

\bibitem[{{Ventura} {et~al.}(2000){Ventura}, {D'Antona}, \&
  {Mazzitelli}}]{Ventura_etal_00}
{Ventura}, P., {D'Antona}, F., \& {Mazzitelli}, I. 2000, \aap, 363, 605.
\newblock \doarXiv{astro-ph/0101374}

\bibitem[{{Ventura} {et~al.}(2018){Ventura}, {Karakas}, {Dell'Agli},
  {Garc{\'\i}a-Hern{\'a}ndez}, \& {Guzman-Ramirez}}]{Ventura_etal_18}
{Ventura}, P., {Karakas}, A., {Dell'Agli}, F., {Garc{\'\i}a-Hern{\'a}ndez},
  D.~A., \& {Guzman-Ramirez}, L. 2018, \mnras, 475, 2282,
  \dodoi{10.1093/mnras/stx3338}

\bibitem[{{Ventura} {et~al.}(2015){Ventura}, {Karakas}, {Dell'Agli}, {Boyer},
  {Garc{\'\i}a-Hern{\'a}ndez}, {Di Criscienzo}, \&
  {Schneider}}]{Ventura_etal_15}
{Ventura}, P., {Karakas}, A.~I., {Dell'Agli}, F., {et~al.} 2015, \mnras, 450,
  3181, \dodoi{10.1093/mnras/stv918}

\bibitem[{{Wagenhuber} \& {Groenewegen}(1998)}]{WagenhuberGroenewegen_98}
{Wagenhuber}, J., \& {Groenewegen}, M.~A.~T. 1998, \aap, 340, 183.
\newblock \doarXiv{astro-ph/9809338}

\bibitem[{{Wagstaff} {et~al.}(2020){Wagstaff}, {Miller Bertolami}, \&
  {Weiss}}]{Wagstaff_etal_20}
{Wagstaff}, G., {Miller Bertolami}, M.~M., \& {Weiss}, A. 2020, \mnras, 493,
  4748, \dodoi{10.1093/mnras/staa362}

\bibitem[{{Weiss} \& {Ferguson}(2009)}]{WeissFerguson_09}
{Weiss}, A., \& {Ferguson}, J.~W. 2009, \aap, 508, 1343,
  \dodoi{10.1051/0004-6361/200912043}

\bibitem[{{Wenger} {et~al.}(2000){Wenger}, {Ochsenbein}, {Egret}, {Dubois},
  {Bonnarel}, {Borde}, {Genova}, {Jasniewicz}, {Lalo{\"e}}, {Lesteven}, \&
  {Monier}}]{simbad}
{Wenger}, M., {Ochsenbein}, F., {Egret}, D., {et~al.} 2000, \aaps, 143, 9,
  \dodoi{10.1051/aas:2000332}

\bibitem[{{Whitelock} {et~al.}(2008){Whitelock}, {Feast}, \& {Van
  Leeuwen}}]{Whitelock_etal_2008}
{Whitelock}, P.~A., {Feast}, M.~W., \& {Van Leeuwen}, F. 2008, \mnras, 386,
  313, \dodoi{10.1111/j.1365-2966.2008.13032.x}

\bibitem[{{Whitelock} {et~al.}(2017){Whitelock}, {Kasliwal}, \&
  {Boyer}}]{Whitelock_etal_2017}
{Whitelock}, P.~A., {Kasliwal}, M., \& {Boyer}, M. 2017, in European Physical
  Journal Web of Conferences, Vol. 152, European Physical Journal Web of
  Conferences, 01009, \dodoi{10.1051/epjconf/201715201009}

\bibitem[{{Wood}(2000)}]{Wood_2000}
{Wood}, P.~R. 2000, \pasa, 17, 18, \dodoi{10.1071/AS00018}

\bibitem[{{Wood}(2015)}]{Wood_15}
---. 2015, \mnras, 448, 3829, \dodoi{10.1093/mnras/stv289}

\bibitem[{{Wood} \& {Nicholls}(2009)}]{WoodNicholls_2009}
{Wood}, P.~R., \& {Nicholls}, C.~P. 2009, \apj, 707, 573,
  \dodoi{10.1088/0004-637X/707/1/573}

\bibitem[{{Wood} \& {Sebo}(1996)}]{Wood_Sebo_1996}
{Wood}, P.~R., \& {Sebo}, K.~M. 1996, \mnras, 282, 958,
  \dodoi{10.1093/mnras/282.3.958}

\bibitem[{{Wood} \& {Zarro}(1981)}]{WoodZarro_81}
{Wood}, P.~R., \& {Zarro}, D.~M. 1981, \apj, 247, 247, \dodoi{10.1086/159032}

\bibitem[{{Wood} {et~al.}(1999){Wood}, {Alcock}, {Allsman}, {Alves}, {Axelrod},
  {Becker}, {Bennett}, {Cook}, {Drake}, {Freeman}, {Griest}, {King}, {Lehner},
  {Marshall}, {Minniti}, {Peterson}, {Pratt}, {Quinn}, {Stubbs}, {Sutherland},
  {Tomaney}, {Vandehei}, \& {Welch}}]{Wood_etal_1999}
{Wood}, P.~R., {Alcock}, C., {Allsman}, R.~A., {et~al.} 1999, in Asymptotic
  Giant Branch Stars, ed. T.~{Le Bertre}, A.~{Lebre}, \& C.~{Waelkens}, Vol.
  191, 151

\bibitem[{{Yu} {et~al.}(2020){Yu}, {Bedding}, {Stello}, {Huber}, {Compton},
  {Gizon}, \& {Hekker}}]{Yu_etal_2020}
{Yu}, J., {Bedding}, T.~R., {Stello}, D., {et~al.} 2020, \mnras, 493, 1388,
  \dodoi{10.1093/mnras/staa300}

\bibitem[{{Zhong} {et~al.}(2020){Zhong}, {Chen}, {Wu}, {Li}, {Bai}, \&
  {Hou}}]{Zhong_etal_20}
{Zhong}, J., {Chen}, L., {Wu}, D., {et~al.} 2020, \aap, 640, A127,
  \dodoi{10.1051/0004-6361/201937131}

\bibitem[{{Zinn}(2021)}]{Zinn21}
{Zinn}, J.~C. 2021, \aj, 161, 214, \dodoi{10.3847/1538-3881/abe936}

\end{thebibliography}
\bibliographystyle{aasjournal}

\acknowledgments
We acknowledge the support from the ERC Consolidator
Grant funding scheme (project STARKEY, grant agreement n. 615604), and from PRD 2021, University of Padova.
This work has made use of data from the European Space Agency (ESA) mission
\gaia\ (\url{https://www.cosmos.esa.int/gaia}), processed by the \gaia\
Data Processing and Analysis Consortium (DPAC,
\url{https://www.cosmos.esa.int/web/gaia/dpac/consortium}). Funding for the DPAC
has been provided by national institutions, in particular the institutions
participating in the \gaia\ Multilateral Agreement.
This research has made use of the Virtual Observatory Sed Analyzer (VOSA), developed under the Spanish Virtual Observatory project supported by the Spanish MINECO through grant AyA2017-84089.
This research has made use of "Aladin sky atlas" developed at CDS, Strasbourg Observatory, France.
DB acknowledges supported by FCT through the research grants UIDB/04434/2020, UIDP/04434/2020 and PTDC/FIS-AST/30389/2017, and by FEDER - Fundo Europeu de Desenvolvimento Regional through COMPETE2020 - Programa Operacional Competitividade e Internacionalização (grant: POCI-01-0145-FEDER-030389).

\software{VOSA \citep{vosa08}, MESA \citep{Paxton_etal_11,mesa13,mesa15,mesa18}, TRILEGAL \citep{Girardi_etal_05}, PARSEC \citep{Bressan_etal_12}, COLIBRI \citep{Marigo_etal_13}, AESOPUS \citep{Marigo_Aringer_09}.}


\end{document}